%% file: 0main.tex
\documentclass[11pt]{report}

\usepackage[utf8]{inputenc}
\usepackage[a4paper, total={418pt,595pt}, headheight=13.6pt]{geometry}
\usepackage{graphicx}
\usepackage{mathtools}
\usepackage{amsmath,amssymb,amsfonts}
\usepackage{siunitx}
\usepackage{calc}
\usepackage{enumitem}
\usepackage{booktabs}
\usepackage[all]{nowidow}
\usepackage{datetime}
\usepackage{epigraph} 
\newdateformat{zadate}{{\THEDAY} \monthname[\THEMONTH] {\THEYEAR}}
\usepackage{xcolor}
\usepackage[ruled,vlined]{algorithm2e}
\usepackage[colorinlistoftodos,textsize=scriptsize,color=cyan!40]{todonotes}
\usepackage[%
	style=phys,%
	biblabel=brackets,%
	pageranges=false,%
	maxbibnames=10,%
	]
	{biblatex}
\usepackage{hyperref}
\usepackage{caption}
\hypersetup{
	colorlinks=true,
	linkcolor=blue,
	citecolor=green!60!black,
	}
\DeclareMathOperator{\spn}{span}

\DeclareMathOperator{\diag}{diag}
\newcommand{\bs}[1]{\boldsymbol{\mathrm{#1}}}
\newcommand{\transpose}{^\mathsf{T}}
\newcommand{\figureref}[1]{\hyperref[#1]{Fig.~\ref*{#1}}}
\newcommand{\chapterref}[1]{\hyperref[#1]{Chapt.~\ref*{#1}}}
\newcommand{\sectionref}[1]{\hyperref[#1]{Sect.~\ref*{#1}}}
\newcommand{\tableref}[1]{\hyperref[#1]{Table~\ref*{#1}}}
\newcommand{\algorithmref}[1]{\hyperref[#1]{Alg.~\ref*{#1}}}
\newcommand{\appendixref}[1]{\hyperref[#1]{Appendix~\ref*{#1}}}

\title{Investigating Hamiltonian Dynamics by the Method of Covariant Lyapunov Vectors}
\author{Jean-Jacq du Plessis}
\date{\today}

\usepackage{fancyhdr}

\begin{document}
	
\include{frontmatter}

\include{introduction}
\include{theory}
\include{results}
\include{dna}
\include{conclusion}

\include{appendix}

\printbibliography[heading=bibintoc,title={Bibliography}]

\end{document}

%% file: frontmatter.tex
\begin{titlepage}
	\begin{center}
		\vspace*{1cm}
		
		\rule{0.95\textwidth}{1pt}
		
		\vspace*{0.1cm}
		\LARGE
		\textbf{Investigating Hamiltonian Dynamics by the Method of Covariant Lyapunov Vectors}
		\vspace*{-0.22cm}
		
		\rule{0.95\textwidth}{1pt}
		
		\vspace{0.45cm}
		\large
		
		\textit{by}
		
		\vspace{0.4cm}
		
		\Large
		\textbf{Jean-Jacq du Plessis}
		
		\vspace{1cm}
		\textit{Supervisor:} A/Prof.\ Haris Skokos\\
		\textit{Co-supervisor:} Dr Malcolm Hillebrand
		
		\vfill
		
		\textit{Dissertation presented for the degree of}\\
		Master of Science
		
		\vspace{0.6cm}
		
		\includegraphics[width=0.3\textwidth]{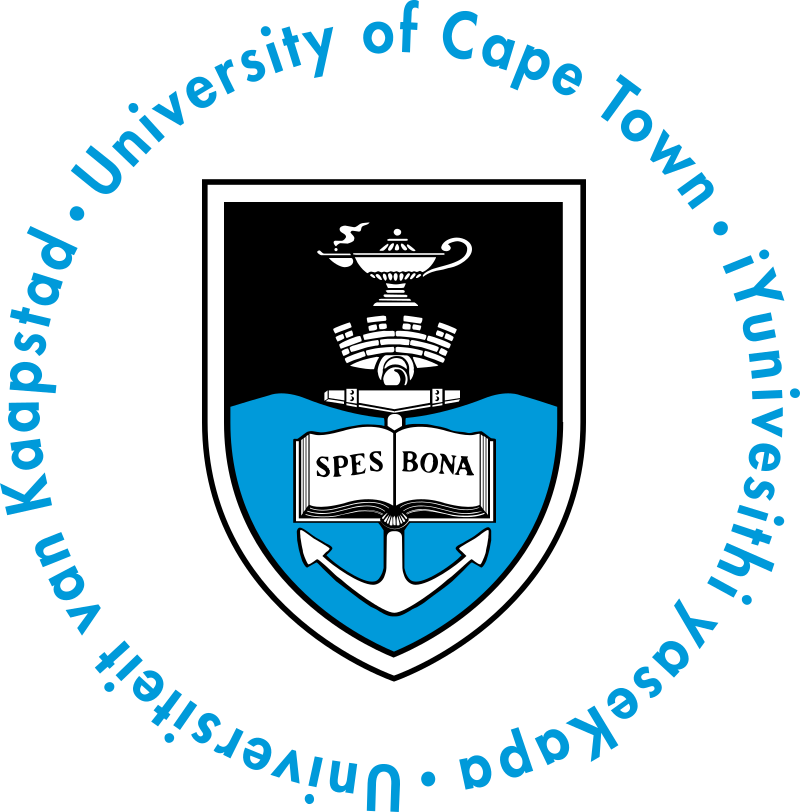}
		
		\vspace{0.35cm}
		
		University of Cape Town\\
		Faculty of Science\\
		Department of Mathematics and Applied Mathematics
		
		\vspace{0.4cm}
		
		16 February 2024
		
	\end{center}
\end{titlepage}

\chapter*{Abstract}
In this thesis, we review the theory of Lyapunov exponents and covariant Lyapunov vectors (CLVs) and use these objects to numerically investigate the dynamics of several autonomous Hamiltonian systems. The algorithm which we use for computing CLVs is the one developed by Ginelli and collaborators (G\&C), which is quite efficient and has been used previously in many numerical investigations. Using two low-dimensional Hamiltonian systems as toy models, we develop a method for measuring the convergence rates of vectors and subspaces computed via the G\&C algorithm, and we use the time it takes for this convergence to occur to determine the appropriate transient time lengths needed when applying this algorithm to compute CLVs. The tangent dynamics of the centre subspace of the H\'enon-Heiles system is investigated numerically through the use of CLVs, and we propose a method that improves the accuracy of the centre subspace computed with the G\&C algorithm. As another application of the method of CLVs to the H\'enon-Heiles system, we find that the splitting subspaces (which form a splitting of the tangent space and define the CLVs) become almost tangent during sticky regimes of motion, an observation which is related to the hyperbolicity of the system. Additionally, we investigate the dynamics of bubbles (i.e.\ thermal openings between base pairs) in homogeneous DNA sequences using the Peyrard-Bishop-Dauxois lattice model of DNA. For the purpose of studying short-lived bubbles in DNA, the notions of instantaneous Lyapunov vectors (ILVs) are introduced in the context of Hamiltonian dynamics. While we find that the size of the opening between base pairs has no clear relationship with the spatial distribution of the first CLV at that site, we do observe a distinct relationship with various ILV distributions.

\chapter*{Acknowledgements}
Firstly, I wish to thank my supervisor, A/Prof.\ Haris Skokos, who never stopped pushing me to do my best. Whenever my adventures in phase space deviated too far from the desired trajectory, you always brought me back to the task at hand. To my co-supervisor, Dr Malcolm Hillebrand, whose thesis first exposed me to the notion of a covariant Lyapunov vector: I sincerely appreciate your invaluable guidance and feedback, whether in person or from Germany. To my other colleagues in the Nonlinear Dynamics and Chaos Group: Arnold, Henok, San\'e, Sebastian, Sam, and Dylan, thank you for your stimulating discussions and friendship. I am especially grateful to Layla and my family for their patience and support during my pursuit of this degree.

Computational resources provided by South Africa's Centre for High Performance Computing (CHPC) were used for the numerical investigations in this thesis. From the CHPC staff, I would like to thank Dr Kevin Colville, Mr Binjamin Barsch, and Mr Inus Scheepers, whose instruction and technical support were most helpful. Last but not least, I am grateful to the David \& Elaine Potter Foundation and the University of Cape Town for funding my MSc studies.

\chapter*{Presentations related to this work}
Some results from this work have been included in the following poster presentation:

\begin{description}[leftmargin=!,labelwidth=\widthof{heo}]
	\item[\,$\,\ \ast$] \textit{Instantaneous Lyapunov vectors in DNA}, International Conference on Statistical Physics 2023 (SigmaPhi 2023), Chania, Crete, Greece, 13 July 2023.
\end{description}

{\hypersetup{linkcolor=black}
\tableofcontents
}

\chapter*{List of acronyms}
\addcontentsline{toc}{chapter}{List of acronyms}
\markboth{List of acronyms}{List of acronyms}
\begin{description}[leftmargin=!,itemsep=0.25\baselineskip,labelwidth=\widthof{$\Delta(U,V)\ $}]
	\item[AT] adenine-thymine
	\item[BLV] backward Lyapunov vector
	\item[CLV] covariant Lyapunov vector
	\item[DNA] deoxyribonucleic acid
	\item[DVD] deviation vector distribution
	\item[FLV] forward Lyapunov vector
	\item[ftLE] finite-time Lyapunov exponent
	\item[ftmLE] finite-time maximum Lyapunov exponent
	\item[G\&C] Ginelli and collaborators (Poggi, Turchi, Chaté, Livi, Politi)
	\item[GC] guanine-cytosine
	\item[GSV] Gram-Schmidt vector
	\item[ILE] instantaneous Lyapunov exponent
	\item[ILV] instantaneous Lyapunov vector
	\item[LE] Lyapunov exponent
	\item[mLE] maximum Lyapunov exponent
	\item[$\boldsymbol{N}$-D] $N$-dimensional
	\item[PBD] Peyrard-Bishop-Dauxois
	\item[PSS] Poincar\'e surface of section
	\item[SALI] smaller alignment index
	\item[SVD] singular value decomposition
	\item[wmLE] windowed maximum Lyapunov exponent
\end{description}

\chapter*{List of symbols}
\addcontentsline{toc}{chapter}{List of symbols}
\markboth{List of symbols}{List of symbols}
\begin{description}[leftmargin=!,labelwidth=\widthof{$\Delta(U,V)\ $},itemsep=0.25\baselineskip]
	\item[$C_i$] computed estimate of the $i$-th splitting subspace $\Omega_i$
	\item[$\hat{\bs c}_i$] computed estimate of the $i$-th covariant Lyapunov vector $\hat{\bs\omega}_i$
	\item[$\mathrm{d}_{\bs x}\Phi^t$] linear propagator which evolves deviation vectors of $T_{\bs x}\mathcal{M}$ in time by $t$
	\item[$E_n$] energy density
	\item[$E_r$] relative energy error
	\item[$G_i$] computed estimate of the $i$-th filtration subspace $\Gamma_i^{-}$
	\item[$\hat{\bs g}_i$] computed estimate of the $i$-th backward Lyapunov vector, i.e.\ the $i$-th Gram-Schmidt vector
	\item[$H$] Hamiltonian function
	\item[$J(\bs x)$] Jacobian matrix of a vector field evaluated at $\bs x$
	\item[$\mathcal{M}$] state space
	\item[$M(\bs x,t)$] fundamental matrix of the variational equations, i.e.\ the matrix of $\mathrm{d}_{\bs x}\Phi^t$ 
	\item[$N$] dimension of the state space
	\item[$n$] degrees of freedom of a Hamiltonian system
	\item[$p_i$] $i$-th generalised momentum coordinate
	\item[$P(\bs w)$] participation number of the distribution of the deviation vector $\bs w$
	\item[$Q$] matrix with orthonormal columns from QR decomposition
	\item[$q_i$] $i$-th generalised position coordinate
	\item[$R$] upper triangular matrix with positive diagonal entries from QR decomposition
	\item[$t;t_b$] time variable; backward time variable
	\item[$T_i$] time marking a boundary of a stage of the G\&C algorithm, where $i=0,1,2,3$
	\item[$T_{\bs x}\mathcal{M}$] tangent space of $\mathcal{M}$ at $\bs x$
	\item[$\bs w;\bs w(t)$] deviation vector; deviation vector at time $t$
	\item[$\hat{\bs w}$] deviation vector of unit magnitude
	\item[$\hat{\bs w}_{\mathrm{flow}}$] direction of the flow
	\item[$\bs x;\bs x(t)$] state vector; state vector at time $t$
	\item[${[\bs x]}$] orbit passing through $\bs x$
	\item[$X_1^{(t_w)}$] windowed maximum Lyapunov exponent with window $t_w$
	\item[$X_i$] $i$-th finite-time Lyapunov exponent
	\item[$\Gamma_i^{\pm}$] $i$-th filtration subspace
	\item[$\Delta(U,V)$] distance between equidimensional subspaces $U$ and $V$
	\item[$\Delta(\bs u,\bs v)$] distance between subspaces $\spn(\bs u)$ and $\spn(\bs v)$
	\item[$\theta_i$] $i$-th principal angle
	\item[$\kappa_i$] multiplicity of the $i$-th distinct Lyapunov exponent
	\item[$\Lambda^{\pm}$] forward/backward Oseledets matrices
	\item[$\lambda_i$] $i$-th distinct Lyapunov exponent
	\item[$\bar\xi_i$] weighted instantaneous Lyapunov vector distribution component at the $i$-th lattice site
	\item[$\xi_i(\bs w)$] distribution component of the deviation vector $\bs w$ at the $i$-th lattice site
	\item[$\sigma(\bs w)$] standard deviation of the distribution of the deviation vector $\bs w$
	\item[$\tau$] time interval between successive (ortho)normalisations of deviation vectors
	\item[$\Phi^t$] flow map which evolves states in time by $t$
	\item[$\chi_i$] $i$-th Lyapunov exponent
	\item[$\chi_i^0$] $i$-th instantaneous Lyapunov exponent
	\item[$\Omega_i$] $i$-th splitting subspace
	\item[$\hat{\bs\omega}_i$] $i$-th covariant Lyapunov vector
	\item[$\hat{\bs\omega}_i^0$] $i$-th instantaneous Lyapunov vector
\end{description}

%% file: introduction.tex
\chapter{Introduction}\label{ch:intro}
In the context of dynamical systems, \textit{Lyapunov exponents} (LEs) are scalar quantities which characterise the growth rates of perturbations applied to a trajectory in different directions of the system's state space. In the 1960s, Oseledets was the first to prove the existence of a full spectrum of LEs for a broad class of dynamical systems \cite{Oseledets1968}. The directions which correspond to these LEs are known as \textit{covariant Lyapunov vectors} (CLVs), the theory of which was introduced by Oseledets and further developed by Ruelle in 1979 \cite{Ruelle1979}. CLVs can be defined for general, possibly aperiodic orbits, but they reduce to the so-called Floquet vectors if the orbit is periodic, and when the orbit is a fixed point then these vectors reduce to the eigenvectors of the Jacobian matrix at that fixed point \cite{WolfeSamelson2007}. In this sense, CLVs generalise the concept of eigenvectors used in stability analysis for fixed points to encompass aperiodic evolution \cites{KuptsovParlitz2012}{Politi2013}. Consequently, LEs and CLVs are invaluable tools for understanding the stability properties of a wide variety of systems.

Due to the generic non-integrability of nonlinear dynamical systems, analytic expressions for LEs and CLVs for a given state do not typically exist (see e.g.\ \cite[p.~7]{PikovskyPoliti2016}) and so must be estimated instead with the help of computers. Methods for numerically computing the full spectrum of LEs were first established in around 1980 \cites{ShimadaNagashima1979}{BenettinEtAl1980}{BenettinEtAl1980a}, but efficient algorithms for computing CLVs were developed far more recently. The first practical CLV algorithms were published independently by Wolfe and Samelson \cite{WolfeSamelson2007} and Ginelli et al.\ \cite{GinelliEtAl2007} in 2007, which subsequently led to the development of newer algorithms such as that of Kuptsov and Parlitz \cite{KuptsovParlitz2012}. With the proliferation of CLV algorithms in recent years, there has been a renewed interest in the subject, with CLVs being applied to a variety of fields. To name a few applications in the atmospheric sciences, CLVs have been used to study ensemble prediction systems \cite{HerreraEtAl2011}, eddy fluxes in geophysical models \cite{SchubertLucarini2015}, and atmospheric circulation over the North Atlantic \cite{QuinnEtAl2021}. CLVs have also been used to classify bifurcations \cite{KamiyamaEtAl2014}, to study wall turbulence in fluid dynamics \cite{InubushiEtAl2015}, to quantify the degree of a dynamical system's hyperbolicity \cite{XuPaul2016}, and to predict catastrophes \cite{BeimsGallas2016}.

In this thesis, we aim to explain the theory of CLVs in a manner that is accessible to applied mathematicians since the topic is often expressed in terms of very abstract mathematics. Furthermore, we strive to compute CLVs along chaotic orbits using the algorithm developed by Ginelli and collaborators (G\&C) \cite{GinelliEtAl2007}, which we refer to hereafter as \textit{the G{\slshape{\&}}C algorithm}, and we apply these vectors to study various autonomous Hamiltonian systems. By numerically investigating these systems with the method of CLVs, we hope to understand their dynamical behaviour better. We also aim to understand the convergence properties of this algorithm and develop a method by which appropriate time intervals for the transients of the G\&C algorithm can be determined.

The contents of this thesis are organised as follows. \hyperref[ch:theory]{Chapter~\ref*{ch:theory}} begins with a short summary of several prerequisite mathematical notions needed for later discussions, followed by an overview of the theory of LEs and CLVs in the context of continuous-time dynamical systems. A detailed description of the numerical methods we employed for LE and CLV computation is also given. CLV computations for chaotic orbits in two low-dimensional autonomous Hamiltonian systems are presented in \chapterref{ch:hh}, and the convergence properties of the G\&C algorithm are numerically analysed. In \chapterref{ch:dna}, a Hamiltonian lattice model of DNA is studied using LEs and CLVs, and new quantities are also introduced for exploring the instantaneous stability properties of the lattice. Finally, the results of this thesis are summarised in \chapterref{ch:conclusion}, alongside our conclusions.

%% file: theory.tex
\chapter{Theory and numerics}\label{ch:theory}
\vspace{-1em}
\epigraph{Eighty percent of mathematics is linear algebra.}{\textit{Raoul Bott}}

\noindent In this chapter, we present the mathematical and numerical methods used in the rest of this thesis. A summary of the necessary mathematics is given in \sectionref{sec:bigN}, followed by the relevant background on continuous dynamical systems in \sectionref{sec:continuoussec}. Autonomous Hamiltonian systems, which are of particular interest to us, are covered in \sectionref{sec:autoham}. Algorithms for the computation of LEs and CLVs are detailed in Sects.~\ref{sec:le} and~\ref{sec:ginelli}, respectively, where pseudocode for each of these algorithms is given. Issues of convergence are discussed in \sectionref{sec:converge}, and finally we summarise this chapter in \sectionref{sec:theorysummary}.

\section{Mathematical preliminaries}\label{sec:bigN}
While we assume a basic understanding of linear algebra and matrix theory, we begin with a brief overview of some key mathematical notions that may be less familiar to the reader. Where relevant, we also discuss some issues of computation. For our discussions involving angles and distances, it is necessary to choose an inner product and norm. Since the vector space in which we work is $\mathbb{R}^N$, we use the standard dot product $\bs u\cdot\bs v = \sum_{i=1}^N u_i v_i$ and its induced norm $\|\bs u\|=\sqrt{\bs u\cdot\bs u}$. Here, $\bs u,\bs v\in\mathbb{R}^N$ are vectors with components $u_i,v_i$ in the standard basis, where $i=1,2,\dots,N$.

\subsection{QR decomposition}\label{sec:qr}
\begin{figure}[htbp]
	\centering
	\includegraphics[height=0.13\linewidth]{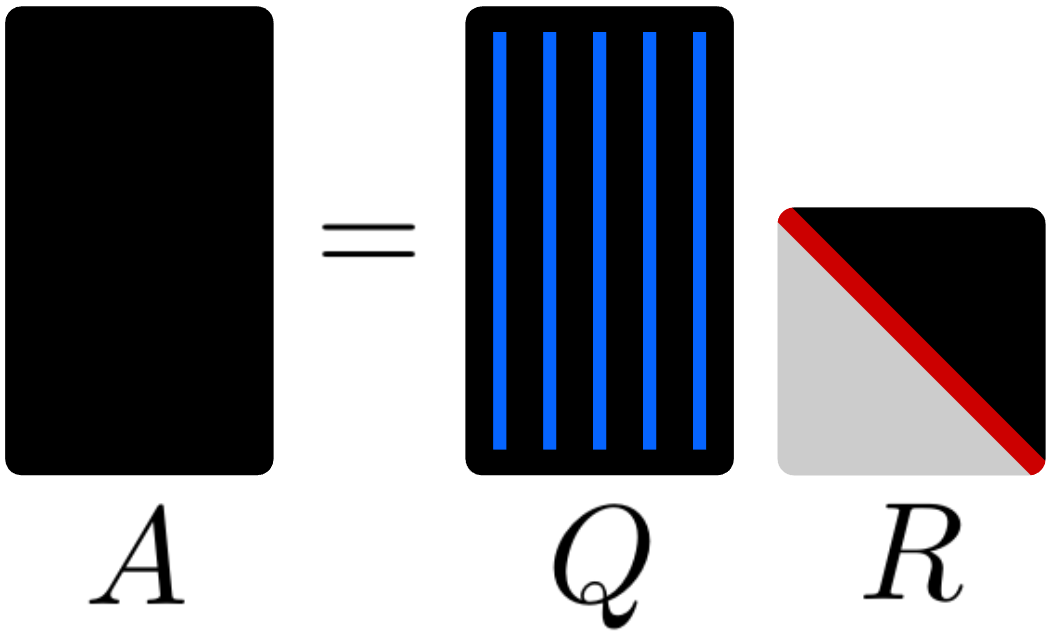}
	\caption{Diagram of QR decomposition. The blue stripes denote the orthonormal columns of $Q$, the red stripe is the positive diagonal of $R$, and the grey region indicates where entries of $R$ are zero.}
	\label{fig:qr}
\end{figure}

One of the most important matrix decompositions, the \textit{QR decomposition}, plays a central role in algorithms for computing LEs and CLVs that we discuss later. Any real matrix $A\in\mathbb{R}^{N\times M}$ ($N\ge M$) with linearly independent columns can be uniquely decomposed into the product
\begin{align}
	A=QR,\label{eq:qr}
\end{align}
where $Q\in\mathbb{R}^{N\times M}$ has orthonormal columns\footnote{By orthonormal columns, we mean that the columns of the matrix constitute a set of orthonormal vectors.} and $R\in\mathbb{R}^{M\times M}$ is upper triangular with positive diagonal entries. We refer to this decomposition simply as QR decomposition, but some authors prefix this title with ``thin'' \cite[p.~230]{GolubVanLoan1996} or ``reduced'' \cite[p.~48]{TrefethenBau1997} to distinguish it from so-called ``full'' QR decomposition, which adds redundant columns and rows to $Q$ and $R$, respectively. The structure of the matrices in decomposition \eqref{eq:qr} is diagrammed in \figureref{fig:qr}. Note that the columns of $Q$ are the same as the vectors produced by the Gram-Schmidt orthonormalisation procedure (see e.g.\ \cite{Skokos2010}).

Many software libraries implement efficient routines for decomposing an arbitrary matrix into its $Q$ and $R$ matrices. However, the numerical routines for QR decomposition provided by some widely used numerical linear algebra libraries yield a decomposition $A=\tilde Q\tilde R$ that is not necessarily the unique QR decomposition in \eqref{eq:qr}, since the $\tilde R$ matrix may have some negative diagonal entries.\footnote{Since we assume $A$ has linearly independent columns, the diagonal entries of $\tilde R$ cannot be zero, and so must either be positive or negative \cite[pp.~48--55]{TrefethenBau1997}.} We have directly observed that the QR decomposition implementations provided by the SciPy \cite{VirtanenEtAl2020} and NumPy \cite{HarrisEtAl2020} libraries give such ``non-unique'' QR decompositions, and this has also been observed in MATLAB \cite[p.~427]{KilianLutkepohl2017} and the LAPACK library \cite{DemmelEtAl2009}. Nevertheless, the unique QR decomposition with positive entries on the diagonal of $R$ is easily recovered by negating the rows of $\tilde R$ that have negative diagonal entries and simultaneously negating the corresponding columns of $\tilde Q$. These elementary row and column operations can be represented by matrix multiplications $S\tilde R$ and $\tilde QS$, respectively, where $S$ is a diagonal matrix with diagonal entries of $\pm1$ and the signs of these diagonal entries are the signs of the diagonal entries of $\tilde R$. Since $S$ is its own inverse, these multiplications with $S$ can be inserted into the decomposition $A=\tilde Q\tilde R$ to produce a new decomposition,
\begin{align}
	\tilde Q\tilde R=\tilde Q(SS)\tilde R=(\tilde QS)(S\tilde R).\label{eq:proc}
\end{align}
Now, $\tilde QS$ has orthonormal columns because negating some orthonormal columns of $S$ does not change the orthonormality of those columns. Furthermore, $S\tilde R$ is upper triangular with positive diagonal entries because negating some rows of $\tilde R$ yields an upper triangular matrix which, by our choice of $S$, has positive diagonal entries. We have therefore demonstrated that the procedure given in \eqref{eq:proc} recovers the unique QR decomposition for $A$, where $Q=\tilde QS$ and $R=S\tilde R$.

\subsection{Singular value decomposition}\label{sec:svd}
\begin{figure}[htbp]
	\centering
	\includegraphics[height=0.13\linewidth]{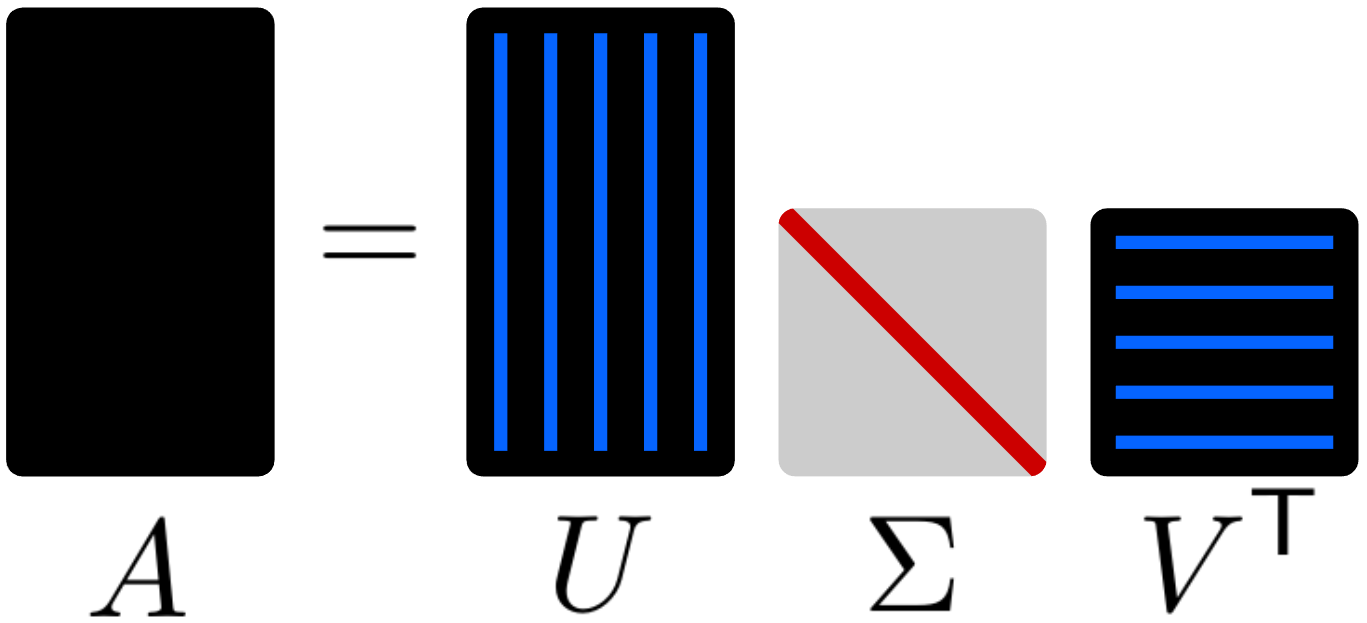}
	\caption{Diagram of singular value decomposition. The blue stripes denote the orthonormal columns of $U$ and $V$, the red stripe is the non-negative diagonal of $\Sigma$, and the grey regions indicate where entries of $\Sigma$ are zero.}
	\label{fig:svd}
\end{figure}

Any real matrix $A\in\mathbb{R}^{N\times M}$ ($N\ge M$) can be decomposed into a product
\begin{align}
	A=U\Sigma V\transpose,\label{eq:svd}
\end{align}
where $(\transpose)$ denotes the transpose, $U\in\mathbb{R}^{N\times M}$ and $V\in\mathbb{R}^{M\times M}$ have orthonormal columns known as \textit{left} and \textit{right singular vectors}, respectively, and $\Sigma\in\mathbb{R}^{M\times M}$ is a diagonal matrix with non-negative diagonal entries, which are known as \textit{singular values}. By convention, the singular values are expressed in descending order along the diagonal of $\Sigma$. We refer to this decomposition simply as \textit{singular value decomposition} (SVD), but (analogous to QR decomposition) some authors prefix this title with ``thin'' \cite[p.~72]{GolubVanLoan1996} or ``reduced'' \cite[p.~27]{TrefethenBau1997} to distinguish it from so-called ``full'' SVD, which adds redundant columns and rows to $U$ and $\Sigma$, respectively. It is easily shown from this decomposition that the right singular vectors of the matrix $A$ are the eigenvectors of the matrix $A\transpose A$. If the singular values are distinct, then decomposition \eqref{eq:svd} is unique up to signs of the columns of $U$ and $V$ \cite[p.~29]{TrefethenBau1997}. The structure of the matrices in the SVD is diagrammed in \figureref{fig:svd}.

\begin{figure}[htbp]
	\centering
	\includegraphics[width=0.8\linewidth]{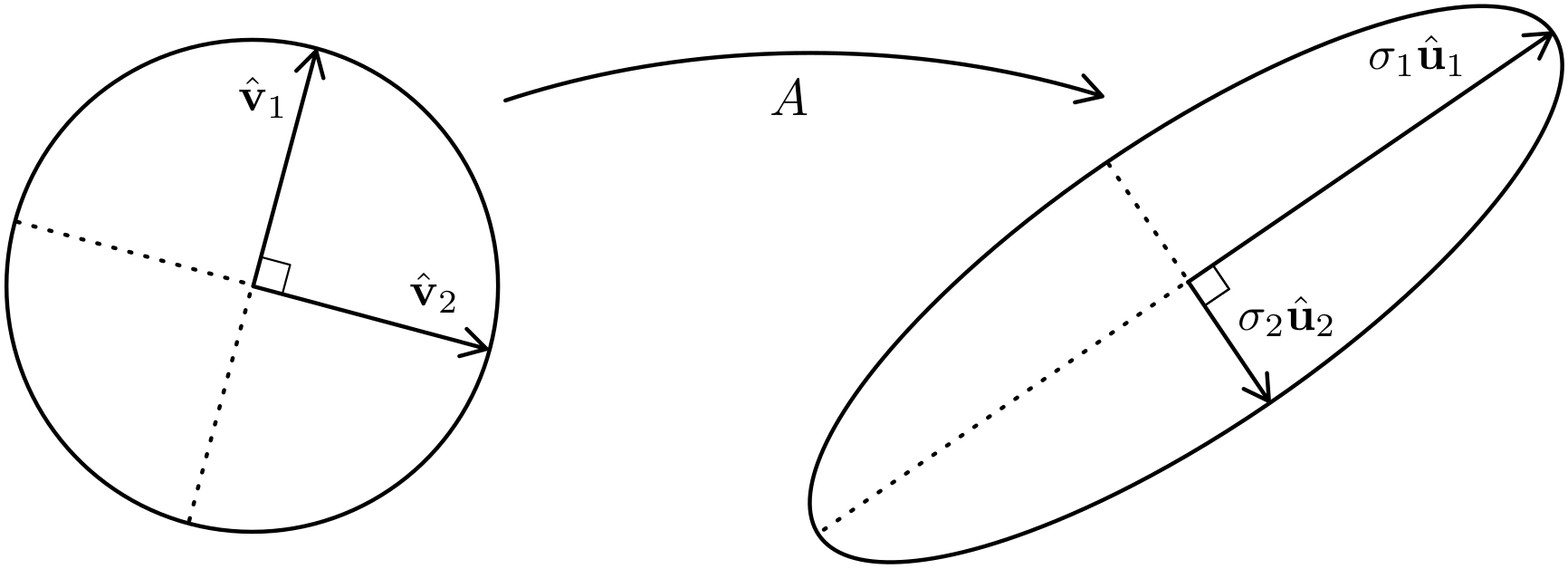}
	\caption{Example of a $2\times 2$ real matrix $A$ which maps a unit circle to an ellipse. The principal semi-axes of the ellipse are denoted by the vectors $\sigma_i\hat{\bs u}_i$, $i=1,2$, and their corresponding pre-images under $A$ are denoted by $\hat{\bs v}_i$.}
	\label{fig:ellipsoid}
\end{figure}

SVD has a simple geometric interpretation rooted in the fact that matrices map unit (hyper-)spheres to (hyper-)ellipsoids \cite[p.~25]{TrefethenBau1997}. For a given $N\times M$ ($N\ge M$) real matrix, the singular values $\sigma_i$ are the lengths of the principal semi-axes of the ellipsoid, the left singular vectors $\hat{\bs u}_i$ are oriented along the principal semi-axes of the ellipsoid, and the right singular vectors $\hat{\bs v}_i$ are the pre-images of the principal semi-axes of the ellipsoid, where $i=1,2,\dots,M$ in each case. In \figureref{fig:ellipsoid}, we diagram this geometric interpretation with an example in $\mathbb{R}^2$. With this understanding, it is clear that the first right singular vector $\hat{\bs v}_1$ is the unit vector which stretches the most under the transformation $A$ because $\hat{\bs v}_1$ is mapped to the longest principal semi-axis $\sigma_1\hat{\bs u}_1$ of the ellipsoid.

\subsection{Principal angles}\label{sec:principal}
\begin{figure}[htbp]
	\centering
	\includegraphics[width=0.36\linewidth]{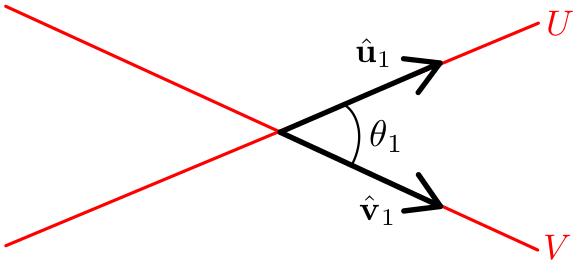}
	\caption{An example of the minimum angle $\theta_1$ between two 1-D subspaces $U$ and $V$ (drawn as red lines), where $\hat{\bs u}_1\in U$ and $\hat{\bs v}_1\in V$ (drawn as black arrows) are representative unit vectors with maximal dot product.}
	\label{fig:r2}
\end{figure}

Let $\hat{\bs u},\hat{\bs v}$ be unit vectors in some real inner product space $W$, where we use $(\,\hat{}\,)$ to denote a unit vector. The angle $\theta\in[0,\pi]$ between $\hat{\bs u}$ and $\hat{\bs v}$ is defined by the equation
\begin{align}
	\cos\theta=\hat{\bs u}\cdot\hat{\bs v}.\label{eq:cos}
\end{align}
Consider now $U=\spn(\hat{\bs u})$ and $V=\spn(\hat{\bs v})$, which are two $1$-dimensional ($1$-D) subspaces of $W$. We can define an angle between these subspaces by choosing a representative unit vector from each subspace and computing the angle between them using \eqref{eq:cos}. However, this angle would depend on the choice of representatives. For example, both $\hat{\bs u}$ and $-\hat{\bs u}$ are unit vectors in $U$, but switching between them will negate the dot product in \eqref{eq:cos} and hence switch the angle $\theta$ between acute and obtuse. If a unique angle between $U$ and $V$ is required, one could use the \textit{minimum angle} $\theta_1$ between any pair of representatives $\hat{\bs u}\in U$ and $\hat{\bs v}\in V$. If $\hat{\bs u}_1\in U$ and $ \hat{\bs v}_1\in V$ are such representatives with a minimal angle $\theta_1$ between them, then $\hat{\bs u}_1\cdot\hat{\bs v}_1$ maximises the dot product $\hat{\bs u}\cdot\hat{\bs v}$. In \figureref{fig:r2}, we give a graphical representation of the minimum angle $\theta_1$ between subspaces $U$ and $V$ of $\mathbb{R}^2$.

\begin{figure}[htbp]
	\centering
	\includegraphics[width=\linewidth]{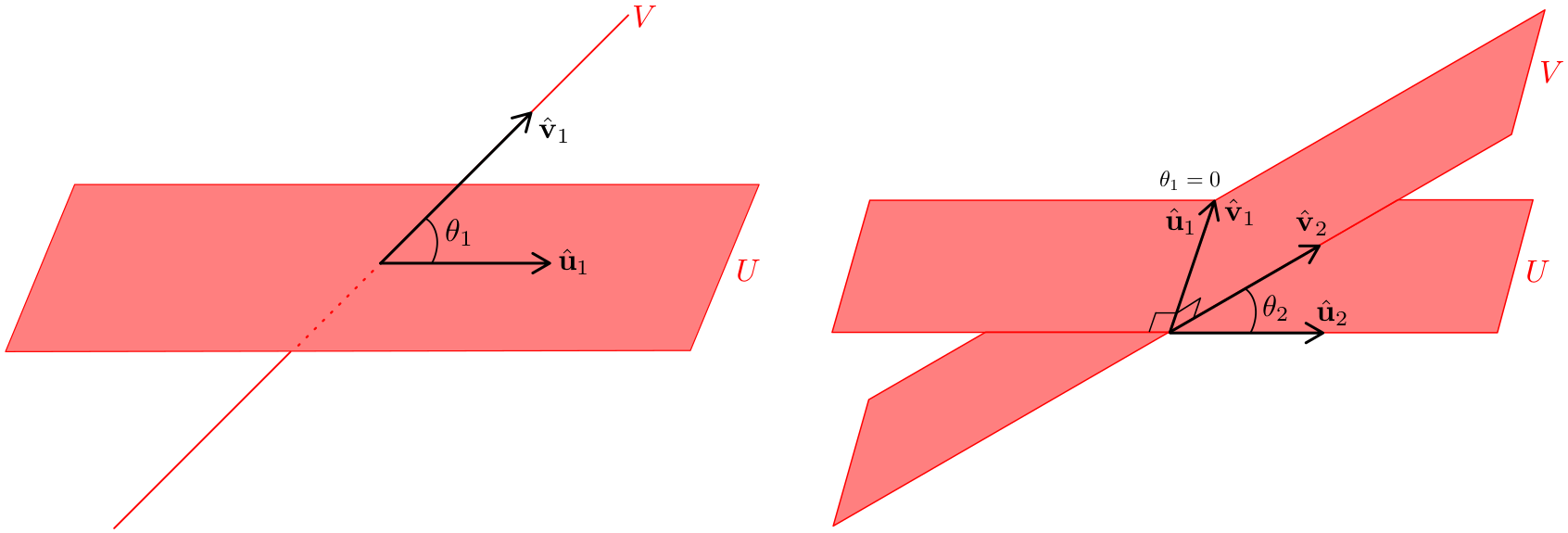}
	\caption{Two examples of principal angles and vectors (drawn as black arrows) between subspaces $U$ and $V$ (drawn as red lines or planes) in $\mathbb{R}^3$. In the example on the left, $\dim V=1$, while $\dim V=2$ on the right. In both cases, $\dim U=2$.}
	\label{fig:r3}
\end{figure}

The minimum angle between $1$-D subspaces can be generalised to so-called \textit{principal angles} between subspaces of arbitrary dimension. Following the definitions given in \cites{BjorckGolub1973}[p.~603]{GolubVanLoan1996}, if $U$ and $V$ are subspaces and $q=\min(\dim U,\dim V)\ge1$, then the principal angles $\theta_k\in[0,\pi/2]$ between $U$ and $V$ are defined recursively for $k=1,\dots,q$ by
\begin{align}
	\cos\theta_k=\max_{\substack{\hat{\bs u}\in U\\\hat{\bs v}\in V}} (\hat{\bs u}\cdot\hat{\bs v})=\hat{\bs u}_k\cdot\hat{\bs v}_k,
\end{align}
where the unit vectors $\hat{\bs u},\hat{\bs v}$ are subject to the constraints
\begin{align}
\begin{split}
	&\hat{\bs u}\cdot\hat{\bs u}_i=0\quad\text{for all}\quad i=1,\dotsc,k-1,\\
	&\hat{\bs v}\cdot\hat{\bs v}_i=0\quad\text{for all}\quad i=1,\dotsc,k-1.
\end{split}
\end{align}
The unit vectors $\hat{\bs u}_1,\dotsc,\hat{\bs u}_q$ and $\hat{\bs v}_1,\dotsc,\hat{\bs v}_q$ are called \textit{principal vectors} between $U$ and $V$. The principal angles satisfy $\theta_1\leq\cdots\leq\theta_q$, and $\theta_1$ is the minimum angle between $U$ and $V$. By this definition, $\theta_1$ in \figureref{fig:r2} is the first (and only) principal angle between the 1-D subspaces $U$ and $V$, and $\hat{\bs u}_1$ and $\hat{\bs v}_1$ are corresponding principal vectors. To provide some intuition for these quantities in higher dimensions, we give two examples in \figureref{fig:r3} of principal angles and vectors between subspaces of $\mathbb{R}^3$. Note that if $\theta_1=0$, then $U$ and $V$ have a non-trivial intersection, i.e.\ $\dim(U\cap V)>0$.\footnote{This observation leads directly to an algorithm for computing the intersection between two subspaces, see e.g.\ \cite[p.~604]{GolubVanLoan1996}.}

Bj\"orck and Golub \cite{BjorckGolub1973} proposed a basic algorithm for computing principal angles and vectors using SVD. Unfortunately, any principal angle $\theta_k$ produced by this algorithm is not accurate when $\theta_k\approx0$ since the algorithm first computes $\cos\theta_k$, which fails to discriminate between very small angles; using double precision, we typically found the cosines of any angles smaller than $10^{-8}$ to be equal. The authors addressed this issue by proposing an alternative algorithm which computes $\sin\theta_k$ instead; unfortunately, this has the new problem of losing accuracy for $\theta_k\approx\pi/2$. Knyazev and Argentati \cite{KnyazevArgentati2002} combined these two approaches into a hybrid algorithm to ensure the most accurate computation of each principal angle is used. For computing principal angles in this thesis, we use the SciPy \cite{VirtanenEtAl2020} implementation of this hybrid algorithm.

\subsection{Distance between subspaces}\label{sec:distanceintro}
In this section, we adapt the definitions and results from \cites[p.~76]{GolubVanLoan1996}[p.~239]{Galantai2004}{YeLim2016}{Noethen2019}. Let $U$ and $V$ be $q$-dimensional ($q$-D) subspaces of a vector space $W$, then the \textit{distance} $\Delta$ between $U$ and $V$ is defined as
\begin{align}
	\Delta(U,V)=\|P_U-P_V\|,\label{eq:proj}
\end{align}
where $P_U$ and $P_V$ are the orthogonal projections of $W$ onto $U$ and $V$, respectively, and the matrix norm is induced by the Euclidean norm (see e.g.\ \cite[p.~15]{Galantai2004}). This distance is also known as the gap \cite[p.~227]{DezaDeza2009} or aperture \cite[p.~69]{AkhiezerGlazman1993} between $U$ and $V$. While \eqref{eq:proj} is the usual definition found in the mathematical literature, this notion of distance can be formulated equivalently in terms of principal angles:
\begin{align}
	\Delta(U,V)=\sin\theta_q,\label{eq:distance}
\end{align}
where $\theta_q$ is the last (i.e.\ largest) principal angle between $U$ and $V$. The advantage of this definition is it provides us with a straightforward method of computing and interpreting $\Delta$ using principal angles. Since $\theta_q\in[0,\pi/2]$ (see \sectionref{sec:principal}), it is clear from \eqref{eq:distance} that $\Delta(U,V)\in[0,1]$, and if $\theta_q$ is small then $\Delta(U,V)\approx\theta_q$.

For convenience, we extend the notion of the distance between subspaces to the distance between vectors $\bs u$ and $\bs v$ and denote it $\Delta(\bs u,\bs v)$, which is simply a shorthand for $\Delta(\spn(\bs u),\spn(\bs v))$. Since $\spn(\bs u)$ and $\spn(\bs v)$ are each $1$-D subspaces, the only principal angle between them is the minimum angle $\theta_1$, so $\Delta(\bs u,\bs v)=\sin\theta_1$.

It turns out that $\Delta$ is a metric on the space of $q$-D subspaces, which is a differentiable manifold known as the Grassmannian. As a metric, it follows that $\Delta(U,V)=0$ if and only if $U=V$. We use this notion of distance between subspaces in later chapters to measure the convergence rates between equidimensional subspaces. It is worth noting that $\Delta$, as defined in \eqref{eq:distance}, is not a metric if we allow $\dim U\ne\dim V$, since $\sin\theta_q$ can be zero if $U\subset V$ or $V\subset U$, but $U\ne V$, so $\Delta$ is not a metric in this case.

\subsection{Functions of matrices}\label{sec:matfunc}
If $A\in\mathbb{R}^{N\times N}$ is a real symmetric matrix, then $A$ is diagonalisable and has eigenvectors which form an orthonormal basis of the vector space \cite[p.~409]{Anton2020}. Therefore, $A$ can be decomposed as
\begin{align}
	A = PDP\transpose,\label{eq:diag}
\end{align}
where the columns of $P\in\mathbb{R}^{N\times N}$ are a set of orthonormal eigenvectors of $A$ and $D\in\mathbb{R}^{N\times N}$ is a diagonal matrix with the eigenvalues of $A$ along the diagonal \cite[p.~423]{Johnston2021}.

Now, given some function $\tilde f:\mathbb R\to\mathbb R$, we can define a corresponding matrix function $f:\mathbb{R}^{N\times N}\to\mathbb{R}^{N\times N}$ on the set of real symmetric matrices by
\begin{align}
	f(A) = P\diag(\tilde f(\lambda_i))P\transpose,\label{eq:func}
\end{align}
where $\lambda_i$, $i=1,2,\dots,N$, are the eigenvalues of $A$ and $\diag(\tilde f(\lambda_i))$ is an $N\times N$ diagonal matrix with $\tilde f(\lambda_i)$ along the diagonal \cite[p.~3]{Higham2008}. Hereafter, we use the same symbol for a real function and its matrix counterpart. The matrix logarithm $\ln A$ (which we will encounter in \sectionref{sec:lyapunov}) is thus defined by \eqref{eq:func} for real symmetric matrices $A$, provided $\ln\lambda_i$ is defined for every eigenvalue $\lambda_i$ of $A$. Therefore, the domain of the matrix logarithm is restricted to positive-definite matrices since these only have positive eigenvalues. The matrix exponential $\exp A$ can be similarly defined but without any additional restriction to its domain.

\section{Continuous dynamical systems}\label{sec:continuoussec}
We begin this section with a review of dynamical systems, adapting the introductions on this topic given in \cites{Meiss2007}{Skokos2010}[Sect.~2.1]{CvitanovicEtAl2020}. Because we are ultimately interested in autonomous Hamiltonian systems, we restrict our discussion of dynamical systems here to ones that are continuous, invertible, and deterministic. Nevertheless, many of the notions we introduce in this section can be applied more broadly, e.g.\ to discrete and stochastic dynamical systems (see e.g.\ \cite{Arnold1998}).

A \textit{continuous dynamical system} is a differentiable manifold $\mathcal M$ together with a continuous map $\Phi:\mathbb R\times\mathcal M\to\mathcal M$ which describes the evolution of points $x\in\mathcal M$ with respect to some variable $t\in\mathbb{R}$, which we refer to as \textit{time}. For convenience, we write $\Phi(t,x)$ as $\Phi^t(x)$, where $\Phi^t$ can be considered as a family of maps $\Phi^t:\mathcal M\to\mathcal M$ parametrised by $t$. We assume that each map $\Phi^t$ is invertible, so the dynamical system is called \textit{invertible}. The $N$-D manifold $\mathcal M$ is known as the \textit{state space}, and a point $x\in\mathcal M$ is called a \textit{state} of the system. Choosing some coordinate system, $x$ can be represented by a vector $\bs x$ whose components are the coordinates of $x$. In this thesis, we always assume that some coordinate system has been chosen and thus refer to $\bs x$ directly as the state instead of the abstract point $x$. Furthermore, we write $\bs x(t)$ to denote the state at time $t$.

If the state of a dynamical system is $\bs x(0)$ at time $0$, then its evolution by time $t$ to the state $\bs x(t)$ is given by
\begin{align}
	\bs x(t) = \Phi^t(\bs x(0)).\label{eq:solution}
\end{align}
The map $\Phi^t$ is called the \textit{flow map}, and we assume it is single-valued (i.e.\ it uniquely determines the consequent state of any initial state), thereby restricting ourselves to so-called \textit{deterministic} dynamical systems. In particular, we only consider such systems that can be expressed as a system of ordinary differential equations $\dot{\bs x} = \bs f(\bs x, t)$, where $(\,\dot{}\,)$ denotes a time derivative and $\bs f$ is a vector field. We further restrict our focus to so-called \textit{autonomous} systems where $\bs f$ has no explicit time dependence, so
\begin{align}
	\dot{\bs x} = \bs f(\bs x).\label{eq:auto}
\end{align}
Given an initial state (or \textit{initial condition}) $\bs x(0)$, the solution of \eqref{eq:auto} at time $t$ is \eqref{eq:solution}. Finally, the set of points traced out by a state $\bs x$ under the evolution of the flow map is a curve known as the \textit{orbit} $[\bs x]$ passing through $\bs x$, defined by
\begin{align}
	[\bs x] = \{\Phi^t(\bs x):t\in\mathbb{R}\}.
\end{align}

We now consider the evolution of a small perturbation to the state $\bs x(t)$. Such a perturbation can be modelled as a \textit{deviation vector} $\bs w(t)$ that lives in a vector space known as the \textit{tangent space} of $\mathcal M$ at $\bs x(t)$, denoted by $T_{\bs x(t)}\mathcal {M}$. Now, consider a deviation vector $\bs w(0)\in T_{\bs x(0)}\mathcal M$, where $\bs x(0)$ is some initial state. The evolution of this deviation vector is described by
\begin{align}
	\bs w(t)=\mathrm{d}_{\bs x(0)}\Phi^t(\bs w(0)),\label{eq:prop}
\end{align}
where $\mathrm{d}_{\bs x}\Phi^t:T_{\bs x}\mathcal M\to T_{\Phi^t(\bs x)}\mathcal M$ is a linear map known as the \textit{linear propagator}. Note that deviation vectors typically evolve through distinct tangent spaces as the underlying state evolves along its orbit. The union of tangent spaces at all points in $\mathcal M$ forms a differentiable manifold $T\mathcal M$ known as the \textit{tangent bundle} \cite[p.~348]{Nakahara2003}, so deviation vectors ultimately live in and evolve through the tangent bundle.

By choosing bases for the relevant vector spaces, we can represent the linear propagator by the matrix $M(\bs x(0),t)$, so \eqref{eq:prop} becomes the matrix equation
\begin{align}
	\bs w(t)=M(\bs x(0),t)\bs w(0).\label{eq:fundamental}
\end{align}
The matrix $M(\bs x(0),t)\in\mathbb{R}^{N\times N}$ is called the \textit{fundamental matrix},\footnote{Authors typically refer to \textit{a} fundamental matrix, but we say \textit{the} fundamental matrix since this is an initial value problem, which specifies the matrix uniquely.} whose columns are independent solutions to the so-called \textit{variational equations}
\begin{align}
	\dot{\bs w}=J(\bs x(0))\bs w,\label{eq:jacob}
\end{align}
where $J(\bs x(0))$ is the Jacobian matrix of $\bs f$ in \eqref{eq:auto} evaluated at $\bs x(0)$. Since fundamental matrices are non-singular (see e.g.\ \cite[p.~352]{ZillEtAl2013}), $M(\bs x(0),t)$ is invertible. We acknowledge our reuse of the symbol $\bs w$ in equations \eqref{eq:prop} and \eqref{eq:fundamental}, but we do so without confusion to avoid introducing additional notation.

The term \textit{dynamics} refers to the collective evolution of all states in $\mathcal M$, while the evolution of deviation vectors in $T\mathcal M$ is called the \textit{tangent dynamics}. Now, the dynamics is governed by typically nonlinear maps $\Phi^t$, but the tangent dynamics is governed by linear maps $\mathrm{d}_{\bs x}\Phi^t$ expressed as matrices $M(\bs x,t)$, hence studying the tangent dynamics usually involves linear algebra and matrix theory.

\subsection{Lyapunov vectors and exponents}\label{sec:lyapunov}
Having reviewed the basics of dynamical systems, we proceed with an overview of the theory of Lyapunov exponents and vectors where we mainly follow the approach of \cite{GinelliEtAl2013} in our presentation with influence from \cites{Skokos2010}{KuptsovParlitz2012}{PikovskyPoliti2016}. We will soon define several subspaces and linear maps which either live in or act on the tangent space, and thus these objects will all depend on an underlying state $\bs x(0)$. In order to simplify the notation, we will simply write $\bs x$ in the remainder of this section with the understanding that $\bs x$ is the state at time $0$. In later sections of this thesis, we will drop the $\bs x$ dependence entirely when there is no confusion.

In the pioneering work of Oseledets, the existence of Lyapunov exponents was proved under fairly general conditions in the \textit{multiplicative ergodic theorem}. It can be shown from this theorem that the two $N\times N$ matrices $\Lambda^{+}(\bs x)$ and $\Lambda^{-}(\bs x)$ (sometimes called the \textit{Oseledets matrices}), which are given by
\begin{align}
	\Lambda^{\pm}(\bs x)=\lim_{t\to\pm\infty}\frac{1}{t}\ln\sqrt{M(\bs x,t)\transpose M(\bs x,t)},\label{eq:matrix}
\end{align}
exist for almost all $\bs x$.\footnote{By \textit{almost} all, we mean there may be states for which the result does not hold, but the set of such states has measure zero, and thus the probability of randomly selecting such an exceptional state is zero.} Note that since $M(\bs x,t)$ is invertible, the argument of the matrix logarithm in \eqref{eq:matrix} is a positive-definite matrix and hence in the domain of the logarithm (see \sectionref{sec:matfunc}). The matrices $\Lambda^{\pm}(\bs x)$ share the same $m\le N$ distinct eigenvalues,
\begin{align}
	\lambda_1([\bs x])>\lambda_2([\bs x])>\cdots>\lambda_m([\bs x]),\label{eq:le}
\end{align}
known as \textit{Lyapunov exponents} (LEs), which are norm-independent quantities and depend only on the orbit $[\bs x]$, so they do not vary with time. The multiplicity of $\lambda_i([\bs x])$ is denoted by $\kappa_i([\bs x])$, so $\sum_{i=1}^{m} \kappa_i([\bs x])=N$. The eigenspace of $\Lambda^{\pm}(\bs x)$ corresponding to $\lambda_i([\bs x])$ is denoted $V_i^{\pm}(\bs x)$ and has dimension $\dim V_i^{\pm}(\bs x)=\kappa_i([\bs x])$.

The \textit{filtration subspaces} $\Gamma_i^{\pm}(\bs x)$ are defined as
\begin{align}
	\begin{split}
		\Gamma_i^+(\bs x) &= V_i^+(\bs x)\oplus V_{i+1}^+(\bs x)\oplus\cdots\oplus V_m^+(\bs x),\\
		\Gamma_i^-(\bs x) &= V_1^-(\bs x)\oplus V_2^-(\bs x)\oplus\cdots\oplus V_i^-(\bs x),
	\end{split}\label{eq:filtration}
\end{align}
where $i=1,2,\dots,m$ and $\oplus$ denotes the direct sum. Extending \eqref{eq:filtration} by defining both $\Gamma_{m+1}^+(\bs x)$ and $\Gamma_0^-(\bs x)$ to be the trivial vector space $\{0\}$, each family of subspaces $\Gamma_i^{\pm}(\bs x)$ forms a nested subspace structure,
\begin{align}
	\begin{split}
		T_{\bs x}\mathcal M&=\Gamma_1^+(\bs x)\supset\Gamma_2^+(\bs x)\supset\cdots\supset\Gamma_m^+(\bs x)\supset\Gamma_{m+1}^+(\bs x)=\{0\},\\
		T_{\bs x}\mathcal M&=\Gamma_m^-(\bs x)\supset\Gamma_{m-1}^-(\bs x)\supset\cdots\supset\Gamma_1^-(\bs x)\supset\Gamma_0^-(\bs x)=\{0\},
	\end{split}
\end{align}
which is known as a \textit{filtration}. The filtration subspaces determine the asymptotic growth rates of the deviation vectors they contain.\footnote{When we refer to the ``growth'' of a vector, we mean the growth of that vector's norm.} In particular,
\begin{align}
	\lim_{t\to\pm\infty}\frac{1}{|t|}\ln\frac{\|M(\bs x,t)\bs w\|}{\|\bs w\|}=\pm\lambda_i([\bs x])\quad\text{if}\quad\bs w\in\Gamma_i^{\pm}(\bs x)\setminus\Gamma_{i\pm1}^{\pm}(\bs x),\label{eq:le_filtration}
\end{align}
where $\setminus$ is the set difference. Loosely speaking, $\lambda_i([\bs x])$ is the exponential growth rate\footnote{We use the term ``exponential growth rate'' even if the rate is negative and thus corresponds to exponential decay.} of almost all deviation vectors in $\Gamma_{i}^{+}(\bs x)$ as $t\to\infty$, while almost all deviation vectors in $\Gamma_{i}^{-}(\bs x)$ have an exponential growth rate of $-\lambda_i([\bs x])$ as $t\to-\infty$. In \appendixref{app:rate}, we explain why \eqref{eq:le_filtration} does not strictly imply that $\lambda_i([\bs x])$ is the exponential growth rate of $M(\bs x,t)\bs w$, but we conclude that it is still a useful characterisation in practice.

It can be shown from \eqref{eq:le_filtration} that each filtration subspace satisfies
\begin{align}
	\mathrm{d}_{\bs x}\Phi^t(\Gamma_i^{\pm}(\bs x))=\Gamma_i^{\pm}(\Phi^t(\bs x)),
\end{align}
so we say the filtration subspaces are \textit{covariant}. While $\Gamma_i^{+}(\bs x)$ and $\Gamma_i^{-}(\bs x)$ are covariant and both correspond to the same LE $\lambda_i([\bs x])$, these subspaces are typically distinct. A natural question to then ask would be if there exists a covariant subspace which corresponds to the same LE, regardless of whether it is evolved forwards or backwards in time. It turns out that such a space can be defined by
\begin{align}
	\Omega_i(\bs x) = \Gamma_i^+(\bs x) \cap \Gamma_i^-(\bs x),\label{eq:omegaintersect}
\end{align}
where $i=1,2,\dots,m$ and $\dim\Omega_i(\bs x)=\kappa_i([\bs x])$ \cite{Ruelle1979}. Together, these subspaces decompose the tangent space,
\begin{align}
	T_{\bs x}\mathcal M=\Omega_1(\bs x)\oplus\Omega_2(\bs x)\oplus\cdots\oplus\Omega_m(\bs x).\label{eq:splitting}
\end{align}
Such a decomposition of the tangent space is known as a \textit{splitting}, and so we refer to $\Omega_i(\bs x)$ as the \textit{splitting subspaces}. Furthermore,
\begin{align}
	\lim_{t\to\pm\infty}\frac{1}{|t|}\ln\frac{\|M(\bs x,t)\bs w\|}{\|\bs w\|}=\pm\lambda_i([\bs x])\quad\text{if}\quad\bs w\in\Omega_i(\bs x)\setminus\{0\},\label{eq:le_splitting}
\end{align}
so we see that the asymptotic growth rates of the deviation vectors in each splitting subspace are governed by the corresponding LEs, both forwards and backwards in time. Like the filtration subspaces, the splitting subspaces are covariant,
\begin{align}
	\mathrm{d}_{\bs x}\Phi^t(\Omega_i(\bs x))=\Omega_i(\Phi^t(\bs x)).\label{eq:covariance}
\end{align}
The splitting subspaces also decompose the filtration subspaces \cite{Noethen2019a},
\begin{align}
	\begin{split}
		\Gamma_i^+(\bs x) &= \Omega_i(\bs x)\oplus\Omega_{i+1}(\bs x)\oplus\cdots\oplus\Omega_m(\bs x),\\
		\Gamma_i^-(\bs x) &= \Omega_1(\bs x)\oplus\Omega_2(\bs x)\oplus\cdots\oplus\Omega_i(\bs x).
	\end{split}\label{eq:noethen}
\end{align}
Note that the subspaces $\Omega_i(\bs x)$ are norm-independent and generically non-orthogonal, whereas the orthogonal eigenspaces $V_i^{\pm}(\bs x)$ depend on the chosen norm \cite{GinelliEtAl2013}.

As a brief aside, some authors refer to $\Gamma_i^{\pm}(\bs x)$ as Oseledets subspaces \cites{KuptsovParlitz2012}{GinelliEtAl2013}{PikovskyPoliti2016}{SharafiEtAl2017}, while Oseledets \cite{Oseledets2008} reserves that name for $\Omega_i(\bs x)$, and Noethen \cite{Noethen2019a} calls $\Omega_i(\bs x)$ the Oseledets spaces. To avoid this ambiguity in the literature, we have chosen to refer to $\Gamma_i^{\pm}(\bs x)$ and $\Omega_i(\bs x)$ descriptively as the filtration and splitting subspaces, respectively.

\begin{figure}[htbp]
	\centering
	\includegraphics[width=0.98\linewidth]{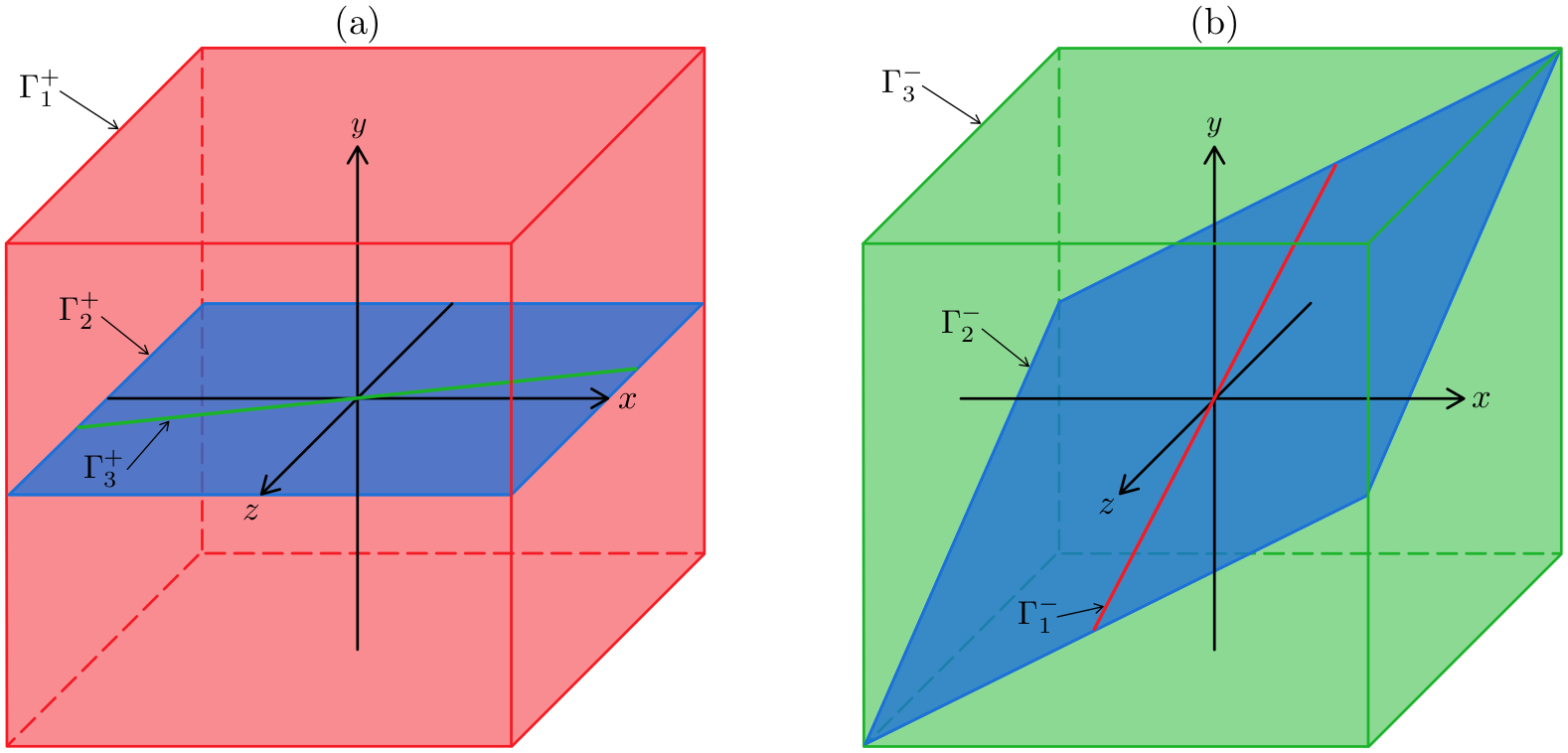}
	\caption{Example of the filtration subspaces $\Gamma_i^{\pm}$, $i=1,2,3$, at a state $\bs x$ for a dynamical system with tangent space $T_{\bs x}\mathcal{M}=\mathbb{R}^3$, where in (a) the subspaces $\Gamma_1^{+}\supset\Gamma_2^{+}\supset\Gamma_3^{+}$ are drawn on a single set of $x$-$y$-$z$ axes and in (b) $\Gamma_1^{-}\subset\Gamma_2^{-}\subset\Gamma_3^{-}$ are drawn on a separate set of axes. The subspaces $\Gamma_1^{\pm}$ are both drawn in red, $\Gamma_2^{\pm}$ in blue, and $\Gamma_3^{\pm}$ in green.}
	\label{fig:visualise}
\end{figure}

For the purpose of visualising the filtration subspaces $\Gamma_i^{\pm}$ and the splitting subspaces $\Omega_i$ in the tangent space at $\bs x$, consider a dynamical system whose state space and tangent space is $\mathbb{R}^3$. For simplicity, we assume that the orbit of $\bs x$ has three distinct LEs, $\lambda_1>\lambda_2>\lambda_3$. In \figureref{fig:visualise}(a), we have drawn the subspaces $\Gamma_1^{+}\supset\Gamma_2^{+}\supset\Gamma_3^{+}$ on a single set of $x$-$y$-$z$ axes, where each subspace has been drawn in a distinct colour. We can use \eqref{eq:le_filtration} to interpret this figure as follows: any deviation vector $\bs w$ in the green line denoting $\Gamma_3^{+}$ will grow at an exponential rate of $\lambda_3$ as $t\to\infty$; if $\bs w$ is in the blue plane denoting $\Gamma_2^{+}$, but not in the green line, then it grows according to $\lambda_2$; and if $\bs w$ is in the red volume denoting $\Gamma_1^{+}$, but not in the blue plane, then its asymptotic growth is governed by $\lambda_1$. Since the dimension of the red volume is greater than that of the blue plane and that of the green line, it is clear from \figureref{fig:visualise}(a) that almost all deviation vectors lie in the red volume, not the blue plane, and hence grow according to $\lambda_1$ as $t\to\infty$. Analogously, we have drawn the subspaces $\Gamma_1^{-}\subset\Gamma_2^{-}\subset\Gamma_3^{-}$ in \figureref{fig:visualise}(b) using the same colours as the corresponding subspaces $\Gamma_i^{+}$ of the same index $i=1,2,3$ in \figureref{fig:visualise}(a). We interpret \figureref{fig:visualise}(b) as follows: any deviation vector $\bs w$ in the red line denoting $\Gamma_1^{-}$ decays at an exponential rate of $\lambda_1$ as $t\to-\infty$; if $\bs w$ is in the blue plane denoting $\Gamma_2^{-}$, but not in the red line, then it decays according to $\lambda_2$; and if $\bs w$ is in the green volume denoting $\Gamma_3^{-}$, but not in the blue plane, then its asymptotic exponential decay rate is $\lambda_3$. Finally, the splitting subspaces $\Omega_i$, $i=1,2,3$, are acquired by simply intersecting the splitting subspaces according to \eqref{eq:omegaintersect}. In particular: $\Omega_1$ is the intersection between the red volume and the red line, which is simply $\Gamma_1^-$ in this case; $\Omega_2$ is the intersection between the blue planes, which is a 1-D line; and $\Omega_3$ is the intersection between the green line and green volume, which is $\Gamma_3^+$. As expected from \eqref{eq:le_splitting}, any deviation vector in $\Omega_i$ grows at an exponential rate of $\lambda_i$ as $t\to\infty$ and decays at an exponential rate of $\lambda_i$ as $t\to-\infty$.

Returning to our theoretical discussion: the tangent space decomposition \eqref{eq:splitting} can be coarsened into the decomposition
\begin{align}
	T_{\bs x}\mathcal M=E^+(\bs x)\oplus E^0(\bs x)\oplus E^-(\bs x),\label{eq:coarse}
\end{align}
where $E^+(\bs x)$, $E^0(\bs x)$, and $E^-(\bs x)$ are called the \textit{unstable}, \textit{centre}, and \textit{stable} subspaces, respectively \cite{Oseledets1968}. The unstable subspace $E^+(\bs x)$ is the direct sum of all those splitting subspaces $\Omega_i$ with positive corresponding LEs $\lambda_i$. Likewise, the centre and stable subspaces are each the direct sum of the splitting subspaces with zero or negative LEs, respectively \cite{GinelliEtAl2013}. For a particular state, these subspaces indicate directions with different stability properties, where these properties are determined by the signs of the corresponding LEs. In fact, the unstable, centre, and stable subspaces are the local linearisations of the so-called unstable, centre, and stable manifolds (see e.g.\ \cite[pp.~14,124]{GuckenheimerHolmes1983}).

In our forthcoming numerical investigations of \chapterref{ch:hh}, it will be convenient to speak in terms of both vectors and subspaces, so it becomes prudent for us to expand the spectrum of $m\leq N$ distinct LEs $\lambda_i([\bs x])$ into a spectrum of $N$ possibly non-distinct LEs in which each LE is repeated according to its multiplicity $\kappa_i([\bs x])$. As an alternative to \eqref{eq:le}, we therefore denote and index the LEs by
\begin{align}
	\chi_1([\bs x])\geq\chi_2([\bs x])\geq\cdots\geq\chi_N([\bs x]),\label{eq:le_non}
\end{align}
where each $\lambda_i([\bs x])$ from \eqref{eq:le} appears precisely $\kappa_i([\bs x])$ times in \eqref{eq:le_non}.

The CLVs are defined as unit vectors which span the splitting \eqref{eq:splitting}. Specifically, if $\chi_i([\bs x])=\lambda_j([\bs x])$ then the $i$-th CLV, denoted by $\hat{\bs\omega}_{i}(\bs x)$, is a unit vector in $\Omega_j(\bs x)$. Each CLV is covariant in the sense that its direction evolves with the tangent dynamics, i.e.
\begin{align}
	M(\bs x,t)\hat{\bs\omega}_{i}(\bs x)=c_i\hat{\bs\omega}_{i}(\bs x(t))\label{eq:clvcov}
\end{align}
for some scalars $c_i$, where typically $c_i\ne1$ since CLVs are defined to be unit vectors. If $\lambda_j([\bs x])$ is non-degenerate (i.e.\ if $\kappa_j([\bs x])=1$), then the corresponding $\hat{\bs\omega}_{i}(\bs x)$ is unique up to a sign since $\dim\Omega_j(\bs x)=1$, otherwise the CLVs $\hat{\bs\omega}_{i}(\bs x)$ which correspond to the same degenerate $\lambda_j([\bs x])$ can be defined as any set of linearly independent covariant unit vectors in $\Omega_j(\bs x)$. It follows from \eqref{eq:splitting} that any choice of $N$ CLVs forms a basis for the tangent space. While \eqref{eq:le_splitting} is stated in terms of $\lambda_i([\bs x])$ and $\Omega_i(\bs x)$, it clearly implies the following in terms of $\chi_i([\bs x])$ and $\hat{\bs\omega}_i(\bs x)$:
\begin{align}
	\lim_{t\to\pm\infty}\frac{1}{|t|}\ln\frac{\|M(\bs x,t)\bs w\|}{\|\bs w\|}=\pm\chi_i([\bs x])\quad\text{if}\quad\bs w\in\spn(\hat{\bs\omega}_i(\bs x))\setminus\{0\}.\label{eq:le_clvs}
\end{align}

The \textit{backward Lyapunov vectors} (BLVs) are an orthonormal set of $N$ eigenvectors of $\Lambda^{-}(\bs x)$. These eigenvectors correspond to the eigenspaces $V_i^-(\bs x)$ of $\Lambda^{-}(\bs x)$ and so can be used to construct the filtration subspaces $\Gamma_i^-(\bs x)$ \eqref{eq:filtration}. Analogously, the \textit{forward Lyapunov vectors} (FLVs) are an orthonormal set of $N$ eigenvectors of $\Lambda^{+}(\bs x)$. BLVs and FLVs are uniquely defined up to a sign if their corresponding LEs are non-degenerate. Note that BLVs and FLVs are not covariant in general, unlike CLVs.

We conclude our theoretical discussion by providing a geometric interpretation of LEs, BLVs, and FLVs in terms of SVD (see \sectionref{sec:svd}). Consider a finite-time version of the matrices $\Lambda^{\pm}$ from \eqref{eq:matrix},
\begin{align}
	\frac{1}{t}\ln\sqrt{M(\bs x,t)\transpose M(\bs x,t)}.\label{eq:matrix_finite}
\end{align}
It follows from our earlier discussions on SVD and matrix functions that the eigenvectors of \eqref{eq:matrix_finite} are the right singular vectors of $M(\bs x,t)$. Furthermore, the singular values of $M(\bs x,t)$ are positive and hence can be written in the form $e^{\lambda_i(t)t}$ for some time-dependent functions $\lambda_i(t)$, which are easily shown to be the eigenvalues of \eqref{eq:matrix_finite}. In the limit as $t\to\infty$, the eigenvalues and eigenvectors of \eqref{eq:matrix_finite} tend to the LEs and FLVs \cite{GoldhirschEtAl1987}, while these eigenvectors in the $t\to-\infty$ limit yield the BLVs. Referring back to the example in $\mathbb{R}^2$ given in \figureref{fig:ellipsoid} and considering $M(\bs x,t)$ as the matrix $A$ in that picture, the FLVs are the limiting case of the vectors $\hat{\bs v}_i$ pointing along the principal semi-axes of the unit circle as $t\to\infty$ in $M(\bs x,t)$, while the lengths $\sigma_i$ of the principal semi-axes of the ellipse can be expressed as $e^{\lambda_i(t)t}$, where the $i$-th LE is the limit of $\lambda_i(t)$ as $t\to\infty$.

\subsection{Asymptotic behaviour of deviation vectors}\label{sec:asymp}
We saw from \eqref{eq:le_filtration} that the asymptotic growth rate of the magnitude of any deviation vector is characterised by one of the LEs, but can we say something about the asymptotic behaviour of a deviation vector's direction? Indeed we can, since any deviation vector $\bs w(0)$ can be expressed as a linear combination of CLVs $\hat{\bs\omega}_i(\bs x(0))$. Now, to simplify the notation, we will drop all $\bs x$ dependence of the CLVs and instead write them as explicit functions of time, so $\hat{\bs\omega}_i(\bs x(0))$ becomes $\hat{\bs\omega}_i(0)$. Therefore,
\begin{align}
	\bs w(0) = \sum_{i=1}^N a_i \hat{\bs\omega}_i(0),
\end{align}
where $a_i$ are some scalars. Evolving this deviation vector to $\bs w(t) = M(\bs x(0),t)\bs w(0)$, it follows from the linearity of the tangent dynamics and \eqref{eq:clvcov} that
\begin{align}
	\bs w(t) = \sum_{i=1}^N a_i M(\bs x(0),t)\hat{\bs\omega}_i(0)
	= \sum_{i=1}^N a_i c_i\hat{\bs\omega}_i(t)\label{eq:exp_grow1}
\end{align}
for some scalars $c_i$. Furthermore, we know from \eqref{eq:le_clvs} that (asymptotically speaking) a deviation vector in the direction of the $i$-th CLV grows at an exponential rate equal to the corresponding LE $\chi_i$, so \eqref{eq:exp_grow1} becomes
\begin{align}
\bs w(t) \approx \sum_{i=1}^N a_i e^{\chi_i t}\hat{\bs\omega}_i(t).\label{eq:exp_grow2}
\end{align}
Assuming non-degeneracy, if $a_1\ne0$ then clearly the $\hat{\bs\omega}_1(t)$ component in \eqref{eq:exp_grow2} dominates as $t\to\infty$ and $\bs w(t)$ converges to $\hat{\bs\omega}_1(t)$, by which we mean $\Delta(\bs w(t),\hat{\bs\omega}_1(t))\to0$ as $t\to\infty$.\footnote{Note that when discussing the convergence of two vectors, we only consider their directions, not their magnitudes.} If instead $a_1=0$, then $\bs w(t)$ will align with $\hat{\bs\omega}_2(t)$ or with the next CLV with a non-zero component in \eqref{eq:exp_grow2}, but this is an exceptional case. Finally, what about the case where the first LE is degenerate? If $\chi_1=\chi_2$, then all we can say is that, as $t\to\infty$, a generic $\bs w(t)$ aligns with the splitting subspace $\Omega_1(\bs x(t))$, which is spanned by the first $\kappa_1=\dim\Omega_1$ CLVs.

A note of caution: CLVs are not constant, fixed vectors in time; they are covariant with the tangent dynamics. The convergence of $\bs w(t)$ to a CLV should not be confused with the false notion that $\bs w(t)$ is converging to some fixed direction. It is only \textit{relative} to the CLVs that we have determined the asymptotic directions of arbitrary deviation vectors in this discussion.

Now that we have addressed the fate of individual deviation vectors, we move on to collections of deviation vectors and consider how their spans evolve. For simplicity, we again assume the LEs are non-degenerate. Our previous conclusion that a generic deviation vector converges to the first CLV can be generalised: loosely speaking, a generic subspace of dimension $d$ converges to the span of the first $d$ CLVs. To be precise, if $G_i(0)$ is a subspace of the tangent space at $\bs x(0)$ and has dimension $d=\dim\Gamma_i^-$, then generically $G_i(t)\to \Gamma_i^-(t)$ as $t\to\infty$ \cite{KuptsovParlitz2012}, by which we mean $\Delta(G_i(t),\Gamma_i^-(t))\to0$ as $t\to\infty$.\footnote{Recall from \eqref{eq:noethen} that $\Gamma_i^-$ is simply the span of the first $d$ CLVs.} Note that we have written $\Gamma_i^-$ here as an explicit function of $t$ instead of $\bs x(t)$ to simplify the notation. A related result by Ershov and Potapov \cite{ErshovPotapov1998} implies that a generic orthonormal set of $N$ deviation vectors, when evolved forwards in time and periodically orthonormalised, converge to the BLVs, assuming those BLVs correspond to non-degenerate LEs and are thus uniquely defined (up to a sign) \cite{KuptsovParlitz2012}.

We can also say something about the speed at which arbitrary subspaces converge to the relevant filtration subspaces. It was shown by Noethen in \cite{Noethen2019} that the exponential rate of convergence of $G_i(t)$ to $\Gamma_i^-(t)$ is given by $\lambda_i-\lambda_{i+1}$ as $t\to\infty$. Note that if we choose an $N$-D subspace $G_m(0)$, where $m$ is the number of distinct LEs in \eqref{eq:le}, then no convergence time is required since the subspace is already equal to the entire tangent space, i.e.\ $G_m(0)=\Gamma_m^-(0)$. Nevertheless, this can be made consistent with the exponential convergence rate of $\lambda_m-\lambda_{m+1}$ by judiciously defining $\lambda_{m+1}=-\infty$, which implies that the convergence is instantaneous. We thus use the convention of \cite{Noethen2019} by defining $\lambda_{0}=\infty$ and $\lambda_{m+1}=-\infty$ to conveniently take care of the edge cases in the spectral gaps. A related convergence result discussed by Legras and Vautard \cite{LegrasVautard1996} holds for BLVs: if a set of generic orthonormal deviation vectors are evolved forwards in time and periodically orthonormalised, the $i$-th vector in such a set converges to the $i$-th BLV at an exponential rate given by $\min(\chi_{i-1}-\chi_i,\chi_i-\chi_{i+1})$. Analogous to the convention used for the distinct LE spectrum, we define $\chi_{0}=\infty$ and $\chi_{N+1}=-\infty$ to take care of the edge cases. As an example where this convention is useful, convergence of a generic deviation vector to the first BLV (which is also the first CLV) happens at an exponential rate of
\begin{align}
	\min(\chi_{0}-\chi_1,\chi_1-\chi_{2}) = \min(\infty,\chi_1-\chi_{2}) = \chi_1-\chi_{2}.
\end{align}

To conclude this discussion, let us finally consider the asymptotic behaviour of deviation vectors during backward evolution. We know from \eqref{eq:le_filtration} that the asymptotic growth rate of any deviation vector is characterised by the negative of some LE. Furthermore, we can see from \eqref{eq:exp_grow2} that as $t\to-\infty$, the component of the last CLV dominates the others (assuming non-degeneracy), hence generic deviation vectors converge to the direction of $\hat{\bs\omega}_N(t)$ as $t\to-\infty$. Similar to the forward evolution case, this can be generalised: if $G_i(0)$ is a generic subspace with dimension $d=\dim\Gamma_i^+$, then $G_i(t)\to \Gamma_i^+(t)$ as $t\to-\infty$ \cite{KuptsovParlitz2012}.

\subsection{Conservation, symmetry, and symplecticity}\label{sec:sym}
Certain features of the LE spectrum have been linked to particular properties of the underlying dynamical system. We summarise some of these properties and their implications in this section, but further details can be found in \cite[pp.~24--27]{PikovskyPoliti2016}.

A dynamical system is said to have a \textit{conserved quantity} $E(\bs x)$ if $\dot E=0$ \cite[p.~160]{Strogatz1994}. Each conserved quantity gives rise to a LE of zero. \textit{Continuous symmetries} are another possible source of zero-valued LEs. For a bounded orbit in an autonomous dynamical system that does not converge to a fixed point, it was shown by Haken that at least one LE vanishes \cite{Haken1983}; this zero LE is due to time translation symmetry since perturbations in the direction of the flow neither grow nor decay asymptotically. By ``direction of the flow'', we mean the direction of the velocity vector $\dot{\bs x}$ \eqref{eq:auto}, and we denote a unit vector in this direction by $\hat{\bs w}_{\mathrm{flow}}(t)$. Since $\hat{\bs w}_{\mathrm{flow}}(t)$ is a solution to the variational equations \cite{Haken1983}, it is a covariant vector. Furthermore, its growth is governed by a zero LE both forwards and backwards in time, so $\hat{\bs w}_{\mathrm{flow}}(t)$ is a CLV.

The final special class of dynamical systems that we discuss here are so-called \textit{symplectic systems}. Let $\Omega\in\mathbb{R}^{N\times N}$ be a block matrix of the form
\begin{align}
	\Omega=\begin{pmatrix}
		0&I\\-I&0
	\end{pmatrix},
\end{align}
where $N=2n$ is even, $0$ denotes the $n\times n$ zero matrix, and $I$ denotes the $n\times n$ identity matrix. A square matrix $M\in\mathbb{R}^{N\times N}$ is said to be \textit{symplectic} if it satisfies
\begin{align}
	M\Omega M\transpose=\Omega\label{eq:defsymplectic}
\end{align}
(see e.g.\ \cite[p.~26]{PikovskyPoliti2016}). A dynamical system is referred to as a symplectic system if its fundamental matrix is symplectic. It has been shown (e.g.\ in \cite[p.~27]{PikovskyPoliti2016}) that the LE spectrum for a symplectic system has the following symmetry:
\begin{align}
	\chi_i=-\chi_{N-i+1}.\label{eq:symplecticsym}
\end{align}

\section{Autonomous Hamiltonian systems}\label{sec:autoham}
In this thesis, we focus on a particular class of autonomous dynamical systems known as \textit{autonomous Hamiltonian systems}, which are frequently used to model physical systems (see e.g.\ \cites[Chapt.~2]{Shankar1994}[Chapt.~8]{GoldsteinEtAl2011}). An autonomous Hamiltonian system is specified by a Hamiltonian function,
\begin{align}
	H(q_1,q_2,\dots,q_n,p_1,p_2,\dots,p_n),
\end{align}
where $q_i$ and $p_i$ are generalised position and momentum coordinates (see e.g.\ \cite{Skokos2010}). The state vector $\bs x$ of an autonomous Hamiltonian system is expressed in terms of generalised coordinates as $\bs x = (q_1\,\ q_2\,\ \cdots\,\ q_n\,\ p_1\,\ p_2\,\ \cdots\,\ p_n)\transpose$. These coordinates evolve in time according to \textit{Hamilton's equations} (also called the \textit{equations of motion}),
\begin{align}
	\dot q_i=\frac{\partial H}{\partial p_i},\qquad\dot p_i=-\frac{\partial H}{\partial q_i}.\label{eq:eom}
\end{align}
The state space of a Hamiltonian system is called the \textit{phase space}. The phase space has even dimension $N=2n$, and we say the system has $n$ degrees of freedom. In later chapters, we will study several autonomous Hamiltonian systems with various numbers of degrees of freedom.

An autonomous Hamiltonian system is an example of a symplectic system with time translation symmetry, so its LE spectrum has the symmetry \eqref{eq:symplecticsym} and it has an LE equal to zero, as discussed in \sectionref{sec:sym}. This zero LE corresponds to a perturbation in the direction of the flow $\hat{\bs w}_{\mathrm{flow}}$. Since $N=2n$ is even, the spectrum symmetry implies that at least two LEs are equal to zero:
\begin{align}
	\chi_n=\chi_{n+1}=0.
\end{align}
The Hamiltonian function $H(\bs x)$ represents the total energy in the physical system and, since the system is autonomous, it is a conserved quantity: $\dot H=0$. This is the source of the other zero LE. Energy conservation further implies that all states with a particular energy evolve on an $(N-1)$-dimensional \textit{energy surface} within the phase space (see e.g.\ \cites[p.~20]{LichtenbergLieberman1992}[p.~494]{GoldsteinEtAl2011}).

\subsection{Chaos and chaos indicators}\label{sec:chaos}
An important concept in the field of nonlinear dynamics is that of \textit{chaos}. While a single definition of chaos has not been universally adopted, a commonly used one is the definition by Devaney \cite[p.~50]{Devaney1989}. Devaney defines a continuous map $f:V\to V$ to be chaotic if it is topologically transitive, its periodic points are dense in $V$, and $f$ has sensitive dependence on initial conditions. While this definition was originally given in the context of discrete dynamical systems, it has been applied to continuous dynamical systems as well \cite[p.~133]{Basener2006}.

In practice, sensitive dependence is often the only criterion from Devaney's definition used to classify a dynamical system as chaotic. For example, it follows from \eqref{eq:le_filtration} that the \textit{maximum Lyapunov exponent} (mLE) is given by
\begin{align}
	\chi_1=\lim_{t\to\infty}\frac{1}{t}\ln\frac{\|\bs w(t)\|}{\|\bs w(0)\|}\label{eq:infinite_mle}
\end{align}
for almost all initial deviation vectors $\bs w(0)$. The mLE is a measure of an orbit's sensitive dependence on initial conditions, so the mLE is frequently used as a \textit{chaos indicator}: if $\chi_1>0$ then the orbit is \textit{chaotic}, if $\chi_1=0$ then the orbit is \textit{regular}. Due to the $t\to\infty$ limit in \eqref{eq:infinite_mle}, the mLE is usually estimated numerically over a finite time interval. The so-called \textit{finite-time mLE} (ftmLE) is given by
\begin{align}
	X_1(t)=\frac{1}{t}\ln\frac{\|\bs w(t)\|}{\|\bs w(0)\|},\label{eq:finite_mle}
\end{align}
where $X_1(t)\to\chi_1$ as $t\to\infty$, so the saturation of $X_1(t)$ to a positive constant indicates chaotic motion and regular motion is indicated by $X_1(t)$ decaying to zero proportional to $t^{-1}$ (see e.g.\ \cite{Skokos2010}). For a detailed overview of various chaos detection techniques, we recommend the book \cite{SkokosEtAl2016}.

One downside to using the ftmLE $X_1(t)$ to detect chaos is that it is a time average of the exponential growth rate of $\bs w(t)$, which is easily seen from \eqref{eq:finite_mle}. As an average, this means that estimates of a positive mLE are often slow to saturate, thus requiring long integration times to distinguish between regular and chaotic orbits. With this in mind, several chaos indicators have been designed to detect chaos quickly, one of which is the \textit{smaller alignment index} (SALI), defined as
\begin{align}
	\text{SALI}(t)=\min\{\|\hat{\bs w}_1(t)+\hat{\bs w}_2(t)\|,\|\hat{\bs w}_1(t)-\hat{\bs w}_2(t)\|\},\label{eq:sali}
\end{align}
where $\hat{\bs w}_1(t),\hat{\bs w}_2(t)$ are generic linearly independent deviation vectors \cites{Skokos2001}{SkokosEtAl2003}{SkokosEtAl2004}{SkokosManos2016}. The SALI discriminates between regular and chaotic motion in autonomous Hamiltonian systems by measuring the alignment or anti-alignment between two randomly selected deviation vectors. When the SALI decays to zero, this indicates chaotic motion since an orbit with a non-degenerate positive mLE will have almost all deviation vectors converge to the first CLV $\hat{\bs\omega}_1$, hence two generic deviation vectors will either align or anti-align. In particular, $\text{SALI}(t)\propto\exp(-(\chi_1-\chi_2)t)$ for generic chaotic orbits, and this quick decay to zero makes the SALI a fast chaos indicator. On the other hand, if the orbit is regular (and assuming the system has at least two degrees of freedom) then the SALI remains approximately constant. As a final note: when $\hat{\bs w}_1$ and $\hat{\bs w}_2$ in \eqref{eq:sali} are nearly parallel or anti-parallel, the SALI is approximately the minimum angle $\theta_1$ between $\spn(\hat{\bs w}_1)$ and $\spn(\hat{\bs w}_2)$, hence $\text{SALI}\approx\Delta(\hat{\bs w}_1,\hat{\bs w}_2)$.

\subsection{Symplectic integration}\label{sec:symplecticsec}
Systems of ordinary differential equations can be solved numerically using a variety of numerical integration schemes, such as the well-known Runge-Kutta fourth-order integrator (see e.g.\ \cite[p.~288]{BurdenFaires2011}). However, numerical solutions typically have some associated error and diverge from the true solution over time. This is particularly evident in chaotic systems where slight changes in the computer's arithmetic (e.g.\ by using a different time step or even a different compiler) can eventually result in a drastically different outcome. This means that a precise prediction of the state of a chaotic system in the distant future is practically unobtainable.

A \textit{symplectic integrator} is a type of numerical integration scheme which can be used to solve autonomous Hamiltonian systems. While numerical errors cannot be escaped, symplectic integrators are designed to conserve the energy of the system by approximating the flow map by a sequence of symplectic maps\footnote{A symplectic map $\Phi:\mathcal{M}\to\mathcal{M}$ is a map whose Jacobian matrix is symplectic \eqref{eq:defsymplectic}.} which can then be iterated exactly. The energy conservation of these integrators ensures that numerically computed orbits are constrained to the (approximately) correct energy surfaces. For a comprehensive text on symplectic integrators, see e.g.\ \cite{HairerEtAl2006}. To keep track of changes to the total energy of the system, we compute the \textit{relative energy error}
\begin{align}
	E_r(t) = \left|\frac{H(\bs x(t)) - H(\bs x(0))}{H(\bs x(0))}\right|,\label{eq:energyerror}
\end{align}
where $H(\bs x(0))$ is the initial value of the Hamiltonian and $H(\bs x(t))$ is its computed value at time $t$ along the orbit. For autonomous Hamiltonian systems, $E_r$ should always be zero, but using many general-purpose numerical integrators (like Runge-Kutta fourth-order) will result in $E_r$ typically increasing over time. On the other hand, symplectic integrators bound $E_r$ from above for all time by some small positive constant that depends on the integration time step used.

We now give a brief overview of symplectic integrators as they apply to the dynamical systems studied in this thesis. The solution to the equations of motion of an autonomous Hamiltonian system can be written as
\begin{align}
	\bs x(t) = e^{tL_H}\bs x(0),\label{eq:expop}
\end{align}
where $L_H$ is a differential operator defined by
\begin{align}
	L_Hf=\sum_{i=1}^n\left(\frac{\partial f}{\partial q_i}\frac{\partial H}{\partial p_i}-\frac{\partial f}{\partial p_i}\frac{\partial H}{\partial q_i}\right)\label{eq:poisson}
\end{align}
(see e.g.\ \cite{LaskarRobutel2001}), which is the so-called Poisson bracket between $f$ and $H$. The Hamiltonians for the systems we study all have the form
\begin{align}
	H(\bs q,\bs p)=A(\bs p)+B(\bs q),\label{eq:hsep}
\end{align}
where $A(\bs p)$ typically represents the kinetic energy and $B(\bs q)$ the potential energy. Similarly to \eqref{eq:poisson}, differential operators $L_A$ and $L_B$ can be defined for $A(\bs p)$ and $B(\bs q)$, respectively, and it follows from \eqref{eq:hsep} that $L_H=L_A+L_B$. The exponential operator in $\eqref{eq:expop}$ can be approximated over a small time step $h$ by the product
\begin{align}
	e^{h L_H}=e^{h (L_A+L_B)}=\prod_{i=1}^k e^{a_ih L_A}e^{b_ih L_B}+O(h^{r+1})\label{eq:prodapprox}
\end{align}
for some constants $a_i,b_i\in\mathbb{R}$ and $k\in\mathbb{N}$, where $r$ is called the \textit{order} of the approximation \cite{Yoshida1990}. Now, given some state $\bs x=\smash{\big(\begin{smallmatrix}\raisebox{0.8pt}{$\scriptstyle\bs q$}\\\raisebox{-0.8pt}{$\scriptstyle\bs p$}\end{smallmatrix}\big)}$, the operators $e^{h L_A}$ and $e^{h L_B}$ evolve this state by time $h$ to the state $\bs x'=\smash{\big(\begin{smallmatrix}\raisebox{-1.5pt}{$\scriptstyle\bs q'$}\\\raisebox{1.5pt}{$\scriptstyle\bs p'$}\end{smallmatrix}\big)}$ according to the following respective maps:
\begin{align}
\begin{split}
	\bs q'&=\bs q + h\frac{\partial A}{\partial\bs p},\\
	\bs p'&=\bs p - h\frac{\partial B}{\partial\bs q},
\end{split}
\end{align}
where $A$ and $B$ come from the separable Hamiltonian \eqref{eq:hsep}. A particular approximation of the form of \eqref{eq:prodapprox} that we use for our computations is the fourth-order symplectic integrator ABA864 introduced by Blanes et al.\ \cite{BlanesEtAl2013}, which is defined as the sequence of maps
\begin{align}
\begin{split}
	ABA864(h) &= e^{a_1 h L_{A}} e^{b_1 h L_{B}} e^{a_2 h L_{A}} e^{b_2 h L_{B}} e^{a_3 h L_{A}} e^{b_3 h L_{B}} e^{a_4 h L_{A}}e^{b_4 h L_{B}}\\ &\qquad\times e^{a_4 h L_{A}} e^{b_3 h L_{B}} e^{a_3 h L_{A}} e^{b_2 h L_{B}} e^{a_2 h L_{A}} e^{b_1 h L_{B}} e^{a_1 h L_{A}},\label{eq:aba}
\end{split}
\end{align}
where the constants $a_i$ and $b_i$ are given in Table~3 of \cite{BlanesEtAl2013}.

Having addressed how symplectic integrators can be used for solving the equations of motion, we conclude with a remark on numerically solving the variational equations. Using the so-called \textit{tangent map method}, the equations of motion and variational equations can be combined into a single system of differential equations and thus solved simultaneously by means of a symplectic integrator while still preserving the energy \cite{SkokosGerlach2010}. For all numerical integration in this thesis, we use this method applied to the ABA864 integrator to solve both the equations of motion and variational equations.

\section{Computation of LEs}\label{sec:le}
We saw from \eqref{eq:finite_mle} that the mLE can be estimated by measuring the exponential rate of stretching of a generic deviation vector over a long time interval. In order to achieve a different growth rate, an initial deviation vector would need to be chosen from one of the lower-dimensional filtration subspaces, but the probability of choosing such an initial condition at random is zero. Even if we knew a deviation vector whose growth would theoretically be governed by a non-maximal LE, errors in the numerical integration would likely result in the deviation vector eventually converging to $\hat{\bs\omega}_1$ and thus still growing according to $\chi_1$ regardless \cite{BenettinEtAl1980a}.

Solutions to these problems were proposed by Shimada and Nagashima \cite{ShimadaNagashima1979} and Benettin et al.\ \cites{BenettinEtAl1980}{BenettinEtAl1980a} who looked at the growth rates of (hyper-)volumes in the tangent space. While the exponential growth rate as $t\to\infty$ of almost all deviation vectors is $\chi_1$, for generic areas it is $\chi_1+\chi_2$, for volumes it is $\chi_1+\chi_2+\chi_3$, and so on. Since the growth of these volumes is related to more LEs than just the first, this leads to a method for computing the full spectrum of LEs known as the \textit{standard method}. The standard method is essentially a procedure of evolving a set of orthonormal deviation vectors forwards in time, periodically orthonormalising the vectors to prevent their exponential growth and alignment (which would lead to numerical issues such as overflow), and tracking the growth of volumes spanned by these deviation vectors at each orthonormalisation time. The factors by which these volumes grow are unaffected by the periodic orthonormalisation, and the standard method uses this fact to determine the growth rates in particular directions and thus estimate the spectrum of LEs. In the remainder of this section, we describe the algorithm that is the standard method in terms of QR decomposition, but we refer the interested reader to \cite{Skokos2010} for a more in-depth treatment of the subject.

Let $Q(0)\in\mathbb{R}^{N\times N}$ be a matrix whose columns are a set of orthonormal deviation vectors. Now evolve the columns of $Q(0)$ forwards by $\tau$ time units by numerically solving the variational equations \eqref{eq:jacob} and denote the resulting matrix by $\tilde Q(\tau)$; in other words, compute $\tilde Q(\tau)=M(0,\tau)Q(0)$, where $M(0,\tau)$ is the fundamental matrix \eqref{eq:fundamental} written as an explicit function of the start and end times $0$ and $\tau$. Next, perform QR decomposition on $\tilde Q(\tau)$ as follows: $\tilde Q(\tau)=Q(\tau)R(\tau)$. Repeat this process by evolving the columns of $Q(\tau)$ to yield $\tilde Q(2\tau)$, upon which QR decomposition is performed to yield $Q(2\tau)$ and $R(2\tau)$. By repeating this procedure some number of times $k$, the matrix $R(i\tau)$ is computed for each $i=1,2,\dots,k$, the diagonal elements $R_{jj}(i\tau)$ of which contain the growth rates of deviation vectors in particular directions and are used to define the so-called \textit{finite-time LEs} (ftLEs). The $j$-th ftLE at time $t=k\tau$, $k\in\mathbb{N}$, is given by
\begin{align}
	X_j(k\tau) = \frac{1}{k\tau}\sum_{i=1}^k\ln R_{jj}(i\tau),\label{eq:rii2}
\end{align}
which approximates the $j$-th LE $\chi_j$ \cite{Skokos2010}, i.e.
\begin{align}
	\chi_j = \lim_{t\to\infty}X_j(t).
\end{align}

The standard method estimates the LEs by computing the ftLEs using \eqref{eq:rii2}. This method is easily adapted to computing only the first $p$ LEs, $1\le p\le N$, instead of all $N$ LEs by simply evolving a set of $p$ deviation vectors instead of $N$, i.e.\ initialise $Q(0)$ as a $p\times N$ (instead of $N\times N$) real matrix. We present pseudocode for computing the first $p$ LEs via the standard method in \algorithmref{alg:le}, which is adapted from Table~3 of \cite{Skokos2010}. The results saved during each iteration of Step~(c) of this algorithm give the time evolution of the first $p$ ftLEs. When an ftLE saturates to a constant value, this value is used as an estimate for the underlying LE. Goldhirsch et al.\ \cite{GoldhirschEtAl1987} showed that an ftLE $X_i(t)$ is related to its respective LE $\chi_i$ by
\begin{align}
	X_i(t) = \chi_i + \frac{b_i+\xi_i(t)}{t},\label{eq:gold}
\end{align}
where $b_i$ is a constant and $\xi_i(t)$ is a noise term with zero mean. If $\chi_i=0$, then this result implies that $X_i(t)$ goes to zero like $t^{-1}$ with some noise.

\begin{table}[htbp]
	\setlength{\tabcolsep}{0pt}
	\small
	\refstepcounter{table}
	\label{alg:le}
	\begin{tabular}{>{\raggedright}p{\dimexpr 0.16\linewidth-2\tabcolsep}p{\dimexpr 0.84\linewidth-2\tabcolsep}}
		\toprule
		\multicolumn{2}{p{\linewidth}}{\textbf{Algorithm \thechapter.1:} The standard method for computing the first $p$ LEs, $1\le p\le N$.}\\
		\midrule
		& \\[-0.75em]
		\textit{Inputs:} &
		\begin{minipage}[t]{\linewidth}\renewcommand{\labelitemi}{\labelitemii}
			\begin{itemize}[leftmargin=1.4em]\raggedright
				\itemsep0em
				\item Initial condition $\bs x(0)$.
				\item Matrix $Q(0)\in\mathbb{R}^{p\times N}$ with orthonormal columns.
				\item Time interval $\tau$ between orthonormalisations.
				\item Total integration time $T$ which is a multiple of $\tau$.
				\item Counter $k=1$.
			\end{itemize}
		\end{minipage}\vspace{0.6em}\\
		\midrule
		\vspace{-0.35em}\textit{Steps:} &
		\vspace{-0.35em}\begin{minipage}[t]{\linewidth}
			\begin{itemize}[leftmargin=1.4em]\raggedright
				\item[] \hspace{-1.5em} While $k\tau\le T$ do: 
				\begin{itemize}[leftmargin=1.8em,topsep=0pt]\raggedright
					\itemsep0em
					\item[(a)] Evolve $\bs x((k-1)\tau)$ and $Q((k-1)\tau)$ forwards by $\tau$ time units to get $\bs x(k\tau)$ and $\tilde Q(k\tau)$, respectively.
					\item[(b)] Perform the QR decomposition $\tilde Q(k\tau)=Q(k\tau)R(k\tau)$.
					\item[(c)] Compute and save $X_j(k\tau) = \smash{\frac{1}{k\tau}\sum_{i=1}^k\ln R_{jj}(i\tau)}$ for each $j=1,2,\dots,p$.
					\item[(d)] Increment $k$ by $1$.
				\end{itemize}
			\end{itemize}
		\end{minipage}\vspace{0.6em}\\
		\midrule
		\vspace{-0.35em}\textit{Output:} &
		\vspace{-0.35em}\begin{minipage}[t]{\linewidth}
			The time evolution of each $X_j(t)$, which estimates the $j$-th LE ($j=1,2,\dots,p$) after an integration time of $t$, where $t=\tau,2\tau,\dots,T$.
		\end{minipage}\vspace{0.5em}\\
		\bottomrule
	\end{tabular}
\end{table}

\section{Computation of CLVs}\label{sec:ginelli}
Having discussed the issue of computing LEs in the previous section, we now address the computation of CLVs, which brings a new set of challenges. Like $\chi_1$, computing an estimate of $\hat{\bs\omega}_1$ is relatively simple: evolve a random deviation vector forwards in time, periodically normalising to avoid overflow, and it will converge to $\hat{\bs\omega}_1(t)$ as $t\to\infty$. The last CLV can be similarly computed by evolving a random deviation backwards in time, but the rest of the CLVs are not so easily found. In this section, we present the G\&C algorithm for computing all CLVs as described in \cite{GinelliEtAl2013}.

\begin{figure}[htbp]
	\centering
	\includegraphics[width=\linewidth]{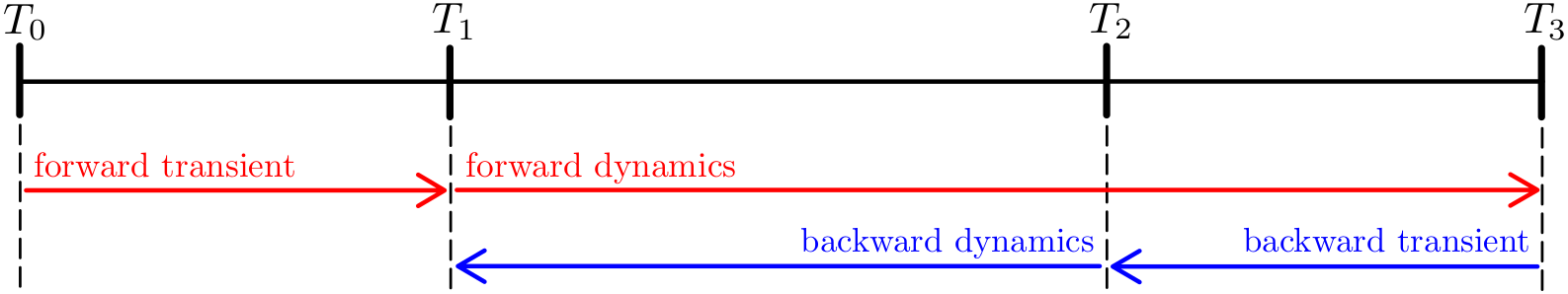}
	\caption{Diagram depicting the four stages of the G\&C algorithm for computing CLVs.}
	\label{fig:alg}
\end{figure}

The initial stages of the G\&C algorithm are very similar to the standard method described in \sectionref{sec:le}: simply take a set of random orthonormal deviation vectors and evolve them each forwards in time, periodically orthonormalising the vectors to prevent numerical issues. As discussed in \sectionref{sec:asymp}, such a set of periodically orthonormalised deviation vectors, which we call the \textit{Gram-Schmidt vectors} (GSVs), will converge to the BLVs.\footnote{It may seem counter-intuitive that the GSVs converge to the BLVs instead of the FLVs when evolving forwards in time, but we note that the BLVs at $\bs x(t)$ are defined from the tangent dynamics along the portion of the orbit $[\bs x(t)]$ in the \textit{past} of time $t$, whereas the FLVs require knowledge of the future. Heuristically, it makes sense that the GSVs would converge to the BLVs instead of the FLVs since the GSVs at time $t$ are computed by evolving some vectors from the \textit{past} of $t$.} With the BLVs acquired over some interval of time, the G\&C algorithm then requires the evolution of deviation vectors backwards in time. Assuming non-degeneracy for now, \eqref{eq:noethen} implies that the first two BLVs span the same subspace as the first two CLVs. By selecting a deviation vector that is a linear combination of the first two BLVs, evolving it backwards in time will result in it converging to the second CLV, not the first! This is due to the fact that the LEs negate when time is reversed, as seen from \eqref{eq:le_clvs}, making $\hat{\bs\omega}_2$ the fastest growing direction in $\spn(\hat{\bs\omega}_1,\hat{\bs\omega}_2)$ under backward evolution. Similarly, $\hat{\bs\omega}_k$ is acquired by taking an arbitrary deviation vector in the subspace spanned by the first $k$ BLVs and evolving it backwards in time. At a high level, this algorithm is summarised by the diagram in \figureref{fig:alg}, which separates the algorithm into four distinct stages whose start and end times are denoted by $T_0,T_1,T_2,T_3$. The first stage, starting at time $T_0$ and ending at time $T_1$, is known as the \textit{forward transient} during which the GSVs converge to the BLVs. The next stage from $T_1$ to $T_3$ is called the \textit{forward dynamics} during which the time evolution of each BLV is computed. The \textit{backward transient} is the interval from $T_3$ to $T_2$ where a random deviation vector in the span of the first $k$ BLVs converges to the $k$-th CLV $\hat{\bs\omega}_k$. Finally, the time evolution of the CLVs is computed over the interval from $T_2$ to $T_1$ during the \textit{backward dynamics}.

This high-level description of the G\&C algorithm might sound straightforward, but a naive attempt at implementing it will reveal a subtle issue: when expressing a deviation vector as a linear combination of BLVs in the coordinate basis of the phase space and evolving that vector backwards in time, small non-zero components of unwanted BLVs will likely be introduced by the numerics, which will increase exponentially (if the corresponding LE is positive) and likely result in the deviation vector converging to the wrong CLV.\footnote{During backward evolution, the LE of $\hat{\bs\omega}_N$ negates and becomes the largest LE, so deviation vectors will typically converge to $\hat{\bs\omega}_N$ as $t\to-\infty$.} Fortunately, the G\&C algorithm provides a neat solution to this problem: perform the backward evolution in a basis formed by the GSVs. By using the GSVs (which we assume have converged to the BLVs) as a basis, we can ensure that unwanted GSV components are exactly zero and remain so under backward evolution, and this is achieved by using QR decomposition as follows. Let $Q(0)\in\mathbb{R}^{N\times N}$ be a matrix with orthonormal columns and evolve this matrix forwards by $\tau$ time units to $\tilde Q(\tau)$, where we have set $T_0=0$ for convenience. Performing QR decomposition on this matrix, $\tilde Q(\tau)=Q(\tau)R(\tau)$, we acquire the matrices $Q(\tau)$ and $R(\tau)$. Note that thus far, this is the same notation and procedure we used for the standard method in \sectionref{sec:le}. Now, $\tilde Q(\tau)=M(0,\tau)Q(0)$, where we have written the fundamental matrix $M$ as an explicit function of times $0$ and $\tau$. It follows that $Q(\tau)R(\tau)=M(0,\tau)Q(0)$, which can be rearranged to yield
\begin{align}
	R(\tau)=Q\transpose(\tau)M(0,\tau)Q(0),\label{eq:rep}
\end{align}
where we have used the fact that square $Q$ matrices are orthogonal, so $Q^{-1}=Q\transpose$. While we already know that, by definition, $M(0,\tau)$ is a matrix representation of the linear propagator $\mathrm{d}_{\bs x(0)}\Phi^{\tau}$, we see from equation \eqref{eq:rep} that $R(\tau)$ is also a representation of $\mathrm{d}_{\bs x(0)}\Phi^{\tau}$. In particular, $M$ is a representation in the coordinate basis, while $R$ is a representation in the GSV basis, where $Q\transpose(\tau)$ and $Q(0)$ perform the necessary coordinate transformations in the tangent spaces $T_{\bs x(\tau)}\mathcal M$ and $T_{\bs x(0)}\mathcal M$, respectively. Therefore, if $\bs c(0)$ is a vector expressed in the basis of GSVs, then its evolution forwards in time by $\tau$ time units is determined by
\begin{align}
	\bs c(\tau)=R(\tau)\bs c(0).
\end{align}
In general, if we wish to evolve a vector $\bs c(k\tau)$, expressed in the GSV basis, backwards by $\tau$ time units then we simply compute
\begin{align}
	\bs c((k-1)\tau)=R^{-1}(k\tau)\bs c(k\tau),
\end{align}
where $R(k\tau)$ is the $R$ matrix from the QR decomposition of $\tilde Q(k\tau)$ acquired during the forward evolution part of the algorithm, so $R^{-1}(k\tau)$ represents the inverse linear propagator $\mathrm{d}_{\bs x(k\tau)}\Phi^{-\tau}:T_{\bs x(k\tau)}\to T_{\bs x((k-1)\tau)}$ over that time interval.

The final practicality we need to discuss before presenting the G\&C algorithm in full is that of initialising deviation vectors for the backward evolution part of the algorithm. If we wish to compute the $j$-th CLV ($j=1,2,\dots,N$), we need simply initialise a deviation vector that is a linear combination of only the first $j$ GSVs and evolve it backwards in time. Doing so in the GSV basis, we denote the initial deviation vector by $\bs c_j(T_3)$, where $T_3$ is the ``initial time'' for backward evolution seen in \figureref{fig:alg}. Evolving $\bs c_j(T_3)$ backwards in time to $\bs c_j(T_2)$, we assume the interval $T_3-T_2$ is long enough such that $\bs c_j(T_2)$ has converged to the $j$-th CLV $\hat{\bs \omega}_j(T_2)$. Since the magnitude of each $\bs c_j(T_2)$ may grow/decay exponentially, we periodically normalise it to avoid issues of overflow/underflow. In fact, we can compute all $N$ CLVs simultaneously by initialising an upper triangular matrix $C(T_3)=(\hat{\bs c}_1(T_3)\ \cdots\ \hat{\bs c}_N(T_3))$, whose columns are the initialised unit deviation vectors $\hat{\bs c}_j(T_3)$ in the GSV basis, then evolving it backwards in time and periodically normalising the columns. We use $\tilde C$ to denote the matrix $C$ before the columns are normalised, in the same spirit as $Q$ and $\tilde Q$. The method we have described for computing all $N$ CLVs can be adapted to compute only the first $p$ CLVs ($1\le p\le N$) by simply initialising $Q(0)\in\mathbb{R}^{p\times N}$ and $C(T_3)\in\mathbb{R}^{p\times p}$, which amounts to only using $p$ deviation vectors during all stages of the algorithm. We present pseudocode for the G\&C algorithm in \algorithmref{alg:clv}. Note that Steps~1--4 in \algorithmref{alg:clv} represent the respective stages of the G\&C algorithm seen in \figureref{fig:alg}: the forward transient, forward dynamics, backward transient, and backward dynamics.

\begin{table}[htbp]
	\setlength{\tabcolsep}{0pt}
	\small
	\refstepcounter{table}
	\label{alg:clv}
	\begin{tabular}{>{\raggedright}p{\dimexpr 0.16\linewidth-2\tabcolsep}p{\dimexpr 0.84\linewidth-2\tabcolsep}}
		\toprule
		\multicolumn{2}{p{\linewidth}}{\textbf{Algorithm \thechapter.2:} The G\&C algorithm for computing the first $p$ CLVs, $1\le p\le N$.}\\
		\midrule
		& \\[-0.75em]
		\textit{Inputs:} &
		\begin{minipage}[t]{\linewidth}\renewcommand{\labelitemi}{\labelitemii}
			\begin{itemize}[leftmargin=1.4em]\raggedright
				\itemsep0em
				\item Initial condition $\bs x(0)$.
				\item Matrix $Q(0)\in\mathbb{R}^{p\times N}$ with orthonormal columns.
				\item Time interval $\tau$ between orthonormalisations.
				\item Times $T_1<T_2<T_3$ which are each multiples of $\tau$.
				\item Upper triangular matrix $C(T_3)\in\mathbb{R}^{p\times p}$ with independent, normalised columns.
				\item Counter $k=0$.
			\end{itemize}
		\end{minipage}\vspace{0.6em}\\
		\midrule
		\vspace{-0.35em}\textit{Step 1:} &
		\vspace{-0.35em}\begin{minipage}[t]{\linewidth}
			\begin{itemize}[leftmargin=1.4em]\raggedright
				\item[] \hspace{-1.3em}While $k\tau<T_1$ do: 
				\begin{itemize}[leftmargin=1.8em,topsep=0pt]\raggedright
					\itemsep0em
					\item[(a)] Evolve $\bs x(k\tau)$ and $Q(k\tau)$ forwards by $\tau$ time units to get $\bs x((k+1)\tau)$ and $\tilde Q((k+1)\tau)$, respectively.
					\item[(b)] Perform the QR decomposition $\tilde Q((k+1)\tau)=Q((k+1)\tau)R((k+1)\tau)$.
					\item[(c)] Increment $k$ by $1$.
				\end{itemize}
			\end{itemize}
		\end{minipage}\vspace{1.1em}\\
		\textit{Step 2:} &
		\begin{minipage}[t]{\linewidth}
			\begin{itemize}[leftmargin=1.4em]\raggedright
				\item[] \hspace{-1.3em}While $k\tau<T_3$ do: 
				\begin{itemize}[leftmargin=1.8em,topsep=0pt]\raggedright
					\itemsep0em
					\item[(a)] Evolve $\bs x(k\tau)$ and $Q(k\tau)$ forwards by $\tau$ time units to get $\bs x((k+1)\tau)$ and $\tilde Q((k+1)\tau)$, respectively.
					\item[(b)] Perform QR decomposition: $\tilde Q((k+1)\tau)=Q((k+1)\tau)R((k+1)\tau)$. Keep $Q((k+1)\tau)$ and $R((k+1)\tau)$ in memory for Steps 3 and 4.
					\item[(c)] Increment $k$ by $1$.
				\end{itemize}
			\end{itemize}
		\end{minipage}\vspace{1.1em}\\
		\textit{Step 3:} &
		\begin{minipage}[t]{\linewidth}
			\begin{itemize}[leftmargin=1.4em]\raggedright
				\item[] \hspace{-1.3em}While $k\tau>T_2$ do: 
				\begin{itemize}[leftmargin=1.8em,topsep=0pt]\raggedright
					\itemsep0em
					\item[(a)] Evolve $C(k\tau)$ backwards by $\tau$ time units to get $\tilde C((k-1)\tau)$, i.e.\ compute $\tilde C((k-1)\tau)=R^{-1}(k\tau)C(k\tau)$.
					\item[(b)] Normalise the columns of $\tilde C((k-1)\tau)$ to get $C((k-1)\tau)$.
					\item[(c)] Decrement $k$ by $1$.
				\end{itemize}
			\end{itemize}
		\end{minipage}\vspace{1.1em}\\
		\textit{Step 4:} &
		\begin{minipage}[t]{\linewidth}
			\begin{itemize}[leftmargin=1.4em]\raggedright
				\item[] \hspace{-1.3em}While $k\tau>T_1$ do: 
				\begin{itemize}[leftmargin=1.8em,topsep=0pt]\raggedright
					\itemsep0em
					\item[(a)] Evolve $C(k\tau)$ backwards by $\tau$ time units to get $\tilde C((k-1)\tau)$, i.e.\ compute $\tilde C((k-1)\tau)=R^{-1}(k\tau)C(k\tau)$.
					\item[(b)] Normalise the columns of $\tilde C((k-1)\tau)$ to get $C((k-1)\tau)$.
					\item[(c)] Compute $Q((k-1)\tau)C((k-1)\tau)$ and save the result.
					\item[(d)] Decrement $k$ by $1$.
				\end{itemize}
			\end{itemize}
		\end{minipage}\vspace{0.6em}\\
		\midrule
		\vspace{-0.35em}\textit{Output:} &
		\vspace{-0.35em}\begin{minipage}[t]{\linewidth}
			The backward time evolution of the matrix $Q(t)C(t)$ whose $j$-th column estimates the $j$-th CLV ($j=1,2,\dots,p$) at time $t$, where $t=T_2-\tau,T_2-2\tau,\dots,T_1$.
		\end{minipage}\vspace{0.5em}\\
		\bottomrule
	\end{tabular}
\end{table}

Performing the G\&C algorithm as described here will produce the time evolution of the first $p$ CLVs ($1\le p\le N$) along an orbit between times $T_1$ and $T_2$, since the BLVs and CLVs are still converging during the forward and backward transients, respectively. From our discussion in \sectionref{sec:asymp}, it follows that the GSVs converge to the BLVs during the forward transient at an exponential rate of $\min(\chi_{i-1}-\chi_i,\chi_i-\chi_{i+1})$. For the backward transient, Ginelli et al.\ argue in \cite{GinelliEtAl2013} that convergence to the $i$-th CLV is exponential with a rate of $\chi_{i-1}-\chi_i$. These exponential convergence rates make the G\&C algorithm an efficient method for computing CLVs.

\subsection{Optimisation of the G\&C algorithm}
The CPU time required for the backward evolution part of the G\&C algorithm can be reduced by taking advantage of the shape of the $R$ matrices in \algorithmref{alg:clv}. Since $R$ is upper triangular, the deviation vectors $\hat{\bs c}_j$ (comprising the columns of the matrix $C$) can each be evolved backwards using a backward substitution algorithm (see e.g.\ \cite[Sect.~6.1]{BurdenFaires2011}) instead of the approach taken in Steps~3(a) and~4(a) of \algorithmref{alg:clv}, which involves explicit matrix inversion and multiplication. This more efficient implementation of the G\&C algorithm by means of backward substitution was proposed in \cite{GinelliEtAl2013}, and pseudocode for this approach can be found in Appendix~B of \cite{PikovskyPoliti2016}.

\begin{figure}[htbp]
	\centering
	\includegraphics[width=\linewidth]{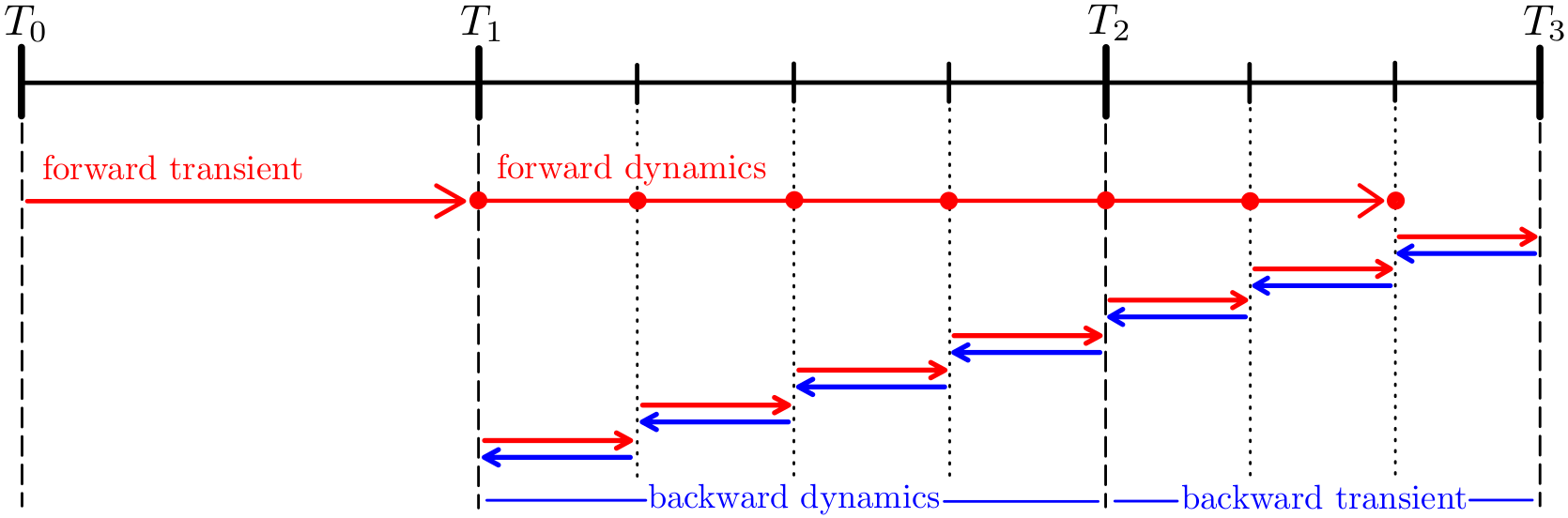}
	\caption{Diagram depicting the G\&C algorithm for computing CLVs using a memory optimisation technique.}
	\label{fig:alg_mem}
\end{figure}

The G\&C algorithm is particularly demanding on memory resources due to the many matrices which need to be stored in order to perform the backward evolution. In \cite{GinelliEtAl2007}, a solution to the memory problem was hinted at, but it is described in more detail in \cite{GinelliEtAl2013}, where the authors also analysed the memory requirements of the algorithm. The solution they proposed is a memory optimisation technique to reduce the memory usage of the G\&C algorithm at the cost of almost doubling the required CPU time. The technique involves breaking up the forward dynamics time interval into many sub-intervals, the boundaries of which are indicated as dotted lines in \figureref{fig:alg_mem}, which diagrams this memory-optimised version of the algorithm (cf.\ \figureref{fig:alg}). The lengths of these sub-intervals must be short enough for the Q and R matrices computed during an individual sub-interval to all be held in memory simultaneously. Then, essentially two computations of the forward dynamics are performed: a first run is performed to compute the state and deviation vectors at the start of each sub-interval and save them in memory, where these points are indicated by big red dots in \figureref{fig:alg_mem}. A second forward dynamics run (denoted by the short red arrows in \figureref{fig:alg_mem}) is done over each sub-interval, starting from the end and alternating with a backward transient/dynamics run (denoted by the short blue arrows). The initial state and deviation vectors used to initialise the forward dynamics run over each sub-interval are those saved during the first forward run, and the upper triangular matrix used to initialise the backward transient/dynamics over each sub-interval is the final upper triangular matrix acquired from the previous backward transient/dynamics run. Thereby, the memory requirements of the G\&C algorithm can be greatly reduced, albeit at the cost of an extra forward dynamics run.

If the angles (e.g.\ the minimum angles) between CLVs or splitting subspaces are all that is required from a particular CLV computation, then additional memory and CPU savings can be made. Since the coordinate and GSV bases are both orthogonal bases, the relevant angles are preserved when changing between bases \cite{GinelliEtAl2013}. Therefore, the CLVs can be saved directly in the GSV basis, from which the relevant angles can then be computed. Referring to our specification of the G\&C algorithm in \algorithmref{alg:clv}, the following changes can be made: in Step~2(b), $Q((k+1)\tau)$ is not stored in memory; in Step 4(c), $Q((k-1)\tau)C((k-1)\tau)$ is not computed, rather $C((k-1)\tau)$ is saved directly and used as the output since its columns are estimates for the CLVs in the GSV basis.

As a final note, recall our discussion in \sectionref{sec:qr} regarding the ``non-unique'' QR decompositions that are implemented by various software, where we described a simple operation which can be performed to recover the unique QR decomposition. In the forward evolution part of the G\&C algorithm, the non-unique decomposition only has the effect of negating some computed GSVs without affecting the subspaces they live in. Therefore, if the sign of the GSVs and CLVs is not of interest, performing this operation to recover the unique decomposition is unnecessary. Removing this step can slightly reduce the CPU time required to run the algorithm.

\subsection{Retroactive computation of BLVs}\label{sec:retro}
In \chapterref{ch:hh}, it will be necessary to compute the CLVs for a given state of a dynamical system. Unfortunately, the G\&C algorithm only prescribes a method for computing the BLVs in the future after some forward transient, hence the CLVs can also only be computed after that transient. In this section, we propose a method of retroactively computing the BLVs that will allow us to then skip the forward transient of the G\&C algorithm and ultimately compute CLVs at that point in time.

Let $Q_b(T_1)$ be an initial $N\times N$ orthogonal matrix, where $T_1$ is the time marking the end of the forward transient (see \figureref{fig:alg}) at which we wish to compute BLVs, and we use the subscript $b$ since we will be moving backwards in time. We now evolve $Q_b(T_1)$ backwards in time, periodically orthonormalising the matrix, until we reach time $T_0$. Using QR decomposition for this orthonormalisation, we acquire a pair of $Q_b$ and $R_b$ matrices at each orthonormalisation time along the way. Recall from our discussion around \eqref{eq:rep} that $R_b$ is a matrix representation of the linear propagator over the relevant interval between successive orthonormalisations, but expressed in the basis of the columns of $Q_b$. So, in an analogous way to the backward evolution part of the G\&C algorithm, we can evolve a generic $N\times N$ orthogonal matrix $Q_f(T_0)$ forwards in time to $T_1$ by successively multiplying by the inverse of the matrices $R_b$ saved in memory, being sure to orthonormalise the columns of the resulting matrix between each multiplication to avoid numerical issues. Unlike in the G\&C algorithm's backward transient/dynamics, we do not constrain $Q_f(T_0)$ to be upper triangular, and we periodically orthonormalise (instead of just normalise) the columns of the matrix in order to converge to the BLVs at time $T_1$. However, the matrix $Q_f(T_1)$ acquired from this procedure is not in the coordinate basis; rather, it is in the basis of the columns of $Q_b(T_1)$. We can convert this back to the coordinate basis by computing $Q_b(T_1)Q_f(T_1)$, the columns of which are the estimated BLVs at $T_1$.

As a final remark on the topic of retroactive computation of BLVs, we emphasise that only the $R_b$ matrices need to be stored during the backward evolution, not all the $Q_b$ matrices. Since we only need to convert the BLVs to the coordinate basis at the end of the computation at time $T_1$, only the matrix $Q_b(T_1)$ needs to be saved in memory.

\section{Convergence of subspaces and vectors}\label{sec:converge}
Using the distance $\Delta$ defined in \eqref{eq:distance}, we can measure how close two subspaces are to each other. If $\Delta(U,V)\to0$ as $t\to\pm\infty$, we say that the subspaces $U$ and $V$ \textit{converge} during forward/backward evolution, respectively. Similarly, we say that two vectors $\bs w_1$ and $\bs w_2$ converge if $\Delta(\bs w_1,\bs w_2)\to0$ as $t\to\pm\infty$.

Many convergence rates we have discussed thus far have been exponential. For example, an arbitrary deviation vector converges to the first CLV (assuming the first LE is non-degenerate) like $e^{-(\chi_1-\chi_2)t}$. However, the first CLV is not known a priori, so instead of directly measuring the convergence rate to the first CLV, we take the approach of evolving two deviation vectors independently and measuring the convergence rate between them as they converge to the first CLV. In general, if generic subspaces of a particular dimension are known to converge exponentially to some as-yet-unknown subspace, we propose using the convergence between the two generic subspaces as a proxy for studying their convergence to the unknown subspace, and we claim that these exponential rates are typically the same.

To motivate this claim, let $P_1(t)$ and $P_2(t)$ be subspaces which both converge exponentially to a subspace $\Pi(t)$ as $t\to\infty$ with an exponential rate $\lambda$, so $\Delta(P_1(t),\Pi(t))\approx c_1 e^{-\lambda t}$ and $\Delta(P_2(t),\Pi(t))\approx c_2 e^{-\lambda t}$ for some constants $c_1,c_2$. Since $\Delta$ is a metric, it obeys the triangle inequality,
\begin{align}
\begin{split}
	\Delta(P_1(t),P_2(t)) &\leq \Delta(P_1(t),\Pi(t)) + \Delta(P_2(t),\Pi(t))\\
	&\approx c_1 e^{-\lambda t} + c_2 e^{-\lambda t}.
\end{split}
\end{align}
Therefore, if we let $c=c_1+c_2$,
\begin{align}
	\Delta(P_1(t),P_2(t)) &\lesssim c e^{-\lambda t}.
\end{align}
So we see that if $P_1(t)$ and $P_2(t)$ are converging exponentially to $\Pi(t)$, then the mutual convergence of $P_1(t)$ and $P_2(t)$ is exponential at the same rate or faster. However, this (approximate) inequality leaves open the possibility that $P_1(t)$ and $P_2(t)$ could mutually converge faster than $e^{-\lambda t}$. To see such a case, we adapt the analysis done in \cite{SkokosEtAl2004} for the SALI, which measures the (anti-)alignment between two generic deviation vectors as they each converge to the first CLV. In particular, we will show in the remainder of this section that some choices for these deviation vectors may result in faster-than-usual convergence, but the probability of choosing such a pair of vectors is zero.

Let $\bs w_1(0)$ and $\bs w_2(0)$ be two generic deviation vectors. We can express each of them as linear combinations of CLVs,
\begin{align}
\begin{split}
	\bs w_1(0) &= \sum_{i=1}^N a_i \hat{\bs\omega}_i(t),\\
	\bs w_2(0) &= \sum_{i=1}^N b_i \hat{\bs\omega}_i(t),
\end{split}
\end{align}
where $a_i,b_i$ are the relevant CLV components. Since CLVs evolve under the tangent dynamics like $M(\bs x(0),t)\hat{\bs \omega}_i(0)\approx e^{\chi_i t}\hat{\bs \omega}_i(t)$ for large $t$, it follows that
\begin{align}
	\begin{split}
		\bs w_1(t) &= M(\bs x(0),t)\bs w_1(0) \approx \sum_{i=1}^N a_i e^{\chi_i t}\hat{\bs\omega}_i(t),\\
		\bs w_2(t) &= M(\bs x(0),t)\bs w_2(0) \approx \sum_{i=1}^N b_i e^{\chi_i t}\hat{\bs\omega}_i(t).
	\end{split}
\end{align}
In order to normalise these deviation vectors, we need to compute their magnitudes. In the CLV expansions of $\bs w_1(t)$ and $\bs w_2(t)$, the leading order terms are $a_1 e^{\chi_1 t}\hat{\bs\omega}_1(t)$ and $b_1 e^{\chi_1 t}\hat{\bs\omega}_1(t)$, respectively, so $\|\bs w_1(t)\| \approx a_1 e^{\chi_1 t}$ and $\|\bs w_2(t)\| \approx b_1 e^{\chi_1 t}$. Note that we have assumed here that $a_1$ and $b_1$ are positive since either initial deviation vector could always be negated to ensure this. Hence,
\begin{align}
	\begin{split}
		\hat{\bs w}_1(t) &= \frac{\bs w_1(t)}{\|\bs w_1(t)\|} \approx \hat{\bs\omega}_1(t) + \sum_{i=2}^N \frac{a_i}{a_1} e^{(\chi_i-\chi_1) t}\hat{\bs\omega}_i(t),\\
		\hat{\bs w}_2(t) &= \frac{\bs w_2(t)}{\|\bs w_2(t)\|} \approx \hat{\bs\omega}_1(t) + \sum_{i=2}^N \frac{b_i}{b_1} e^{(\chi_i-\chi_1) t}\hat{\bs\omega}_i(t).
	\end{split}
\end{align}
Therefore, the difference between these unit vectors is
\begin{align}
	\hat{\bs w}_1(t) - \hat{\bs w}_2(t) \approx \sum_{i=2}^N \left(\frac{a_i}{a_1}-\frac{b_i}{b_1}\right) e^{(\chi_i-\chi_1) t}\hat{\bs\omega}_i(t).\label{eq:leading}
\end{align}
Assuming $a_2/a_1\ne b_2/b_1$, the leading order term in this expansion is the $\hat{\bs\omega}_2(t)$ term, so the magnitude of this vector can be approximated by the absolute value of that term's component. Furthermore, $\Delta(\hat{\bs w}_1(t),\hat{\bs w}_2(t))\approx\|\hat{\bs w}_1(t)-\hat{\bs w}_2(t)\|$ for large $t$, since both quantities reduce to simply the angle between $\hat{\bs w}_1(t)$ and $\hat{\bs w}_2(t)$ when the angle is small. Therefore,
\begin{align}
	\Delta(\hat{\bs w}_1(t),\hat{\bs w}_2(t))\approx \|\hat{\bs w}_1(t)-\hat{\bs w}_2(t)\|\approx \left|\frac{a_2}{a_1}-\frac{b_2}{b_1}\right| e^{-(\chi_1-\chi_2) t}.
\end{align}
From this result, we see that the exponential rate of convergence between the two deviation vectors is $\chi_1-\chi_2$, which is the same as the rate at which they each converge to the first CLV. However, in the probability zero case where $a_2/a_1=b_2/b_1$, the next leading order term in \eqref{eq:leading} would be the $\hat{\bs\omega}_3(t)$ term, and so the mutual convergence between our two deviation vectors would instead have an exponential rate of $\chi_1-\chi_3$. This supports our claim that mutual convergence faster than the exponential rate $\chi_1-\chi_2$ is possible, but not generic. We therefore assume in subsequent chapters that measuring mutual exponential convergence rates between two generic subspaces is a valid proxy for measuring their exponential convergence rates to some other subspace.

\section{Summary}\label{sec:theorysummary}
In this chapter, we gave an overview of the mathematics and numerical methods used in other parts of this thesis. Beginning in \sectionref{sec:bigN}, we summarised several notions from linear algebra and matrix theory; of particular importance is the distance $\Delta$ between vectors/subspaces introduced in \sectionref{sec:distanceintro}, which we use in \chapterref{ch:hh} to measure convergence rates between different computations of the BLVs, CLVs, filtration subspaces, and splitting subspaces. We then discussed the theory of continuous dynamical systems in \sectionref{sec:continuoussec}, with emphasis on Lyapunov vectors and exponents in \sectionref{sec:lyapunov}. Since the dynamical systems we study in the remainder of this thesis are all autonomous Hamiltonian systems, we discussed such systems in \sectionref{sec:autoham} and the use of symplectic integrators for numerically solving them. We described efficient algorithms for computing LEs and CLVs in Sections~\ref{sec:le} and~\ref{sec:ginelli}, before finally motivating in \sectionref{sec:converge} the approach which we use in \chapterref{ch:hh} for measuring the convergence rates of vectors and subspaces computed via the G\&C algorithm.

%% file: results.tex
\chapter{Numerical investigation of low-dimensional systems}\label{ch:hh}
The first autonomous Hamiltonian system which we study in this thesis is the H\'enon-Heiles system, which is a simple model of a star's motion in a planar galaxy \cite{HenonHeiles1964}. The H\'enon-Heiles system is a prototypical model of Hamiltonian dynamics which has been used in many numerical investigations (e.g.\ \cites{SkokosManos2016}{Barrio2016}{HillebrandEtAl2022}). This system has two degrees of freedom with a Hamiltonian function $H_2$, given by
\begin{align}
	H_2=\frac12(p_x^2+p_y^2)+\frac12(x^2+y^2)+x^2y-\frac13y^3,\label{eq:HH}
\end{align}
where $x,y$ are position coordinates and $p_x,p_y$ are the conjugate momenta. In \sectionref{sec:hhsection}, we use the G\&C algorithm to compute the CLVs for chaotic orbits in the H\'enon-Heiles system, and we discuss the convergence of these vectors and the angles between the splitting subspaces they define.

In \sectionref{sec:3d}, we repeat our numerical investigation into the convergence properties of the G\&C algorithm for a system with three degrees of freedom with the Hamiltonian
\begin{align}
	H_3 = \frac{\omega_x}{2}(x^2+p_x^2) + \frac{\omega_y}{2}(y^2+p_y^2) + \frac{\omega_z}{2}(z^2+p_z^2) + x^2(y + z),\label{eq:ham3d}
\end{align}
where $\omega_x=1$, $\omega_y=\sqrt{2}$, and $\omega_z=\sqrt{3}$. The position coordinates are $x,y,z$, and the conjugate momenta are $p_x,p_y,p_z$. This 3-D system, which exhibits both regular and chaotic dynamics, was introduced by Contopoulos et al.\ in 1978 \cite{ContopoulosEtAl1978} and has been used in numerous numerical studies since (e.g.\ \cites{BenettinEtAl1980a}{SkokosEtAl2007}).

\section{The H\'enon-Heiles system}\label{sec:hhsection}
We begin our study of the dynamics of the H\'enon-Heiles system by computing a \textit{Poincar\'e surface of section} (PSS) from which we select an orbit that we demonstrate to be chaotic by computing its ftLEs. Proceeding with a numerical investigation into the convergence properties of the G\&C algorithm for computing CLVs, we see that our numerical results agree with theoretical results from the literature. Our discussion then moves to the dynamics of the $2$-D centre subspace $\Omega_2$ of the H\'enon-Heiles system and the inaccuracies which arise when computing that subspace using the G\&C algorithm, and we propose a modification of the algorithm to correct for this. We also examine the angles between the splitting subspaces for a sticky chaotic orbit, which switches between regular and chaotic regimes of motion, and we find that these regimes correlate with finite-time mLE computed over short windows.

The energy we use for the H\'enon-Heiles system is $H_2=1/8$, which gives rise to a mixture of regular and chaotic dynamics. For all numerical integration of the equations of motion \eqref{eq:eom} and variational equations \eqref{eq:jacob}, we use the ABA864 symplectic integrator \eqref{eq:aba}, and we orthonormalise any computed deviation vectors every $\tau=1$ time unit. We use an integration time step of 0.025 in Sections~\ref{sec:les}--\ref{sec:cent}, which we find keeps the relative energy error $E_r$ \eqref{eq:energyerror} below $10^{-10}$. However, we use a smaller time step of 0.02 in \sectionref{sec:prado} for improved accuracy.

\subsection{Lyapunov exponent spectrum}\label{sec:les}
\begin{figure}[htbp]
	\centering
	\includegraphics[width=0.55\linewidth]{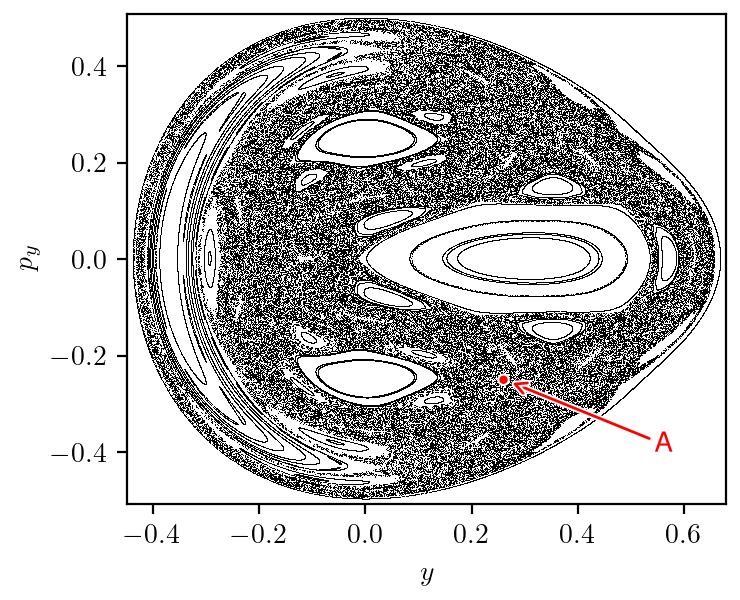}
	\caption{The $x=0$, $p_x>0$ PSS of the H\'enon-Heiles system \eqref{eq:HH} with energy $H_2=1/8$, where many orbits were used in the construction of this PSS. The red point labelled ``A'' has coordinates $(y,p_y)=(0.26,-0.25)$.}
	\label{fig:pss}
\end{figure}

Let us begin by computing the PSS (see e.g.\ \cite[p.~17]{LichtenbergLieberman1992}) defined by $x=0,$ $p_x>0$ for the H\'enon-Heiles system, which we present in \figureref{fig:pss}. This PSS is computed by selecting a $13\times12$ grid of $(y,p_y)$ coordinates that we use as initial conditions, where we set $x=0$ and choose $p_x>0$ such that the energy is $H_2=1/8$. We then evolve these initial states forwards in time and compute each intersection of their orbits with $x=0$ (when $p_x>0$) using H\'enon's approach \cite{Henon1982}, where we use a single step of the Runge-Kutta fourth-order integrator to accurately determine the point of intersection. The orbit for each initial condition is numerically integrated until 2000 such crossings have been computed, which typically occurs in about $10^4$ time units. Many islands of stability consisting of closed curves are visible in \figureref{fig:pss}; these closed curves indicate where the PSS intersects invariant tori upon which regular motion occurs. These islands are surrounded by a chaotic sea (which resembles randomly distributed points) where chaotic orbits intersect the PSS.

\begin{figure}[htbp]
	\centering
	\includegraphics[width=0.55\linewidth]{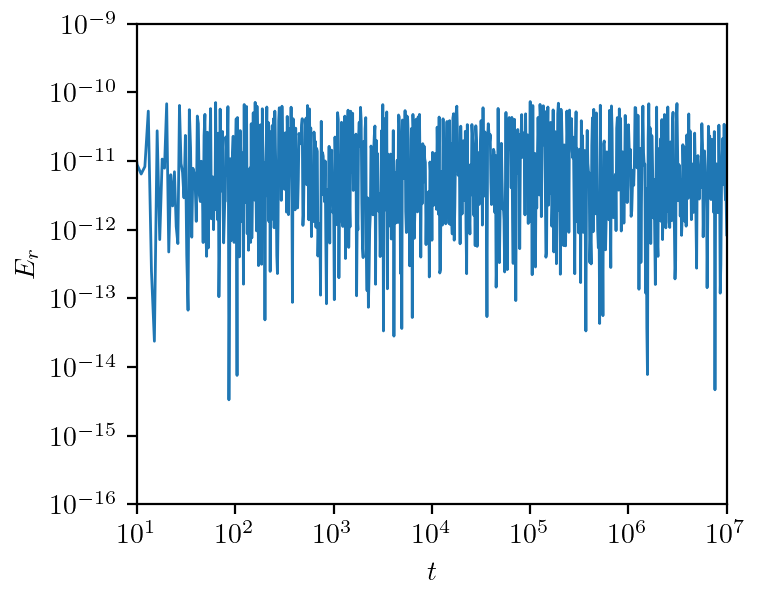}
	\caption{The relative energy error $E_r$ as a function of time $t$ for the chaotic orbit with initial condition \eqref{eq:ic} in the H\'enon-Heiles system \eqref{eq:HH}. The figure is in log-log scale.}
	\label{fig:error}
\end{figure}

\begin{figure}[htbp]
	\centering
	\includegraphics[width=\linewidth]{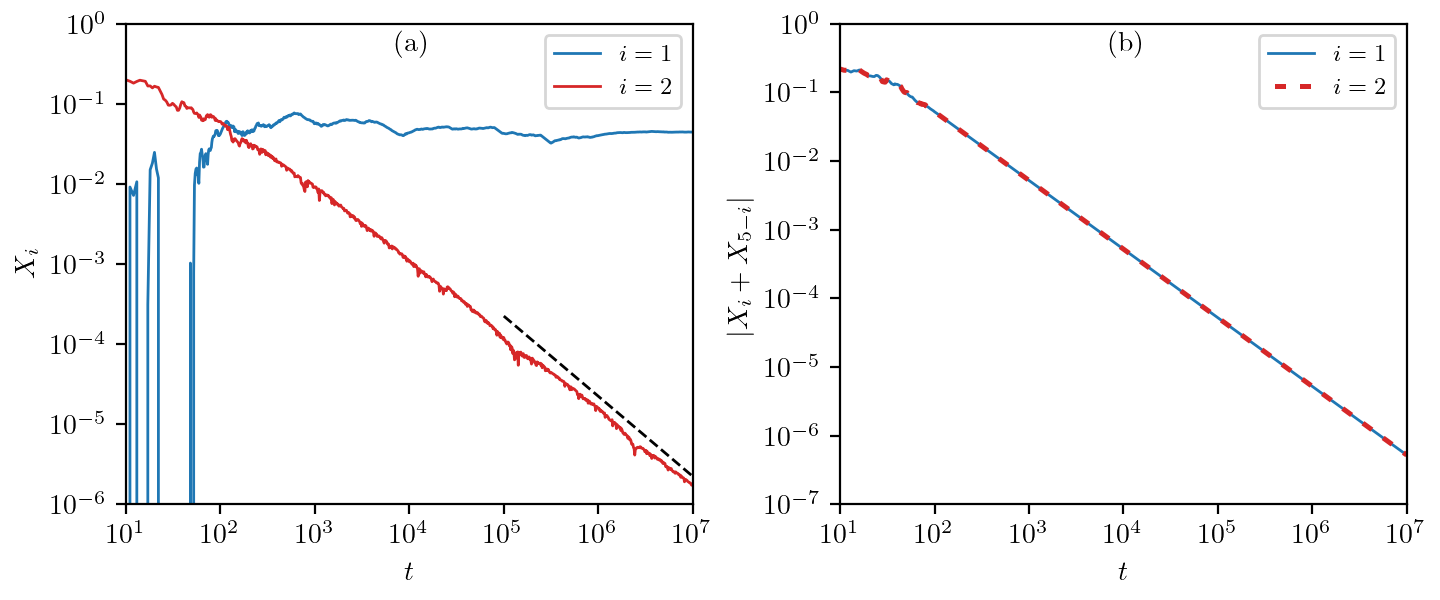}
	\caption{The time evolution of the largest ftLE $X_1$ and second-largest ftLE $X_2$ is given in (a) for the orbit with initial condition \eqref{eq:ic} in the H\'enon-Heiles system \eqref{eq:HH}. The black dashed line in (a) denotes a function $\propto t^{-1}$. The time evolution of $|X_i+X_{5-i}|$ is given in panel (b) for $i=1,2$. Both panels are in log-log scale.}
	\label{fig:lces}
\end{figure}

The red point labelled ``A'' in \figureref{fig:pss} has coordinates
\begin{align}
	(y,p_y)=(0.26,-0.25)\label{eq:ic}
\end{align}
and appears to lie in the chaotic sea of the PSS. To demonstrate that the orbit of this initial condition is chaotic, we estimate the orbit's LEs $\chi_i$, $i=1,2,3,4$, by computing its ftLEs $X_i(t)$ using the standard method described in \sectionref{sec:le}. We compute the ftLEs over an integration time of $10^7$ time units, using a time step of $0.025$ and an orthonormalisation interval of $\tau=1$. During this ftLE computation, we calculate the relative energy error $E_r$ as a function of time $t$ and plot the result in \figureref{fig:error}, where we see that $E_r<10^{-10}$ and that, on average, $E_r$ is approximately constant over time, as expected from our use of a symplectic integrator. The time evolution of the first two ftLEs, $X_1$ and $X_2$, are given in \figureref{fig:lces}(a). We see from this figure that $X_1$ saturates by the end of the integration time to a value of $X_1(10^7)=0.044$, which is our estimate for the mLE $\chi_1$. On the other hand, $X_2(t)$ decays to zero $\propto t^{-1}$, which is consistent with the LE $\chi_2$ being zero (see the discussion at the end of \sectionref{sec:le}). Now, recall from \eqref{eq:symplecticsym} that $\chi_i+\chi_{N-i+1}=0$ for the LEs, so the following holds for the ftLEs:
\begin{align}
	\lim_{t\to\infty}|X_i(t)+X_{N-i+1}(t)|=0.
\end{align}
Since the dimension of the phase space for the H\'enon-Heiles system is $N=4$, the time evolution of $|X_i+X_{5-i}|$ is given in \figureref{fig:lces}(b) for $i=1,2$, from which we see that this quantity tends to zero for each $i$, so our computed estimate of the LE spectrum has the expected symmetry.

Our estimates of the LEs $\chi_i$ \eqref{eq:le_non}, $i=1,2,3,4$, acquired from the computed ftLEs $X_i(t)$ evaluated at $t=10^7$, are summarised in \tableref{tab:vector} together with the spectral gaps between neighbouring LEs. We give this table for reference since the GSVs converge to the BLVs at an exponential rate of $\min(\chi_{i-1}-\chi_i,\chi_i-\chi_{i+1})$, which is given in the last column of the table. Furthermore, convergence to the CLVs during the backward transient of the G\&C algorithm happens at an exponential rate of $\chi_{i-1}-\chi_i$, which is given in the third column of \tableref{tab:vector}. Note that we use the convention that $\chi_0=\infty$ and $\chi_5=-\infty$, as discussed in \sectionref{sec:asymp}. We also give the distinct LEs $\lambda_i$ \eqref{eq:le}, $i=1,2,3$, and their spectral gaps in \tableref{tab:subspace} since generic subspaces $G_i$ converge to the filtration subspaces $\Gamma_i^-$ of the same dimension at an exponential rate of $\lambda_i-\lambda_{i+1}$. Note that we use the convention that $\lambda_0=\infty$ and $\lambda_4=-\infty$. In particular, we conclude from our positive estimate of the mLE that the orbit passing through initial condition \eqref{eq:ic} is chaotic.

\begin{table}[htbp]
	\caption{The $i$-th LE $\chi_i$, $i=1,2,3,4$, and its spectral gaps with neighbouring LEs $\chi_{i-1},\chi_{i+1}$ for the H\'enon-Heiles system \eqref{eq:HH}, computed using the chaotic orbit with initial condition \eqref{eq:ic}.}
	\label{tab:vector}
	\centering
	\setlength{\tabcolsep}{9pt}
	\begin{tabular}{crrrr}
		$i$ & $\chi_i$ & $\chi_{i-1}-\chi_i$ & $\chi_{i}-\chi_{i+1}$ & $\min(\chi_{i-1}-\chi_i,\chi_i-\chi_{i+1})$ \\
		\midrule
		$1$ & $0.044$  & $\infty$            & $0.044$               & $0.044$                                     \\
		$2$ & $0.000$  & $0.044$             & $0.000$               & $0.000$                                     \\
		$3$ & $0.000$  & $0.000$             & $0.044$               & $0.000$                                     \\
		$4$ & $-0.044$ & $0.044$             & $\infty$               & $0.044$                                     \\                            
	\end{tabular}
\end{table}

\begin{table}[htbp]
	\caption{The $i$-th distinct LE $\lambda_i$, $i=1,2,3$, and its spectral gaps with neighbouring distinct LEs $\lambda_{i-1},\lambda_{i+1}$, computed using the same chaotic orbit as in \tableref{tab:vector}.}
	\label{tab:subspace}
	\centering
	\setlength{\tabcolsep}{9pt}
	\begin{tabular}{crrr}
		$i$ & $\lambda_i$ & $\lambda_{i-1}-\lambda_i$ & $\lambda_{i}-\lambda_{i+1}$ \\
		\midrule
		$1$ & $0.044$     & $\infty$                  & $0.044$                     \\
		$2$ & $0.000$     & $0.044$                   & $0.044$                     \\
		$3$ & $-0.044$    & $0.044$                   & $\infty$                   
	\end{tabular}
\end{table}

\subsection{Convergence properties of the G\&C algorithm}\label{sec:hhconv}
We now proceed with a numerical investigation of the convergence properties of the G\&C algorithm for computing CLVs (see \sectionref{sec:ginelli}), beginning with the forward transient part of the algorithm, which computes estimates of the BLVs and the filtration subspaces they define, then moving on to the backward transient part, which uses those subspaces to compute the splitting subspaces and their associated CLVs. For this investigation, we use the initial condition \eqref{eq:ic} belonging to a chaotic orbit. Despite only presenting results in this section for a particular initial condition, our findings are representative of typical chaotic orbits since we repeated this investigation for several different initial conditions of chaotic orbits and obtained similar results.

\subsubsection{Forward transient}\label{sec:forward}
As discussed in \sectionref{sec:ginelli}, we compute the BLVs by evolving a set of linearly independent vectors forwards in time while periodically orthonormalising the vectors in order to prevent alignment to the first CLV. These orthonormal vectors (which converge to the BLVs) are the GSVs, and waiting for this convergence to occur is the purpose of the forward transient part of the G\&C algorithm.

\begin{figure}[htbp]
	\centering
	\includegraphics[width=\linewidth]{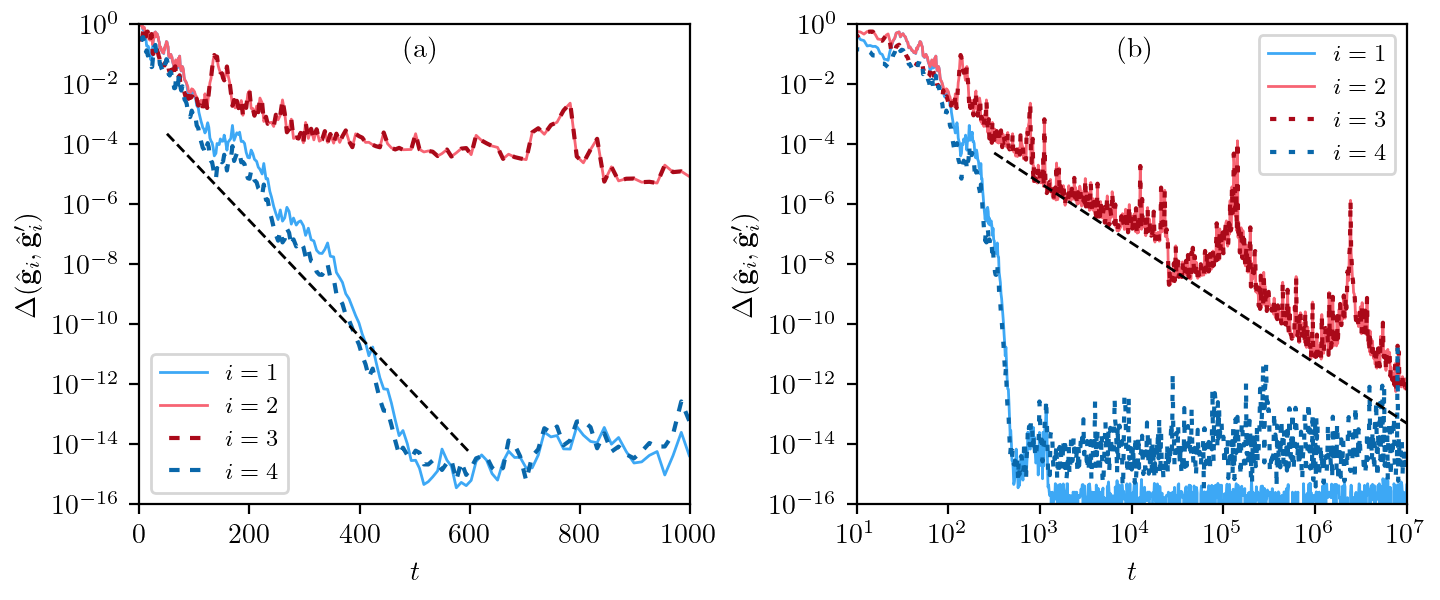}
	\caption{The time evolution of the distance $\Delta$ between two versions $\hat{\bs g}_i$ and $\hat{\bs g}_i'$ of the $i$-th GSV ($i=1,2,3,4$) for the chaotic orbit with initial condition \eqref{eq:ic} in the H\'enon-Heiles system \eqref{eq:HH}. In panel (a) the black dashed line denotes a function $\propto e^{-0.044t}$, while in (b) the black dashed line is $\propto t^{-2}$. Panel (a) is in log-linear scale, while (b) is in log-log scale.}
	\label{fig:gsv_vec_converge}
\end{figure}

In order to measure the rates of convergence of GSVs to BLVs, we take the pragmatic, albeit indirect, approach of measuring the rates of convergence between two different sets of GSVs computed using different initial deviation vectors. In other words, we compute two versions of the $i$-th GSV, $\hat{\bs g}_i$ and $\hat{\bs g}_i'$, for $i=1,2,3,4$, then we compute the distance \eqref{eq:distance} between them, $\Delta(\hat{\bs g}_i,\hat{\bs g}_i')$, and the time evolution of this distance function gives us a view of how independent computations of GSVs converge. As discussed in \sectionref{sec:converge}, we expect the exponential convergence rate between $\hat{\bs g}_i$ and $\hat{\bs g}_i'$ to be equal to their exponential convergence rates to the $i$-th BLV, i.e.\ $\min(\chi_{i-1}-\chi_i,\chi_i-\chi_{i+1})$. We give the time evolution of $\Delta(\hat{\bs g}_i,\hat{\bs g}_i')$ in \figureref{fig:gsv_vec_converge} over a forward transient of $10^7$, where we have parametrised time such that $t=0$ corresponds to $T_0$ in the G\&C algorithm (see \figureref{fig:alg}). We save the value of $\Delta(\hat{\bs g}_i,\hat{\bs g}_i')$ at evenly spaced points in $\log t$ (up to the resolution of our time step) such that we have 150 data points per decade of the $t$-axis in log scale. We see from \figureref{fig:gsv_vec_converge}(a) that $\Delta(\hat{\bs g}_1,\hat{\bs g}_1')$ and $\Delta(\hat{\bs g}_4,\hat{\bs g}_4')$ follow an approximate exponential decay rate of $0.044$ (indicated by the black dashed line), which agrees with the expected convergence rates of each of these vectors to their corresponding BLVs given in the last column of \tableref{tab:vector}. Note that we use double precision in all our computations, so we assume that the saturation of $\Delta(\hat{\bs g}_1,\hat{\bs g}_1')$ and $\Delta(\hat{\bs g}_4,\hat{\bs g}_4')$ to approximately $10^{-15}$ seen in \figureref{fig:gsv_vec_converge}(a) is an artefact of the finite precision of the numerics and can be ignored. On the other hand, we see from \figureref{fig:gsv_vec_converge}(b) that $\hat{\bs g}_2$ and $\hat{\bs g}_2'$ converge sub-exponentially $\propto t^{-2}$, and the same is true for $\hat{\bs g}_3$ and $\hat{\bs g}_3'$. We interpret this sub-exponential rate of convergence as being consistent with an exponential convergence rate of zero (as predicted by \tableref{tab:vector}) in the sense that $t^{-2}$ is asymptotically bounded between $e^{0t}$ and $e^{-\epsilon t}$ for arbitrarily small $\epsilon>0$. Considering that the BLVs associated with these GSVs are not uniquely defined due to the degeneracy in the LE spectrum, it is interesting that these GSVs converge at all.

\begin{figure}[htbp]
	\centering
	\includegraphics[width=0.55\linewidth]{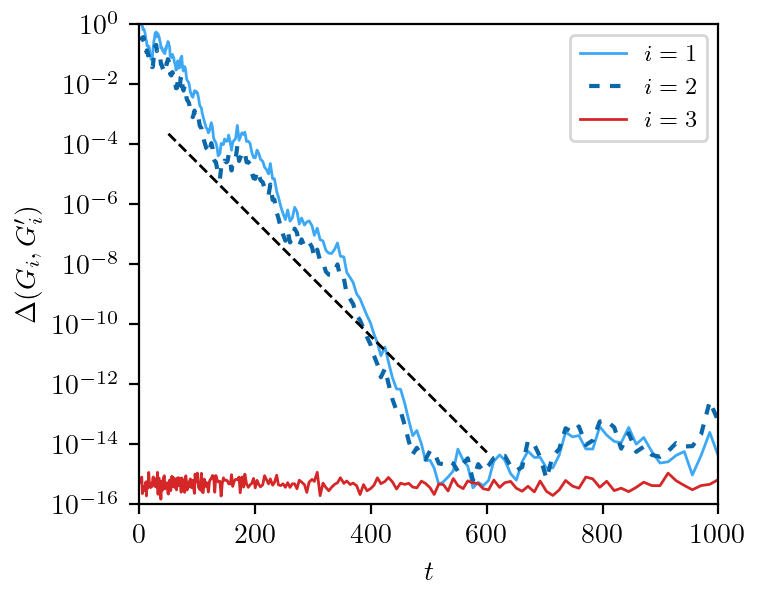}
	\caption{The time evolution of the distance $\Delta$ between two generic subspaces $G_i$ and $G_i'$, $i=1,2,3$, each of dimension $\dim\Gamma_i^-$ for the chaotic orbit with initial condition \eqref{eq:ic} in the H\'enon-Heiles system \eqref{eq:HH}. The black dashed line denotes a function $\propto e^{-0.044t}$. The figure is in log-linear scale.}
	\label{fig:gsv_sub_converge}
\end{figure}

We now move on to study the convergence rates of generic subspaces to the backward filtration subspaces $\Gamma_i^-$ \eqref{eq:filtration}, $i=1,2,3$. Since the multiplicities of the distinct LEs $\lambda_1$, $\lambda_2$, and $\lambda_3$ are $1$, $2$, and $1$ (respectively) for this orbit, it follows from \eqref{eq:filtration} that $\dim\Gamma_1^-=1$, $\dim\Gamma_2^-=3$, and $\dim\Gamma_3^-=4$. With this in mind, we define the following subspaces using the computed GSVs:
\begin{align}
\begin{split}
	G_1&=\spn(\hat{\bs g}_1),\\
	G_2&=\spn(\hat{\bs g}_1,\hat{\bs g}_2,\hat{\bs g}_3),\\
	G_3&=\spn(\hat{\bs g}_1,\hat{\bs g}_2,\hat{\bs g}_3,\hat{\bs g}_4).
\end{split}
\end{align}
Since $\dim G_i=\dim\Gamma_i^-$ for each $i$, we know from \sectionref{sec:asymp} that $G_i\to\Gamma_i^-$ as $t\to\infty$. Therefore, if we compute the subspaces $G_i$ using a set of GSVs $\hat{\bs g}_i$, and we similarly compute a second set of subspaces $G_i'$ using an independent computation of GSVs $\hat{\bs g}_i'$, then each pair $G_i$ and $G_i'$ should converge. We compute these subspaces $G_i$ and $G_i'$ using the same data from \figureref{fig:gsv_vec_converge} and present the time evolution of $\Delta(G_i,G_i')$ in \figureref{fig:gsv_sub_converge}. We see from this figure that $\Delta(G_i,G_i')$ decays approximately $\propto e^{-0.044t}$ when $i=1,2$ since the blue curves in \figureref{fig:gsv_sub_converge} roughly follow the black dashed line. For $i=3$, note that $G_3$ and $G_3'$ have the same dimension as the entire tangent space, so $G_3=G_3'$ and no convergence time is required; indeed, we see from this figure that $\Delta(G_3,G_3')$ is practically zero throughout the computation. These estimates of the exponential rates of convergence between generic subspaces $G_i$ and $G_i'$ are in agreement with their expected rates of convergence to the $i$-th filtration subspace $\Gamma_i^-$ as given in the last column of \tableref{tab:subspace}, where we interpret the convergence rate of $\infty$ in the $i=3$ case as indicating that the convergence happens instantaneously. The fact that $\Delta(G_3,G_3')$ is near zero but not identically zero in \figureref{fig:gsv_sub_converge}, despite $G_3$ and $G_3'$ being the same subspace, is simply an artefact of our numerics when calculating $\Delta(G_3,G_3')$ via principal angles.

Since the goal of the forward transient is to compute accurate estimates of the filtration subspaces $\Gamma_i^-$, it is sufficient to wait until $G_i$ and $G_i'$ have each converged to $\Gamma_i^-$. We see from \figureref{fig:gsv_sub_converge} that the distance $\Delta(G_i,G_i')$ is practically zero up to machine error after only 500 time units. Since we use this proxy method to indirectly measure $\Delta(G_i,\Gamma_i^-)$, we conclude that $10^3$ time units is more than sufficient for the forward transient part of the G\&C algorithm in this case.

\subsubsection{Backward transient}\label{sec:back_trans}
We turn our attention to the backward transient part of the G\&C algorithm, where deviation vectors chosen from the computed estimates of each $\Gamma_i^-$, $i=1,2,3$, are evolved backwards in time in order to converge to the CLVs in $\Omega_i$ \eqref{eq:splitting}. Since we will be moving backwards through time, we define a backward time variable $t_b$, which we parametrise such that $t_b=0$ corresponds to $T_3$ in \figureref{fig:alg}. To initialise the backward transient, we randomly choose a non-singular upper triangular matrix with normalised columns which represent deviation vectors $\hat{\bs c}_i$ ($i=1,2,3,4$) expressed in the basis of GSVs $\hat{\bs g}_i$. During the backward transient, the vectors $\hat{\bs c}_i$ converge to the CLVs $\hat{\bs\omega}_i$, so we initialise a second set of deviation vectors $\hat{\bs c}_i'$ and measure the convergence rate between $\hat{\bs c}_i$ and $\hat{\bs c}_i'$ as a proxy for their convergence rate to $\hat{\bs\omega}_i$. Before performing the backward transient, we compute a forward transient of $10^4$ time units in order to converge to the BLVs. Note: we concluded from \figureref{fig:gsv_sub_converge} that a forward transient of $10^3$ is sufficient, but we use a longer transient to be safe. For the forward dynamics, we use an interval of $10^7$ time units to prepare the GSV basis over a suitably long interval for studying CLV convergence during the backward transient.

\begin{figure}[htbp]
	\centering
	\includegraphics[width=\linewidth]{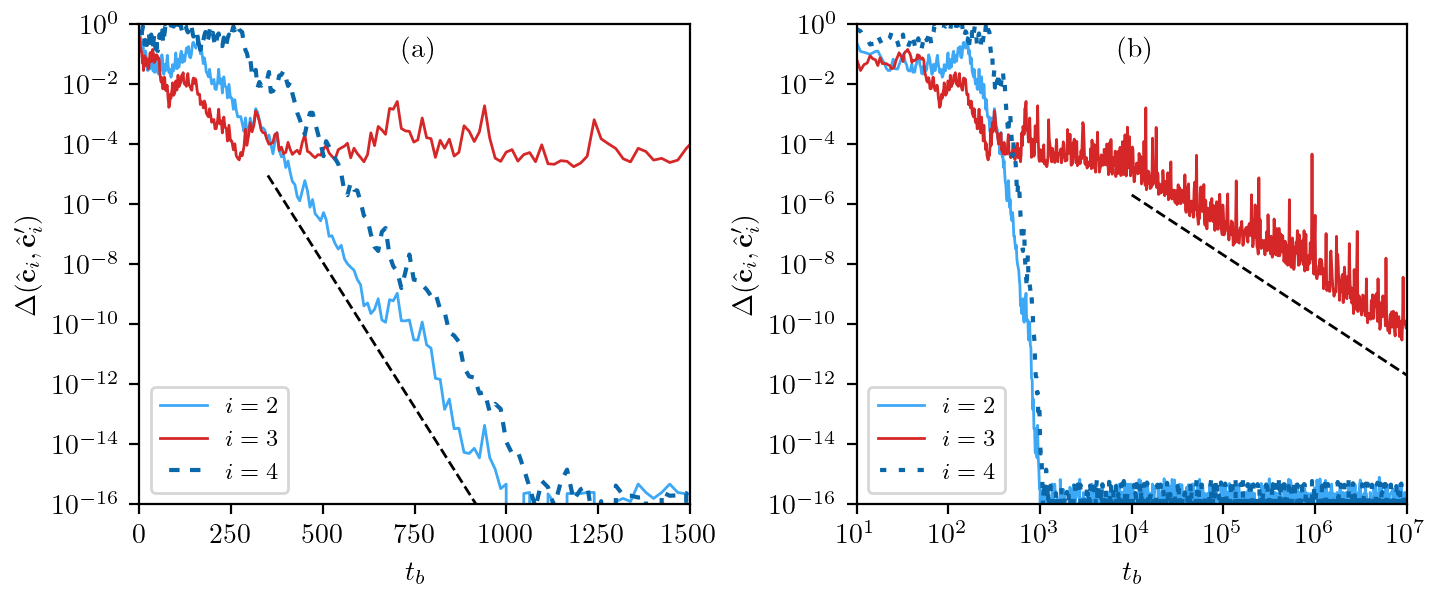}
	\caption{The distance $\Delta$ between estimates $\hat{\bs c}_i$ and $\hat{\bs c}_i'$ of the $i$-th CLV ($i=2,3,4$) as a function of backward time $t_b$, computed during the backward transient of the G\&C algorithm for the chaotic orbit with initial condition \eqref{eq:ic} in the H\'enon-Heiles system \eqref{eq:HH}. The black dashed line in panel (a) denotes a function $\propto e^{-0.044t_b}$, while in (b) it denotes a function $\propto t_b^{-2}$. Panel (a) is in log-linear scale and (b) is in log-log scale.}
	\label{fig:clv_vec_converge}
\end{figure}

In \figureref{fig:clv_vec_converge}, we give the time evolution of the distance $\Delta(\hat{\bs c}_i,\hat{\bs c}_i')$ over a backward transient of $10^7$ time units, where we save the value of $\Delta(\hat{\bs c}_i,\hat{\bs c}_i')$ at evenly spaced points in $\log t_b$ such that we have 150 data points per decade of the $t_b$-axis in log scale. We see from \figureref{fig:clv_vec_converge}(a) that $\Delta(\hat{\bs c}_i,\hat{\bs c}_i')$ decays approximately $\propto e^{-0.044t_b}$ when $i=2,4$ since the blue curves in this figure roughly follow the black dashed line. On the other hand, we see from \figureref{fig:clv_vec_converge}(b) that $\Delta(\hat{\bs c}_3,\hat{\bs c}_3')$ decays sub-exponentially $\propto t_b^{-2}$. Since $\hat{\bs c}_1$ and $\hat{\bs c}_1'$ are both initialised from the same subspace $\spn(\hat{\bs g}_1)$, the vectors $\hat{\bs c}_1$ and $\hat{\bs c}_1'$ are identical (up to a sign), so we omit the $i=1$ case from \figureref{fig:clv_vec_converge} since $\Delta(\hat{\bs c}_1,\hat{\bs c}_1')=0$. These exponential rates of convergence between $\hat{\bs c}_i$ and $\hat{\bs c}_i'$, $i=1,2,3,4$, are in agreement with their expected rates of convergence to the $i$-th CLVs of $\chi_{i-1}-\chi_i$ given in \tableref{tab:vector} in the third column.\footnote{Here, as in the case of the forward transient, we interpret instantaneous convergence as having an exponential convergence rate of $\infty$ and we interpret a sub-exponential rate as being consistent with an exponential rate of zero.}

Despite the degeneracy of the splitting subspace $\Omega_2$, it is unsurprising that the CLV estimates $\hat{\bs c}_2$ and $\hat{\bs c}_2'$ converge since they are initialised for the backward transient of the G\&C algorithm as linear combinations of the first two BLV estimates, which each have distinct LEs. However, considering that $\hat{\bs c}_3$ and $\hat{\bs c}_3'$ are initialised from the span of the first three BLV estimates, which only correspond to two distinct LEs, it is surprising that these CLV estimates converge at all (even sub-exponentially) in the presence of this degeneracy. Since $\hat{\bs c}_3$ and $\hat{\bs c}_3'$ are randomly initialised from a superset of the centre subspace $\Omega_2$, and since both vectors converge to $\Omega_2$, this suggests that generic deviation vectors in the centre subspace converge to each other under backward evolution proportional to $t_b^{-2}$. We provide an explanation for this observation in \sectionref{sec:implic} after discussing the dynamics of the centre subspace.

We now move on to study the convergence rates of subspaces to the splitting subspaces $\Omega_i$ during backward evolution. Informed by the degeneracy in the LE spectrum, we use the computed estimates of the CLVs to define the following subspaces:
\begin{align}
\begin{split}
	C_1&=\spn(\hat{\bs c}_1),\\
	C_2&=\spn(\hat{\bs c}_2,\hat{\bs c}_3),\\
	C_3&=\spn(\hat{\bs c}_4).\label{eq:cdefs}
\end{split}
\end{align}
Since the vectors used to define these subspaces converge to their respective CLVs, each $C_i$ converges to the splitting subspace $\Omega_i$, $i=1,2,3$, as $t_b\to\infty$. While we are not aware of any theoretical studies which determine these convergence rates for the backward transient alone,\footnote{Noethen analyses these convergence rates, but only in conjunction with the convergence rates of the forward transient, not the backward transient alone \cites{Noethen2019}{Noethen2019a}.} if $\dim\Omega_i=1$ holds for each splitting subspace, then it trivially follows from the convergence rates of CLVs that $C_i$ converges to $\Omega_i$ with an exponential rate of $\lambda_{i-1}-\lambda_i$ during the backward transient. However, if $\dim\Omega_i>1$ for some $i$, then it is reasonable to conjecture that a $\lambda_{i-1}-\lambda_i$ rate still holds in this case since one could take the argument used in \cite{Noethen2019}, which shows that a generic subspace of dimension $\dim\Gamma_i^-$ converges to $\Gamma_i^-$ with an exponential rate of $\lambda_{i}-\lambda_{i+1}$ as $t\to\infty$, and apply it to the restricted system with tangent space $\Gamma_i^-$ as $t\to-\infty$ \cite{noethen}. We therefore conjecture that a generic subspace of $\Gamma_i^-$ with dimension $\dim\Omega_i$ converges to $\Omega_i$ with an exponential rate of $\lambda_{i-1}-\lambda_i$ as $t\to-\infty$, which is given in the third column of \tableref{tab:subspace}.

\begin{figure}[htbp]
	\centering
	\includegraphics[width=\linewidth]{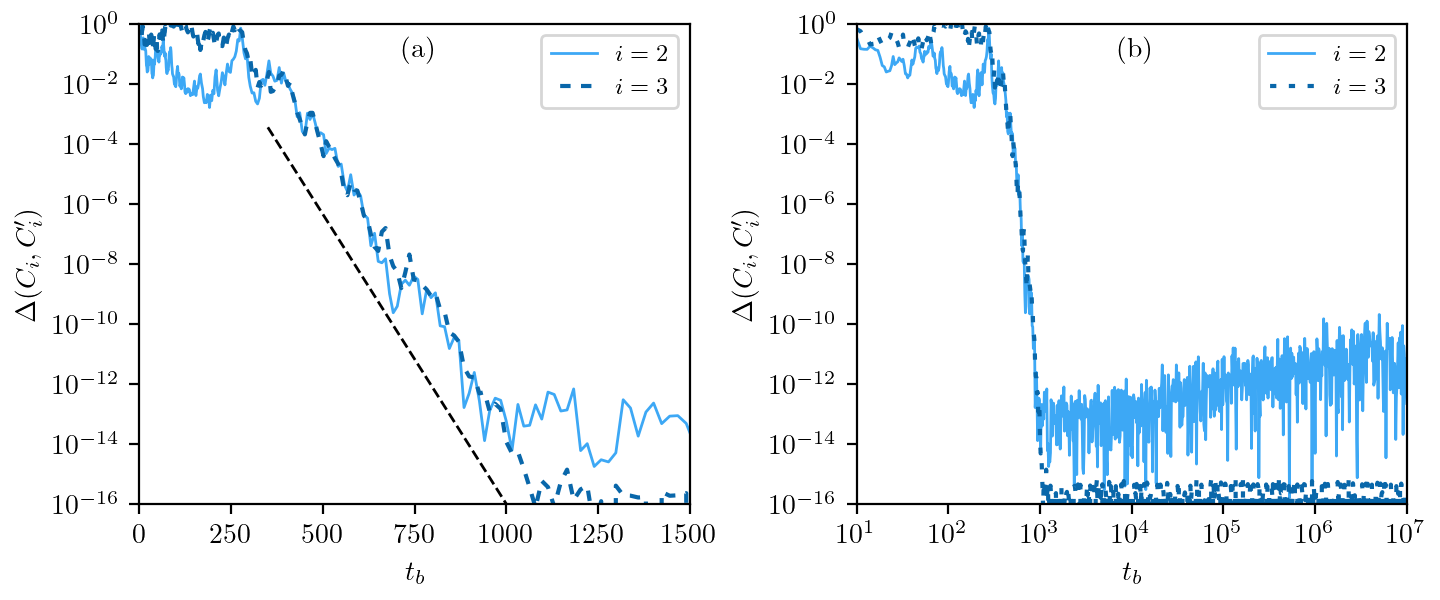}
	\caption{The distance $\Delta$ between estimates $C_i$ and $C_i'$ of the $i$-th splitting subspace ($i=2,3$) as a function of backward time $t_b$, computed during the backward transient of the G\&C algorithm for the chaotic orbit with initial condition \eqref{eq:ic} in the H\'enon-Heiles system \eqref{eq:HH}. In panel (a) the black dashed line denotes a function $\propto e^{-0.044t_b}$. Panel (a) is in log-linear scale and (b) is in log-log scale.}
	\label{fig:clv_sub_converge}
\end{figure}

In the same way that we defined the subspaces $C_i$, we construct $C_i'$ from the vectors $\hat{\bs c}_i'$, and we measure how $C_i$ and $C_i'$ converge as a proxy for measuring their convergence to $\Omega_i$. Applying the same computations of $\hat{\bs c}_i$ and $\hat{\bs c}_i'$ used to produce \figureref{fig:clv_vec_converge}, we show the time evolution of $\Delta(C_i,C_i')$ in \figureref{fig:clv_sub_converge} for $i=2,3$. We see from panel (a) of the figure that $\Delta(C_i,C_i')$ decays $\propto e^{-0.044t_b}$ for each $i$. Note that $C_1$ and $C_1'$ are identical as they are both initialised from the same 1-D subspace, so $\Delta(C_1,C_1')=0$ throughout the computation and is not shown in the figure. These exponential convergence rates between each $C_i,C_i'$ pair are all in agreement with $\lambda_{i-1}-\lambda_i$ for each $i=1,2,3$, which we conjectured to be the rate at which these subspaces converge to $\Omega_i$ during the backward transient. However, in \figureref{fig:clv_sub_converge}(b) we see that, after the initial fast convergence between $C_2$ and $C_2'$, these subspaces slowly diverge. We argue in the remainder of this section that when the G\&C algorithm is used to compute the centre subspace $\Omega_2$, the numerical error of this estimation increases over time and that this explains the observed divergence between $C_2$ and $C_2'$. We then propose a method to improve the accuracy of the $\Omega_2$ estimation.

\begin{figure}[htbp]
	\centering
	\includegraphics[width=\linewidth]{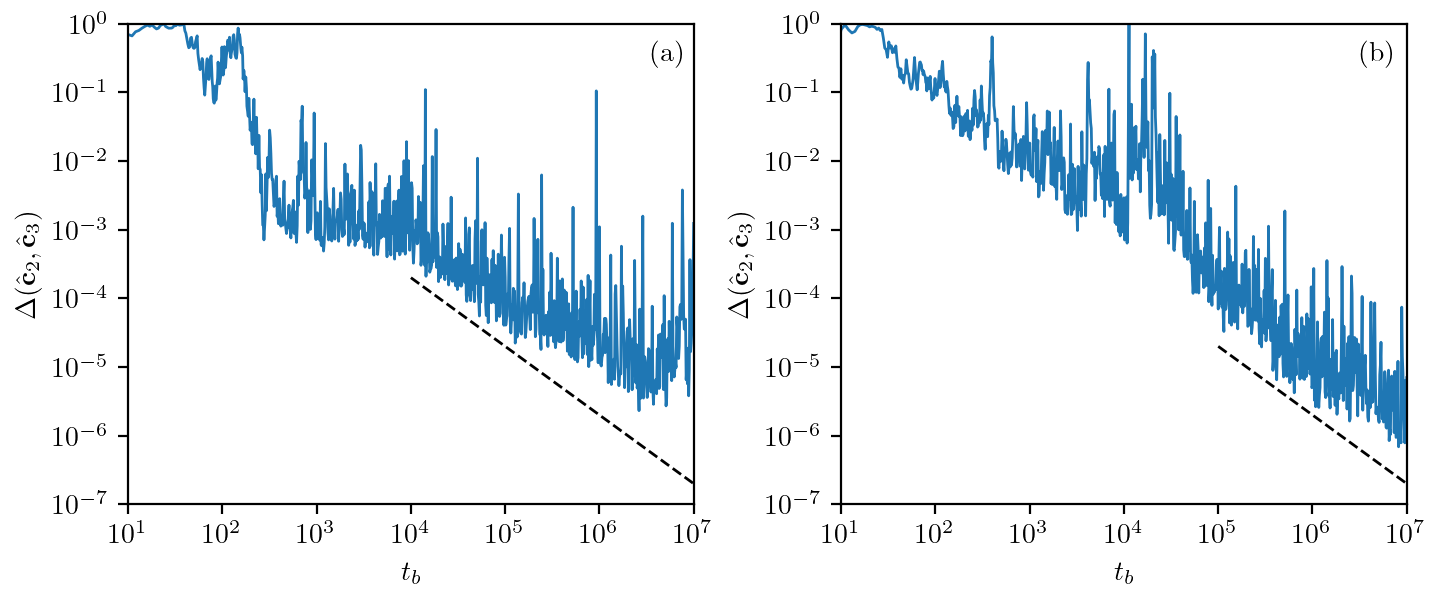}
	\caption{The distance $\Delta$ between CLV estimates $\hat{\bs c}_2$ and $\hat{\bs c}_3$ as a function of backward time $t_b$, computed during the backward transient of the G\&C algorithm for the chaotic orbit with initial condition \eqref{eq:ic} in the H\'enon-Heiles system \eqref{eq:HH}. The forward transient lengths used in the computations of (a) and (b) are $10^4$ and $10^7$ time units, respectively. The black dashed lines in both panels denote functions $\propto t_b^{-1}$, and both panels are in log-log scale.}
	\label{fig:clv_cen_long_converge}
\end{figure}

For the purpose of the backward transient, both $\hat{\bs c}_2$ and $\hat{\bs c}_3$ are initialised as vectors in $G_2$, which is an estimate of $\Gamma_2^-$. Each of these vectors converges exponentially fast to the centre subspace during backward evolution, in the sense that the minimum angle between each vector and the centre subspace decays exponentially, so the dynamics of the covariant subspace $\Omega_2$ governs the asymptotic behaviour of $\hat{\bs c}_2$ and $\hat{\bs c}_3$ during backward evolution. \hyperref[fig:clv_cen_long_converge]{Figure~\ref*{fig:clv_cen_long_converge}}(a) shows the evolution of $\Delta(\hat{\bs c}_2,\hat{\bs c}_3)$ during the backward transient, from which we see that $\hat{\bs c}_2$ and $\hat{\bs c}_3$ converge $\propto t_b^{-1}$ before beginning to diverge as seen in the final decade of the $t_b$-axis. We only explain this divergence in \sectionref{sec:cent} after discussing the dynamics of the centre subspace; for now, we simply remark that this divergent behaviour is eliminated by performing a new run with a longer forward transient, as we do in \figureref{fig:clv_cen_long_converge}(b) using a forward transient of $10^7$ time units, and from this figure we see that the $t_b^{-1}$ decay of $\Delta(\hat{\bs c}_2,\hat{\bs c}_3)$ is reaffirmed. The alignment of CLVs $\hat{\bs c}_2$ and $\hat{\bs c}_3$ computed using the G\&C algorithm for a chaotic orbit in the H\'enon-Heiles system was noted in \cite{PradoReynosoEtAl2021}, albeit for a different energy. This alignment of the CLVs defining the centre subspace implies that numerical estimations of $\Omega_2$, using the CLVs computed from the G\&C algorithm, become increasingly inaccurate as $t_b$ increases. This explains the worsening agreement between $C_2$ and $C_2'$ for large $t_b$ which we saw in \figureref{fig:clv_sub_converge}(b), since both $C_2$ and $C_2'$ become worse estimations of $\Omega_2$ over time.

\begin{figure}[htbp]
	\centering
	\includegraphics[width=\linewidth]{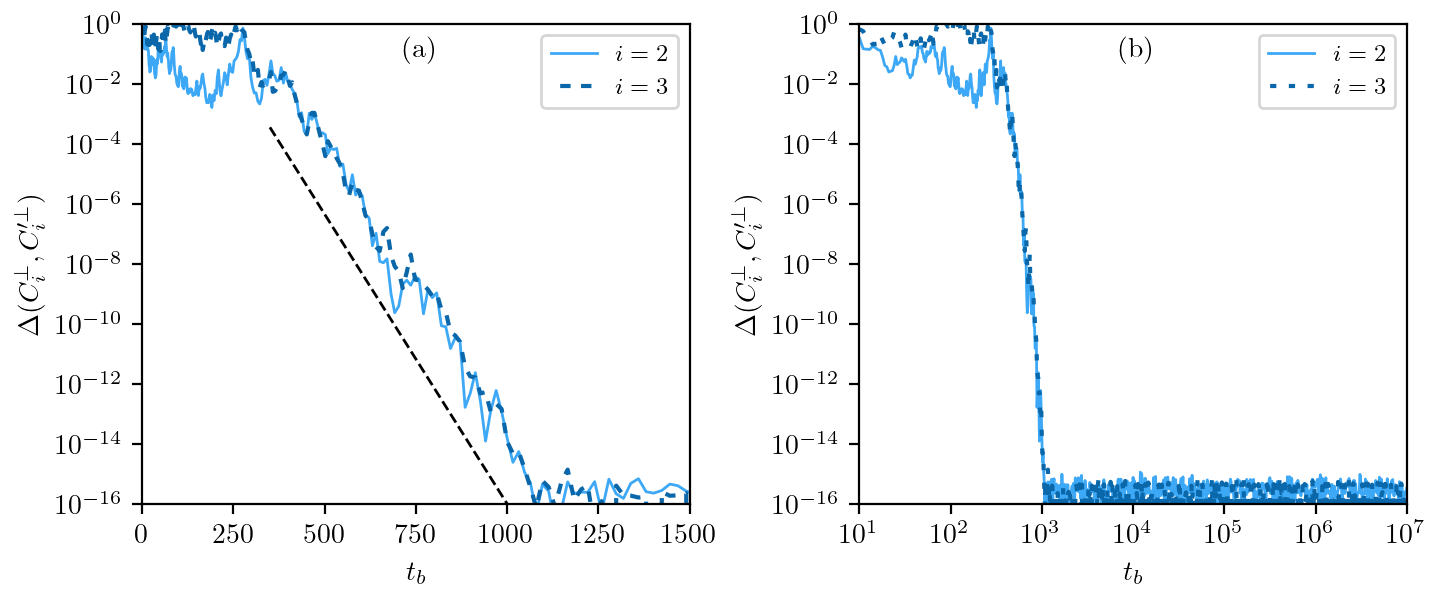}
	\caption{Similar to \figureref{fig:clv_sub_converge}, but the distance $\Delta$ is between estimates $C_i^{\perp}$ and $C_i^{\prime\perp}$ of the $i$-th splitting subspace ($i=1,2$), computed using the G\&C algorithm with the centre correction.}
	\label{fig:clv_sub_converge_corr}
\end{figure}

In light of this minor issue of estimating $\Omega_2$ using the G\&C algorithm, we propose the following solution which improves the accuracy of the $\Omega_2$ estimate by adapting the backward evolution part of the algorithm. After each backward propagation of the upper triangular matrix by $\tau$ time units and subsequent normalisation of its columns, we orthonormalise the vectors $\hat{\bs c}_2$ and $\hat{\bs c}_3$. This orthonormalisation has no effect on $\hat{\bs c}_2$, but $\hat{\bs c}_3$ becomes $\hat{\bs c}_3^{\perp}$ which is orthogonal to $\hat{\bs c}_2$. In theory, $C_2=\spn(\hat{\bs c}_2,\hat{\bs c}_3)=\spn(\hat{\bs c}_2,\hat{\bs c}_3^{\perp})$, so $C_2$ should not be affected by this operation. Numerically, however, this method allows for a more accurate computation of $\Omega_2$, making $C_2^{\perp}=\spn(\hat{\bs c}_2,\hat{\bs c}_3^{\perp})$ a more accurate estimate of the centre subspace than $C_2$. Similarly, $C_2^{\prime\perp}=\spn(\hat{\bs c}_2',\hat{\bs c}_3^{\prime\perp})$ is the modified version of $C_2'$. We refer to this modified algorithm as the G\&C algorithm with the \textit{centre correction}. To demonstrate the effectiveness of our proposed solution, we reproduce \figureref{fig:clv_sub_converge} using the same method as before, including the same initial condition and initial deviation vectors, except we now use the centre correction during the backward evolution. These results are given in \figureref{fig:clv_sub_converge_corr}, where we introduce the notation $C_i=C_i^{\perp}$ and $C_i'=C_i^{\prime\perp}$ for each $i\ne2$ (this is merely a convenience for the figure). As desired, $C_2^{\prime\perp}$ and $C_2^{\prime\perp}$ do not diverge over long times, thus improving the numerical stability of our estimate for $\Omega_2$. As a final remark, we point out that while $\hat{\bs c}_3^{\perp}$ is still an estimate of a vector in $\Omega_2$, it is no longer a \textit{covariant} vector since its evolution is not governed purely by the dynamics. The purpose of computing $\hat{\bs c}_3^{\perp}$ is merely to improve the accuracy of our $\Omega_2$ estimate, but it is not a CLV.

The goal of the backward transient is to estimate the splitting subspaces $\Omega_i$, from which the CLVs are determined. We see from \figureref{fig:clv_sub_converge_corr}(a) that $C_i^{\perp}$ and $C_i^{\prime\perp}$ converge practically up to machine error after $t_b\approx10^3$. Using this proxy method to indirectly measure the convergence of these subspaces to the relevant splitting subspaces, we conclude that a forward transient length of $10^4$ time units for the G\&C algorithm is more than sufficient in this case.

\subsection{Dynamics of the centre subspace}\label{sec:cent}
In \sectionref{sec:hhconv}, we discussed the convergence of the relevant vectors and subspaces computed during the transients of the G\&C algorithm. Furthermore, we showed that the two middle CLVs spanning the centre subspace of the H\'enon-Heiles system converge during backward evolution when computed via this algorithm. In this section, we argue that generic deviation vectors in the centre subspace typically converge to the direction of the flow $\hat{\bs w}_{\mathrm{flow}}$ at a sub-exponential rate, both forwards and backwards in time.

\begin{figure}[htbp]
	\centering
	\includegraphics[width=\linewidth]{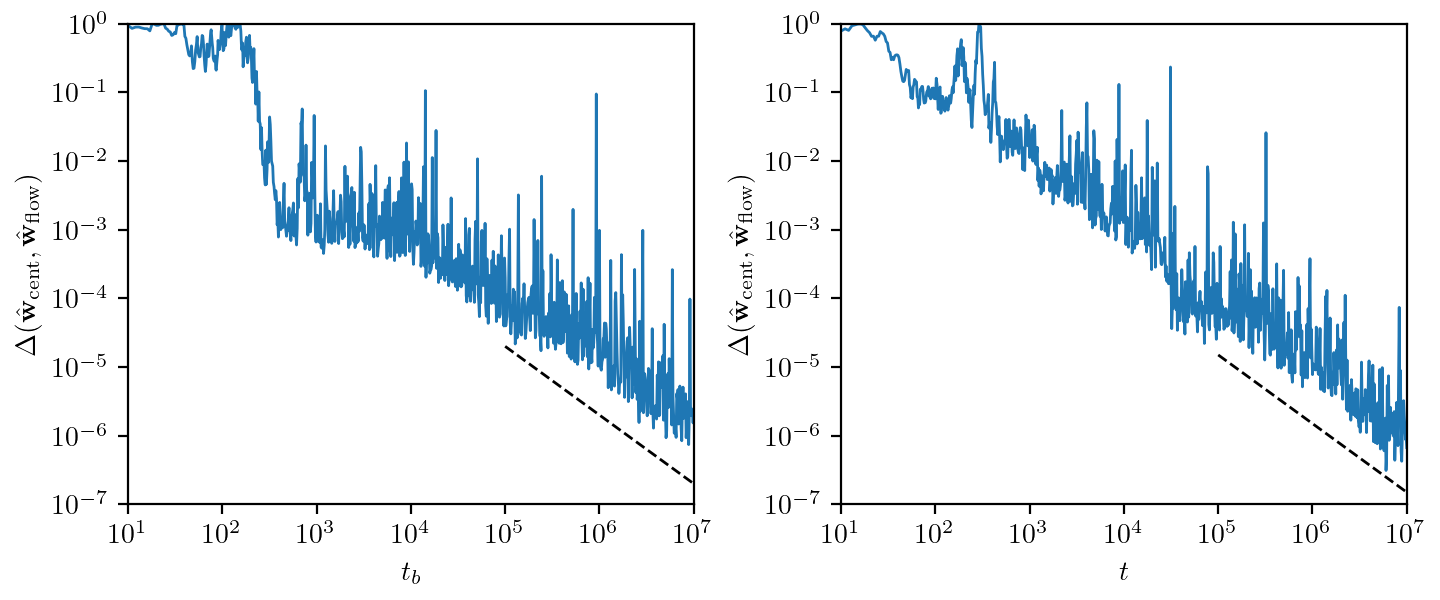}
	\caption{The time evolution of the distance $\Delta$ between a generic deviation vector $\hat{\bs w}_{\mathrm{cent}}$ in the centre subspace and the direction of the flow $\hat{\bs w}_{\mathrm{flow}}$, computed during the backward dynamics of the G\&C algorithm with the centre correction for the chaotic orbit with initial condition \eqref{eq:ic} in the H\'enon-Heiles system \eqref{eq:HH}. The time variable in (a) is backward time $t_b$ and in (b) it is forward time $t$. The dashed line in (a) denotes a function $\propto t_b^{-1}$, and in (b) it denotes a function $\propto t^{-1}$. Both panels are in log-log scale.}
	\label{fig:flow_convergence}
\end{figure}

Starting with initial condition \eqref{eq:ic}, we use the G\&C algorithm with the centre correction to estimate CLVs over an interval of $10^7$ time units, with forward and backward transients of $10^4$ time units.\footnote{This means that the ``forward dynamics'' part of the algorithm is $10^7+10^4$ time units.} After the backward transient is complete, we initialise a generic unit vector $\hat{\bs w}_{\mathrm{cent}}$ in the computed centre subspace by taking a random linear combination of $\hat{\bs c}_2$ and $\hat{\bs c}_3^{\perp}$; we also compute $\hat{\bs w}_{\mathrm{flow}}$, converted to the GSV basis so that it can be compared with $\hat{\bs w}_{\mathrm{cent}}$. In \figureref{fig:flow_convergence}(a), we give the backward evolution of $\Delta(\hat{\bs w}_{\mathrm{cent}},\hat{\bs w}_{\mathrm{flow}})$. Since the evolution of $\Delta(\hat{\bs w}_{\mathrm{cent}},\hat{\bs w}_{\mathrm{flow}})$ in this figure follows the trend of the black dashed line with slope $-1$, we conclude that generic vectors in the centre subspace converge to the direction of the flow $\propto t_b^{-1}$.

We wish to complement the previous result by computing the forward evolution of a generic $\hat{\bs w}_{\mathrm{cent}}$ to see if it also converges to $\hat{\bs w}_{\mathrm{flow}}$ as $t\to\infty$. However, our approach only worked in the backward evolution case for the following reason. Recall that initially $\hat{\bs w}_{\mathrm{cent}}\in\spn(\hat{\bs c}_2,\hat{\bs c}_3)$, which implies $\hat{\bs w}_{\mathrm{cent}}\in\spn(\hat{\bs c}_1,\hat{\bs c}_2,\hat{\bs c}_3)=\spn(\hat{\bs g}_1,\hat{\bs g}_2,\hat{\bs g}_3)$. Now, since we evolve $\hat{\bs w}_{\mathrm{cent}}$ in the GSV basis, the G\&C algorithm ensures that this vector remains in the subspace $\spn(\hat{\bs g}_1,\hat{\bs g}_2,\hat{\bs g}_3)$ under backward evolution because the $\hat{\bs g}_4$ component is exactly zero and cannot inadvertently be changed by the numerics. So if the numerical backward evolution of $\hat{\bs w}_{\mathrm{flow}}$ introduces any unwanted perturbation away from $\spn(\hat{\bs c}_2,\hat{\bs c}_3)$, then that perturbation necessarily has a $\hat{\bs c}_1$ component (and not a $\hat{\bs c}_4$ component) which will exponentially decay under backward evolution, meaning our computation is stable. Unfortunately, simply using the relevant $R$ matrices to propagate a generic $\hat{\bs w}_{\mathrm{cent}}$ in the forward time direction would not work, since any small component of $\hat{\bs c}_1$ would grow exponentially and $\hat{\bs w}_{\mathrm{flow}}$ would quickly escape the centre subspace. As an alternative approach, we simply perform the exact same computation as in \figureref{fig:flow_convergence}(a) but using a negative time step, so the ``forward'' parts of the G\&C algorithm are now technically moving backwards in time, and the ``backward'' part moves forwards in time. This method produces CLVs in the past of our initial condition instead of the future. Note that the direction of the flow simply changes sign when reversing the direction of time, so $\spn(\hat{\bs w}_{\mathrm{flow}})$ is invariant under time reversal. This approach allows us to reliably evolve a generic $\hat{\bs w}_{\mathrm{cent}}$ during the ``backward dynamics'' of the algorithm (which is actually forwards in time). We use this method to compute the forward time evolution of $\Delta(\hat{\bs w}_{\mathrm{cent}},\hat{\bs w}_{\mathrm{flow}})$ and present the results in \figureref{fig:flow_convergence}(b). From this figure, we conclude that generic vectors in the centre subspace converge to the direction of the flow $\propto t^{-1}$ during forward evolution since $\Delta(\hat{\bs w}_{\mathrm{cent}},\hat{\bs w}_{\mathrm{flow}})$ follows the black dashed line.

These results show that arbitrary vectors in the centre subspace converge to $\hat{\bs w}_{\mathrm{flow}}$ like $t^{-1}$ under forward evolution and $t_b^{-1}$ under backward evolution. This conclusion is not unique to the initial condition \eqref{eq:ic}, as we have found the same results for several chaotic orbits with different initial conditions. We can now use this result to explain the strange divergence between $\hat{\bs c}_2$ and $\hat{\bs c}_3$ seen for large $t_b$ in \figureref{fig:clv_cen_long_converge}(a). During the forward transient, $\spn(\hat{\bs g}_1,\hat{\bs g}_2)$ converges to a subspace of $\Gamma_2^-$, and also note that $\Gamma_2^-$ contains $\Omega_2$. If we assume that $\spn(\hat{\bs g}_1,\hat{\bs g}_2)$ has completely converged and thus is a subspace of $\Gamma_2^-$, then it follows that $\dim(\spn(\hat{\bs g}_1,\hat{\bs g}_2)+\Omega_2)\leq3$ since both $\spn(\hat{\bs g}_1,\hat{\bs g}_2)$ and $\Omega_2$ are subspaces of the $3$-D subspace $\Gamma_2^-$. It then follows from an elementary dimension-counting theorem of linear algebra (see e.g.\ \cite[p.~47]{Axler2015}) that
\begin{align}
	\dim(\spn(\hat{\bs g}_1,\hat{\bs g}_2)\cap\Omega_2)&=\dim(\spn(\hat{\bs g}_1,\hat{\bs g}_2))+\dim(\Omega_2)-\dim(\spn(\hat{\bs g}_1,\hat{\bs g}_2)+\Omega_2)\notag\\
	&\geq 2 + 2 - 3\\
	&\geq 1.\notag
\end{align}
Therefore, $\spn(\hat{\bs g}_1,\hat{\bs g}_2)$ contains a non-trivial vector from $\Omega_2$, but recall from the discussion around \figureref{fig:flow_convergence}(b) that generic vectors in $\Omega_2$ converge to $\hat{\bs w}_{\mathrm{flow}}$ like $t^{-1}$. We therefore conclude that the subspace $\spn(\hat{\bs g}_1,\hat{\bs g}_2)$ computed during the forward transient converges to a subspace that contains $\hat{\bs w}_{\mathrm{flow}}$, but only at a sub-exponential rate due to the slow dynamics of the centre subspace, meaning that $\spn(\hat{\bs g}_1,\hat{\bs g}_2)$ only contains a poor approximation of $\hat{\bs w}_{\mathrm{flow}}$ when these GSVs are evolved over a short forward transient of $10^4$. Meanwhile during the backward dynamics, $\hat{\bs c}_3$ is free to converge to $\hat{\bs w}_{\mathrm{flow}}$ but $\hat{\bs c}_2\in\spn(\hat{\bs g}_1,\hat{\bs g}_2)$, so $\hat{\bs c}_2$ is constrained by the poor convergence of $\spn(\hat{\bs g}_1,\hat{\bs g}_2)$ to a subspace containing $\hat{\bs w}_{\mathrm{flow}}$, which only gets worse for large $t_b$. This explains the divergence seen between $\hat{\bs c}_2$ and $\hat{\bs c}_3$ in \figureref{fig:clv_cen_long_converge}(a) and also why the issue was remedied by using a longer transient of $10^7$ in \figureref{fig:clv_cen_long_converge}(b), since $\spn(\hat{\bs g}_1,\hat{\bs g}_2)$ would then contain a good approximation of $\hat{\bs w}_{\mathrm{flow}}$.

There is, however, an easier approach to ensure that $\spn(\hat{\bs g}_1,\hat{\bs g}_2)$ contains a vector that well approximates $\hat{\bs w}_{\mathrm{flow}}$ and avoids the need for an extremely long forward transient. Since we have already established that $\spn(\hat{\bs g}_1,\hat{\bs g}_2)$ converges to a subspace containing $\hat{\bs w}_{\mathrm{flow}}$, we can simply set $\hat{\bs g}_2=\hat{\bs w}_{\mathrm{flow}}$ at initialisation of the GSVs, orthonormalising the GSVs as usual, and thus bypass the slow centre dynamics.

\subsubsection{Shear dynamics}\label{sec:implic}
Our results from \figureref{fig:flow_convergence} have shown that generic deviation vectors in the centre subspace converge to the direction of the flow like $t^{-1}$ when evolved forwards in time and like $t_b^{-1}$ when evolved backwards in time. Here, we examine the restricted linear map which evolves the centre subspace in time; in particular, we compute the matrix of this linear map in a convenient basis. In doing so, we will see the $t^{-1}$ and $t_b^{-1}$ convergence behaviour emerge, and we will also explain the $t_b^{-2}$ mutual convergence of $\hat{\bs c}_3$ and $\hat{\bs c}_3'$ seen in \figureref{fig:clv_vec_converge}(b).

Since any chaotic orbit in the H\'enon-Heiles system has a 2-D covariant centre subspace, there is a linear map which evolves deviation vectors between centre subspaces at different times, and this linear map can be represented as a $2\times2$ matrix. However, such a matrix representation could be expressed using any basis vectors from the relevant centre subspaces. Since $\hat{\bs w}_{\mathrm{flow}}$ is covariant and we wish to uncover the dynamics of generic deviation vectors in the centre subspace (hereafter, \textit{centre vectors}) in relation to $\hat{\bs w}_{\mathrm{flow}}$, we choose $\hat{\bs w}_{\mathrm{flow}}$ to be our first basis vector. For the other basis vector of the centre subspace, a reasonable choice would be one which preserves angles in the centre subspace from the coordinate basis, hence we choose the second basis vector to be the unit centre vector orthogonal to $\hat{\bs w}_{\mathrm{flow}}$, which is unique up to a sign. We denote this vector by $\hat{\bs w}_{\mathrm{flow}}^{\perp}$.

Before proceeding, a quick note on notation and terminology. Let $\Omega$ and $\Omega'$ be the $2$-D centre subspaces of tangent spaces $T_{\bs x(k\tau)}\mathcal {M}$ and $T_{\bs x((k+1)\tau)}\mathcal {M}$, respectively, for some $k\in\mathbb{N}$. The \textit{restriction} of the linear propagator $\mathrm{d}_{\bs x(k\tau)}\Phi^{\tau}$ to $\Omega$ and $\Omega'$ is defined by
\begin{align}
\begin{split}
	\mathrm{d}_{\bs x(k\tau)}\Phi^{\tau}|_{\Omega}^{\Omega'}:\Omega&\longrightarrow \Omega'\\
	\bs\omega&\longmapsto\mathrm{d}_{\bs x(k\tau)}\Phi^{\tau}(\bs\omega),
\end{split}\label{eq:cool}
\end{align}
where this definition only makes sense if $\mathrm{d}_{\bs x(k\tau)}\Phi^{\tau}(\Omega)\subseteq \Omega'$ \cite[p.~75]{SteffenEtAl2018}, which is the case here due to the covariance of the centre subspace. Finally, we denote the \textit{matrix} of $\mathrm{d}_{\bs x(k\tau)}\Phi^{\tau}$ with respect to the ordered bases\footnote{We assume all bases are ordered in this discussion.} $\mathcal B=(\beta_1,...,\beta_N)$ and $\mathcal B'=(\beta_1',...,\beta_N')$ by
\begin{align}
	\big[\mathrm{d}_{\bs x(k\tau)}\Phi^{\tau}\big]_{\mathcal B}^{\mathcal B'}.\label{eq:bracketdefinition}
\end{align}

\begin{figure}[htbp]
	\centering
	\includegraphics[width=0.7\linewidth]{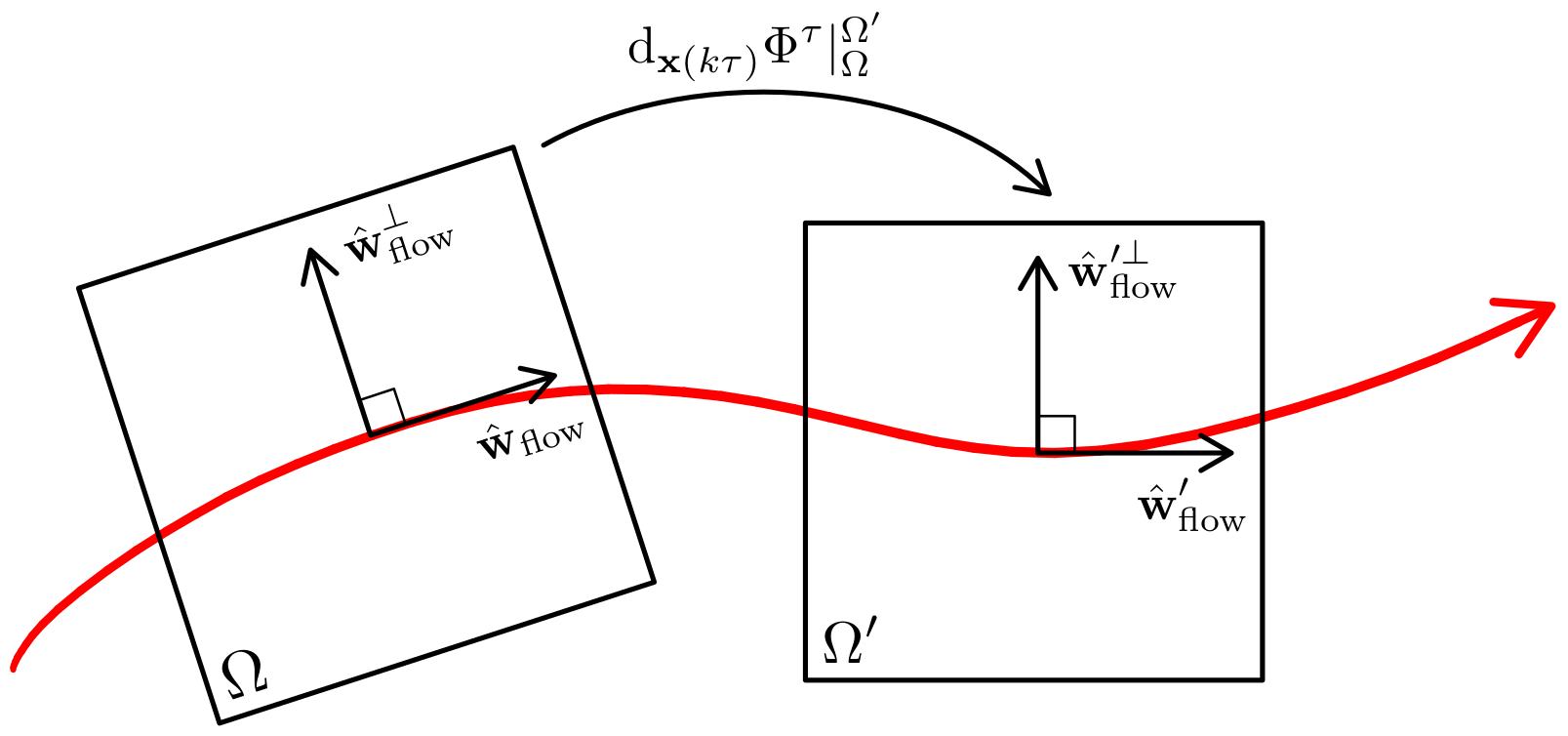}
	\caption{A diagram representing the linear propagator $\smash{\mathrm{d}_{\bs x(k\tau)}\Phi^{\tau}|_{\Omega}^{\Omega'}}$ \eqref{eq:cool} restricted to the $2$-D centre subspaces $\Omega$ and $\Omega'$ (drawn as black squares) along an orbit (drawn as a red curve). The perpendicular arrows in $\Omega$ and $\Omega'$ represent a particular orthonormal basis \eqref{eq:particularbasis} of each centre subspace.}
	\label{fig:restrict}
\end{figure}

Assume that we are able to compute the matrix of $\mathrm{d}_{\bs x(k\tau)}\Phi^{\tau}$ with respect to some orthonormal bases. We now wish to compute the matrix of $\smash{\mathrm{d}_{\bs x(k\tau)}\Phi^{\tau}|_{\Omega}^{\Omega'}}$ with respect to the orthonormal bases $\overline{\mathcal B}=(\beta_1,\beta_2)$ and $\overline{\mathcal B}'=(\beta_1',\beta_2')$. Motivated by our earlier discussion, we choose the following bases of $\Omega$ and $\Omega'$, respectively:
\begin{align}
\begin{split}
	\overline{\mathcal B}&=(\hat{\bs w}_{\mathrm{flow}},\hat{\bs w}_{\mathrm{flow}}^{\perp}),\\
	\overline{\mathcal B}'&=(\hat{\bs w}_{\mathrm{flow}}',\hat{\bs w}_{\mathrm{flow}}^{\prime\perp}).\label{eq:particularbasis}
\end{split}
\end{align}
A diagram of this setup is given in \figureref{fig:restrict}. Now, how do we compute these proposed basis vectors? Recall from \sectionref{sec:back_trans} that we introduced the G\&C algorithm with the centre correction where we compute the two orthonormal centre vectors $\hat{\bs c}_2$ and $\hat{\bs c}_3^{\perp}$. Since generic centre vectors converge to the direction of the flow (assuming sufficiently long transients), it follows that $\hat{\bs c}_2$ converges to $\hat{\bs w}_{\mathrm{flow}}$ during backward evolution. Therefore, $\hat{\bs c}_2$ and $\hat{\bs c}_3^{\perp}$ can be used as our respective computed estimates of $\hat{\bs w}_{\mathrm{flow}}$ and $\hat{\bs w}_{\mathrm{flow}}^{\perp}$ in the GSV basis. Similarly, we use $\hat{\bs c}_2'$ and $\hat{\bs c}_3^{\prime\perp}$ as estimates for $\hat{\bs w}_{\mathrm{flow}}'$ and $\hat{\bs w}_{\mathrm{flow}}^{\prime\perp}$.

Recall from \sectionref{sec:ginelli} that the matrix $R((k+1)\tau)$ computed during the forward dynamics of the G\&C algorithm is a matrix representation of the linear propagator $\mathrm{d}_{\bs x(k\tau)}\Phi^{\tau}$ over a single orthonormalisation interval $\tau$ with respect to the GSV bases. Given this matrix of the linear propagator, how do we compute the matrix of the restricted linear propagator $\smash{\mathrm{d}_{\bs x(k\tau)}\Phi^{\tau}|_{\Omega}^{\Omega'}}$ with respect to the bases $\overline{\mathcal B}$ and $\overline{\mathcal B}'$, that is, $\smash{\big[\mathrm{d}_{\bs x(k\tau)}\Phi^{\tau}|_{\Omega}^{\Omega'}\big]_{\raisebox{-0.5pt}{$\scriptstyle\overline{\mathcal B}$}}^{\raisebox{-2pt}{$\scriptstyle\overline{\mathcal B}'$}}}$? To simplify the notation, we denote this matrix by $\overline{M}(k\tau,(k+1)\tau)$ and we claim that
\begin{align}
	\overline{M}(k\tau,(k+1)\tau)=(\hat{\bs c}_2'\,\ \hat{\bs c}_3^{\prime\perp})\transpose R((k+1)\tau)(\hat{\bs c}_2\,\ \hat{\bs c}_3^{\perp}),\label{eq:bases}
\end{align}
where $(\hat{\bs c}_2'\,\ \hat{\bs c}_3^{\prime\perp})$ is an $N\times2$ matrix with orthonormal columns $\hat{\bs c}_2'$ and $\hat{\bs c}_3^{\prime\perp}$, and $(\hat{\bs c}_2\,\ \hat{\bs c}_3^{\perp})$ is an $N\times2$ matrix with orthonormal columns $\hat{\bs c}_2$ and $\hat{\bs c}_3^{\perp}$. We show in \appendixref{app:basis} that \eqref{eq:bases} holds. By iteratively multiplying these matrices together for successive values of $k$, we can compute $\overline{M}(0,t)$ for an arbitrary time $t=K\tau$, $K\in\mathbb{N}$, as follows:
\begin{align}
	\overline{M}(0,t)=\overline{M}((K-1)\tau,K\tau)\cdots\overline{M}(\tau,2\tau)\overline{M}(0,\tau).\label{eq:comb}
\end{align}

\begin{figure}[htbp]
	\centering
	\includegraphics[width=0.55\linewidth]{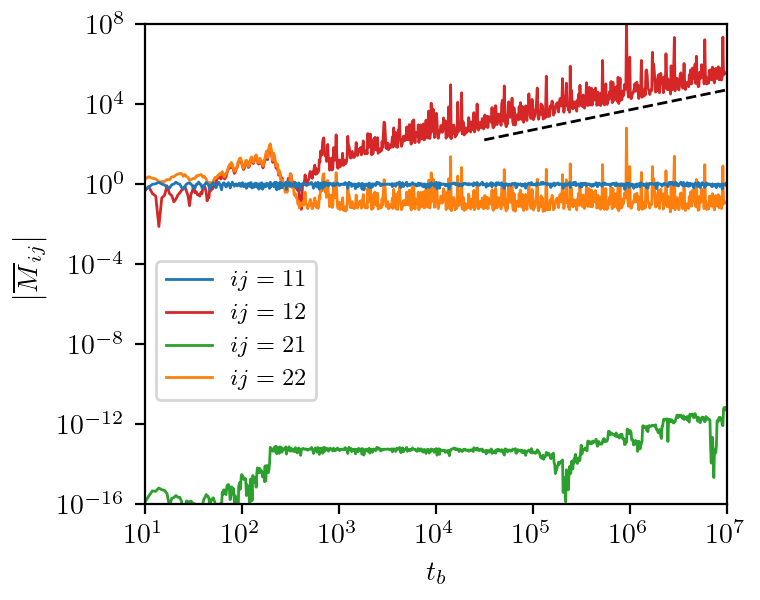}
	\caption{The absolute values of entries $\overline{M}_{ij}$ of the matrix \eqref{eq:comb} of the restricted linear propagator as a function of backward time $t_b$, computed during the backward dynamics of the G\&C algorithm with the centre correction for the chaotic orbit with initial condition \eqref{eq:ic} in the H\'enon-Heiles system \eqref{eq:HH}. The dashed line denotes a function $\propto t_b$. The figure is in log-log scale.}
	\label{fig:shear}
\end{figure}

Choosing the initial condition \eqref{eq:ic}, we compute the matrices $\overline{M}(k\tau,(k+1)\tau)$ according to \eqref{eq:bases} after each orthonormalisation interval $\tau$ during the backward dynamics of the G\&C algorithm with the centre correction. As suggested earlier in \sectionref{sec:cent}, we initially set $\hat{\bs g}_2=\hat{\bs w}_{\mathrm{flow}}$ so that a forward transient of $10^4$ is sufficient to ensure that $\hat{\bs c}_2$ is an accurate estimate of $\hat{\bs w}_{\mathrm{flow}}$. Note that this choice of $\hat{\bs g}_2$ is not essential, but it is convenient; alternatively, a long transient of $10^7$ can be used, but we use the first approach to save CPU time. For the backward transient, we again use a length of $10^4$ time units. Using \eqref{eq:comb}, we iteratively compute $\overline{M}(0,t_b)$ after each orthonormalisation interval during the backward dynamics, where $t_b=0$ corresponds to $T_2$ (see \figureref{fig:alg}). The time evolution of the entries of $\overline{M}(0,t_b)$, which we denote by $\overline{M}_{ij}(0,t_b)$, are given in \figureref{fig:shear}. Note that we take the absolute values of the matrix entries $\overline{M}_{ij}$ in the figure so that we can give the results in log scale even when some entries are negative. In particular, $M_{21}$ is a small value oscillating around zero, its magnitude remaining under $10^{-11}$, while $M_{12}$ becomes negative after an initial transient and remains negative throughout. From \figureref{fig:shear}, we see that $\overline{M}_{11}$ and $\overline{M}_{22}$ remain roughly constant over time, fluctuating near a value of $1$, while $\overline{M}_{21}\approx0$, and $\overline{M}_{12}(0,t_b)\approx kt_b$ for some constant $k$. Hence,
\begin{align}
	\overline{M}(0,t_b)\approx\begin{pmatrix}1&kt_b\\0&1\end{pmatrix}.\label{eq:shear}
\end{align}
Since this matrix is a simple shear matrix with shear factor $kt_b$, we conclude that centre vectors shear along the direction of the flow.\footnote{Note that some centre vectors will align with $\hat{\bs w}_{\mathrm{flow}}$ and some will anti-align.} Since the inverse of a shear matrix of the form \eqref{eq:shear} is also a shear matrix with a negated shear factor, it follows that centre vectors also shear along the direction of the flow during backward evolution, just in the opposite direction. The form \eqref{eq:shear} of the matrix of the restricted linear propagator is not particular to \eqref{eq:ic}, since we obtained the same result when using several different chaotic orbits.

As an informal aside, the dynamics of the centre subspace which we have discussed has an interesting parallel to regular motion. In autonomous Hamiltonian systems with $n$ degrees of freedom, regular orbits evolve on an $n$-D invariant torus, and generic deviation vectors converge to the tangent space of the torus $\propto t^{-1}$ \cites{SkokosEtAl2003}{Skokos2010}. In particular, consider a periodic orbit in a system with a single degree of freedom (such as the oscillation of a \mbox{$1$-D} pendulum) where each invariant torus is \mbox{$1$-D} and the centre subspace is the entire \mbox{$2$-D} tangent space at each state since the system is not chaotic. Generic deviation vectors converge to the direction of the flow $\propto t^{-1}$, and the linear transformation governing this tangent dynamics can also be shown to be a shear. Therefore, it appears that the tangent dynamics of the centre subspace for a generic aperiodic, chaotic orbit in the H\'enon-Heiles system behaves like the tangent dynamics of a regular periodic orbit for a $1$-D pendulum, where the orbit acts like a $1$-D invariant torus of infinite length.

We now demonstrate that the $t^{-1}$ convergence of centre vectors to the direction of the flow as $t\to\pm\infty$ (see \figureref{fig:flow_convergence}) emerges from the form of the restricted linear propagator given in \eqref{eq:shear}. Consider the shear matrix presented in \eqref{eq:shear}. For convenience, we assume $k$ is positive, but the result is the same for negative $k$. In the bases chosen for that matrix, $(1,0)\transpose$ represents the direction of the flow $\hat{\bs w}_{\mathrm{flow}}$ and orthonormal to that is $\hat{\bs w}_{\mathrm{flow}}^{\perp}$, represented by $(0,1)\transpose$. Considering the evolution of these vectors under the action of \eqref{eq:shear}, the vector $(1,0)\transpose$ is unchanged, while $(0,1)\transpose$ gets mapped to $(kt,1)\transpose$, i.e.\ it shears in the direction of $(1,0)\transpose$. Therefore, vectors do indeed converge to the direction of the flow under this transformation, but at what rate? To address this, we normalise these converging vectors, $(1,0)\transpose$ and $(kt,1)\transpose$, and consider the magnitude of their difference,
\begin{align}
	\left\lVert\frac{1}{\sqrt{k^2t^2+1}}\begin{pmatrix}kt\\1\end{pmatrix}-\begin{pmatrix}1\\0\end{pmatrix}\right\rVert=\sqrt{\frac{\left(kt-\sqrt{k^2t^2+1}\right)^2+1}{k^2t^2+1}}.\label{eq:presurd}
\end{align}
When the angle $\theta$ between the unit vectors is small, the norm of this difference is approximately equal to $\theta$. Since $\theta\approx\sin\theta$ for small $\theta$, the asymptotic behaviour of \eqref{eq:presurd} is the same as the distance $\Delta$ defined in \eqref{eq:distance} in terms of principal angles, of which there is only one angle in this case. Simplifying the contents of the square root in \eqref{eq:presurd}, it follows that for large $t$,
\begin{align}
	\Delta\left(\begin{pmatrix}1\\0\end{pmatrix},\begin{pmatrix}kt\\1\end{pmatrix}\right)\approx\sqrt{2-\frac{2kt}{\sqrt{k^2t^2+1}}}.\label{eq:surd}
\end{align}
In \appendixref{app:conv}, we show that
\begin{align}
	\sqrt{2-\frac{2kt}{\sqrt{k^2t^2+1}}}\sim \frac{1}{kt},\label{eq:showequiv}
\end{align}
where the symbol $\sim$ denotes asymptotic equivalence (see e.g.\ \cite[p.~4]{Murray1984}), so the vectors $(1,0)\transpose$ and $(kt,1)\transpose$ converge $\propto t^{-1}$. Indeed, we observed such convergence rates between centre vectors and the direction of the flow in \figureref{fig:flow_convergence}.

Recall that our earlier observation from \figureref{fig:clv_vec_converge}(b) suggested that arbitrary vectors in the centre subspace converge $\propto t_b^{-2}$ during backward evolution. Using the result that generic deviation vectors converge to the direction of the flow $\propto t^{-1}$, we will now show that generic deviation vectors converge \textit{to each other} $\propto t^{-2}$ during forward evolution.\footnote{Similarly, this convergence during backward evolution is $\propto t_b^{-2}$ since the inverse of \eqref{eq:shear} is also a shear matrix.} It is worth noting that there is no contradiction between this result we are about to show and \sectionref{sec:converge} where we argued that two deviation vectors which converge exponentially to a particular direction will converge to each other at the \textit{same} exponential rate since in the current discussion we are dealing with \textit{sub}-exponential rates.

\begin{figure}[htbp]
	\centering
	\includegraphics[width=0.75\linewidth]{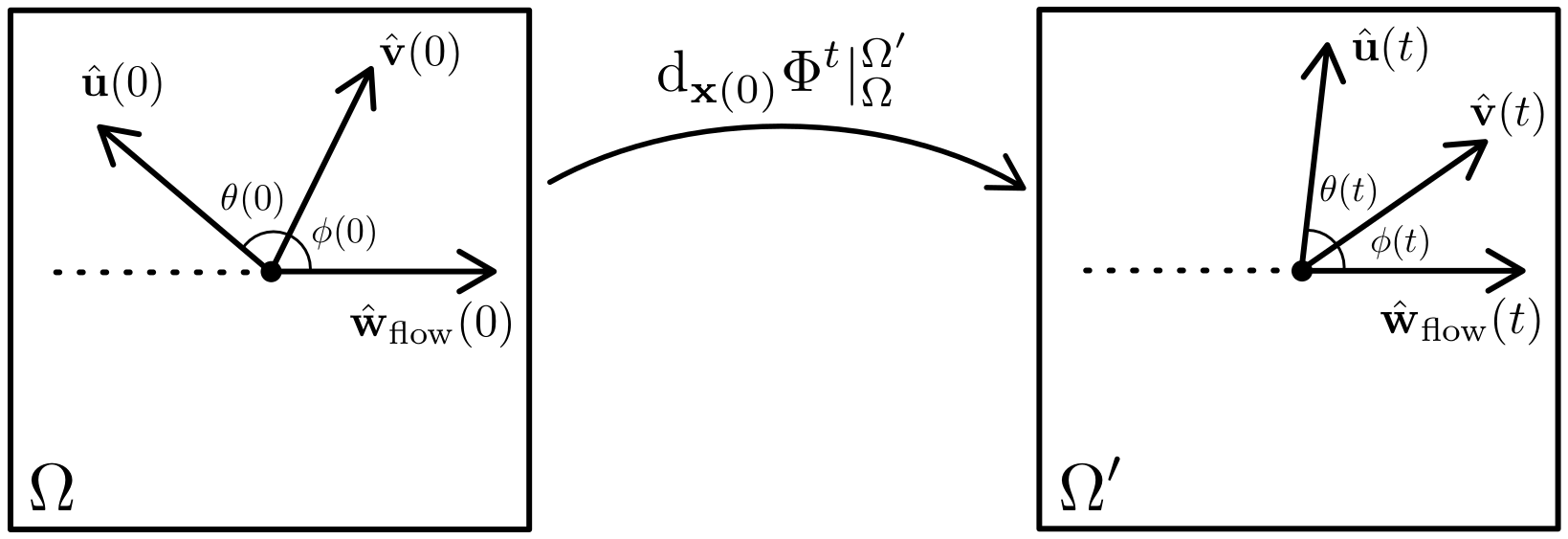}
	\caption{A diagram representing the linear propagator $\mathrm{d}_{\bs x(0)}\Phi^{t}|_{\Omega}^{\Omega'}$ restricted to the $2$-D centre subspaces $\Omega$ and $\Omega'$ (drawn as black squares) at times $0$ and $t$, respectively. The angle between unit vectors $\hat{\bs u}$ and $\hat{\bs v}$ is $\theta$, and the angle between $\hat{\bs v}$ and $\hat{\bs w}_{\mathrm{flow}}$ is $\phi$.}
	\label{fig:restricted_convergence}
\end{figure}

Consider the evolution of any two generic centre vectors $\hat{\bs u}(0)$ and $\hat{\bs v}(0)$ under the restricted linear propagator, which can be represented by the shear matrix \eqref{eq:shear}. For convenience, we keep the evolving vectors $\hat{\bs u}(t)$ and $\hat{\bs v}(t)$ normalised at all points in time, since we are only interested in the evolution of the 1-D subspaces they each span. We know that $\hat{\bs u}(t)$ and $\hat{\bs v}(t)$ converge to $\hat{\bs w}_{\mathrm{flow}}(t)$ proportional to $t^{-1}$ (see \figureref{fig:flow_convergence}). We assume that the $\hat{\bs w}_{\mathrm{flow}}^{\perp}(t)$ components of $\hat{\bs u}(0)$ and $\hat{\bs v}(0)$ are positive; if they are not, we can always negate the vectors to get the desired vectors without affecting their spans. Furthermore, we assume without loss of generality that the minimum angle between $\hat{\bs v}(t)$ and $\hat{\bs w}_{\mathrm{flow}}(t)$ is less than the minimum angle between $\hat{\bs u}(t)$ and $\hat{\bs w}_{\mathrm{flow}}(t)$. A diagram of this setup is given in \figureref{fig:restricted_convergence}, where $\Omega$ and $\Omega'$ are the centre subspaces at times $0$ and $t$, respectively.

Since the shear matrix \eqref{eq:shear} represents the restricted linear propagator, and assuming that the shear factor is positive, both $\hat{\bs u}(t)$ and $\hat{\bs v}(t)$ align with $\hat{\bs w}_{\mathrm{flow}}(t)$ as $t\to\infty$. Since both vectors are converging to the same direction, it follows that $\hat{\bs u}(t)$ is simply a lagged version of $\hat{\bs v}(t)$, so $\hat{\bs u}(t)=\hat{\bs v}(t-a)$ where $a$ is the time by which $\hat{\bs u}(t)$ is lagged behind $\hat{\bs v}(t)$ as they both converge to the direction of the flow. Now consider the magnitude
\begin{align}
	\left\lVert\hat{\bs u}(t)-\hat{\bs v}(t)\right\rVert=\left\lVert\hat{\bs v}(t-a)-\hat{\bs v}(t)\right\rVert
\end{align}
of the difference between $\hat{\bs v}(t)$ and $\hat{\bs u}(t)=\hat{\bs v}(t-a)$, which is approximately equal to the angle $\theta(t)$ between the two vectors when the angle is small, i.e.\ when $t$ is large. Let $\phi(t)$ be the angle between $\hat{\bs v}(t)$ and $\hat{\bs w}_{\mathrm{flow}}(t)$, then $\phi(t-a)$ is the angle between $\hat{\bs u}(t)=\hat{\bs v}(t-a)$ and $\hat{\bs w}_{\mathrm{flow}}(t)$, from which it follows that
\begin{align}
	\theta(t)=\phi(t-a)-\phi(t).
\end{align}
But generic centre vectors converge to $\hat{\bs w}_{\mathrm{flow}}(t)$ like $t^{-1}$, so $\phi(t)\approx bt^{-1}$ for large $t$ and some constant $b$, hence
\begin{align}
	\left\lVert\hat{\bs v}(t-a)-\hat{\bs v}(t)\right\rVert \approx\theta(t)=\phi(t-a)-\phi(t)\approx\frac{b}{t-a}-\frac{b}{t} = \frac{ab}{t(t-a)}\sim \frac{ab}{t^2}.\label{eq:sim}
\end{align}
Since $\theta(t)\approx\sin\theta(t)$ for large $t$, this implies that $\Delta(\hat{\bs u}(t),\hat{\bs v}(t))\sim ab/t^2$, so the generic centre vectors $\hat{\bs u}(t)$ and $\hat{\bs v}(t)$ converge to each other $\propto t^{-2}$.

\subsection{Angles between splitting subspaces}\label{sec:prado}
The minimum angles between pairs of CLVs were computed in \cite{PradoReynosoEtAl2021} for the H\'enon-Heiles system \eqref{eq:HH} with an energy of $H_2=1/6$. The authors of that paper did this in order to detect violations of the system's hyperbolicity. A system is said to be \textit{hyperbolic} if, for each state $\bs x$, the tangent space $T_{\bs x}\mathcal{M}$ can be decomposed as a direct sum of the unstable, centre, and stable subspaces \cite{GinelliEtAl2013}, as given in \eqref{eq:coarse}. Note that some authors (e.g.\ \cites{HasselblattPesin2011}[p.~22]{Kuznetsov2012}) use the term \textit{partially hyperbolic} to describe such systems, but we simply refer to them as hyperbolic. For hyperbolic systems, the minimum angle between the stable and unstable subspaces is bounded away from zero \cite{XuPaul2016}, since any non-trivial intersection between these subspaces implies the tangent space cannot be fully decomposed in the sense of \eqref{eq:coarse}. Violations of hyperbolicity therefore occur when the minimum angle between the stable and unstable subspaces vanishes and usually indicates the presence of so-called homoclinic tangencies \cite{GinelliEtAl2013}. We extend the numerical investigation of \cite{PradoReynosoEtAl2021} by computing the minimum angles between the splitting subspaces along a chaotic orbit in the H\'enon-Heiles system with an energy of $H_2=1/8$. We choose to study angles between splitting subspaces instead of angles between CLVs because the 2-D centre subspace is uniquely defined, unlike the CLVs contained therein. Note that, as mentioned at the start of \sectionref{sec:hhsection}, we use a time step of 0.02 in this section.

\begin{figure}[htbp]
	\centering
	\includegraphics[width=\linewidth]{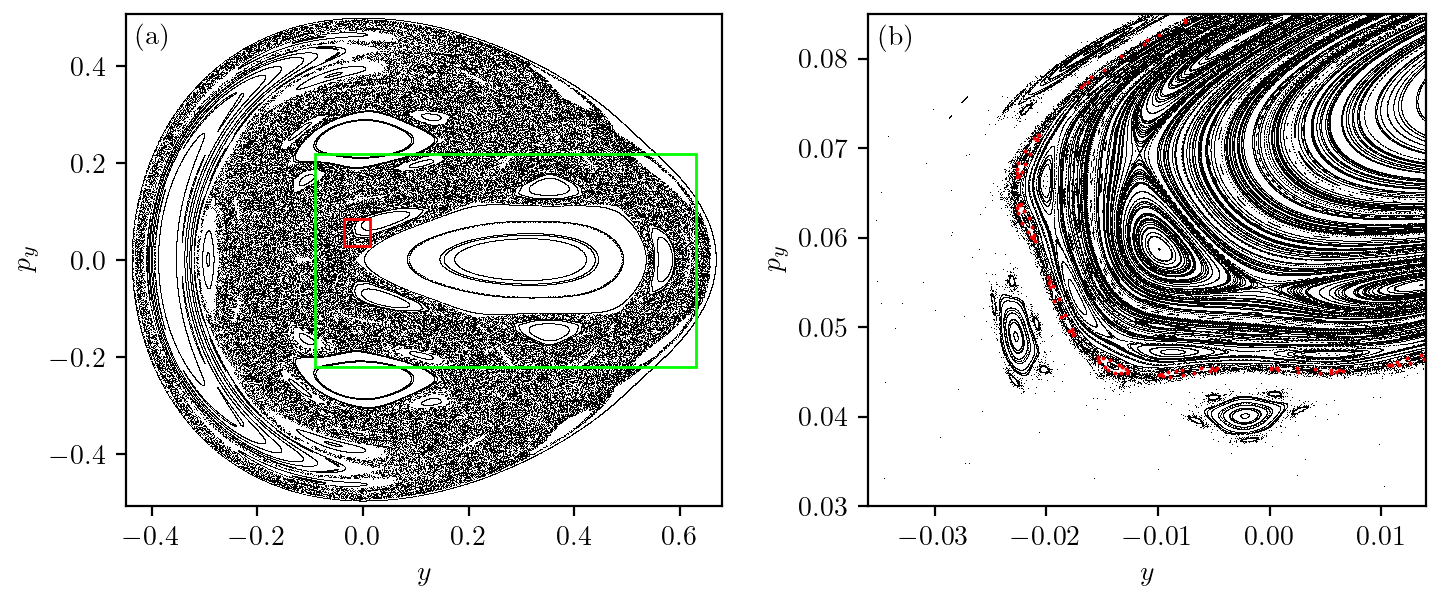}
	\caption{Panel (a) is the $x=0$, $p_x>0$ PSS of the H\'enon-Heiles system \eqref{eq:HH} with energy $H_2=1/8$. Panel (b) is a zoomed-in PSS of the region bordered by the red rectangle in (a), where only those points that do not leave the green rectangular region of (a) over an interval of $10^5$ time units are plotted in (b). The red dots in (b) denote initial condition \eqref{eq:supersticky} and its orbit's consequent intersections with the PSS over the $10^5$ interval.}
	\label{fig:zoom}
\end{figure}

We begin by selecting a \textit{sticky orbit}, i.e.\ an orbit which spends some time very close to an island of stability and behaves almost like a regular orbit (see e.g.\ \cite{ContopoulosHarsoula2010}). In \figureref{fig:zoom}(a), we reproduce the PSS from \figureref{fig:pss} but with two rectangular regions outlined in red and green. From the PSS, we see a large island of stability in the region bordered by the green rectangle, and that large island of stability is surrounded by five smaller islands. We wish to select a sticky orbit which stays near the borders of the five small islands for a long time. Since regular orbits crossing any one of these small islands of stability also cross the other four small islands, we search for a sticky initial condition in the region surrounded by the red rectangle in \figureref{fig:zoom}(a) whose orbit remains near the five small islands (and so inside the green rectangle) for the duration of its stickiness phase. In particular, we try to find a sticky orbit which remains inside the green rectangular region for at least $10^5$ time units. In \figureref{fig:zoom}(b), we give a higher-resolution PSS of the red rectangular region from \figureref{fig:zoom}(a) using a $24\times24$ grid of initial conditions in that region, but we only plot those points on the PSS which belong to orbits which remain inside the green rectangular region in \figureref{fig:zoom}(a) for at least $10^5$ time units. By doing so, we find that most of the chaotic orbits from the red rectangular region are excluded in \figureref{fig:zoom}(b) because they diffuse to the large chaotic sea outside the red rectangular region. However, along the border of the largest island of stability in \figureref{fig:zoom}(b), there are a few points which would make good candidates for sticky initial conditions. We pick the following initial condition which we will show to be sticky:
\begin{align}
	(y,p_y)=(-0.00991304348, 0.0826086957).\label{eq:supersticky}
\end{align}
This initial condition and the consequent intersections of its orbit with the PSS are shown as red dots in \figureref{fig:zoom}(b).

\begin{figure}[htbp]
	\centering
	\includegraphics[width=\linewidth]{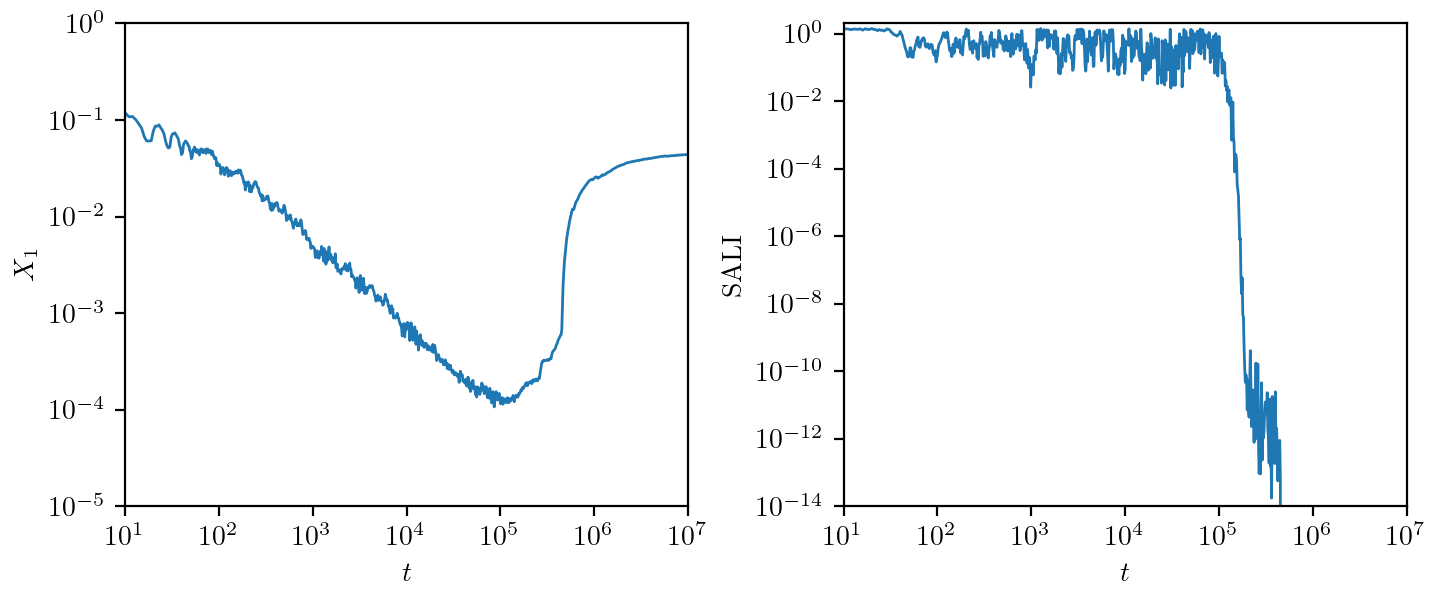}
	\caption{The time evolution of (a) the ftmLE $X_1$ \eqref{eq:finite_mle} and (b) the SALI \eqref{eq:sali} using the initial condition \eqref{eq:supersticky} for a sticky orbit in the H\'enon-Heiles system \eqref{eq:HH}. The ftmLE in (a) is computed for an integration time of $10^7$, while the SALI is only computed until it falls below a threshold of $10^{-14}$. Both panels are in log-log scale.}
	\label{fig:sali_sticky}
\end{figure}

In order to confirm that this proposed sticky orbit is indeed chaotic, we utilise two chaos indicators: the ftmLE $X_1$ \eqref{eq:finite_mle} and the SALI \eqref{eq:sali}. We compute the ftmLE over an integration time of $10^7$ time units, but we only compute the SALI until its value falls below a value of $10^{-14}$; this small threshold was chosen for classifying an orbit as chaotic since the SALI goes to zero for such orbits. The time evolution of the ftmLE and the SALI are presented in panels (a) and (b) of \figureref{fig:sali_sticky}, respectively. We see from \figureref{fig:sali_sticky}(a) that the ftmLE starts off decreasing steadily towards zero until around $t=10^5$ time units when it starts increasing and eventually saturates to a value of $X_1(10^7)=0.044$, which coincides with our mLE estimate in \sectionref{sec:les} for a different chaotic orbit. In \figureref{fig:sali_sticky}(b), the SALI remains approximately constant for over $10^5$ time units before finally decaying to zero. Up until $10^5$ units, neither of the chaos indicators suggests that the orbit is chaotic, which is a consequence of the orbit being sticky. Nevertheless, both chaos indicators correctly identify the sticky orbit as chaotic instead of regular when computed for longer times.

\begin{figure}[htbp]
	\centering
	\includegraphics[width=\linewidth]{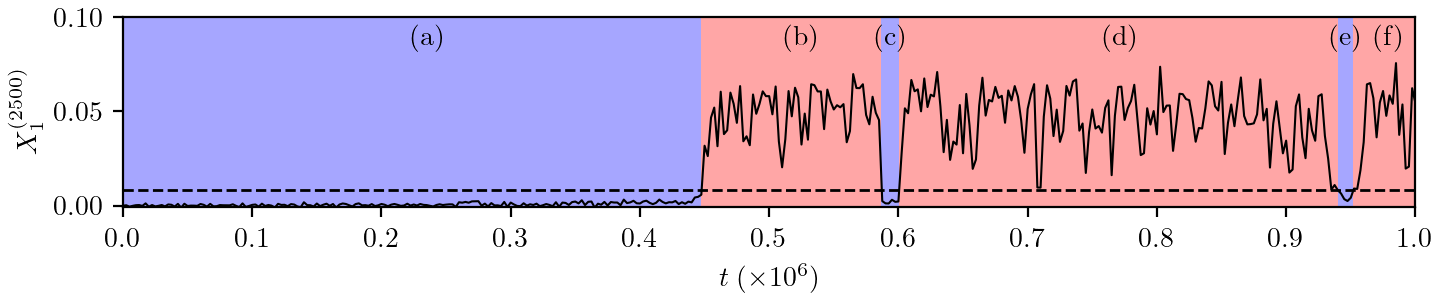}
	\caption{The time evolution of the wmLE $X_1^{(2500)}$ \eqref{eq:wmledef} with a time window of $t_w=2500$, using the initial condition \eqref{eq:supersticky} in the H\'enon-Heiles system \eqref{eq:HH}. The alternating blue and red intervals, labelled (a)--(f), indicate regular and chaotic regimes, respectively, and are distinguished by the threshold $X_1^{(2500)}=0.0083$, denoted by the dashed line.}
	\label{fig:ftle}
\end{figure}

We know the initial condition \eqref{eq:supersticky} is sticky, but we now wish to determine over which time intervals the orbit exhibits sticky behaviour. To do this, we compute \textit{windowed mLEs} (wmLEs) which are given by
\begin{align}
	X_1^{(t_w)}(t)=\frac{1}{t_w}\ln\frac{\|\bs w(t+t_w)\|}{\|\bs w(t)\|},\label{eq:wmledef}
\end{align}
Note that unlike the ftmLE $X_1(t)$, which is the time average of the logarithm of stretching factors up to time $t$, the wmLE $X_1^{(t_w)}(t)$ is the time average over some short window $t_w$ and thus gives insight into the stability properties of the tangent space during that short time. Using our sticky initial condition \eqref{eq:supersticky}, we compute $X_1^{(2500)}(t)$ over a time interval of $10^6$ time units and present the results in \figureref{fig:ftle}. We use a time window of $t_w=2500$ because we find it is small enough to capture the varied dynamics and large enough to avoid significant noise, which appears in the time evolution of the wmLE when using short time windows. The choice of initial deviation vector for computing $X_1^{(2500)}(t)$ is important since the deviation vector would typically undergo some transient phase as it converges to the first CLV, and this transient is particularly long in this case because our choice of a sticky orbit will slow down the convergence rate to the first CLV. We chose the first CLV as our initial deviation vector for this computation, and we will explain how we computed it retroactively in a few paragraphs. Returning to \figureref{fig:ftle}, we see that the wmLE is approximately zero on the interval from $t=0$ to around $t=0.45\times10^6$, after which $X_1^{(2500)}$ fluctuates around $0.05$ for some time.\footnote{Note: this wmLE value of $0.05$ roughly equals our mLE estimate of $0.044$ from \figureref{fig:sali_sticky}(a).} We see from \figureref{fig:ftle} that the wmLE exhibits different behaviour over different time intervals, where either $\smash{X_1^{(2500)}}\approx0$ or $\smash{X_1^{(2500)}}\approx0.05$. Taking a similar approach to that of \cite{PradoReynosoEtAl2021}, we set a threshold in order to distinguish between these two regimes of motion: a \textit{regular regime} when the wmLE is below the threshold and a \textit{chaotic regime} when the wmLE is above the threshold. We choose this arbitrary threshold to be $\smash{X_1^{(2500)}}=0.0083$, which is given as a dashed line in \figureref{fig:ftle}. The regular and chaotic regimes are highlighted in blue and red (respectively) in the figure.

\begin{figure}[htbp]
	\centering
	\includegraphics[width=\linewidth]{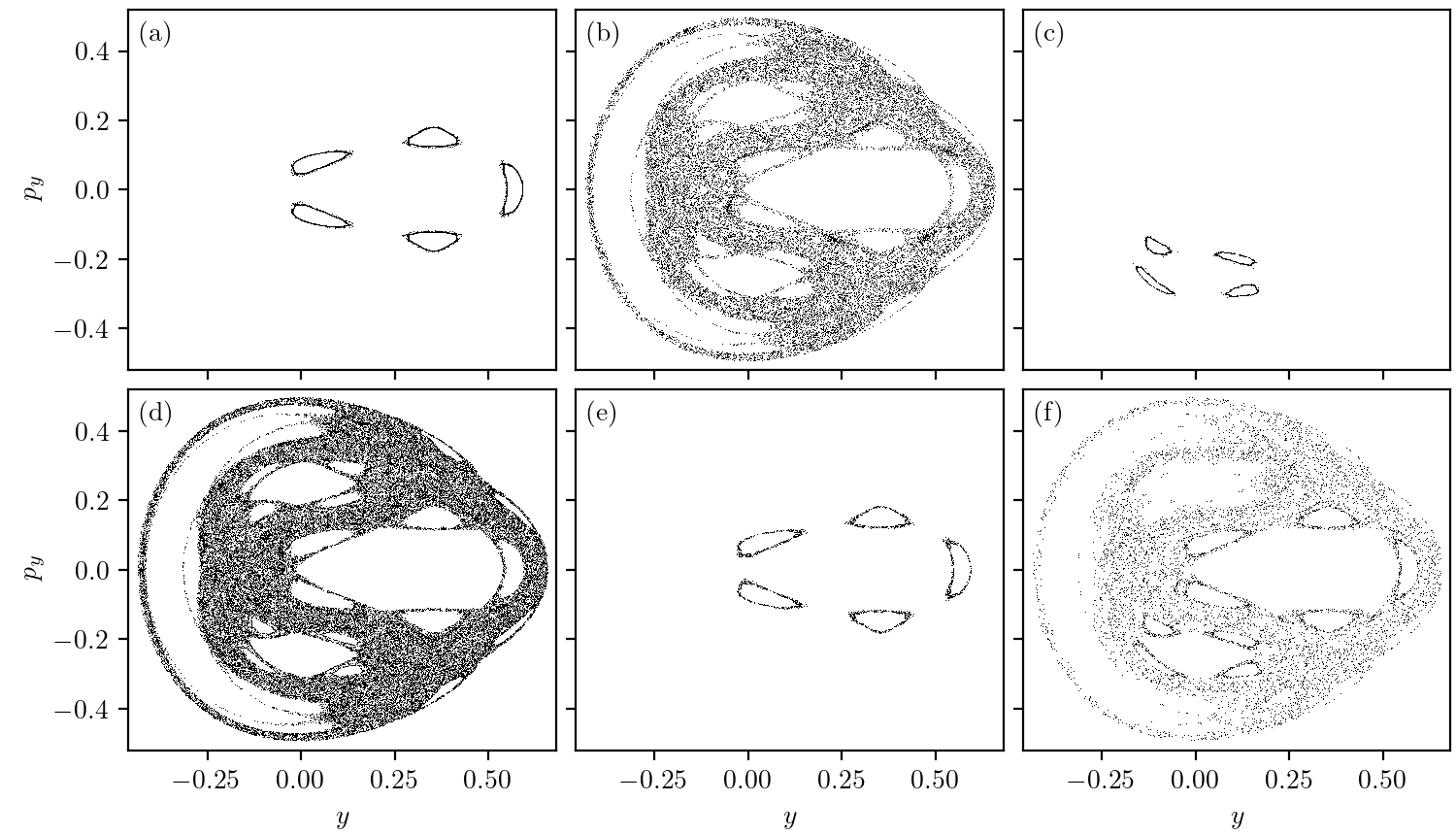}
	\caption{The $x=0$, $p_x>0$ PSS of the H\'enon-Heiles system \eqref{eq:HH} with energy $H_2=1/8$ constructed from a single orbit with initial condition \eqref{eq:supersticky}. The segments of this orbit used to construct panels (a)--(f) correspond to the time intervals labelled (a)--(f) in \figureref{fig:ftle}.}
	\label{fig:psss}
\end{figure}

To visualise the orbit of initial condition \eqref{eq:supersticky} during the regular and chaotic regimes labelled (a)--(f) from \figureref{fig:ftle}, we compute the $x=0$, $p_x>0$ PSS over each of these regimes using only this orbit and present each PSS in the respective panels (a)--(f) of \figureref{fig:psss}. Comparing these figures, we see that the orbit's intersections with the PSS remain near some islands of stability during regular regimes, indicating that the orbit is close to an invariant torus during such times. During chaotic regimes, however, the orbit's intersections with the PSS spread throughout the chaotic sea.

\begin{figure}[htb]
	\centering
	\includegraphics[width=\linewidth]{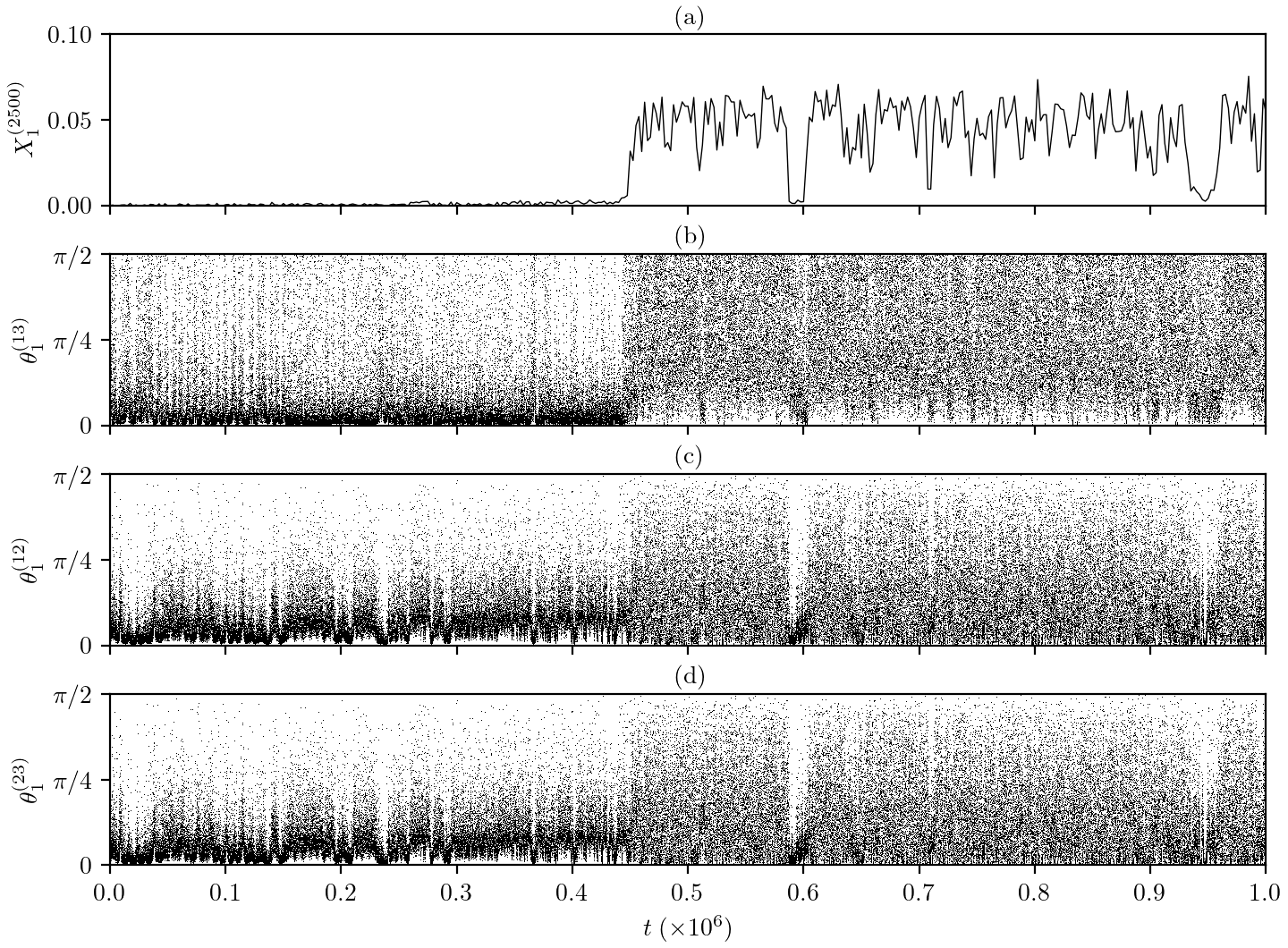}
	\caption{The time evolution of (a) the wmLE $X_1^{(2500)}$ reproduced from \figureref{fig:ftle}, (b) $\smash{\theta_1^{(13)}}$, (c) $\smash{\theta_1^{(12)}}$, and (d) $\smash{\theta_1^{(23)}}$, where $\smash{\theta_1^{(ij)}}$ is the minimum angle between estimates $C_i$ and $C_j$ of the respective splitting subspaces $\Omega_i$ and $\Omega_j$ for the H\'enon-Heiles system \eqref{eq:HH} using the orbit with initial condition \eqref{eq:supersticky}. Note: the data in panels (b)--(d) are plotted as points without connecting lines. All panels share the same horizontal axis.}
	\label{fig:clv_angles}
\end{figure}

Using the method described in \sectionref{sec:retro}, we compute the GSVs retroactively with a transient length of $5\times10^5$ time units.\footnote{We could not use short transient of $10^4$, which we used in \sectionref{sec:hhconv}, due to the stickiness of this orbit, so the GSVs fail to converge in such a short time.} Note that we used the first GSV from this computation as the initial deviation vector for the wmLE computation in \figureref{fig:ftle}. With an initial set of BLV estimates, we can perform the G\&C algorithm with the centre correction while skipping the forward transient; doing this, we compute the CLVs over a $10^6$ time interval, saving the CLVs every $10$ time units. Using these CLVs, we compute the subspaces $C_i$ ($i=1,2,3$) defined in \eqref{eq:cdefs}, which estimate the splitting subspaces $\Omega_i$. Calculating the minimum angle $\smash{\theta_1^{(ij)}}$ between each pair of distinct subspaces $C_i$ and $C_j$, we present the time evolution of these angles in panels (b)--(d) of \figureref{fig:clv_angles}. For convenience of comparison, we have reproduced the time evolution of the wmLE $\smash{X_1^{(2500)}}$ in \figureref{fig:clv_angles}(a). We see from \figureref{fig:clv_angles} that when $\smash{X_1^{(2500)}}\approx0$, the angles $\smash{\theta_1^{(ij)}}$ typically remain near zero, while the values of $\smash{\theta_1^{(ij)}}$ are distributed more homogeneously in the interval $[0,\pi/2]$ when the wmLE fluctuates around $\smash{X_1^{(2500)}}\approx0.05$. We therefore conclude that each of these minimum angles tends to be close to zero during regular regimes and away from zero during chaotic regimes. In particular, this conclusion for the minimum angle $\smash{\theta_1^{(13)}}$ between the stable and unstable subspaces suggests that more frequent violations of hyperbolicity occur in the H\'enon-Heiles system during regular regimes, indicating the presence of nearby homoclinic tangencies. Similar observations and conclusions regarding hyperbolicity were made in \cite{PradoReynosoEtAl2021} for the H\'enon-Heiles system with a different energy.

\section{A 3-D Hamiltonian system}\label{sec:3d}
In \sectionref{sec:hhconv}, we studied the convergence of various vectors and subspaces computed using the G\&C algorithm for the H\'enon-Heiles system to determine the transient lengths needed when computing CLVs. In this section, we repeat that numerical investigation for an autonomous Hamiltonian system with three degrees (hereafter, \textit{the 3-D system}) whose Hamiltonian function $H_3$ is given in \eqref{eq:ham3d}.

For our computations throughout this section, we use the ABA864 integrator \eqref{eq:aba} with a time step of 0.025, which we find keeps the relative energy error $E_r$ \eqref{eq:energyerror} below $3\times10^{-10}$, and orthonormalise any computed deviation vectors every $\tau=1$ time unit. We use the following initial condition,
\begin{align}
	x=y=z=0,\quad p_x=\sqrt{\frac{0.06}{\omega_x}},\quad p_y=\sqrt{\frac{0.06}{\omega_y}},\quad p_z=\sqrt{\frac{0.06}{\omega_z}},\label{eq:3dic}
\end{align}
which sets the energy to $H_3=0.09$. This initial condition has been used in numerical investigations where it was found to correspond to a chaotic orbit of this system \cites{BenettinEtAl1980a}[p.~114]{BountisSkokos2012}. Although we use this single initial condition \eqref{eq:3dic} in our computations, we obtained the same results when using several different initial conditions for chaotic orbits.

\begin{figure}[htbp]
	\centering
	\includegraphics[width=0.55\linewidth]{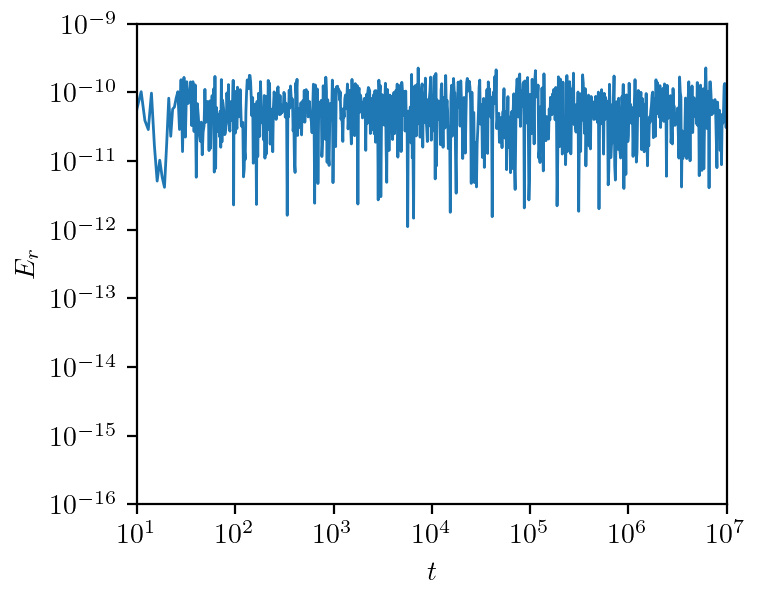}
	\caption{The relative energy error $E_r$ as a function of time $t$ for the chaotic orbit with initial condition \eqref{eq:3dic} in the 3-D system \eqref{eq:ham3d}. The figure is in log-log scale.}
	\label{fig:error_3d}
\end{figure}

\begin{figure}[htbp]
	\centering
	\includegraphics[width=\linewidth]{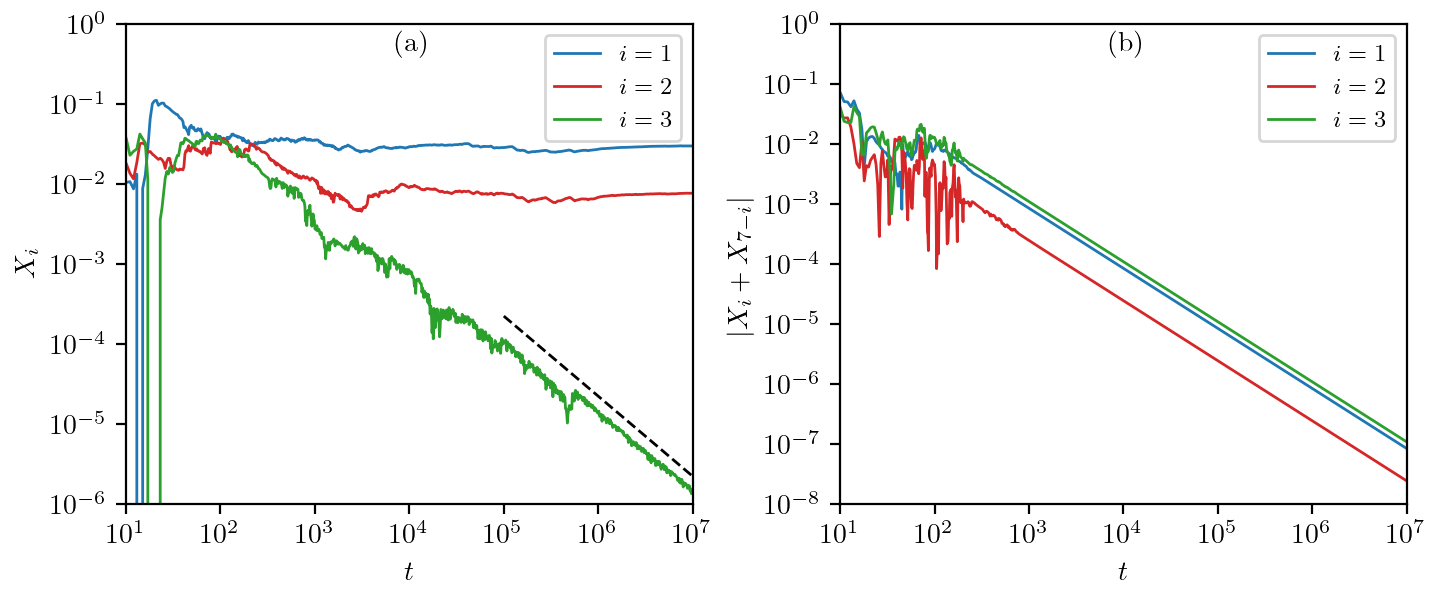}
	\caption{The time evolution of the ftLEs $X_i$, $i=1,2,3$, is given in (a) for the orbit with initial condition \eqref{eq:3dic} in the 3-D system \eqref{eq:ham3d}. The black dashed line in (a) denotes a function $\propto t^{-1}$. The time evolution of $|X_i+X_{7-i}|$ is given in panel (b) for $i=1,2,3$. Both panels are in log-log scale.}
	\label{fig:lces_3d}
\end{figure}

In order to estimate the LE spectrum of the orbit with initial condition \eqref{eq:3dic}, we compute the ftLEs $X_i(t)$, $i=1,2,\dots,6$, over an integration time of $10^7$ time units. During this computation, we calculate the time evolution of the relative energy error $E_r$ and plot the result in \figureref{fig:error_3d}, where we see that $E_r<3\times 10^{-10}$ and that $E_r$ is approximately constant over time. The time evolution of the first three ftLEs and of the quantity $|X_i+X_{7-i}|$ are given in panels (a) and (b) of \figureref{fig:lces_3d}, respectively. From \figureref{fig:lces_3d}(a), we see that the first two ftLEs saturate to values of $X_1(10^7)=0.030$ and $X_2(10^7)=0.008$, while $X_3(t)$ decays to zero $\propto t^{-1}$, which suggests $\chi_3=0$. Furthermore, since $|X_i+X_{7-i}|$ in \figureref{fig:lces_3d}(b) tends to zero for each $i=1,2,3$, our computed estimate of the LE spectrum has the expected symmetry.

Using our ftLE computations as estimates for the LEs, we present the full spectrum of LEs in \tableref{tab:vector_3D} together with the spectral gaps between neighbouring LEs. We give this table for reference, since the GSVs converge to the BLVs at an exponential rate of $\min(\chi_{i-1}-\chi_i,\chi_i-\chi_{i+1})$, while convergence to the CLVs during the backward transient happens at an exponential rate of $\chi_{i-1}-\chi_i$. We also give the distinct LEs $\lambda_i$ and their spectral gaps in \tableref{tab:subspace_3D}, since generic subspaces $G_i$ converge to the filtration subspaces $\Gamma_i^{-}$ of the same dimension at an exponential rate of $\lambda_i-\lambda_{i+1}$, and we conjecture that convergence to the $i$-th splitting subspace $\Omega_i$ happens at an exponential rate of $\lambda_{i-1}-\lambda_i$.

\begin{table}[htbp]
	\caption{The $i$-th LE $\chi_i$, $i=1,2,\dots,6$, and its spectral gaps with neighbouring LEs $\chi_{i-1},\chi_{i+1}$ for the 3-D system \eqref{eq:ham3d}, computed using the chaotic orbit with initial condition \eqref{eq:3dic}.}
	\label{tab:vector_3D}
	\centering
	\setlength{\tabcolsep}{9pt}
	\begin{tabular}{crrrr}
		$i$ & $\chi_i$ & $\chi_{i-1}-\chi_i$ & $\chi_{i}-\chi_{i+1}$ & $\min(\chi_{i-1}-\chi_i,\chi_i-\chi_{i+1})$ \\
		\midrule
		$1$ & $0.030$  & $\infty$            & $0.022$               & $0.022$                                     \\
		$2$ & $0.008$  & $0.022$             & $0.008$               & $0.008$                                     \\
		$3$ & $0.000$  & $0.008$             & $0.000$               & $0.000$                                     \\
		$4$ & $0.000$  & $0.000$             & $0.008$               & $0.000$                                     \\
		$5$ & $-0.008$ & $0.008$             & $0.022$               & $0.008$                                     \\
		$6$ & $-0.030$ & $0.022$             & $\infty$              & $0.022$                                    
	\end{tabular}
\end{table}

\begin{table}[htbp]
	\caption{The $i$-th distinct LE $\lambda_i$, $i=1,2,\dots,5$, and its spectral gaps with neighbouring distinct LEs $\lambda_{i-1},\lambda_{i+1}$, computed using the same chaotic orbit as in \tableref{tab:vector_3D}.}
	\label{tab:subspace_3D}
	\centering
	\setlength{\tabcolsep}{9pt}
	\begin{tabular}{crrr}
		$i$ & $\lambda_i$ & $\lambda_{i-1}-\lambda_i$ & $\lambda_{i}-\lambda_{i+1}$ \\
		\midrule
		$1$ & $0.030$     & $\infty$                  & $0.022$                     \\
		$2$ & $0.008$     & $0.022$                   & $0.008$                     \\
		$3$ & $0.000$     & $0.008$                   & $0.008$                     \\
		$4$ & $-0.008$    & $0.008$                   & $0.022$                     \\
		$5$ & $-0.030$    & $0.022$                   & $\infty$                   
	\end{tabular}
\end{table}

\subsection{Forward transient of the G\&C algorithm}
\begin{figure}[htbp]
	\centering
	\includegraphics[width=\linewidth]{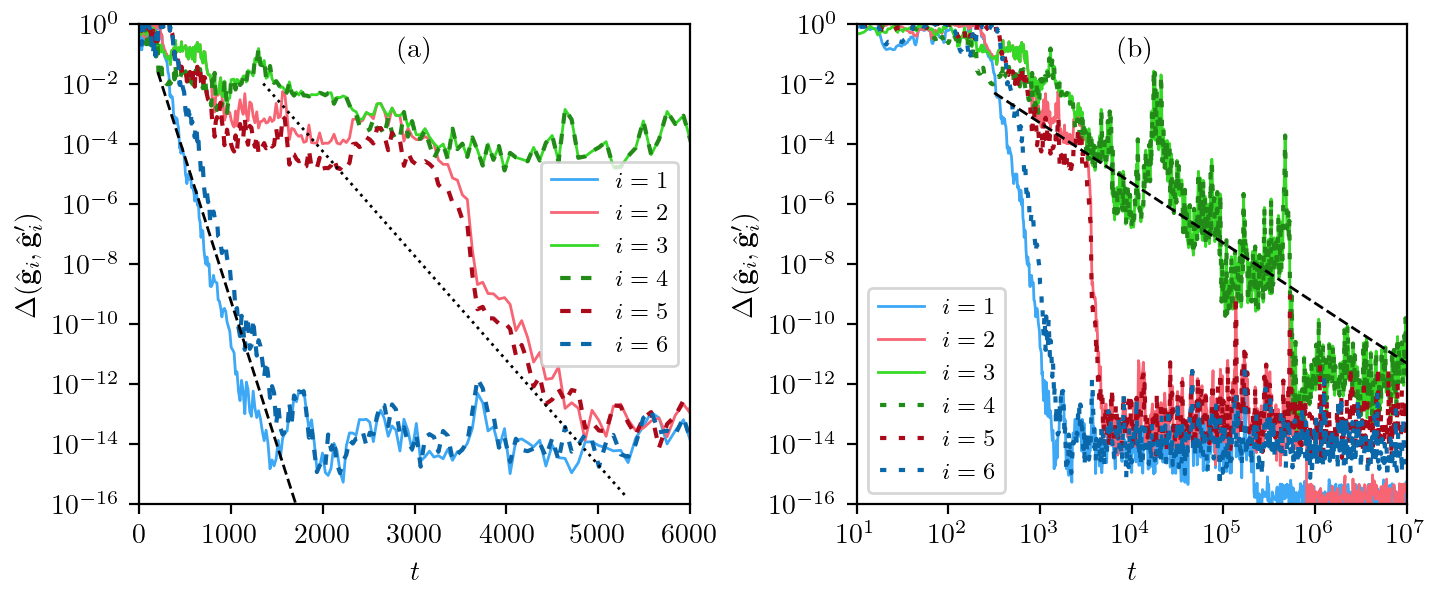}
	\caption{The distance $\Delta$ between two versions $\hat{\bs g}_i$ and $\hat{\bs g}_i'$ of the $i$-th GSV ($i=1,2,\dots,6$) for the chaotic orbit with initial condition \eqref{eq:3dic} in the 3-D system \eqref{eq:ham3d}. In panel (a) the black dashed line denotes a function $\propto e^{-0.022t}$ and the black dotted line is $\propto e^{-0.008t}$, while in (b) the black dashed line denotes a function $\propto t^{-2}$. Panel (a) is in log-linear scale and (b) is in log-log scale.}
	\label{fig:gsv_vec_converge_3D}
\end{figure}

Taking the same approach as in \sectionref{sec:forward}, we compute the BLVs by evolving a set of linearly independent vectors (the GSVs) forwards in time while periodically orthonormalising them. By performing this computation twice, we generate two sets of GSVs that both converge to their respective BLVs and hence to each other. We give the time evolution of the distance $\Delta$ between the $i$-th GSVs $\hat{\bs g}_i$ and $\hat{\bs g}_i'$, $i=1,2,\dots,6$, in \figureref{fig:gsv_vec_converge_3D} over a forward transient of $10^7$. We see from this figure that $\Delta(\hat{\bs g}_i,\hat{\bs g}_i')$ decays roughly $\propto e^{-0.022t}$ for $i=1,6$ and $\propto e^{-0.008t}$ for $i=2,5$. For $i=3,4$, the distance $\Delta(\hat{\bs g}_i,\hat{\bs g}_i')$ decays $\propto t^{-2}$, which is sub-exponential. These exponential rates of convergence between $\hat{\bs g}_i$ and $\hat{\bs g}_i'$ are in agreement with their expected rates of convergence to the $i$-th BLV as given in the last column of \tableref{tab:vector_3D}.

\begin{figure}[htbp]
	\centering
	\includegraphics[width=0.55\linewidth]{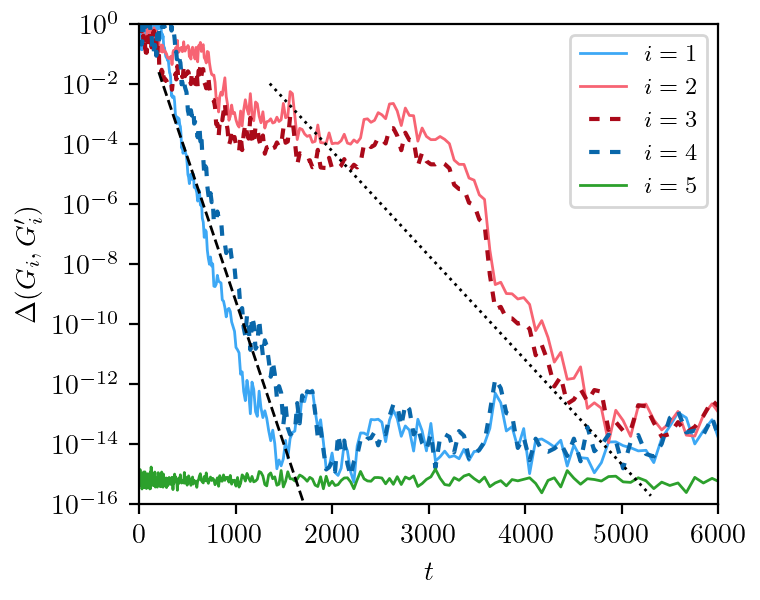}
	\caption{The time evolution of the distance $\Delta$ between two generic subspaces $G_i$ and $G_i'$, $i=1,2,\dots,5$, each of dimension $\dim\Gamma_i^-$ for the chaotic orbit with initial condition \eqref{eq:3dic} in the 3-D system \eqref{eq:ham3d}. The black dashed line denotes a function $\propto e^{-0.022t}$ and the black dotted line is $\propto e^{-0.008t}$. The figure is in log-linear scale.}
	\label{fig:gsv_sub_converge_3D}
\end{figure}

We now move on from the convergence between vectors to the convergence between subspaces. Informed by the degeneracy in the LE spectrum, we define
\begin{align}
\begin{split}
	G_1 &= \spn(\hat{\bs g}_1),\\
	G_2 &= \spn(\hat{\bs g}_1,\hat{\bs g}_2),\\
	G_3 &= \spn(\hat{\bs g}_1,\hat{\bs g}_2,\hat{\bs g}_3,\hat{\bs g}_4),\\
	G_4 &= \spn(\hat{\bs g}_1,\hat{\bs g}_2,\hat{\bs g}_3,\hat{\bs g}_4,\hat{\bs g}_5),\\
	G_5 &= \spn(\hat{\bs g}_1,\hat{\bs g}_2,\hat{\bs g}_3,\hat{\bs g}_4,\hat{\bs g}_5,\hat{\bs g}_6).
\end{split}
\end{align}
Since each generic subspace $G_i$, $i=1,2,\dots,5$, satisfies $\dim G_i=\dim\Gamma_i^-$, it follows that $G_i\to\Gamma_i^-$ as $t\to\infty$. Computing two sets of these subspaces $G_i$ and $G_i'$ from the two previously computed sets of GSVs $\hat{\bs g}_i$ and $\hat{\bs g}_i'$, respectively, we present the time evolution of the distance $\Delta$ between $G_i$ and $G_i'$ in \figureref{fig:gsv_sub_converge_3D}. We see from this figure that $\Delta(G_i,G_i')$ decays approximately $\propto e^{-0.022t}$ for $i=1,4$ and $\propto e^{-0.008t}$ for $i=2,3$. For $i=5$, the distance $\Delta(G_5,G_5')$ is practically zero throughout the computation. These exponential rates of convergence between $G_i$ and $G_i'$ are in agreement with their expected convergence rates to the $i$-th filtration subspace $\Gamma_i^-$ given in the last column of \tableref{tab:subspace_3D}.

\subsection{Backward transient of the G\&C algorithm}
\begin{figure}[htbp]
	\centering
	\includegraphics[width=\linewidth]{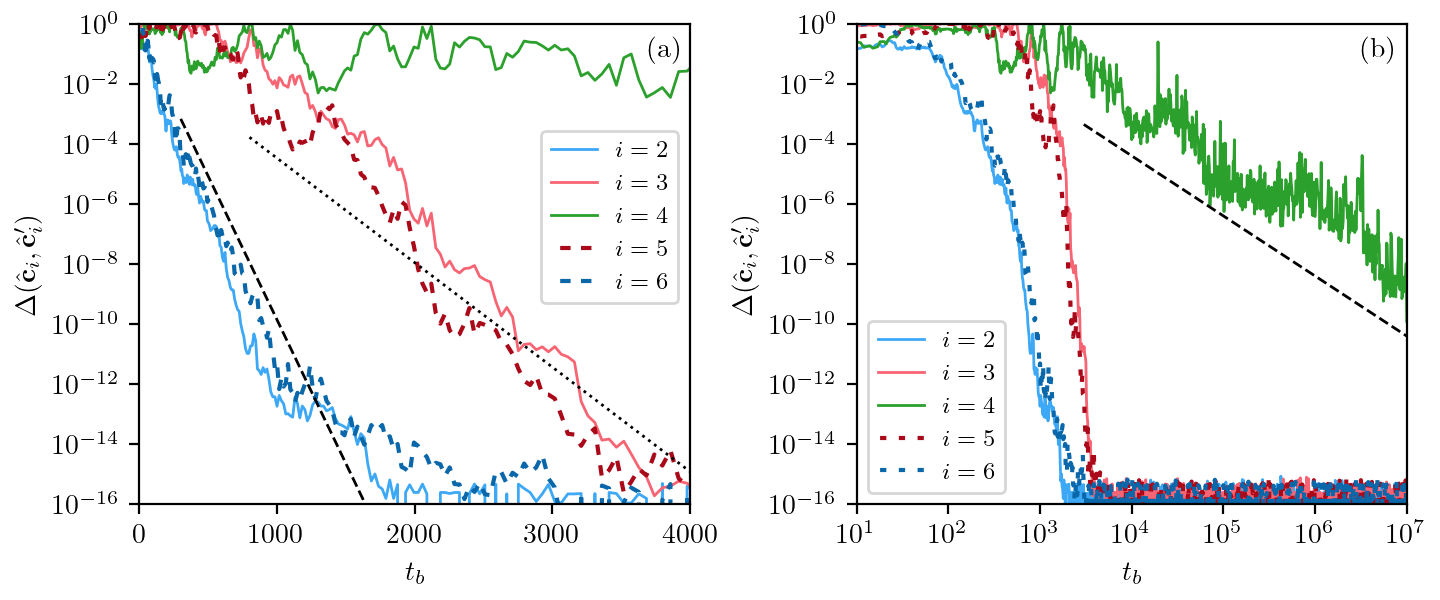}
	\caption{The distance $\Delta$ between estimates $\hat{\bs c}_i$ and $\hat{\bs c}_i'$ of the $i$-th CLV ($i=2,3,4,5,6$) as a function of backward time $t_b$, computed during the backward transient of the G\&C algorithm for the chaotic orbit with initial condition \eqref{eq:3dic} in the 3-D system \eqref{eq:ham3d}. In panel (a) the black dashed line denotes a function $\propto e^{-0.022t_b}$ and the black dotted line is $\propto e^{-0.008t_b}$, while in (b) the black dashed line denotes a function $\propto t_b^{-2}$. Panel (a) is in log-linear scale and (b) is in log-log scale.}
	\label{fig:clv_vec_converge_3D}
\end{figure}

Let us now consider the backward transient part of the G\&C algorithm, where deviation vectors chosen from each of the computed estimates of $\Gamma_i^-$ are evolved backwards in time in order to converge to the splitting subspaces $\Omega_i$ and hence the CLVs. Using the same method as in \sectionref{sec:back_trans} for the H\'enon-Heiles system, we compute two estimates $\hat{\bs c}_i$ and $\hat{\bs c}_i'$ of the $i$-th CLV ($i=1,2,\dots,6$) over a backward transient of $10^7$ time units. To set up this computation, we first calculate the BLVs during a forward transient of $10^4$ time units, which we deem sufficient because of the high degree of convergence between each $G_i,G_i'$ pair seen in \figureref{fig:gsv_sub_converge_3D} during this time interval. In \figureref{fig:clv_vec_converge_3D} we give the backward time evolution of the distance $\Delta$ between corresponding CLV estimates. We see from the figure that $\Delta(\hat{\bs c}_i,\hat{\bs c}_i')$ decays approximately $\propto e^{-0.022t_b}$ for $i=2,6$ and $\propto e^{-0.008t_b}$ for $i=3,5$. For $i=4$, the distance $\Delta(\hat{\bs c}_4,\hat{\bs c}_4')$ decays sub-exponentially $\propto t_b^{-2}$, which is the same sub-exponential decay rate we computed for $\Delta(\hat{\bs c}_3,\hat{\bs c}_3')$ in the case of the H\'enon-Heiles system (see \figureref{fig:clv_vec_converge}) where we showed in \sectionref{sec:cent} that this rate is a consequence of these CLV estimates converging to the direction of the flow like $t_b^{-1}$ and hence to each other like $t_b^{-2}$; we assume that the same argument holds in this case for the 3-D system. Note that we do not show the distance $\Delta$ for the $i=1$ case (similarly to \figureref{fig:clv_vec_converge} for the H\'enon-Heiles system), since it is identically zero throughout the computation. These exponential rates of convergence between $\hat{\bs c}_i$ and $\hat{\bs c}_i'$, $i=1,2,\dots,6$, are in agreement with their expected rates of convergence to the $i$-th CLVs of $\chi_{i-1}-\chi_i$ given in the third column of \tableref{tab:vector_3D}.

\begin{figure}[htbp]
	\centering
	\includegraphics[width=\linewidth]{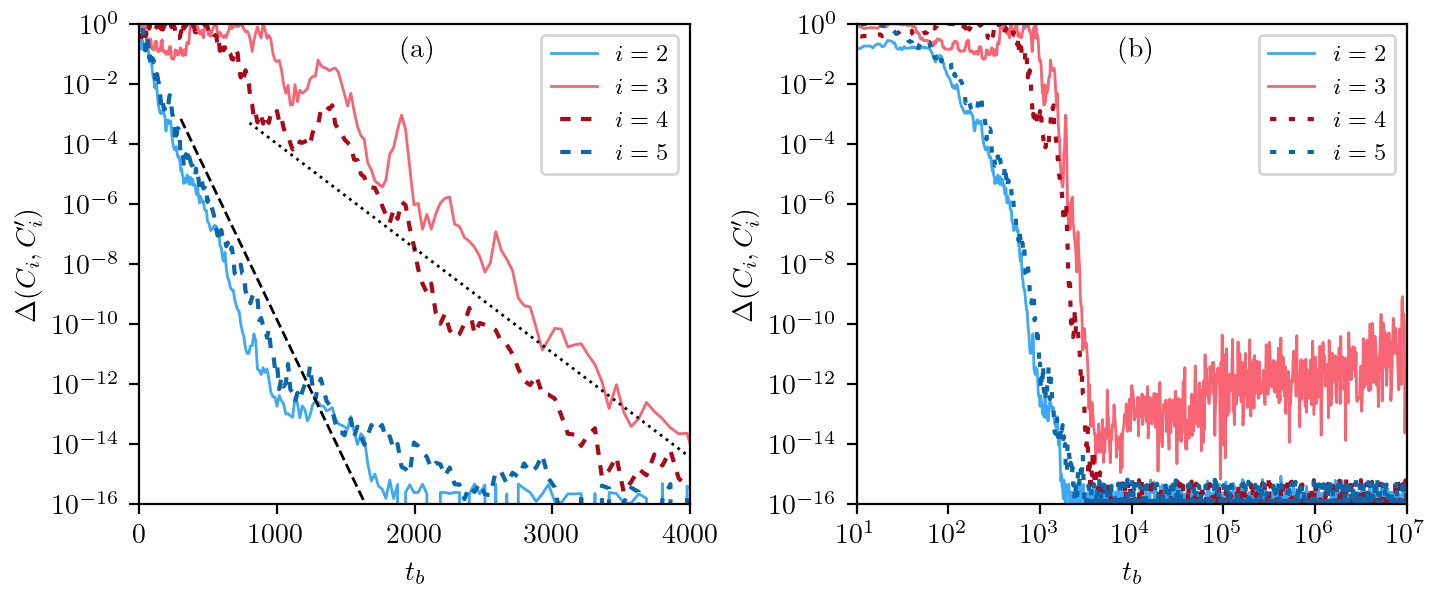}
	\caption{The distance $\Delta$ between estimates $C_i$ and $C_i'$ of the $i$-th splitting subspace ($i=2,3,4,5$) as a function of backward time $t_b$, computed during the backward transient of the G\&C algorithm for the chaotic orbit with initial condition \eqref{eq:3dic} in the 3-D system \eqref{eq:ham3d}. In panel (a) the black dashed line denotes a function $\propto e^{-0.022t_b}$ and the black dotted line is $\propto e^{-0.008t_b}$. Panel (a) is in log-linear scale and (b) is in log-log scale.}
	\label{fig:clv_sub_converge_3D}
\end{figure}

\begin{figure}[htbp]
	\centering
	\includegraphics[width=\linewidth]{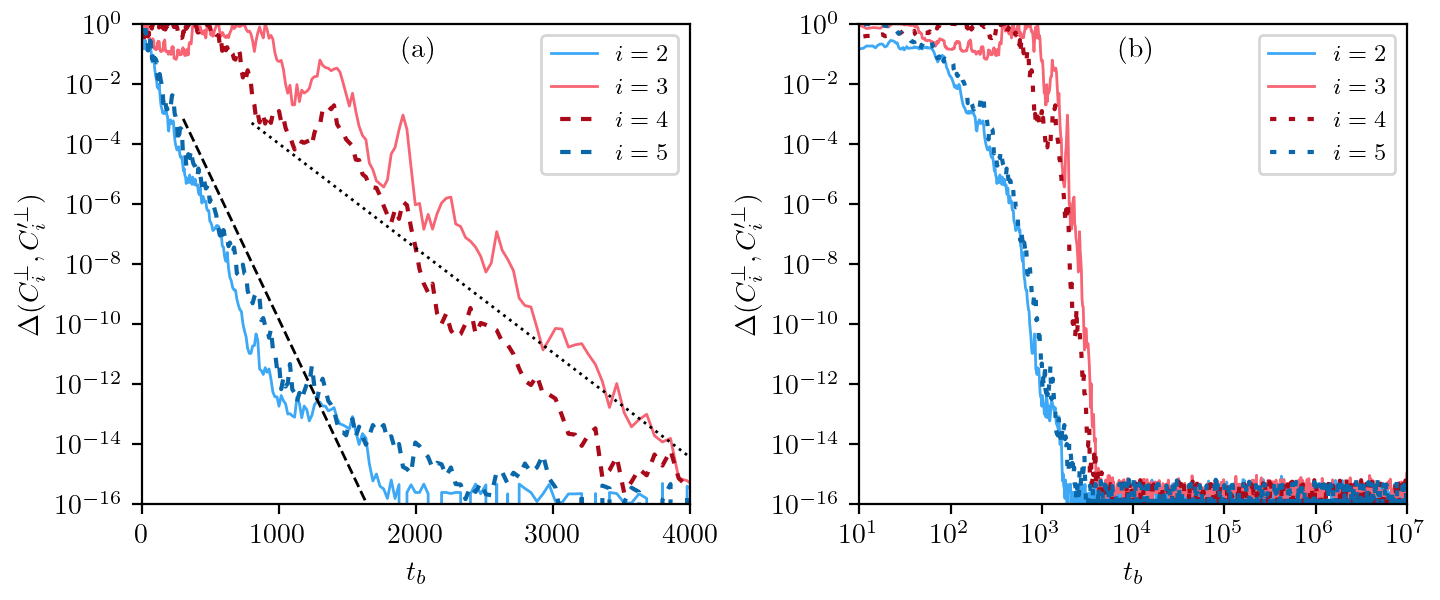}
	\caption{Similar to \figureref{fig:clv_sub_converge_3D}, but the distance $\Delta$ is between estimates $C_i^{\perp}$ and $C_i^{\prime\perp}$ of the $i$-th splitting subspace ($i=2,3,4,5$), computed using the G\&C algorithm with the centre correction.}
	\label{fig:clv_sub_converge_corr_3D}
\end{figure}

Finally, we investigate the convergence between subspaces that converge to the splitting subspaces $\Omega_i$ during the backward transient. We define
\begin{align}
\begin{split}
	C_1 &= \spn(\hat{\bs c}_1),\\
	C_2 &= \spn(\hat{\bs c}_2),\\
	C_3 &= \spn(\hat{\bs c}_3,\hat{\bs c}_4),\\
	C_4 &= \spn(\hat{\bs c}_5),\\
	C_5 &= \spn(\hat{\bs c}_6).
\end{split}
\end{align}
Computing two sets of these subspaces $C_i$ and $C_i'$ from the two previously computed sets of CLV estimates $\hat{\bs c}_i$ and $\hat{\bs c}_i'$ (respectively), we present the backward time evolution of the distance $\Delta$ between $C_i$ and $C_i'$ in \figureref{fig:clv_sub_converge_3D}. We see from \figureref{fig:clv_sub_converge_3D}(a) that $\Delta(C_i,C_i')$ decays approximately $\propto e^{-0.022t_b}$ for $i=2,5$ and $\propto e^{-0.008t_b}$ for $i=3,4$. Note that the $i=1$ case is excluded since $\Delta(C_1,C_1')$ is identically zero throughout the computation. These exponential rates of convergence between $C_i$ and $C_i'$, $i=1,2,\dots,5$, are in agreement with their expected convergence rates to the $i$-th splitting subspace $\Omega_i$ of $\lambda_{i-1}-\lambda_i$ given in the third column of \tableref{tab:subspace_3D}. However, we see from \figureref{fig:clv_sub_converge_3D}(b) that our estimates $C_3$ and $C_3'$ for the centre subspace $\Omega_3$ diverge slowly over the $10^7$ interval. This divergent behaviour is not merely due to errors in the numerical integration, since we checked that it persists when using smaller time steps. As in the H\'enon-Heiles case (see \sectionref{sec:hhsection}), we find that the CLVs of the centre subspace computed using the G\&C algorithm ($\hat{\bs c}_3$ and $\hat{\bs c}_4$ in the 3-D system case) tend to converge during the backward dynamics, resulting in the centre subspace becoming poorly defined in the numerics as we move further in backward time $t_b$. This is easily remedied by computing the G\&C algorithm with the centre correction (see \sectionref{sec:back_trans}), where we periodically orthonormalise the CLV estimates in the centre subspace. Specifically, the CLV estimate $\hat{\bs c}_4$ is replaced with $\hat{\bs c}_4^{\perp}\in\spn(\hat{\bs c}_3,\hat{\bs c}_4)$, which is orthogonal to $\hat{\bs c}_3$, and we denote the splitting subspace estimates using this new set of CLV estimates by $C_i^{\perp}$ to distinguish them from $C_i$. The subspaces $C_i^{\prime\perp}$ are similarly constructed from $C_i'$. The backward time evolution of the distance $\Delta$ between $C_i^{\perp}$ and $C_i^{\prime\perp}$ is given in \figureref{fig:clv_sub_converge_corr_3D}. We see that \figureref{fig:clv_sub_converge_corr_3D}(a) is practically unchanged from \figureref{fig:clv_sub_converge_3D}(a), but we see from \figureref{fig:clv_sub_converge_corr_3D}(b) that the centre subspace estimates $C_3^{\perp}$ and $C_3^{\prime\perp}$ no longer diverge in the long run.

\section{Summary}
In this chapter, we performed a numerical investigation into the convergence properties of the G\&C algorithm applied to the H\'enon-Heiles system \eqref{eq:HH} and the 3-D system \eqref{eq:ham3d}, both for the forward transient and the backward transient parts of the algorithm. In particular, we studied the rates at which relevant vectors/subspaces converge to the BLVs, backward filtration subspaces, CLVs, and splitting subspaces. Instead of directly computing the convergence rates of generic vectors/subspaces to these a priori unknown objects, we measured the convergence rates between two sets of independently computed vectors/subspaces and found them to match the expected rates of convergence to the unknown objects. These findings validate our approach of indirectly measuring rates of convergence when those rates are exponential, which we recommend as an efficient means of determining the transient lengths needed when using the G\&C algorithm.

We also studied the dynamics of the centre subspace of the H\'enon-Heiles system after the intriguing observation that the two CLVs in the centre subspace converge when computed using the G\&C algorithm. We found that generic centre vectors converge to the direction of the flow $\propto t^{-1}$ during forward evolution and $\propto t_b^{-1}$ during backward evolution. As a consequence of this convergence, we showed that generic centre vectors also converge to each other, but only $\propto t^{-2}$. By computing the matrix of the linear propagator restricted to the centre subspaces with respect to particular orthonormal bases, we determined that the mechanism by which deviation vectors in the centre subspace converge to the direction of the flow is through a shear transformation, which explains the aforementioned convergence rates in the centre subspace, as well as how convergence to a single direction can occur in both directions of time. These results provide new insights into the dynamics of the centre subspace.

Finally, as an application of the method of CLVs to study hyperbolicity, we computed the angles between the splitting subspaces for a very sticky orbit in the H\'enon-Heiles system, which alternates between regular and chaotic regimes. We found that the minimum angle between each pair of splitting subspaces is distributed close to zero during regular regimes, indicating more frequent violations of hyperbolicity than during chaotic regimes where the minimum angles are distributed more uniformly.

%% file: dna.tex
\chapter{Numerical investigation of a DNA lattice model}\label{ch:dna}
We now shift focus from our investigation of low-dimensional systems to investigate a high-dimensional physical system of particular importance to living organisms: \textit{deoxyribonucleic acid} (DNA). DNA is a molecule composed of two sugar-phosphate chains that are twisted around each other in the form of a double helix and held together by \textit{base pairs}. Each base in a base pair is one of four compounds: adenine, thymine, guanine, or cytosine. The only bases which bond together in DNA to form base pairs are adenine-thymine (AT) and guanine-cytosine (GC), and each base pair is held together by hydrogen bonds. Biological ``instructions'' for the synthesis of proteins are encoded in sequences of these AT and GC base pairs. Since we will be studying DNA from a dynamical systems point of view, this gross over-simplification of the subject will suffice, but we refer the interested reader to Chapt.~1 of \cite{Calladine2004} for a general introduction to DNA from a molecular biology standpoint.

\section{The Peyrard-Bishop-Dauxois model of DNA}\label{sec:pbdsec}
In this chapter, we study the dynamics of a DNA model. We are particularly interested in thermal openings, known as \textit{bubbles}, which occur between base pairs. Such dynamical openings have biological importance as there is evidence that they may direct transcription \cites{ChoiEtAl2004}{AlexandrovEtAl2010}, a mechanism by which a sequence of base pairs in a DNA segment is ``read'' and stored temporarily in a ribonucleic acid (RNA) molecule, which in turn relays instructions for biological processes such as protein synthesis. Extending the work of \cite{HillebrandEtAl2019}, we investigate how bubbles contribute to chaos in DNA using CLVs and a new type of vector, the so-called instantaneous Lyapunov vector. As a first step in our research approach, we focus exclusively on homogeneous sequences of DNA, where each lattice consists of only AT or only GC base pairs.

To study the dynamical behaviour of DNA, we use the \textit{Peyrard-Bishop-Dauxois} (PBD) model, which describes the displacement $x_i$ from equilibrium and momentum $p_i$ of the $i$-th base pair, modelling DNA as a $1$-D lattice with $n$ sites \cite{DauxoisEtAl1993}. For a detailed explanation of the PBD model and the physical interpretation of the variables $x_i$ and $p_i$ in the context of a DNA molecule, see e.g.\ Chapt.~2 of \cite{Hillebrand2021}. The PBD model is an autonomous Hamiltonian system, the Hamiltonian function for which is given by
\begin{align}
	H=\sum_{i=1}^n\Bigg[\hspace{3pt}\underbrace{\frac{1}{2m\vphantom{m_2}}p_i^2}_{\substack{\text{kinetic}\\\text{energy}}}\ +\ \underbrace{\vphantom{\frac{K}{m_2}}D_i(e^{-a_ix_i}-1)^2}_{\substack{\text{on-site Morse}\\\text{potential}}}\ +\ \underbrace{\frac{K}{2\vphantom{m_2}}\left(1+\rho e^{-b(x_i+x_{i-1})}\right)(x_i-x_{i-1})^2}_{\text{coupling potential}}\ \Bigg],\label{eq:dnahamiltonian}
\end{align}
where $m$, $D_i$, $a_i$, $K$, $\rho$, and $b$ are parameters. The number of sites $n$ is also the number of degrees of freedom for this Hamiltonian model. Note that by calling this lattice model $1$-D, we are simply referring to the structure of the lattice, not to the dimension of the system's phase space which, of course, has $N=2n$ dimensions. In our study of this model, we use periodic boundary conditions, so $x_0=x_n$ and $p_0=p_n$. As indicated in \eqref{eq:dnahamiltonian}, the Hamiltonian consists of a kinetic energy term, an on-site potential for which a Morse potential is used, and a nonlinear coupling potential between neighbouring sites.

\begin{figure}[htbp]
	\centering
	\includegraphics[width=0.55\linewidth]{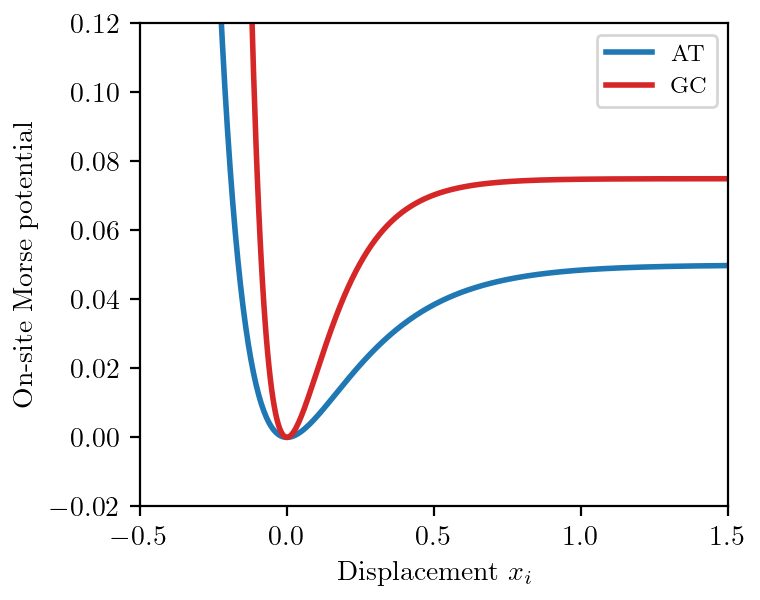}
	\caption{The on-site Morse potential from the PBD Hamiltonian \eqref{eq:dnahamiltonian} as a function of displacement $x_i$ at the $i$-th lattice site, for both the pure AT and pure GC cases.}
	\label{fig:potentials}
\end{figure}

In our numerical investigation, we use the PBD model parameters from \cite{CampaGiansanti1998} which were found to fit empirical data and have been used in numerous works since, see e.g.\ \cites{KalosakasEtAl2004}{AresEtAl2005}{KalosakasAres2009}{HillebrandEtAl2019}. These parameters are $D_{\scriptscriptstyle AT}=\qty{0.05}{eV}$, $D_{\scriptscriptstyle GC}=\qty{0.075}{eV}$, $a_{\scriptscriptstyle AT}=\qty{4.2}{\angstrom^{-1}}$, $a_{\scriptscriptstyle GC}=\qty{6.9}{\angstrom^{-1}}$, $K=\qty{0.025}{eV.\angstrom^{-2}}$, $\rho=2$, and $b=\qty{0.35}{\angstrom^{-1}}$. To show the effect that these different parametrisations for AT and GC have on the on-site Morse potential in \eqref{eq:dnahamiltonian}, we plot this potential as a function of displacement $x_i$ in \figureref{fig:potentials}. We see from this figure that the potential well for GC is deeper and narrower than in the AT case. For the mass parameter $m$, we use $m=\qty{300}{amu}$, which has been used in previous studies of the PBD model (e.g.\ in \cites{KalosakasEtAl2004}{HillebrandEtAl2019}). Note that the parameters with subscript $AT$ are used in the potential energy terms of \eqref{eq:dnahamiltonian} to simulate AT base pairs, and similarly, the parameters with $GC$ subscripts are used for simulating GC base pairs. We use the angstrom (\AA) as our length unit, the picosecond (ps) as our time unit, and the electronvolt (eV) as our energy unit. In this system of units, the natural mass unit is \unit{eV.ps^2.\angstrom^{-2}}. While the mass parameter $m$ was given above in atomic mass units (amu), converting this to our natural mass unit yields $m=\qty{0.031}{eV.ps^2.\angstrom^{-2}}$ (rounded to two significant figures), which we use in all of our computations.

When selecting an initial condition for the PBD model, we have the freedom to choose the total energy $H$ of the system \eqref{eq:dnahamiltonian} or, equivalently, the energy density $E_n=H/n$. A nonlinear relationship between the energy density $E_n$ and the temperature $T$ (which is proportional to the average kinetic energy of the system) of the PBD lattice was proposed and fitted to numerically simulated data in \cite{Hillebrand2021}. This relationship is given by
\begin{align}
	E_n = k_B T + \gamma\hspace{1pt} T^3,\label{eq:relate}
\end{align}
where $k_B$ is Boltzmann's constant in units of \unit{eV/K} and $\gamma$ is a fitted parameter which depends linearly on the proportion of AT and GC base pairs in the lattice. For pure AT and pure GC lattices, these fitted parameters $\gamma_{\scriptscriptstyle AT}$ and $\gamma_{\scriptscriptstyle GC}$ are, respectively,
\begin{align}
	\gamma_{\scriptscriptstyle AT} = \qty{4.97e-10}{eV/K^3},\qquad\gamma_{\scriptscriptstyle GC} = \qty{6.15e-10}{eV/K^3}.\label{eq:param}
\end{align}
We choose to base our decision of which energy density to use on the temperature $T$. Choosing a physiological temperature of \qty{310}{K}, we determine the related energy densities for pure AT and pure GC lattices using \eqref{eq:relate} with parameters \eqref{eq:param}. For pure AT and pure GC, we therefore use the following respective energy densities per site,
\begin{align}
	E_n^{\scriptscriptstyle AT} = \qty{0.0415}{eV},\qquad E_n^{\scriptscriptstyle GC} = \qty{0.0450}{eV},\label{eq:energydensity}
\end{align}
rounded to three significant figures in each case.

\section{Spatial distribution of deviation vectors}
The PBD lattice model which we study is a spatially extended system, so it is prudent for us to have some measure of the spatial distribution of deviation vectors (including CLVs) in the lattice. We define the \textit{distribution} $\xi_i$ of a deviation vector $\bs w$ at the $i$-th site of a Hamiltonian lattice model as
\begin{align}
	\xi_i(\bs w)=\frac{w_i^2+w_{i+n}^2}{\sum_{k=1}^n(w_k^2+w_{k+n}^2)},\label{eq:dvd}
\end{align}
where $w_i$ and $w_{i+n}$ are the respective displacement and momentum components of $\bs w$ at the $i$-th site, and $n$ is the number of sites in the lattice. This quantity measures how a particular deviation vector is distributed across the lattice. Note that $\xi_i\in [0,1]$ and $\sum_{i=1}^n\xi_i=1$, so $\xi_i$ as a function of the site index $i$ can be thought of as a probability mass function (see e.g.\ \cite[p.~94]{UnderhillBradfield2013}) to which all the usual statistical notions of location and spread of the distribution apply.

For a randomly-chosen deviation vector $\bs w$, the distribution $\xi_i(\bs w)$ \eqref{eq:dvd} is the same as the so-called \textit{deviation vector distribution} (DVD), which has been used to track chaos within various lattice models by determining which degrees of freedom depend most sensitively on initial conditions \cites{SkokosEtAl2013}{SenyangeEtAl2018}. Since generic deviation vectors converge exponentially fast to the first CLV $\hat{\bs\omega}_1$, the DVD is an accurate approximation of $\xi_i(\hat{\bs\omega}_1)$ once the random deviation vector $\bs w$ has evolved for a short transient.

\subsection{Measures of location and spread}\label{sec:locspread}
\begin{figure}[htbp]
	\centering
	\includegraphics[width=\linewidth]{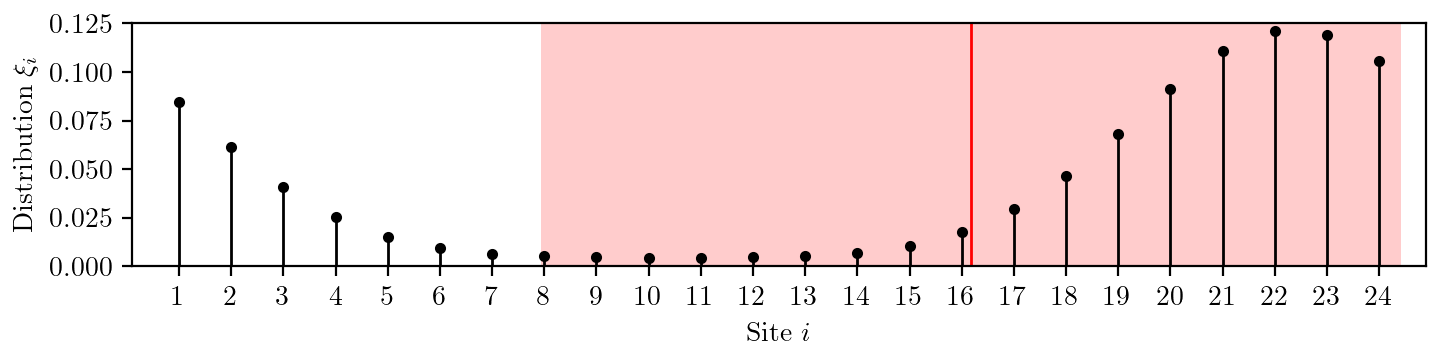}
	\caption{A plot of an example distribution $\xi_i$ as a function of the site $i$ for a periodic lattice of 24 sites. Each black dot represents the value of $\xi_i$ at the $i$-th lattice site. The red line is the location of the mean $\bar{i}$ and the region within one standard deviation $\sigma$ of the mean is shaded red.}
	\label{fig:polar_linear}
\end{figure}

Given a distribution $\xi_i$ \eqref{eq:dvd} of some deviation vector $\bs w$, how do we determine its central location and quantify its spread? For distributions on a line, a sensible measure of location is the \textit{mean} $\bar{i}$ of the distribution, defined as
\begin{align}
	\bar{i}=\sum_{i=1}^n i\,\xi_i(\bs w),\label{eq:mean}
\end{align}
while the \textit{standard deviation} $\sigma$ is a commonly used measure of spread given by
\begin{align}
	\sigma(\bs w)=\sqrt{\sum_{i=1}^n (i-\bar{i})^2\xi_i(\bs w)}.\label{eq:sd}
\end{align}
However, $\xi_i$ is not a distribution on a line due to our use of periodic boundary conditions where lattice sites $i$ and $i+n$ are identified. To demonstrate why the usual notions of mean and standard deviation can be problematic in the presence of periodicity, we give an example of a distribution $\xi_i$ over a 24-site periodic lattice in \figureref{fig:polar_linear}. We naively compute the mean $\bar i$ and standard deviation $\sigma$ of the distribution, where the red line in that figure denotes the mean and the red region represents the interval within one standard deviation of the mean. With the understanding that the underlying lattice is periodic, it is clear from the figure that the mean and standard deviation do not accurately describe the location and spread of the distribution. These measures tend to poorly describe the underlying periodic distribution when that distribution is concentrated near the boundaries of the lattice, as in the case of \figureref{fig:polar_linear}.

\begin{figure}[htbp]
	\centering
	\includegraphics[width=0.7\linewidth]{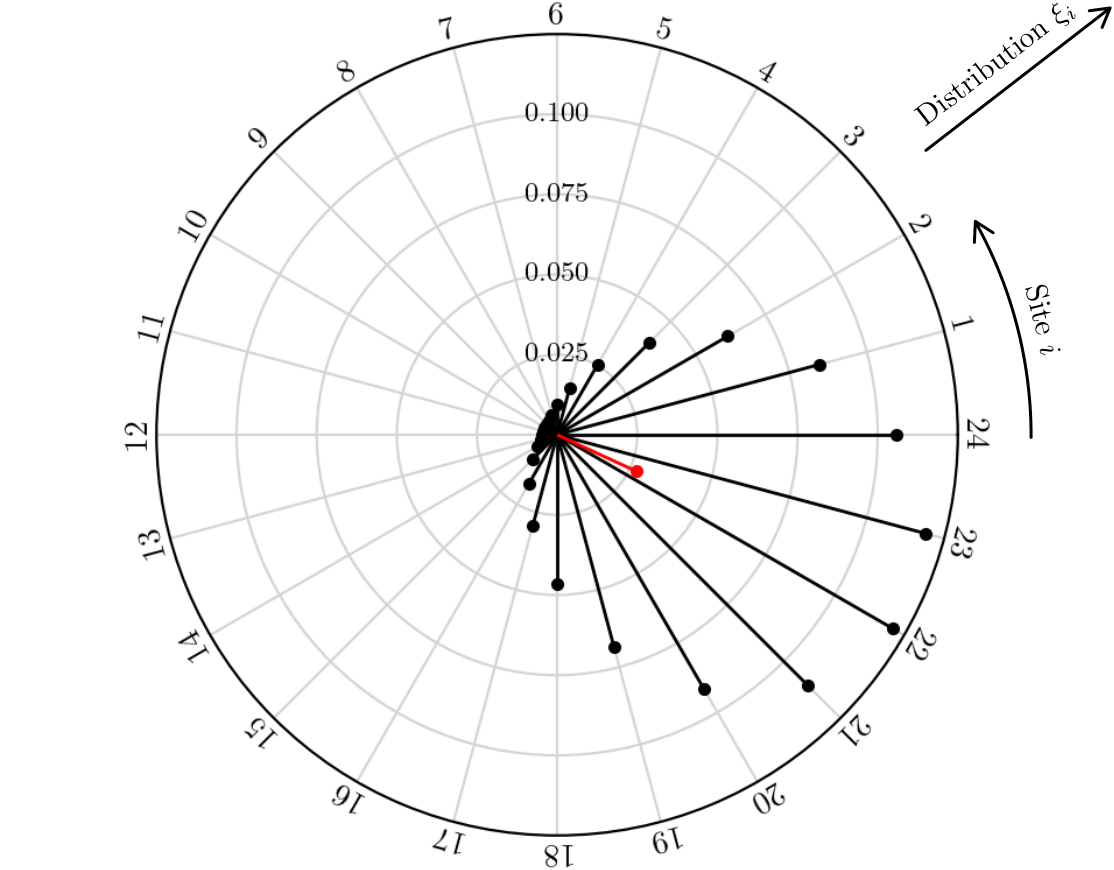}
	\caption{A polar plot of the distribution $\xi_i$ from \figureref{fig:polar_linear} as a function of the site $i$ for a periodic lattice of 24 sites. Each black dot represents the value of $\xi_i$ (the radial variable) at the $i$-th lattice site (the angular variable). The average of the black dots in the plane $\mathbb{R}^2$ is represented by the red dot, the direction of which is the circular mean of the distribution $\xi_i$.}
	\label{fig:polar}
\end{figure}

While $\xi_i$ should not be thought of as a distribution on a line, it can be thought of as a distribution on a \textit{circle}, and in this framework, a sensible measure of location can be defined. Such problems are formalised in the field of directional statistics or circular statistics, and we refer the interested reader to several introductory texts on the subject like \cites{Mardia1972}{JammalamadakaSengupta2001}{PewseyEtAl2013}. A circular distribution can be depicted by the use of polar coordinates: taking the distribution $\xi_i$ represented in \figureref{fig:polar_linear} using rectangular axes, we alternatively represent $\xi_i$ using a polar plot in \figureref{fig:polar} where each site $i$ is represented by a particular angle and $\xi_i$ is described by the radial variable. The periodicity of the distribution is accounted for by the angular variable in this representation, and a sensible measure of the distribution's location is determined as follows: considering the black dots in \figureref{fig:polar} as points in $\mathbb{R}^2$, compute the average of these points in the plane, then the direction of this average point is known as the \textit{mean direction} or \textit{circular mean} of the distribution. This average point is shown in \figureref{fig:polar} as a red dot, and its direction in the polar plot is the circular mean.\footnote{We forgo presenting a specific formula for the circular mean, which involves several conversions between Cartesian and polar coordinates, angles and lattice sites. For further details, see \cites{Mardia1972}{JammalamadakaSengupta2001}{PewseyEtAl2013}.} Comparing Figs.~\ref{fig:polar_linear} and~\ref{fig:polar}, it is clear that the circular mean, which lies between sites~22 and~23 in \figureref{fig:polar}, more accurately describes the location (or central tendency) of $\xi_i$ than the usual mean $\bar{i}$, which lies between sites~16 and~17 in \figureref{fig:polar_linear}. As a final remark, the circular mean is undefined if the aforementioned average point is at the origin where the angular variable is not uniquely defined, such as in the case of a uniform distribution.

\begin{figure}[htbp]
	\centering
	\includegraphics[width=\linewidth]{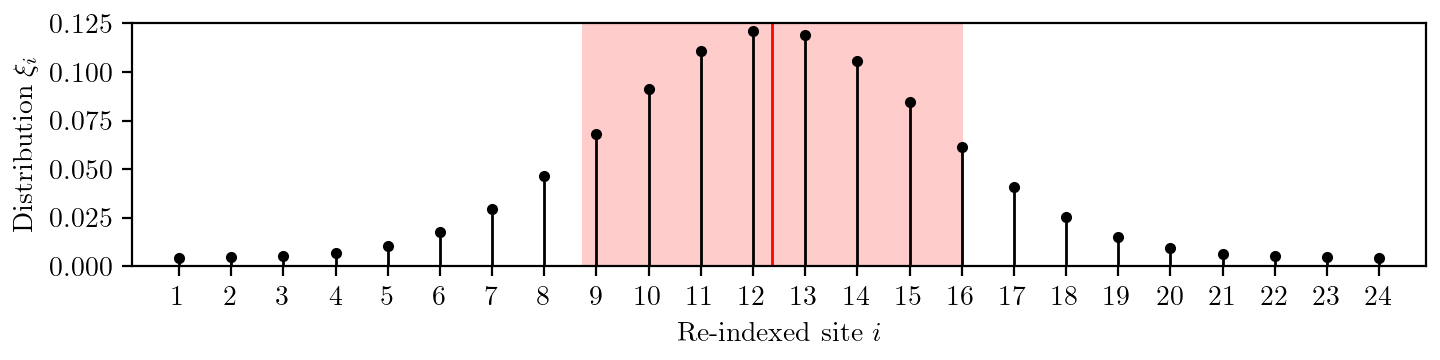}
	\caption{A plot of the distribution $\xi_i$ from \figureref{fig:polar_linear} as a function of the re-indexed site $i$ for a periodic lattice of 24 sites. Each black dot represents the value of $\xi_i$ at the $i$-th lattice site. The red line is the location of the mean $\bar{i}$ and the region within one standard deviation $\sigma$ of the mean is shaded red.}
	\label{fig:polar_linear_corr}
\end{figure}

Finally, we return to the issue of the spread of the distribution $\xi_i$. While there is an established definition of the circular standard deviation (see e.g.\ \cite[p.~41]{PewseyEtAl2013}), the explanation, calculation, and interpretation thereof each warrant lengthy discussions. Instead of broaching this relatively obscure topic, we opt for a simpler approach resulting from the following observation: the choice of site indices of a periodic lattice of $n$ sites is arbitrary since they can always be cyclically re-indexed by replacing the index of each site $i$ with $i+k\ (\text{mod}\ n)$ for some integer $k$. We therefore aim to remove the undesired boundary effects from the usual mean $\bar i$ and standard deviation $\sigma$ observed in \figureref{fig:polar_linear} by re-indexing the lattice such that the distribution is concentrated at the middle of the lattice and away from the boundaries. To determine the necessary re-indexing, we compute the circular mean of $\xi_i$, then re-index the lattice such that the circular mean is nearest to the centre of the lattice. In the case of our example distribution, the circular mean is located between sites~22 and~23, but after re-indexing, it is located between sites~12 and~13 in the middle of the 24-site lattice. This re-indexed version of the distribution from \figureref{fig:polar_linear} is displayed in \figureref{fig:polar_linear_corr}. Now that the distribution is concentrated far from the boundaries,\footnote{We assume a unimodal distribution, which suffices for our use case.} we compute the usual mean and standard deviation and indicate them in red in the exact same manner as \figureref{fig:polar_linear}. These measures of location and spread indicated in \figureref{fig:polar_linear_corr} now appear to accurately describe the periodic distribution. We use the method described here to compute the standard deviation of $\xi_i$ in this chapter.

Another measure of a distribution's spatial extent that complements the standard deviation is the \textit{participation number}, which estimates the number of sites participating strongly in that distribution. Given some deviation vector $\bs w$ with a distribution $\xi_i(\bs w)$, the participation number is defined as
\begin{align}
	P(\bs w) = \frac{1}{\sum_{i=1}^n (\xi_i(\bs w))^2},\label{eq:partip}
\end{align}
where $P(\bs w)\in[1,n]$. Note that $P=1$ if $\bs w$ only has position and momentum components at a single site (i.e.\ the distribution $\xi_i$ is non-zero for only a single site) and $P=n$ if $\bs w$ is distributed evenly across all lattice sites. In the case of the example distribution $\xi_i$ of this section (depicted in Figs.~\ref{fig:polar_linear}--\ref{fig:polar_linear_corr}), the participation number turns out to be $P\approx12.1$, which can be interpreted as approximately 12 sites contributing significantly to the distribution $\xi_i$. The participation number has been used in previous studies to study the spatial extent of various quantities in Hamiltonian lattices, such as the energy distribution and DVD in lattices of one \cite{SenyangeEtAl2018} and two \cite{ManyMandaEtAl2020} spatial dimensions, as well as the normal mode distributions of the disordered discrete nonlinear Schrödinger model and the disordered Klein-Gordon model \cites{FlachEtAl2009}{SenyangeEtAl2020}.

\section{Instantaneous Lyapunov exponents and vectors}\label{sec:ilv}
CLVs are defined from the matrices $\Lambda^{\pm}$ in \eqref{eq:matrix}, which are time averages of a function of the fundamental matrix $M(\bs x(0),t)$ as $t\to\pm\infty$. However, we wish to probe the stability properties of DNA bubbles, which are very short-lived phenomena, as seen in the study of bubble lifetimes in \cite{HillebrandEtAl2020}. Due to the averaging inherent in the CLV definition, it is reasonable to suspect that CLVs might not be optimal tools for studying bubbles at a particular time when these bubbles typically do not persist far into the past or future. We therefore suggest an alternative tool for studying such phenomena, which is based on the $N\times N$ matrix
\begin{align}
	\Lambda^0(\bs x(0))=\lim_{t\to0}\frac{1}{t}\ln\sqrt{M(\bs x(0),t)\transpose M(\bs x(0),t)}.\label{eq:lambda0}
\end{align}
This matrix is identical to the matrices $\Lambda^{\pm}$ in \eqref{eq:matrix} except that $t\to\pm\infty$ has been replaced with $t\to0$. Note that the superscript $0$ in this section does not denote an exponent; it is simply a reminder that the relevant quantity is defined instantaneously instead of over an infinite time interval in the future or past. To further simplify the notation, we now drop all explicit $\bs x$ and $t$ dependencies. We define the \textit{instantaneous Lyapunov exponents} (ILEs) as the eigenvalues of $\Lambda^0$. Furthermore, we define the \textit{instantaneous Lyapunov vectors} (ILVs) as an orthonormal set of eigenvectors of $\Lambda^0$. ILEs and ILVs have been previously defined and applied in the context of fluid dynamics \cite{NolanEtAl2020}. We have not shown the existence of \eqref{eq:lambda0}, but later in this section, we show that this matrix can be expressed in terms of the Jacobian matrix, which does exist.

We denote the $j$-th ILE by $\chi_j^0$ and its corresponding ILV by $\hat{\bs\omega}_j^0$, in accordance with our notation for LEs and CLVs given in \sectionref{sec:lyapunov}. We choose to order the $N$ (possibly non-distinct) ILEs in descending order,
\begin{align}
	\chi_1^0 \geq \chi_2^0 \geq \cdots \geq \chi_N^0,
\end{align}
so $\chi_1^0$ is the \textit{maximum} ILE.\footnote{We could define a distinct spectrum of ILEs as done for the distinct LEs $\lambda_i$ in \eqref{eq:le}, but this is unnecessary because we only use eigenvectors in this discussion, not their eigenspaces.} The ILV $\hat{\bs\omega}_1^0$ corresponding to the maximum ILE is the deviation vector whose length grows the fastest at that instant in time since $\hat{\bs\omega}_1^0$ is actually the first right singular vector of $M(\bs x(0),t)$ as $t\to0$. Therefore, the system is most sensitive to perturbations in the direction of $\hat{\bs\omega}_1^0$ over an infinitesimal time interval, so we propose using the first ILV as an indicator of which degrees of freedom are the contributing most to the system's chaoticity at that instant in time. Despite the similarities between the definitions of CLVs and ILVs, note that CLVs are covariant while ILVs, in general, are not, so we should not expect these vectors to share all the same properties.

ILEs and ILVs can be estimated by computing the singular vectors of $M$ over a very small time interval $t$, but we can derive a simpler form of \eqref{eq:lambda0} that will allow us to accurately compute $\Lambda^0$ without the use of finite-time approximations. Since $M$ is a fundamental matrix, it satisfies the equation
\begin{align}
	\dot M = J M
\end{align}
(see e.g.\ \cite[p.~352]{ZillEtAl2013}), where $J$ is the Jacobian matrix from \eqref{eq:jacob}. Therefore, a first-order series expansion of $M$ as $t\to0$ is given by $M=I+tJ+O(t^2)$, hence
\begin{align}
	M\transpose M = I+t(J\transpose+J) + O(t^2).\label{eq:mmt}
\end{align}
Substituting \eqref{eq:mmt} into \eqref{eq:lambda0}, a series expansion of the logarithm and root functions then yields
\begin{align}
	\Lambda^0 &= \lim_{t\to0}\frac{1}{t}\ln\sqrt{I+t(J\transpose+J) + O(t^2)}\notag\\
	&= \lim_{t\to0}\frac{1}{t}\left(\frac{t}{2}(J\transpose+J) + O(t^2)\right)\notag\\
	&= \frac{1}{2}\left(J\transpose+J\right).\label{eq:symjac1}
\end{align}
We conclude that ILEs and ILVs are the respective eigenvalues and eigenvectors of the symmetric part of the Jacobian matrix $J$ evaluated at the relevant state. This result provides a simple means of computing the ILEs and ILVs of a state using the Jacobian (which is known a priori) instead of the fundamental matrix (which must be computed).

We show in \appendixref{app:appsymmetry} that the following symmetries exist in the ILE spectrum and in the distributions of ILVs for autonomous Hamiltonian systems:
\begin{align}
	\chi_j^0&=-\chi_{N-j+1}^0,\label{eq:specsym}\\
	\xi_i(\hat{\bs\omega}_j^0)&=\xi_i(\hat{\bs\omega}_{N-j+1}^0).\label{eq:specvecsym}
\end{align}
Note that this ILE symmetry \eqref{eq:specsym} is analogous to the LE symmetry \eqref{eq:symplecticsym} for symplectic systems. These symmetries \eqref{eq:specsym} and \eqref{eq:specvecsym} allow us to only study the first half of the ILEs and ILVs in our numerical investigations.

\section{Numerical results for a pure AT lattice}\label{sec:all_at}
In this section, we present numerical results for a 40-site ($n=40$) PBD lattice  of which every site is initialised as an AT base pair. When numerically integrating this system, we use the ABA864 integrator \eqref{eq:aba} with a time step of \qty{0.01}{ps}, which we find bounds the relative energy error $E_r\lesssim10^{-7}$. Finally, we use initial conditions with position coordinates of zero (i.e.\ $x_i=0$ for $i=1,\dots,n$) and with momentum coordinates $p_i$ drawn randomly from a normal distribution of mean zero and variance one, after which the initial state vector is scaled to achieve an energy density of $E_n^{\scriptscriptstyle AT}=\qty{0.0415}{eV}$ per site which, as discussed in \sectionref{sec:pbdsec}, corresponds to a physiological temperature.

\subsection{Lyapunov exponents and the SALI}\label{sec:dna_lesali}
\begin{figure}[htbp]
	\centering
	\includegraphics[width=\linewidth]{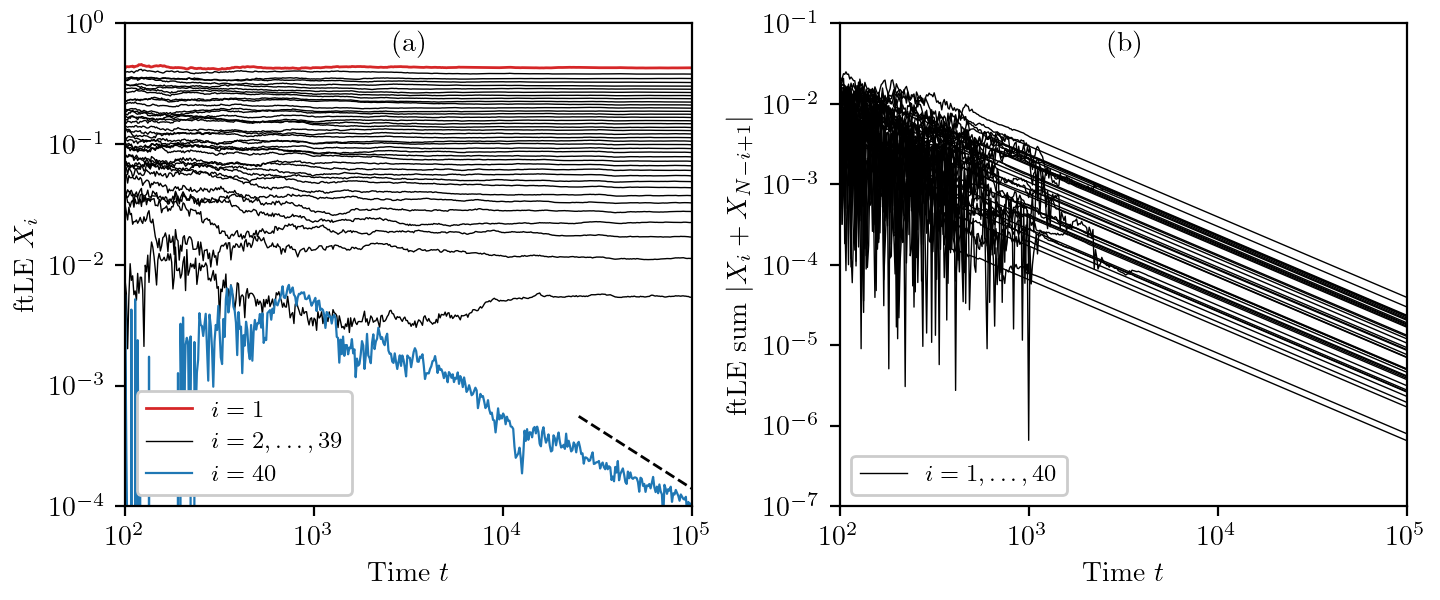}
	\caption{The time evolution the ftLEs $X_i$, $i=1,\dots,40$, is given in (a) for an orbit with a random initial condition in a 40-site pure AT lattice. The ftmLE $X_1$ is drawn in red, $X_{40}$ is drawn in blue, and the other ftLEs are drawn in black. The dashed line denotes a function $\propto t^{-1}$. The time evolution of $|X_i+X_{N-i+1}|$ is shown in (b) for $N=80$, $i=1,\dots,40$. Both panels are in log-log scale.}
	\label{fig:dna_lces}
\end{figure}

Choosing a random initial set of $N=80$ orthonormal deviation vectors, we compute the full spectrum of ftLEs $X_i(t)$, $i=1,\dots,80$, over an integration time of $10^5$ time units using the standard method (see \sectionref{sec:le}). In \figureref{fig:dna_lces}(a), we give the time evolution of the first 40 ftLEs. Note that we only compute and report ftLE values at evenly spaced points in $\log t$ such that we have at most 150 data points per decade of the $t$-axis in log scale. Furthermore, we use an orthonormalisation interval of $\tau=\qty{0.1}{ps}$ to prevent significant alignment of the deviation vectors. We see from \figureref{fig:dna_lces}(a) that the first 39 ftLEs saturate to positive constants, while $X_{40}$ decays to zero approximately $\propto t^{-1}$. The time evolution of $|X_i+X_{N-i+1}|$, $i=1,\dots,40$, is given in \figureref{fig:dna_lces}(b), from which we see that this quantity tends to zero for each $i$, so the computed LE spectrum has the expected symmetry \eqref{eq:symplecticsym}. From this symmetry, together with the roughly $\propto t^{-1}$ decay of $X_{40}$ seen in \figureref{fig:dna_lces}(a), we conclude from these numerical results that all LEs are non-zero except for the middle two, $\chi_{40}=\chi_{41}=0$. The ftmLE saturates to a positive value of $X_1\approx0.43$ after $t=\qty{e5}{ps}$. Since this estimate of the mLE is positive, we conclude that the orbit is chaotic.

\begin{figure}[htbp]
	\centering
	\includegraphics[width=0.55\linewidth]{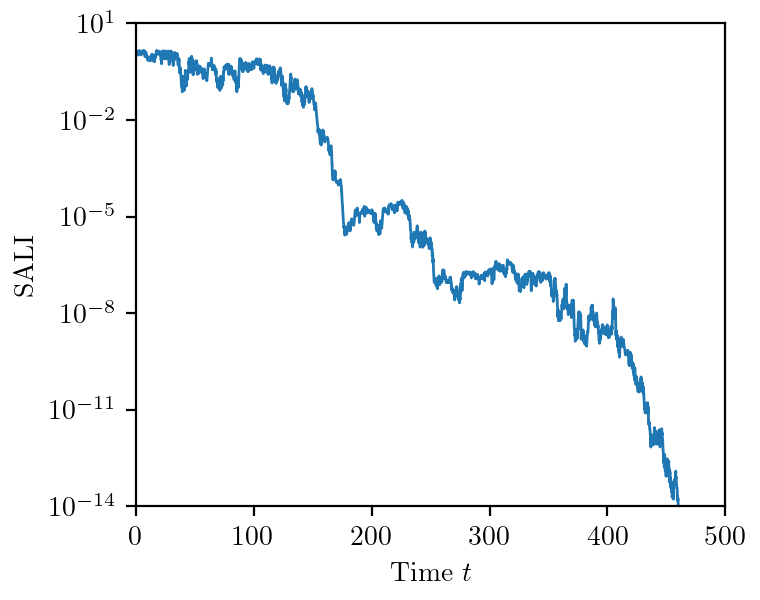}
	\caption{The time evolution of the SALI \eqref{eq:sali} using the same random initial condition as in \figureref{fig:dna_lces} for a 40-site pure AT lattice. We stop computing the SALI once its value falls below a threshold of $10^{-14}$. The figure is in log-linear scale.}
	\label{fig:dna_sali}
\end{figure}

Recall from \sectionref{sec:chaos} that the SALI chaos indicator \eqref{eq:sali} measures the exponential rate of convergence between two generic deviation vectors and that, when the two vectors are nearly aligned, the SALI is approximately equal to the distance $\Delta$ \eqref{eq:distance} between the two vectors. Since we concluded in the previous chapters that the rate of convergence between two random deviation vectors is a suitable proxy for measuring their convergence rates to the first CLV, we compute the SALI to determine the forward transient length needed in order to later compute the first CLV.\footnote{Here we use SALI because it is simple and cheap to compute, but $\Delta$ can be used instead. Either approach yields almost identical results since $\text{SALI}\approx\Delta$ for nearly (anti)-parallel deviation vectors.} In \figureref{fig:dna_sali}, we give the time evolution of the SALI, where we use a normalisation interval of $\tau=\qty{0.1}{ps}$ and we stop the computation once the SALI falls below a threshold of $10^{-14}$. Again we conclude that the orbit is chaotic. Furthermore, it is clear from the figure that the two converging deviation vectors are separated by a distance of only $10^{-14}$ after $t\approx\qty{460}{ps}$, so using a forward transient of \qty{e3}{ps} should be more than sufficient to accurately compute the first CLV, which we do next.

\subsection{Distributions of the first CLV and ILV}\label{sec:firstilv}
\begin{figure}[htbp]
	\centering
	\includegraphics[width=\linewidth]{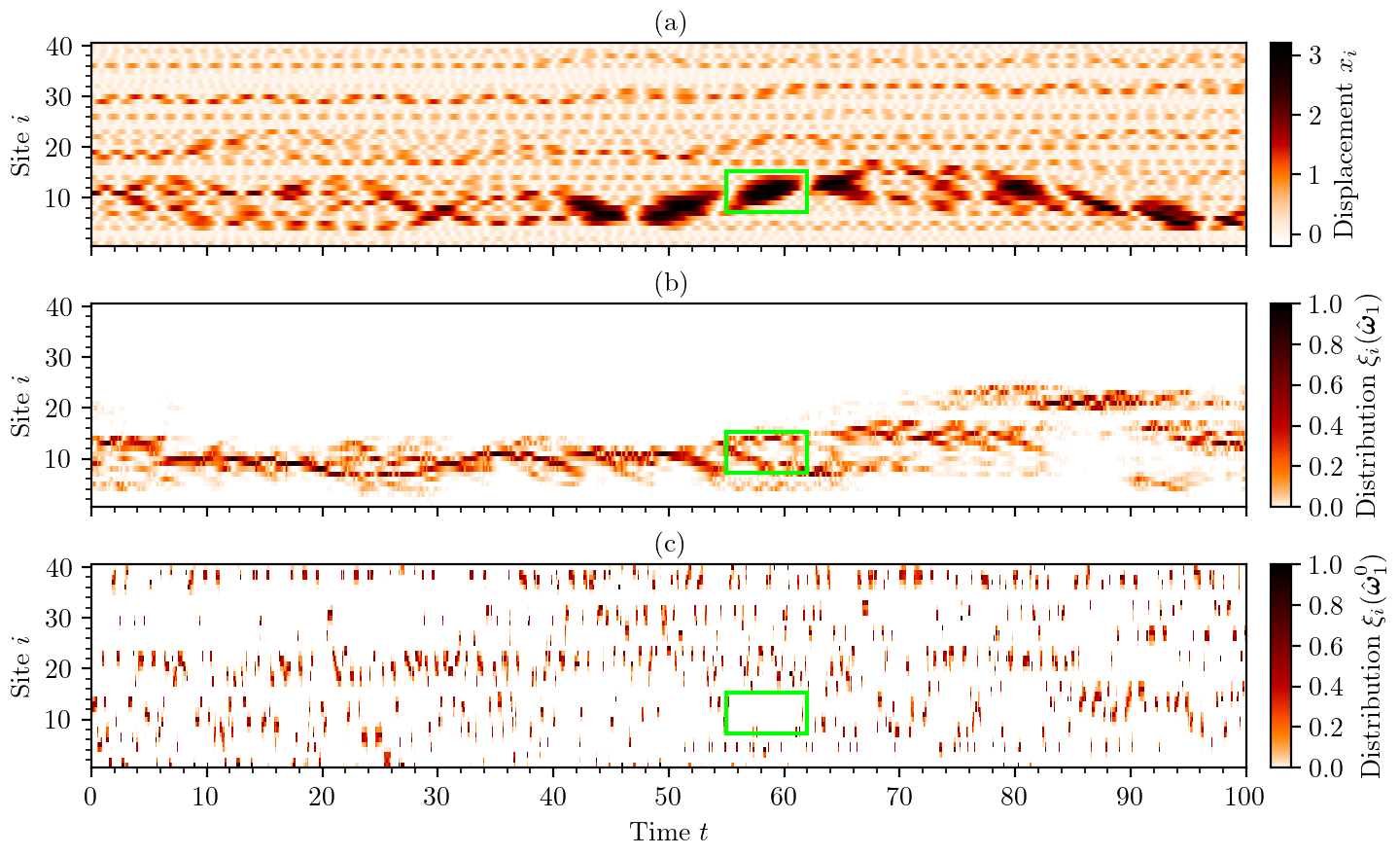}
	\caption{For a 40-site pure AT lattice: the spatiotemporal evolution of (a) the displacement $x_i$, (b) the distribution $\xi_i(\hat{\bs\omega}_1)$ of the first CLV (see \sectionref{sec:lyapunov}), and (c) the distribution $\xi_i(\hat{\bs\omega}_1^0)$ of the first ILV (see \sectionref{sec:ilv}). The values of $x_i$, $\xi_i(\hat{\bs\omega}_1)$, and $\xi_i(\hat{\bs\omega}_1^0)$ are given in the relevant colour scales. The green rectangle in panel (a) highlights a particular bubble, and the corresponding regions in (b) and (c) are also indicated by green rectangles. All panels share the same horizontal axis.}
	\label{fig:heat}
\end{figure}

Using the same 40-site AT lattice and initial condition as in \sectionref{sec:dna_lesali}, we evolve the state and a random deviation vector for \qty{e4}{ps} to allow for thermalisation of the lattice and for the deviation vector to converge to the first CLV. Based on \figureref{fig:dna_sali}, we expect satisfactory CLV convergence to occur in under \qty{e3}{ps}, but we use a longer transient of \qty{e4}{ps} to be safe and to ensure thermalisation of the lattice. Over a subsequent \qty{100}{ps} interval, we further evolve the state and deviation vector while computing the site displacements and the distributions of the first CLV $\hat{\bs\omega}_1$ and first ILV $\hat{\bs\omega}_1^0$ at each time step. The spatiotemporal evolution of the displacement $x_i$ and the distributions $\xi_i(\hat{\bs\omega}_1)$ and $\xi_i(\hat{\bs\omega}_1^0)$ are given in \figureref{fig:heat}. Note that time $t=0$ in this figure corresponds to the end of the transient. We see the formation of several bubbles in \figureref{fig:heat}(a) involving multiple sites, such as the one surrounded by a green rectangle in this figure.

As a brief aside, let us provide some physical context of the time scales used in our results before proceeding. Looking at quiet regions of the lattice in \figureref{fig:heat}(a) where sites oscillate with very small displacements (e.g.\ sites 1--3 for the entire \qty{100}{ps} interval), we see that these sites oscillate regularly with a period of roughly \qty{1}{ps}. If we ignore the effect of the coupling potential in \eqref{eq:dnahamiltonian} and assume the displacements at these sites are sufficiently small, then we can approximate these sites as uncoupled simple harmonic oscillators. In this approximation, the on-site Morse potential of \eqref{eq:dnahamiltonian} is given by $\tilde V(x)\approx D_{\scriptscriptstyle AT}a_{\scriptscriptstyle AT}^2 x^2$. Using the same mass parameter $m=\qty{0.031}{eV.ps^2.\angstrom^{-2}}$ as in \sectionref{sec:pbdsec} to model the mass of this oscillator, the period of such a harmonic oscillator can be shown to be
\begin{align}
	\sqrt{\frac{2\pi^2m}{D_{\scriptscriptstyle AT}a_{\scriptscriptstyle AT}^2}}=\qty{0.83}{ps},\label{eq:at_osc}
\end{align}
which agrees with our rough \qty{1}{ps} estimate from \figureref{fig:heat}(a). Therefore, our choice of time unit, the picosecond, physically corresponds to the approximate period of individual AT base pairs undergoing low-amplitude oscillations. Repeating this exercise for a pure GC lattice (which we study later in \sectionref{sec:gcc}), the period of low-amplitude oscillations of GC base pairs is only \qty{0.41}{ps}. This understanding provides some physical context to the time scales discussed in this chapter for PBD lattices.

We now return to the task of interpreting the remaining panels of \figureref{fig:heat}. Inspecting \figureref{fig:heat}(b), we see that the distribution of $\hat{\bs\omega}_1$ is quite localised\footnote{When we say that a distribution is localised, we mean that it is almost zero everywhere except for a few neighbouring sites.} and is usually located near the most energetic regions of the lattice where sites have large displacements, such as the region surrounded by the green rectangle in \figureref{fig:heat}(a) which highlights a bubble. The same behaviour was observed in \cite{HillebrandEtAl2019} for the DVD in a heterogeneous PBD lattice (i.e.\ consisting of both AT and GC base pairs), which is relevant here since the DVD approximates the distribution of $\hat{\bs\omega}_1$. On the other hand, the distribution of $\hat{\bs\omega}_1^0$ also appears very localised but moves rapidly throughout the lattice, as seen in \figureref{fig:heat}(c). A careful comparison between panels (a) and (c) in \figureref{fig:heat} reveals that the distribution of $\hat{\bs\omega}_1^0$ often avoids sites of maximal displacement, e.g.\ the bubble surrounded by the green rectangle in \figureref{fig:heat}(a) has a near-zero distribution $\xi_i(\hat{\bs\omega}_1^0)$, as seen by the corresponding region in \figureref{fig:heat}(c), indicating that the first ILV avoids sites inside this bubble.

\begin{figure}[htb]
	\centering
	\includegraphics[width=\textwidth]{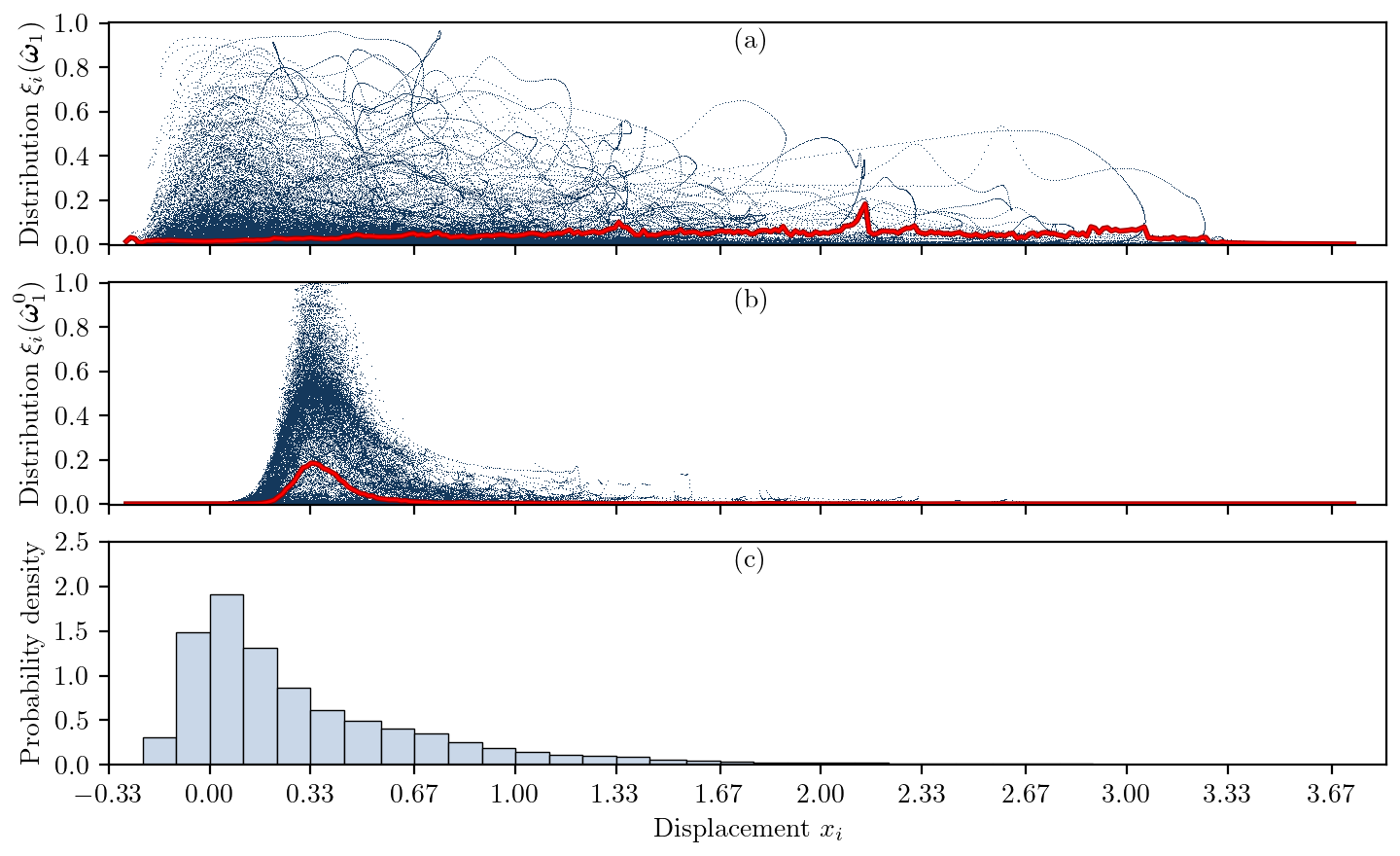}
	\caption{Scatter plots showing the relationship between displacement $x_i$ and the distributions $\xi_i$ used in \figureref{fig:heat} of (a) the first CLV $\hat{\bs\omega}_1$ and (b) the first ILV $\hat{\bs\omega}_1^0$ for a 40-site pure AT lattice. Binning the data into small displacement intervals, the average distribution in each bin is given by the red curves in (a) and (b). A normalised histogram of displacement is given in (c). All panels share the same horizontal axis.}
	\label{fig:scatter}
\end{figure}

To further investigate the previous observations that there may be a relationship between the displacement $x_i$ and the distributions $\xi_i(\hat{\bs\omega}_1)$ and $\xi_i(\hat{\bs\omega}_1^0)$, we present scatter plots in panels (a) and (b) of \figureref{fig:scatter} in which the coordinates of each plotted point are the respective displacement and distribution values from \figureref{fig:heat} at a common time and lattice site. We then bin these data into small displacement intervals and compute the mean of the distribution values in each interval, which is shown by the red curves in panels (a) and (b) of \figureref{fig:scatter}. A normalised histogram of the displacement data used in these scatter plots is presented in \figureref{fig:scatter}(c) where we see that near-zero displacements are most frequent, which is expected since this is the location of the bottom of the well of the Morse potential (see \figureref{fig:potentials}). The histogram of \figureref{fig:scatter}(c) has a long right tail, indicating that sites with very large displacements are rare. Returning to \figureref{fig:scatter}(a), we see no clear relationship between displacement and the distribution of the first CLV. There is a small spike in the red curve near $\qty{2.15}{\angstrom}$ in this figure, but this is not significant due to the scarcity of sites with such large displacements, and we find that additional runs using different initial conditions typically do not exhibit a similar spike in the distribution. On the other hand, it is clear from \figureref{fig:scatter}(b) that the distribution of the first ILV $\hat{\bs\omega}_1^0$ is greatest at sites with displacements near \qty{0.33}{\angstrom}, indicating that these sites are the most sensitively dependent on initial conditions at instants in time.

\begin{figure}[htbp]
	\centering
	\includegraphics[width=\linewidth]{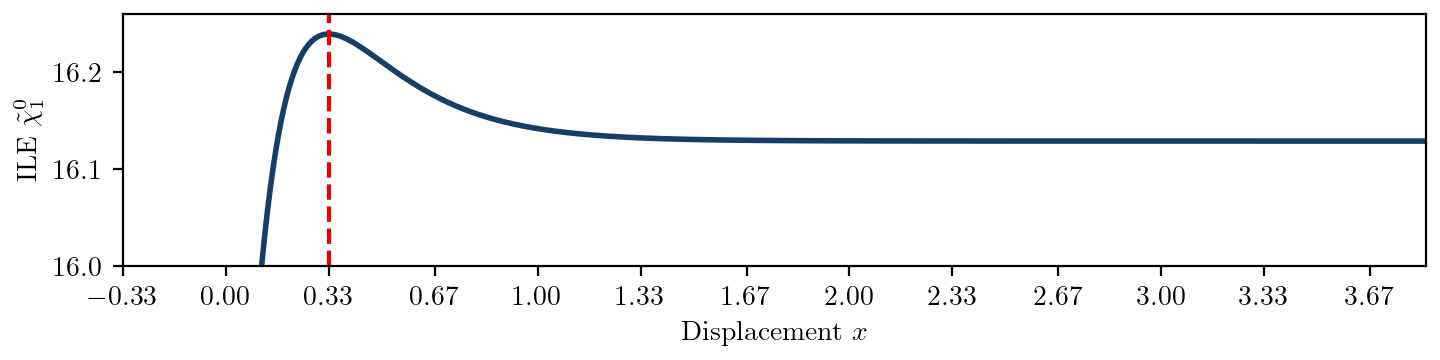}
	\caption{The first ILE $\tilde\chi_1^0$ \eqref{eq:maxstiff} of a single AT oscillator as a function of displacement $x$. A maximum of $\tilde\chi_1^0$ occurs at $x=\qty{0.33}{\angstrom}$, which is indicated by the red dashed line.}
	\label{fig:ile}
\end{figure}

Why does the first ILV prefer sites with a particular displacement? To answer this, we consider a single AT oscillator whose Hamiltonian $\tilde H(x,p)$ consists of a single kinetic term and on-site potential term from \eqref{eq:dnahamiltonian} without any coupling. In particular,
\begin{align}
	\tilde H(x,p)=\tilde T(p)+\tilde V(x),
\end{align}
where $\tilde T(p)$ is the kinetic energy as a function of only the momentum coordinate $p$ and $\tilde V(x)$ is the potential energy as a function of only the displacement coordinate $x$. Note that we use (\,$\tilde{}$\,) to distinguish quantities relating to a single oscillator from ones relating to the full lattice. In our case, for a single AT oscillator,
\begin{align}
\begin{split}
	\tilde T(p)&=\frac{p^2}{2m},\\
	\tilde V(x)&=D_{\scriptscriptstyle AT}(e^{-a_{\scriptscriptstyle AT}x}-1)^2.
\end{split}
\end{align}
The Jacobian matrix for this single-oscillator system is given by
\begin{align}
	\tilde J=\begin{pmatrix}
		0 & \tilde T''(p) \\
		-\tilde V''(x) & 0
	\end{pmatrix},
\end{align}
where $('')$ denotes the second derivative with respect to the relevant variable ($x$ or $p$). Therefore, the symmetric part of the Jacobian $\tilde J$ is
\begin{align}
	\frac{\tilde J\transpose + \tilde J}{2} = \begin{pmatrix}0& \frac12(\tilde T''(p)-\tilde V''(x))\\\frac12(\tilde T''(p)-\tilde V''(x))&0\end{pmatrix},
\end{align}
whose two eigenvalues are $\pm\frac12(\tilde T''(p)-\tilde V''(x))$, so the first (i.e.\ maximum) ILE is
\begin{align}
	\tilde\chi_1^0=\left|\frac{\tilde T''(p)-\tilde V''(x)}{2}\right|.\label{eq:maxstiff}
\end{align}
Since $\tilde T''(p)=1/m$ is a constant, $\tilde\chi_1^0$ is a function of only displacement $x$, which we plot in \figureref{fig:ile}. We see from the figure that $\tilde\chi_1^0$ is maximised on the given displacement interval when $x\approx\qty{0.33}{\angstrom}$. Since an AT lattice in the PBD model consists of many coupled Morse oscillators, each with various displacements at any particular time, we would expect the distribution of the first ILV $\hat{\bs\omega}_1^0$ (for the full lattice) to be located at sites with displacements near \qty{0.33}{\angstrom}. Indeed, this is what we observe from \figureref{fig:scatter}(b) where the distribution of $\hat{\bs\omega}_1^0$ is typically maximised when the displacement is near \qty{0.33}{\angstrom}. This argument relies on the assumption that ILV distributions are localised so that they can be ``located'' at sites of a particular displacement. This localisation of the first ILV distribution can be seen qualitatively by close inspection of \figureref{fig:heat}(c), but we discuss in more detail the spread of ILV distributions in \sectionref{sec:character}.

\begin{figure}[htbp]
	\centering
	\includegraphics[width=\linewidth]{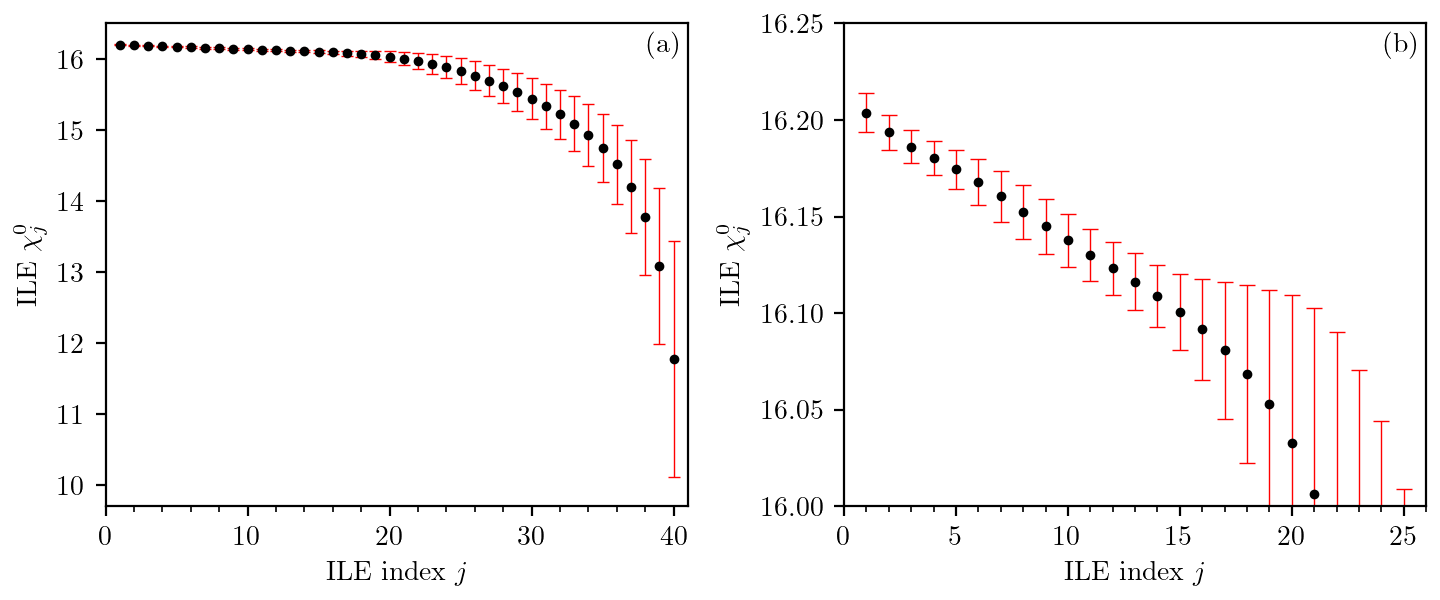}
	\caption{The ILEs $\chi_j^0$ (see \sectionref{sec:ilv}), $i=j,\dots,40$, computed over a \qty{100}{ps} interval for a 40-site pure AT lattice are given in (a), using the same initial condition as \figureref{fig:heat}. The time average of each ILE is plotted as a black dot and the red error bars denote one standard deviation from the mean. Panel (b) is the same as (a), except for the scales of the axes.}
	\label{fig:ile_spectrum_wide}
\end{figure}

While there is no strong correlation between the first CLV distribution and displacement, we have seen that the first ILV is typically distributed at sites with displacements near \qty{0.33}{\angstrom}, hence these sites are instantaneously the most sensitively dependent on initial conditions. However, if the ILEs turn out to be nearly degenerate, then the meaning of the first ILV as the deviation vector whose length grows the fastest at that instant in time is somewhat diminished, since other ILVs may grow practically equally as fast if their corresponding ILEs are nearly maximal. Utilising the same initial condition and the same \qty{100}{ps} interval which were used to generate Figs.~\ref{fig:heat} and~\ref{fig:scatter}, we therefore compute the spectrum of ILEs at each point in time, and the time averages of each ILE are presented in \figureref{fig:ile_spectrum_wide} together with their respective standard deviations (red error bars). Note that we only present the first half of the 80 ILEs in the spectrum due to the symmetry \eqref{eq:specsym}. We observe from the figure that approximately the first 20 ILEs are nearly degenerate,\footnote{Due to the spectrum symmetry, the last 20 ILEs are also nearly degenerate.} which tempers the significance of our analysis of the first ILV in isolation. Therefore, in \sectionref{sec:character} we analyse the remaining ILVs to obtain a more holistic view of which sites depend most sensitively on initial conditions at an instant.

\subsection{Characteristics of ILV distributions}\label{sec:character}
Now that we have analysed the first ILV distribution, we move on to a discussion of the other ILV distributions, which we begin by quantifying their spread. Applying the same method used in \sectionref{sec:firstilv} to compute the first ILV distribution for a 40-site AT lattice over a \qty{100}{ps} interval, we compute the distributions of the first 40 ILVs over the same interval. As with the ILEs, it is unnecessary to compute all 80 ILV distributions due to the symmetry \eqref{eq:specvecsym}. Previously we used only a single initial condition to analyse the distribution of the first ILV, but now we perform our computations for 10 random initial conditions to ensure that our results are not particular to a single run. We use many runs for this investigation since we find the relevant features of these ILV distributions are more easily distinguished when more data is used.

\begin{figure}[htbp]
	\centering
	\includegraphics[width=\linewidth]{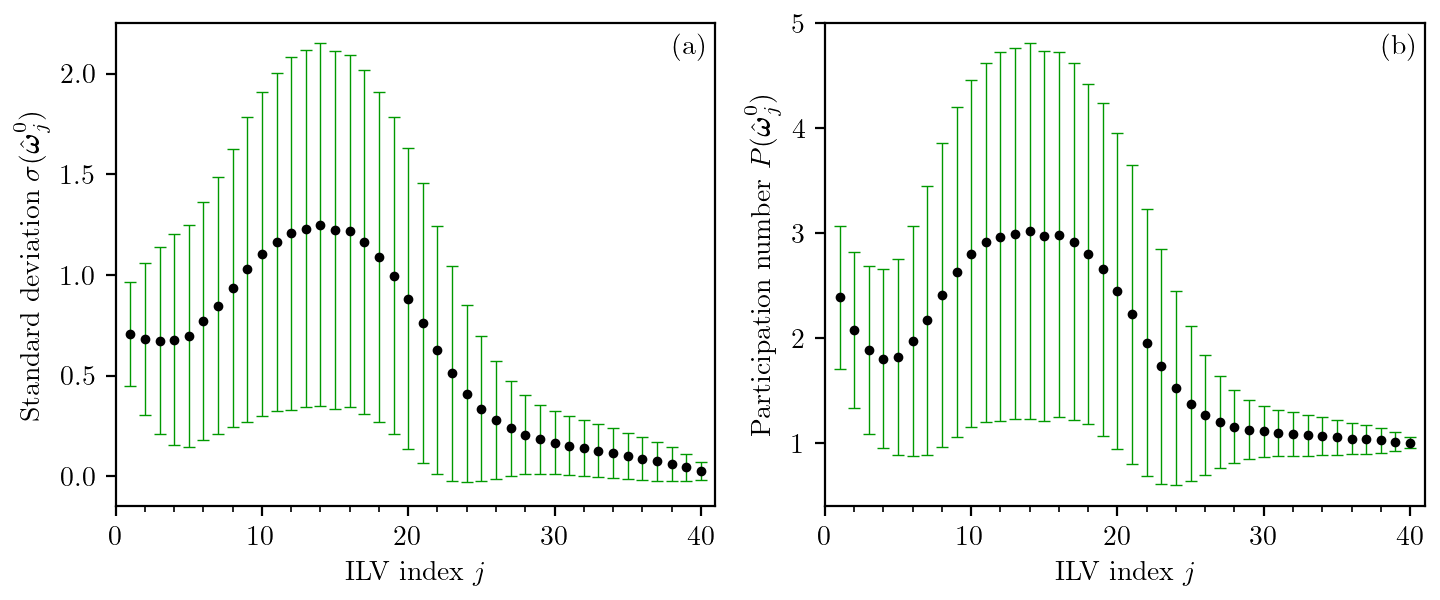}
	\caption{The (a) standard deviation $\sigma$ \eqref{eq:sd} and (b) participation number $P$ \eqref{eq:partip} of the distribution of each ILV $\hat{\bs\omega}_j^0$ (see \sectionref{sec:ilv}), $i=j,\dots,40$, computed at every time step in a \qty{100}{ps} interval for each of 10 random initial conditions for a 40-site pure AT lattice. The black dots represent the mean of the respective quantities $\sigma$ and $P$, and the error bars represent one standard deviation from the mean.}
	\label{fig:deviation}
\end{figure}

In order to quantify the spread of the ILV distributions, we compute the standard deviation $\sigma$ and participation number $P$ using the method described in \sectionref{sec:locspread} for periodic lattices. We denote the standard deviation of the distribution of the $j$-th ILV $\hat{\bs\omega}_j^0$ by $\sigma(\hat{\bs\omega}_j^0)$, and similarly the participation number by $P(\hat{\bs\omega}_j^0)$. For each of the 40 ILVs indexed by $j$, we compute $\sigma(\hat{\bs\omega}_j^0)$ and $P(\hat{\bs\omega}_j^0)$ at each time step over the \qty{100}{ps} interval for each of the 10 different initial conditions, then we compute the average of each of these quantities and plot the results in \figureref{fig:deviation}. Furthermore, we also compute the standard deviation of the $\sigma(\hat{\bs\omega}_j^0)$ and $P(\hat{\bs\omega}_j^0)$ values and represent them by the error bars in the figure. We see from \figureref{fig:deviation} that the distribution of the 40-th ILV is the most localised with a standard deviation of $\sigma(\hat{\bs\omega}_{40}^0)\approx0$ and a participation number of $P(\hat{\bs\omega}_{40}^0)\approx1$ on average, so both quantities indicate that this ILV distribution is typically located at only a single site with zero spread. On the other hand, the 14-th ILV distribution is the least localised, with $\sigma(\hat{\bs\omega}_{14}^0)\approx1.2$ and $P(\hat{\bs\omega}_{14}^0)\approx3.1$ on average, so its distribution is usually localised to a site and its immediate neighbours. In conclusion, it is clear from these measures of spread that typical ILV distributions are spatially localised within the lattice to 1--3 neighbouring sites.

\begin{figure}[htbp]
	\centering
	\includegraphics[width=\linewidth]{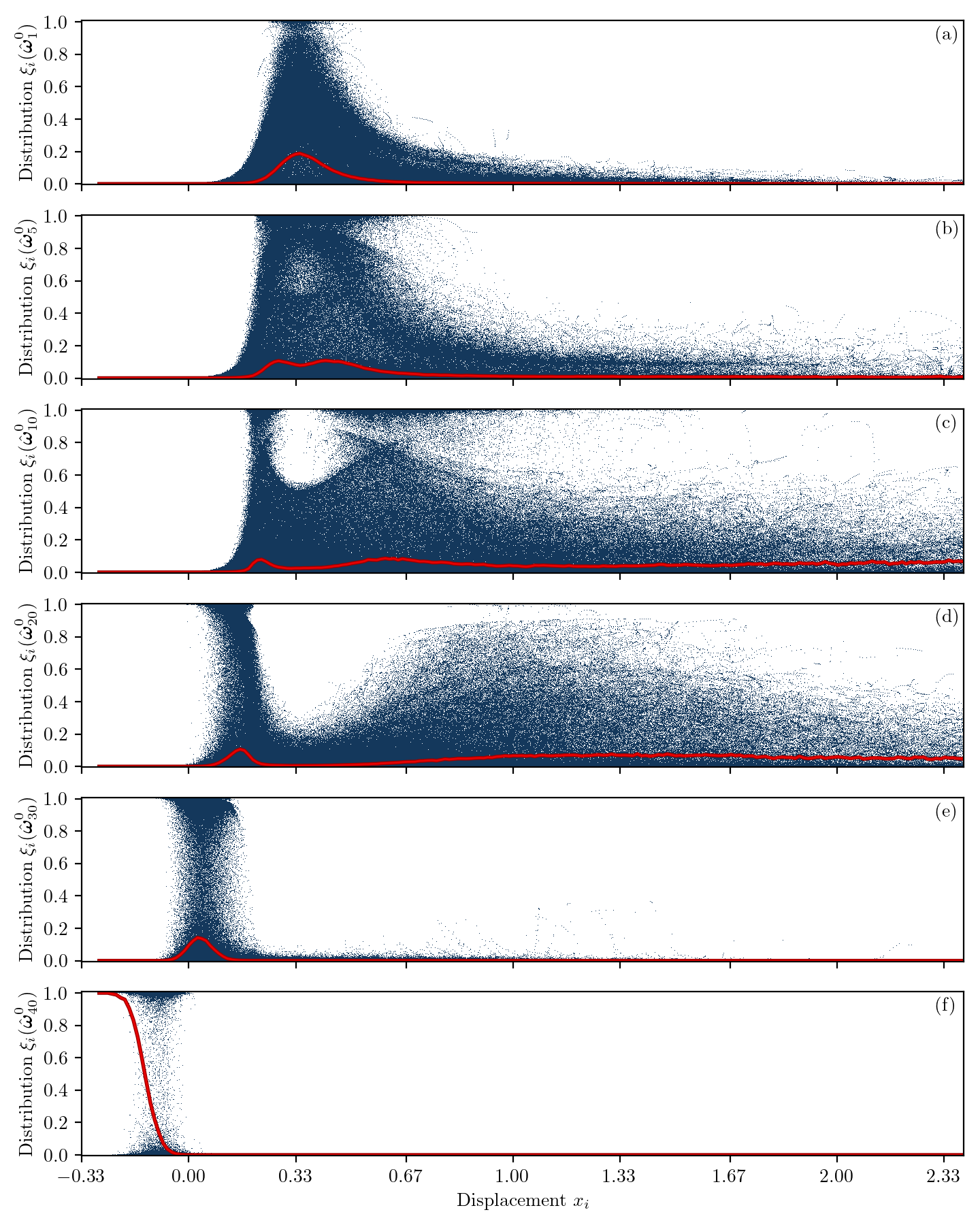}
	\caption{Scatter plots showing the relationship between displacement $x_i$ and the distributions $\xi_i$ of the ILVs (a) $\hat{\bs\omega}_1^0$, (b) $\hat{\bs\omega}_5^0$, (c) $\hat{\bs\omega}_{10}^0$, (d) $\hat{\bs\omega}_{20}^0$, (e) $\hat{\bs\omega}_{30}^0$, and (f) $\hat{\bs\omega}_{40}^0$ for a 40-site pure AT lattice, using data from 10 independent runs over a \qty{100}{ps} interval. Binning the data into small displacement intervals, the average distribution value in each bin is given by the red curves. All panels share the same horizontal axis.}
	\label{fig:all_ilvs}
\end{figure}

Taking the set of ILVs used in \figureref{fig:deviation} over a \qty{100}{ps} interval for 10 runs using different initial conditions in each case, we now study the relationship between their distributions and displacements. Applying the method used to produce \figureref{fig:scatter}(b) for the first ILV of a single run, we present scatter plots in \figureref{fig:all_ilvs} for several different ILVs using data from all 10 runs. Recalling the long right tail of the histogram in \figureref{fig:scatter}(c), we truncate the horizontal axes of \figureref{fig:all_ilvs} at \qty{2.39}{\angstrom}, which we determine to be the 99-th displacement percentile of the data for the 10 runs, thus we exclude any points with very large displacements from our presentation due to the sparse data. Examining \figureref{fig:all_ilvs}(a), we see that the distribution of $\hat{\bs\omega}_1^0$ peaks at around \qty{0.33}{\angstrom} as seen previously for the single run in \sectionref{sec:firstilv}. In \figureref{fig:all_ilvs}(b), however, the distribution of $\hat{\bs\omega}_{5}^0$ appears to have two peaks, one on either side of \qty{0.33}{\angstrom}. We again see two peaks in the distribution of $\hat{\bs\omega}_{10}^0$ in panel (c), but these are further from \qty{0.33}{\angstrom} than in the previous case, and likewise for $\hat{\bs\omega}_{20}^0$ in panel (d). While the left peak (corresponding to a small displacement) keeps moving leftward as we transition to the respective distributions of $\hat{\bs\omega}_{30}^0$ and $\hat{\bs\omega}_{40}^0$ in panels (e) and (f), the right peak vanishes altogether. In summary, the progression shown in \figureref{fig:all_ilvs} demonstrates how the relationship between the ILV distributions and displacement changes as the index $j$ of the ILV increases.

We offer a heuristic explanation of these observations. Recall from \figureref{fig:ile} that the maximum ILE $\tilde\chi_1^0$ for a single AT oscillator gets its largest value at \qty{0.33}{\angstrom}. However, if displacements at \qty{0.33}{\angstrom} are inaccessible, then the largest attainable $\tilde\chi_1^0$ would occur at \textit{two} displacement values on either side of the maximum at \qty{0.33}{\angstrom}. Concentrating on the full lattice system again, once the first few ILVs in the AT lattice have claimed all sites with displacements near \qty{0.33}{\angstrom} at any particular time, what remains are mostly sites with displacements other than \qty{0.33}{\angstrom}. As confirmed by \figureref{fig:all_ilvs}(b), an ILV such as $\hat{\bs\omega}_{5}^0$ is then mainly located\footnote{Recall from \figureref{fig:deviation} that ILV distributions are typically very localised, so it is meaningful to speak of their location in the lattice.} at sites with displacements slightly larger or smaller than \qty{0.33}{\angstrom} since such sites contribute the most to the ILE $\chi_{5}^0$ in the absence of sites with displacements of \qty{0.33}{\angstrom}. As the index $j$ of the ILV $\hat{\bs\omega}_{j}^0$ increases, the displacements of sites in the distribution move further away from \qty{0.33}{\angstrom}. Eventually, what remains for the higher-indexed ILVs are sites with displacements which contribute minimally to the corresponding ILE; since the left tail of the curve in \figureref{fig:ile} extends lower than the right tail, we would expect that such sites which contribute the least to the overall lattice ILE have only small displacements, and indeed we see this behaviour in \figureref{fig:all_ilvs} with ILVs $\hat{\bs\omega}_{30}^0$ and $\hat{\bs\omega}_{40}^0$. We assume in this argument that many ILVs cannot all be concentrated at the same sites, thereby excluding sites with displacements such as \qty{0.33}{\angstrom} from the distributions of higher-indexed ILVs. This assumption is valid since ILVs are spatially localised and form a basis of the tangent space, so each coordinate must have a significant non-zero component in some ILV(s) and hence all lattice sites feature prominently in the distribution of some ILV(s).

In totality, our observations from \figureref{fig:all_ilvs} indicate that different lattice sites exhibit different levels of instability at an instant when perturbed, depending on the displacement of those sites. We interpret this level of instability of a site as its contribution to the system's overall instability, and hence chaoticity, at that moment in time. Since many ILV distributions contributed to this conclusion, we propose measuring a site's contribution to chaos at an instant by taking an average of the ILV distributions, weighted by their corresponding ILEs. We therefore define the \textit{weighted ILV distribution} $\bar{\xi}_i$ at the $i$-th lattice site ($i=1,\dots,n$) as
\begin{align}
	\bar{\xi}_i = \frac{\sum_{j=1}^n \chi_j^0\, \xi_i(\hat{\bs\omega}_j^0)}{\sum_{j=1}^n \chi_j^0},\label{eq:weight}
\end{align}
where $\bar{\xi}_i\in[0,1]$ and $\sum_{i=1}^n\bar{\xi}_i=1$. Note that we only include the first $n$ ILEs and ILV distributions in this definition due to the symmetries \eqref{eq:specsym} and \eqref{eq:specvecsym}, since if we instead used all $N=2n$ ILEs and ILVs then the denominator in \eqref{eq:weight} would be zero. It is this weighted ILV distribution $\bar\xi_i$ that we propose as an instantaneous measure of the $i$-th site's contribution to chaos in the lattice.

\begin{figure}[htbp]
	\centering
	\includegraphics[width=\linewidth]{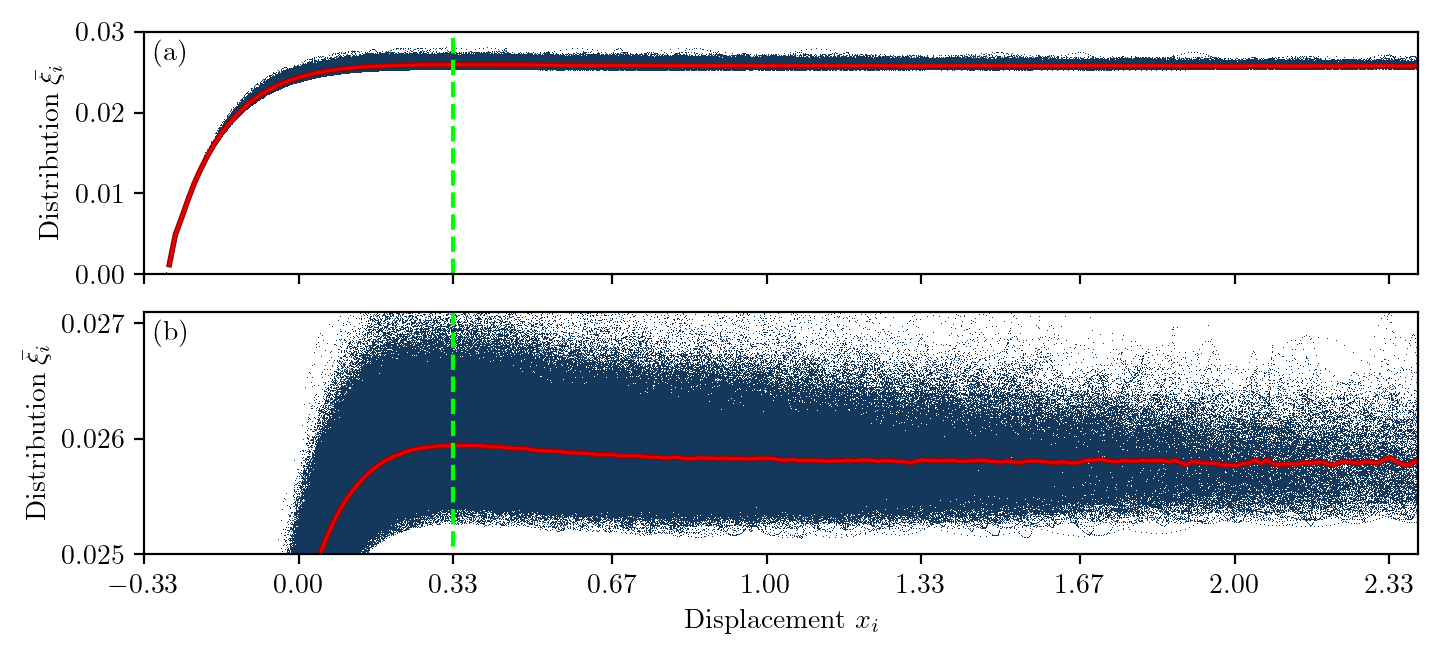}
	\caption{Scatter plots showing the relationship between displacement $x_i$ and the weighted distribution $\bar{\xi}_i$ \eqref{eq:weight} of ILVs for a 40-site pure AT lattice, using data from 10 independent runs over a \qty{100}{ps} interval. Binning the data into small displacement intervals, the average distribution value in each bin is given by the red curves. Panels (a) and (b) are plots of the same data using different scales of the vertical axis. The green dashed line in both panels denotes $x_i=\qty{0.33}{\angstrom}$. Both panels share the same horizontal axis.}
	\label{fig:weight}
\end{figure}

Using all 40 ILVs computed previously over a \qty{100}{ps} interval for 10 different runs, as well as their corresponding ILEs, we compute the weighted ILV distribution $\bar{\xi}_i$. The value of $\bar{\xi}_i$ is relevant to the $i$-th site at a particular time which has a displacement $x_i$. We present a scatter plot relating the weighted ILV distribution and displacement in \figureref{fig:weight}, where we have once again truncated the displacement axis at the 99-th percentile. From \figureref{fig:weight}(a), we see that the weighted distribution increases with increasing displacement until $x_i\approx\qty{0.15}{\angstrom}$ after which the weighted distribution remains approximately constant. However, we give a new perspective of this plot in \figureref{fig:weight}(b) by rescaling the vertical axis, and from this, we can see that the weighted distribution has a small peak which occurs at approximately \qty{0.33}{\angstrom} (indicated by the green dashed line). As expected from our analysis of the first ILV distribution alone (see \figureref{fig:scatter}), the weighted distribution is maximised at sites with a displacement of \qty{0.33}{\angstrom} since the first ILV distribution receives the greatest weighting in $\bar{\xi}_i$. However, sites with displacements greater than approximately $\qty{0.15}{\angstrom}$ have very similar values of $\bar{\xi}_i$, which can be explained by the near-degeneracy of the first 20 ILEs seen previously in \figureref{fig:ile_spectrum_wide}, whose corresponding ILV distributions are weighted almost equally in $\bar{\xi}_i$. Since the first 20 ILVs are usually located at sites with medium-to-large displacements $x_i\gtrsim\qty{0.15}{\angstrom}$ and the higher-indexed ILV distributions (which have smaller weights in $\bar{\xi}_i$) are mainly located at sites with small displacements $x_i\lesssim\qty{0.15}{\angstrom}$ (see \figureref{fig:all_ilvs}), it makes sense that the weighted distribution is almost constant at sites with medium-to-large displacements.

\begin{figure}[htbp]
	\centering
	\includegraphics[width=\linewidth]{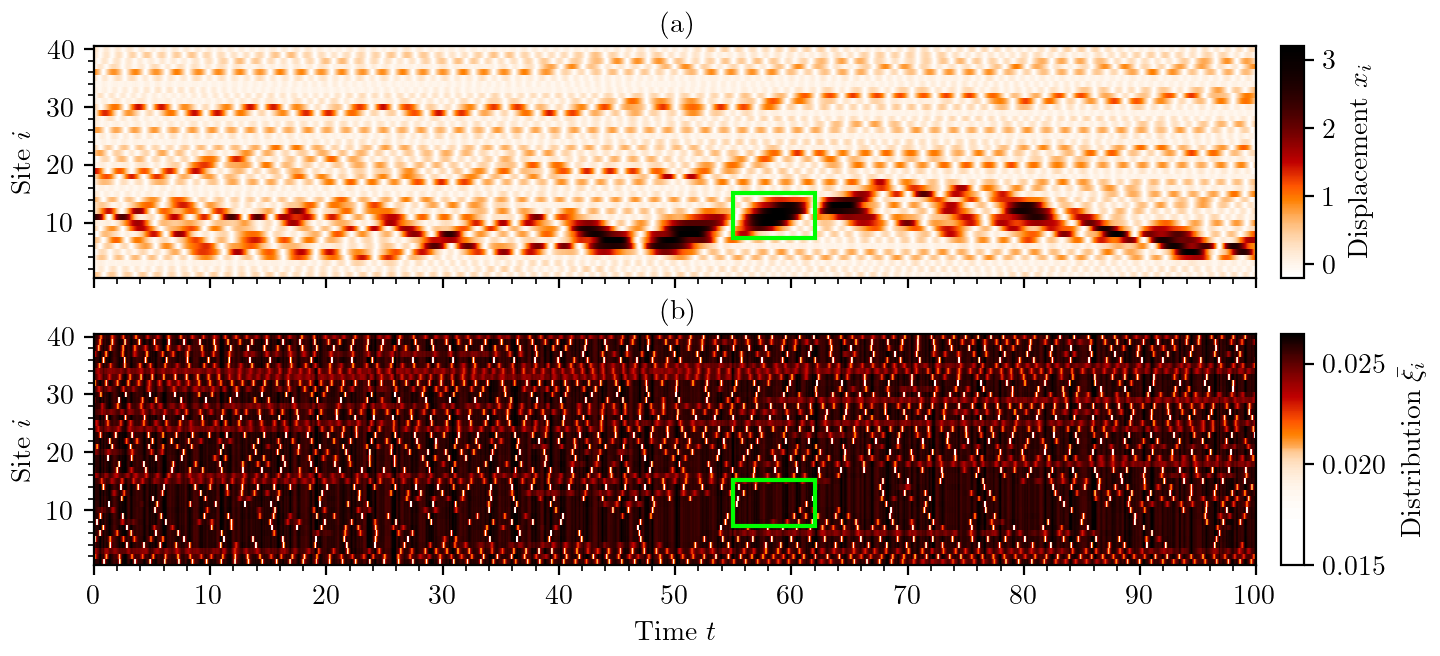}
	\caption{For a 40-site pure AT lattice: the spatiotemporal evolution of (a) the displacement $x_i$ and (b) the weighted ILV distribution $\bar{\xi}_i$ \eqref{eq:weight} using the same initial condition as in \figureref{fig:heat}. The green rectangle in panel (a) highlights a particular bubble, and the corresponding region in (b) is also indicated by a green rectangle. Both panels share the same horizontal axis.}
	\label{fig:weighted_heat}
\end{figure}

Making use of the same simulation used to produce \figureref{fig:heat}, we present the spatiotemporal evolution of the displacement $x_i$ and the weighted ILV distribution $\bar{\xi_i}$ in \figureref{fig:weighted_heat}(b). Note that panel (a) in Figs.~\ref{fig:heat} and~\ref{fig:weighted_heat} are the same. Comparing the location of bubbles in \figureref{fig:weighted_heat}(a), such as the one inside the green rectangle, to the value of the weighted ILV distribution $\bar{\xi_i}$ in \figureref{fig:weighted_heat}(b), it appears that large values of $\bar{\xi_i}$ typically occur at sites with large displacements, including those inside bubbles. However, sites with large displacements can be shown to linearise the potential of \eqref{eq:dnahamiltonian}, so we would not expect bubbles to contribute strongly to the chaoticity of the system, which is in opposition to our results from Figs.~\ref{fig:weight} and~\ref{fig:weighted_heat}. This suggests that further work is needed in order to understand the correct interpretation of the weighted distribution $\bar{\xi}_i$ and its relationship with chaos.

\section{Numerical results for a pure GC lattice}\label{sec:gcc}
In this section, we present numerical results for a 40-site pure GC lattice using the same methodology as we used for a pure AT lattice in \sectionref{sec:all_at}. As before when numerically integrating this system, we use the ABA864 \eqref{eq:aba} integrator, but now with a smaller time step of \qty{0.005}{ps} which we find bounds the relative energy error $E_r\lesssim10^{-7}$. We choose random initial conditions using the same method as in the AT case, except we use the energy density from \eqref{eq:energydensity} of $E_n^{\scriptscriptstyle GC}=\qty{0.0450}{eV}$ per site, which corresponds to a physiological temperature for a pure GC lattice. This section follows the same structure as \sectionref{sec:all_at}, but we avoid repeating some discussions here which are completely analogous to the pure AT case.

\subsection{Lyapunov exponents and the SALI}\label{sec:dna_gc_lesali}
\begin{figure}[htbp]
	\centering
	\includegraphics[width=\linewidth]{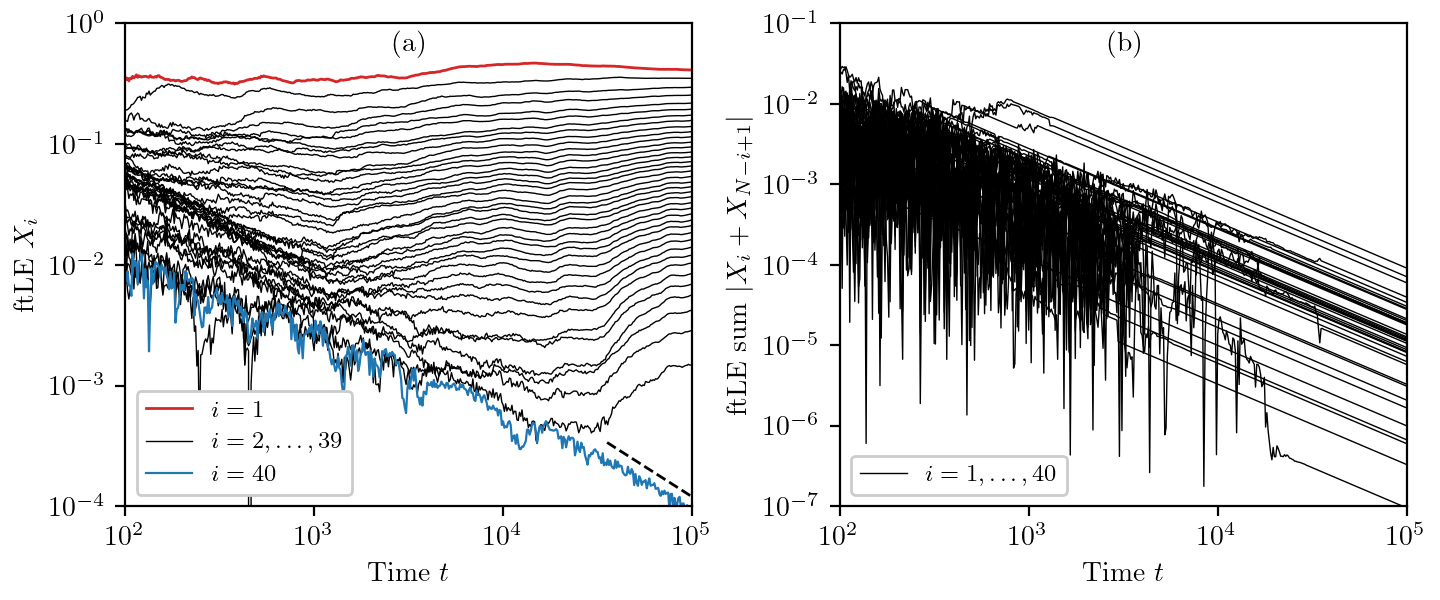}
	\caption{Similar to \figureref{fig:dna_lces}, but for a 40-site pure GC lattice in this case.}
	\label{fig:dna_gc_lces}
\end{figure}

Using the same method as in \sectionref{sec:dna_lesali}, we compute the full spectrum of ftLEs over an integration time of $10^5$ units for a 40-site pure GC lattice with an orthonormalisation interval of $\tau=\qty{0.05}{ps}$. We present the time evolution of the first 40 ftLEs in \figureref{fig:dna_gc_lces}(a). From the figure, we see that the first 39 ftLEs saturate to positive constants, while $X_{40}$ decays to zero following $\propto t^{-1}$. The time evolution of $|X_i+X_{N-i+1}|$, $i=1,\dots,40$, $N=80$, is given in \figureref{fig:dna_gc_lces}(b) where we see this quantity tends to zero for each $i$, thus the computed LE spectrum has the expected symmetry. Our results show that all LEs are non-zero except for the middle two, $\chi_{40}=\chi_{41}=0$. The ftmLE saturates to a positive value of $X_1=0.41$ after \qty{e5}{ps}. Since our estimate for the mLE is positive, we understand that the orbit is chaotic. These results for the LE spectrum are similar to the pure AT case in \sectionref{sec:dna_lesali}.

\begin{figure}[htbp]
	\centering
	\includegraphics[width=0.55\linewidth]{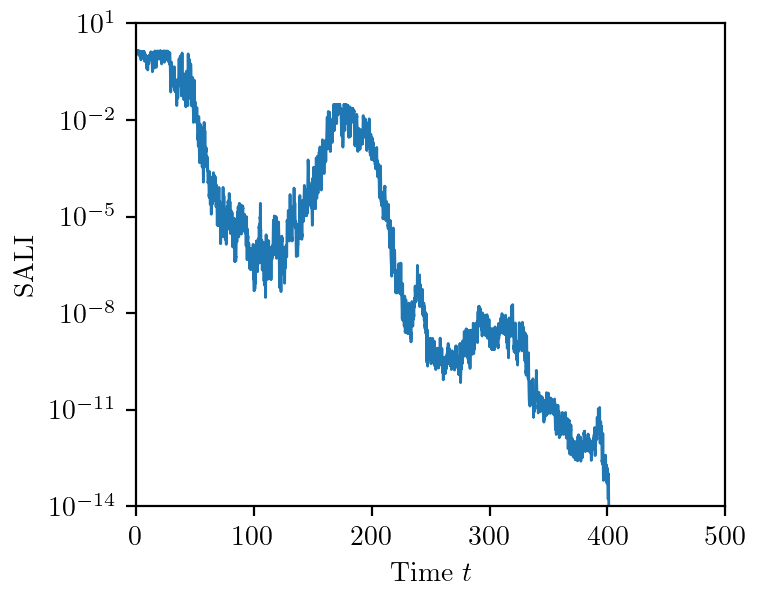}
	\caption{Similar to \figureref{fig:dna_sali}, but for a 40-site pure GC lattice in this case.}
	\label{fig:dna_gc_sali}
\end{figure}

We now compute the SALI for a pure GC lattice using the same initial condition that was used for the ftLE computation. In \figureref{fig:dna_gc_sali}, we give the time evolution of the SALI, where we use a normalisation interval of $\tau=\qty{0.05}{ps}$ and we stop the computation once the SALI falls below a threshold of $10^{-14}$. It is clear from the figure that the two converging deviation vectors are separated by a distance (approximately equal to the SALI) of only $10^{-14}$ after $t\approx\qty{400}{ps}$, so using a forward transient of at least \qty{400}{ps} should be sufficient to accurately compute the first CLV. Note that SALI decays to zero slightly faster than the pure AT case, which can be explained by the larger spectral gap between the first two LEs in the GC case: we estimate from \figureref{fig:dna_lces} for AT that $\chi_1-\chi_2=0.05$, while from \figureref{fig:dna_gc_lces} we find that $\chi_1-\chi_2=0.06$ for GC.

\subsection{Distributions of the first CLV and ILV}\label{sec:firstilv_gc}
\begin{figure}[htbp]
	\centering
	\includegraphics[width=\linewidth]{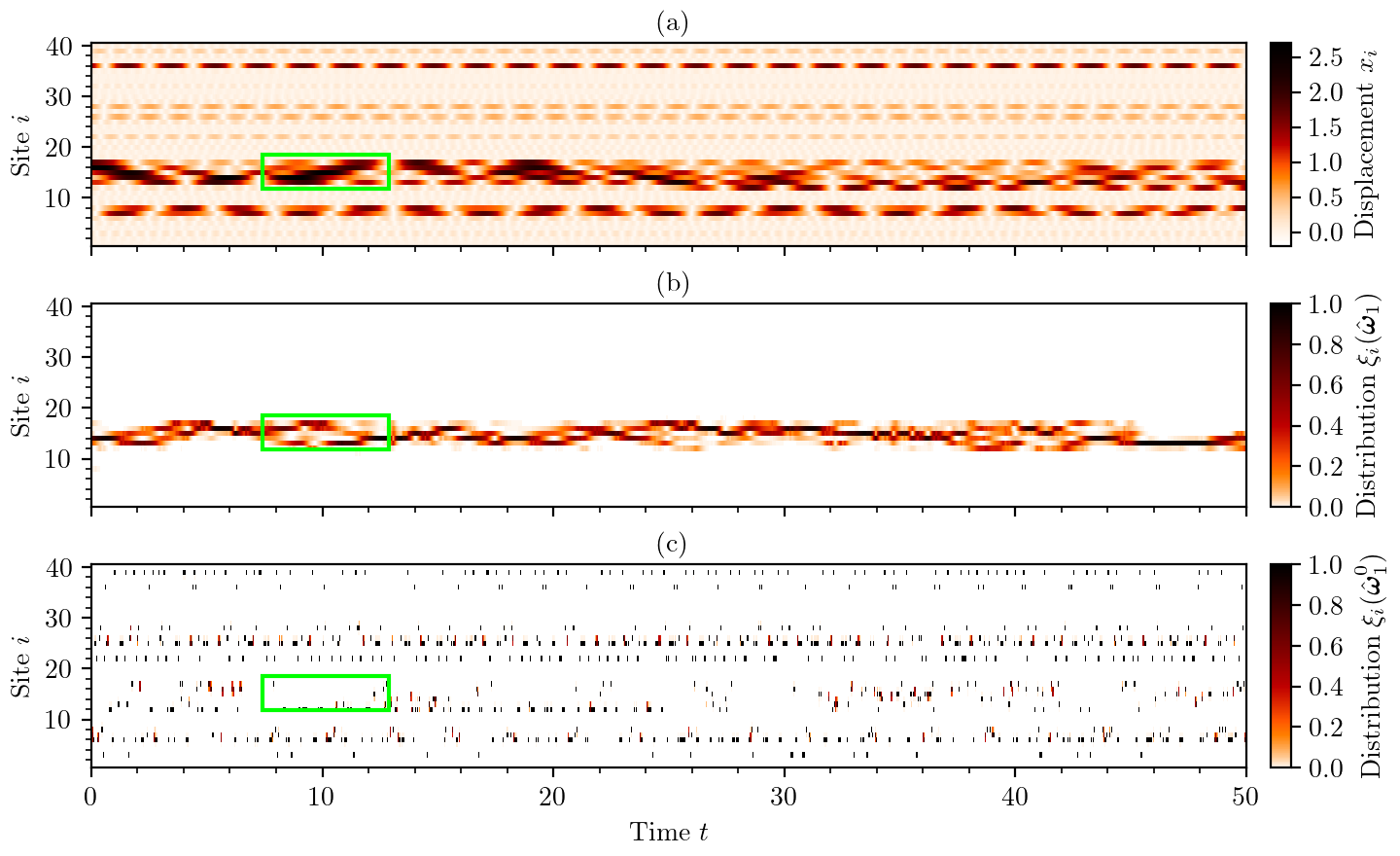}
	\caption{Similar to \figureref{fig:heat}, but for a 40-site pure GC lattice over a \qty{50}{ps} interval, and the green rectangle highlights a different region in this case.}
	\label{fig:heat_gc}
\end{figure}

Using the same 40-site GC lattice and initial condition as in \sectionref{sec:dna_gc_lesali}, we evolve the state and a random deviation vector for \qty{e4}{ps} to allow for thermalisation and for the deviation vector to converge to the first CLV. Over a subsequent \qty{50}{ps} interval, we further evolve the state and deviation vector while computing the site displacements and the first CLV and ILV distributions, $\hat{\bs\omega}_1$ and $\hat{\bs\omega}_1^0$, at each time step. These results are given in \figureref{fig:heat_gc}. Note that we choose a shorter time interval of \qty{50}{ps} to integrate over than in the AT case due to the faster dynamics of GC which was discussed around \eqref{eq:at_osc} in the context of low-amplitude oscillations of individual AT and GC base pairs. We see from \figureref{fig:heat_gc}(b) that the distribution of $\hat{\bs\omega}_1$ is usually localised near energetic regions of the lattice where nearby sites have large displacements. On the other hand, the distribution of $\hat{\bs\omega}_1^0$ appears very localised but moves rapidly throughout the lattice, as seen in \figureref{fig:heat_gc}(c). A careful comparison between panels (a) and (c) in \figureref{fig:heat_gc} reveals that the distribution of $\hat{\bs\omega}_1^0$ typically avoids sites of maximal displacement, e.g.\ the bubble inside the green rectangle of \figureref{fig:heat_gc}(a) corresponds to a region in \figureref{fig:heat_gc}(c) where typically $\xi_i(\hat{\bs\omega}_1^0)=0$. This result is the same as in the pure AT case.

\begin{figure}[htbp]
	\centering
	\includegraphics[width=\textwidth]{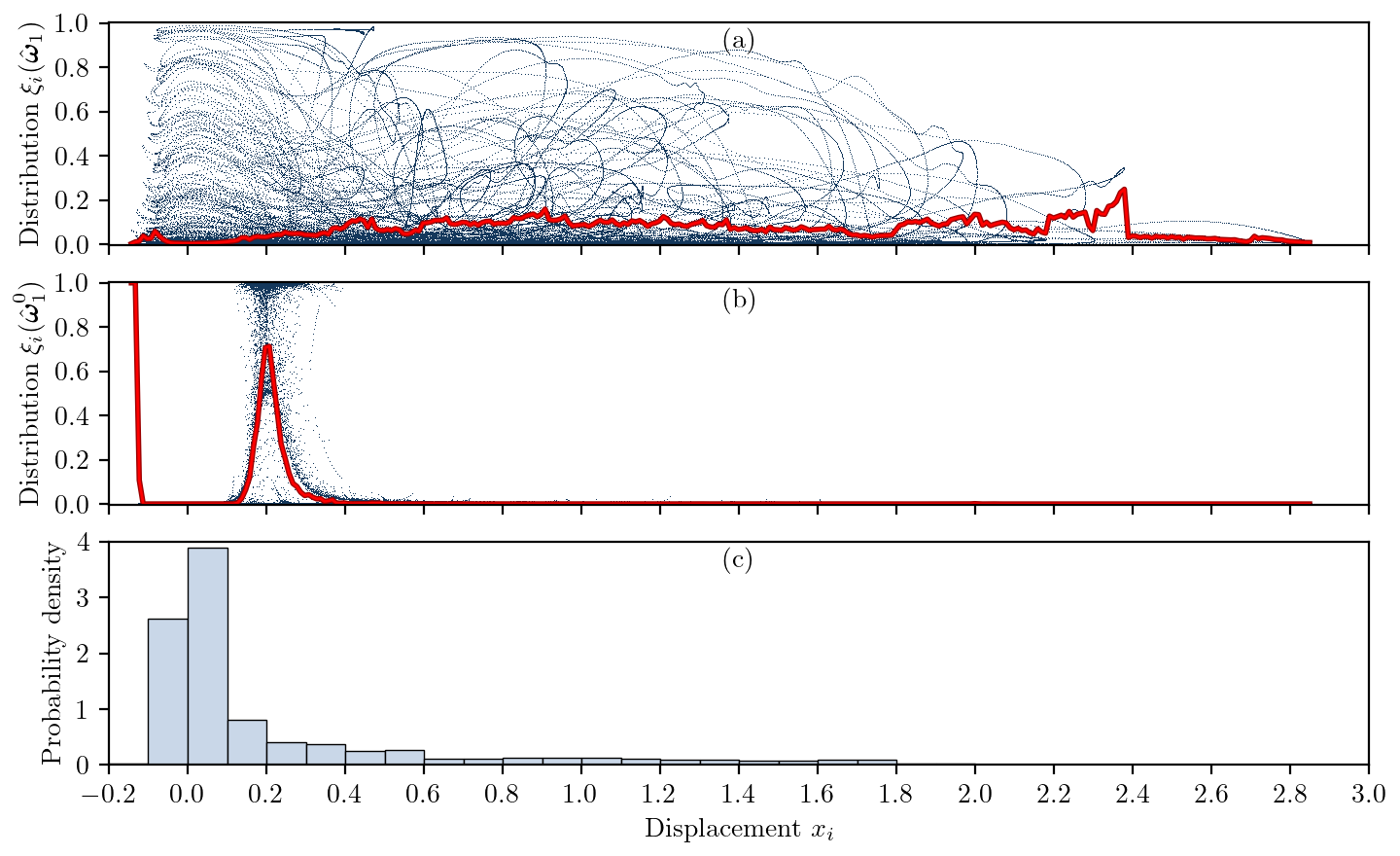}
	\caption{Similar to \figureref{fig:scatter}, but for a 40-site pure GC lattice over a \qty{50}{ps} interval, using the data for distributions $\xi_i(\hat{\bs\omega}_1)$ and $\xi_i(\hat{\bs\omega}_1^0)$ from \figureref{fig:heat_gc}.}
	\label{fig:scatter_gc}
\end{figure}

To further examine the previous observations, we present scatter plots in panels (a) and (b) of \figureref{fig:scatter_gc} in which the coordinates of each plotted point are the respective displacement and distribution values from \figureref{fig:heat_gc}, where each coordinate in the pair corresponds to a common time and lattice site. We bin these data into small displacement intervals and compute the average of the distribution values in each interval, and we show these averages by the red curves in panels (a) and (b) of \figureref{fig:scatter_gc}. We see no clear relationship between displacement and the distribution of the first CLV from \figureref{fig:scatter_gc}(a), similar to \figureref{fig:scatter}(a). However, we see from \figureref{fig:scatter_gc}(b) that the distribution of $\hat{\bs\omega}_1^0$ peaks at sites with displacements near \qty{0.2}{\angstrom} and the smallest possible displacements which, in this case, are around \qty{-0.13}{\angstrom}. This second peak at negative displacements differs qualitatively from the AT case where we only saw a single peak. We also present a normalised histogram of the displacement data used in these scatter plots in \figureref{fig:scatter_gc}(c) where we see that near-zero displacements are most frequent, while the long right tail of the histogram indicates that sites with large displacements are rare.

\begin{figure}[htbp]
	\centering
	\includegraphics[width=\linewidth]{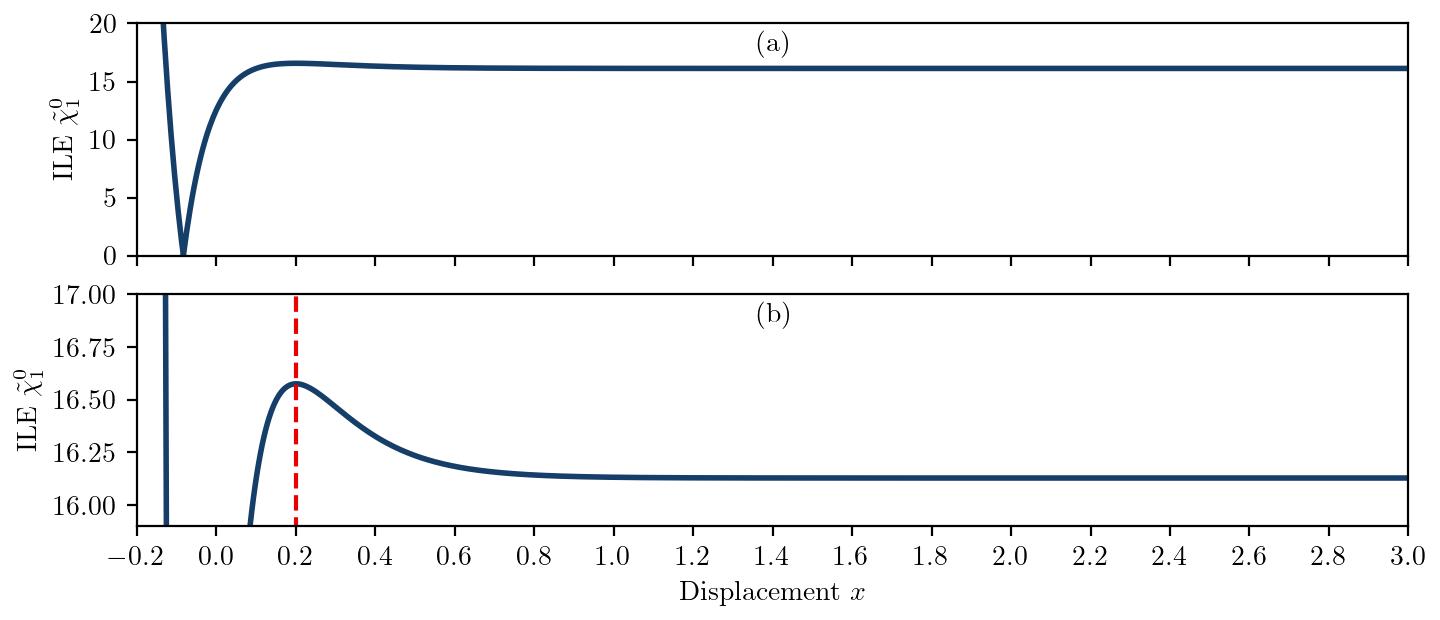}
	\caption{The first ILE $\tilde\chi_1^0$ \eqref{eq:maxstiff} of a single GC oscillator as a function of displacement $x$. Panels (a) and (b) are plots of the same function using different scales of the vertical axis. A local maximum of $\tilde\chi_1^0$ occurs at $x=\qty{0.2}{\angstrom}$ and is indicated by the red dashed line in panel (b). Both panels share the same horizontal axis.}
	\label{fig:ile_gc}
\end{figure}

As seen in the pure AT case, we can explain the observation that ILV distributions prefer sites with particular displacements by considering a single GC oscillator. In this case, the first (i.e.\ maximum) ILE $\tilde\chi_1^0$ is also given by \eqref{eq:maxstiff}, except the potential energy $V$ for GC is given by
\begin{align}
	\tilde V(x)=D_{\scriptscriptstyle GC}(e^{-a_{\scriptscriptstyle GC}x}-1)^2.
\end{align}
The first ILE is a function of only displacement $x$, which we plot in \figureref{fig:ile_gc} using different vertical scales in the two panels. We see from \figureref{fig:ile_gc}(b) that $\tilde\chi_1^0$ achieves a local maximum when $x\approx\qty{0.2}{\angstrom}$, so we would expect the distribution of $\hat{\bs\omega}_1^0$ (for the full lattice) to often be located at sites with displacements near \qty{0.2}{\angstrom}, which is what we observe in \figureref{fig:scatter_gc}(b). However, we also see from \figureref{fig:ile_gc}(a) that $\tilde\chi_1^0\to\infty$ as $x\to-\infty$. Therefore, sites with very small displacements should also be found in the distribution of $\hat{\bs\omega}_1^0$, which we see in \figureref{fig:scatter_gc}(b). Note that the shape of the $\tilde\chi_1^0$ curve is actually the same in the pure AT and pure GC cases; the left arm of the curve was not shown in \figureref{fig:ile} because such small displacements were not attained in our simulations and so irrelevant for that particular discussion. For the parameters used in the pure GC case, however, this feature is relevant and results in the peak seen at the smallest attained displacements in \figureref{fig:scatter_gc}(b).

\begin{figure}[htbp]
	\centering
	\includegraphics[width=\linewidth]{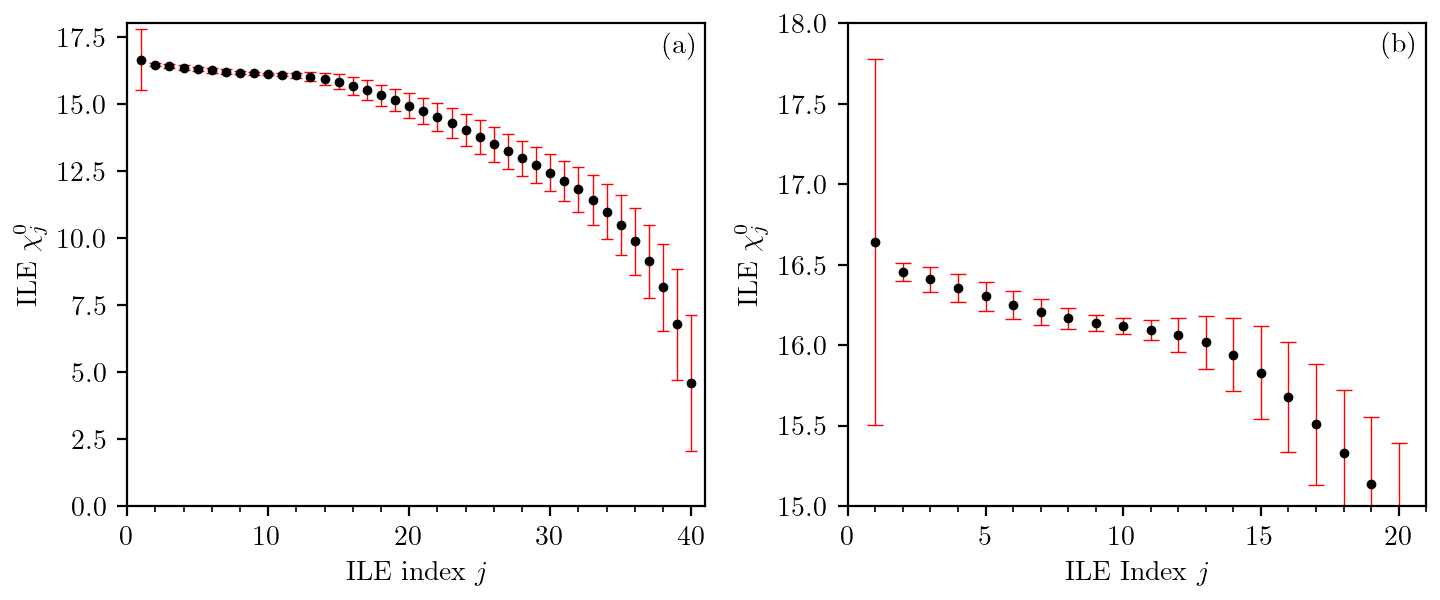}
	\caption{Similar to \figureref{fig:ile_spectrum_wide}, but for a 40-site pure GC lattice over a \qty{50}{ps} interval.}
	\label{fig:ile_spectrum_wide_gc}
\end{figure}

Using the same initial condition and the same \qty{50}{ps} time interval which were used to generate Figs.~\ref{fig:heat_gc} and~\ref{fig:scatter_gc}, we compute the first half of the ILE spectrum at each point in time, and the time averages of each ILE are presented in \figureref{fig:ile_spectrum_wide_gc} together with their respective standard deviations. We observe from the figure that approximately the first 15 ILEs are nearly degenerate, which tempers the significance of our analysis of the first ILV in isolation. Therefore, in \sectionref{sec:character_gc}, we incorporate the remaining ILVs in our analysis. It is interesting that the number of nearly degenerate ILEs in this case is fewer than in the pure AT case (cf.\ \figureref{fig:ile_spectrum_wide}). The large standard deviation of the first ILE $\chi_1^0$ in \figureref{fig:ile_spectrum_wide_gc} is also noteworthy. Recall from \figureref{fig:ile_gc} that $\tilde{\chi}_1^0$ is maximised at the smallest attained displacement, but such displacements occur infrequently, so the local maximum at \qty{0.2}{\angstrom} in \figureref{fig:ile_gc} is prominently featured in \figureref{fig:scatter_gc}(b) for the ILV distribution since such displacements occur regularly in our simulations. When a very negative displacement is on occasion attained, however, we see from \figureref{fig:ile_gc} that the $\tilde{\chi}_1^0$ (for the single GC oscillator) value can be significantly higher than when the displacement is \qty{0.2}{\angstrom}. These outlying values for the first ILE skew the averages of $\chi_1^0$ (for the lattice system) given in \figureref{fig:scatter_gc}(b) and are the cause of the large error bar in \figureref{fig:ile_spectrum_wide_gc} for the first ILE.

\subsection{Characteristics of ILV distributions}\label{sec:character_gc}
We turn our attention to the distributions of ILVs other than the first (cf.\ \sectionref{sec:character}). Analogous to the pure AT case, we compute the distributions of the first 40 ILVs over a \qty{50}{ps} interval using 10 random initial conditions. Note that this is in contrast to \sectionref{sec:firstilv_gc} where only a single initial condition was used.

\begin{figure}[htbp]
	\centering
	\includegraphics[width=\linewidth]{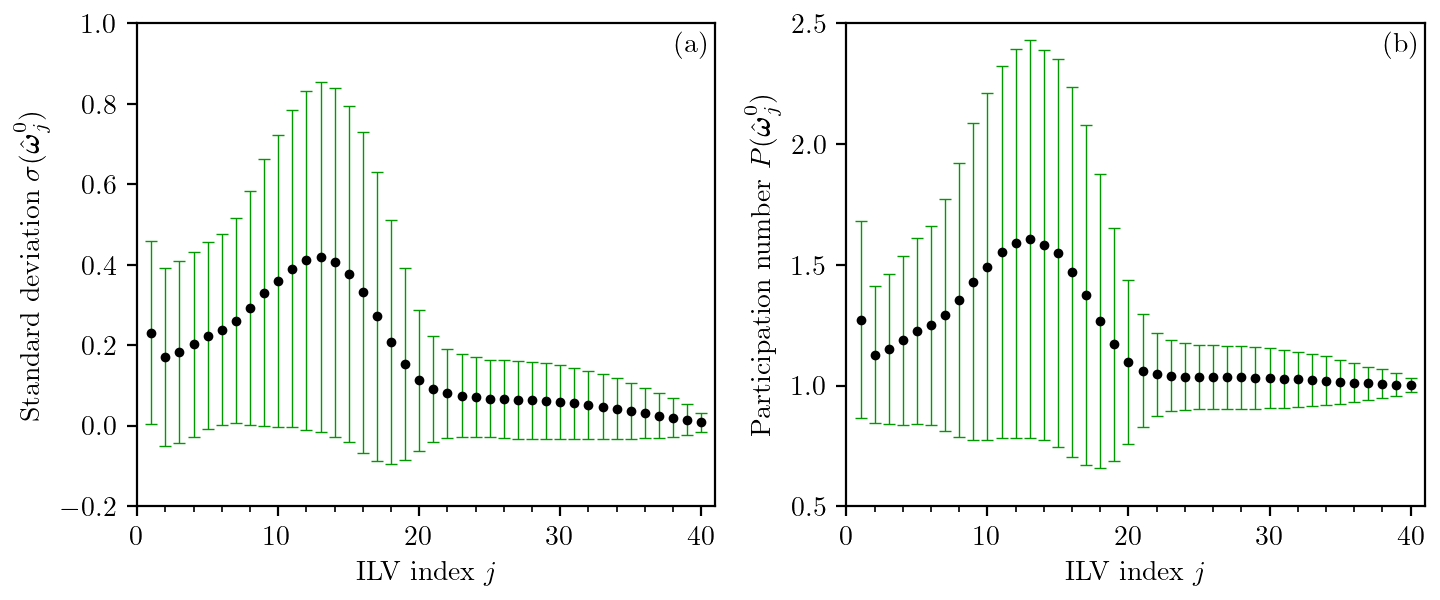}
	\caption{Similar to \figureref{fig:deviation}, but using data from 10 simulations of a 40-site pure GC lattice over a \qty{50}{ps} interval.}
	\label{fig:deviation_gc}
\end{figure}

We begin by quantifying the spread of each ILV distribution $\xi_i(\hat{\bs\omega}_j^0)$ by computing the standard deviation $\sigma$ and participation number $P$. For each of the 40 ILVs $\hat{\bs\omega}_1^0,\dots,\hat{\bs\omega}_{40}^0$, we compute $\sigma(\hat{\bs\omega}_j^0)$ and $P(\hat{\bs\omega}_j^0)$ at each time step over the \qty{50}{ps} interval for each of the 10 different initial conditions, then we compute the average and standard deviation of each of these quantities and plot the results in \figureref{fig:deviation_gc}. We also compute the standard deviation of the $\sigma(\hat{\bs\omega}_j^0)$ and $P(\hat{\bs\omega}_j^0)$ values and represent them by the error bars in the figure. We see from \figureref{fig:deviation_gc} that the distribution of the 40-th ILV is the most localised with a standard deviation of $\sigma(\hat{\bs\omega}_{40}^0)\approx0$ and a participation number of $P(\hat{\bs\omega}_{40}^0)\approx1$ on average, indicating that this ILV distribution is typically located at a single site with zero spread. On the other hand, the 13-th ILV is the least localised with $\sigma(\hat{\bs\omega}_{13}^0)\approx0.4$ and $P(\hat{\bs\omega}_{13}^0)\approx1.6$ on average. From these measures of spread, we conclude that typical ILV distributions are spatially localised to 1--2 neighbouring lattice sites. This is similar to the result for the pure AT case from \figureref{fig:deviation}, but we find that the ILV distributions are localised to fewer sites in the pure GC case.

\begin{figure}[htbp]
	\centering
	\includegraphics[width=\linewidth]{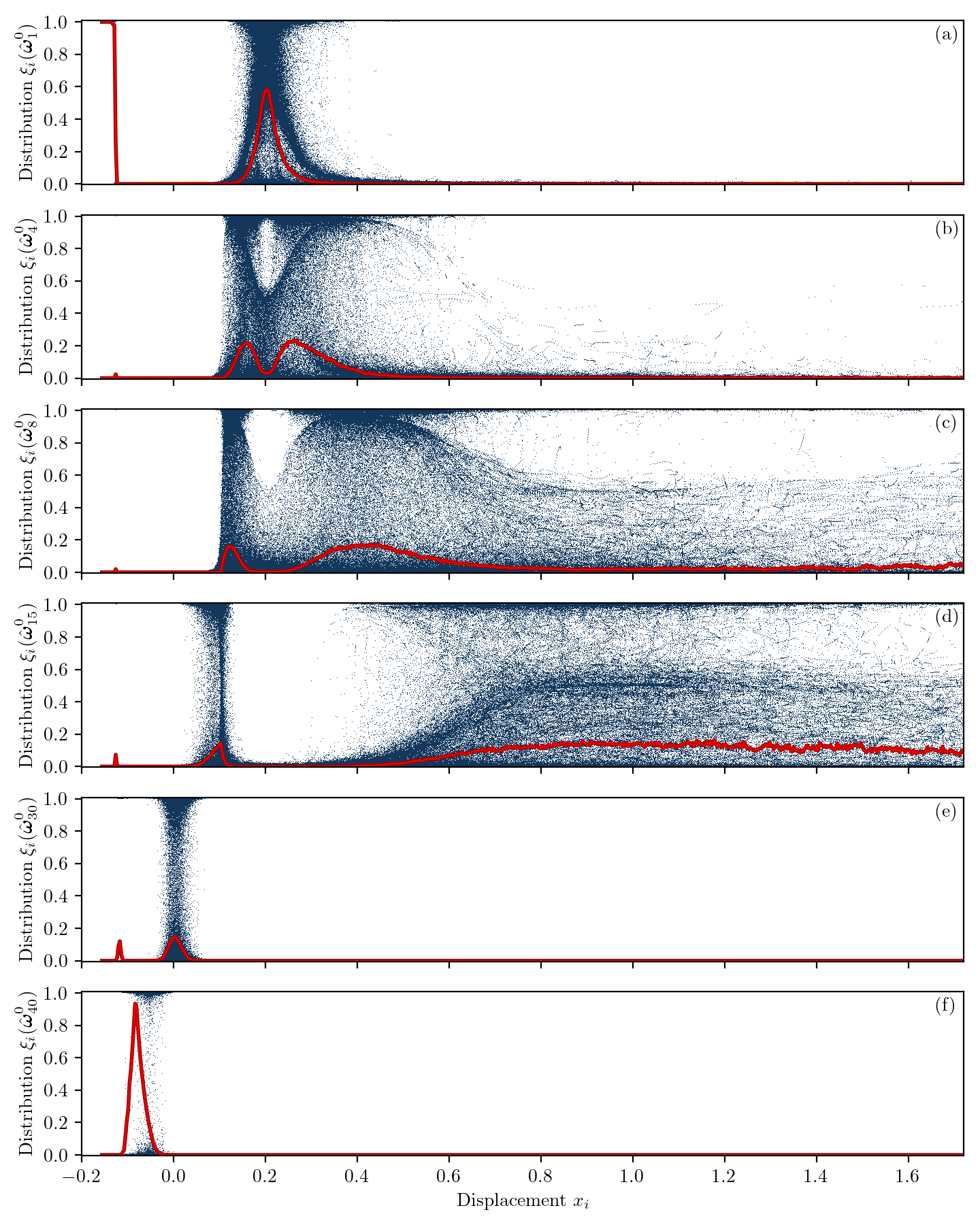}
	\caption{Similar to \figureref{fig:all_ilvs}, but showing the distributions $\xi_i$ of the ILVs (a) $\hat{\bs\omega}_1^0$, (b) $\hat{\bs\omega}_4^0$, (c) $\hat{\bs\omega}_{8}^0$, (d) $\hat{\bs\omega}_{15}^0$, (e) $\hat{\bs\omega}_{30}^0$, and (f) $\hat{\bs\omega}_{40}^0$ for a 40-site pure GC lattice, using data from 10 independent simulations over a \qty{50}{ps} interval.}
	\label{fig:all_ilvs_gc}
\end{figure}

Taking the same approach as the one used in \sectionref{sec:character} to produce \figureref{fig:all_ilvs} in the AT case, we present scatter plots in \figureref{fig:all_ilvs_gc} for several different ILVs using data from all 10 pure GC simulations. Due to the long right tail of the histogram in \figureref{fig:scatter_gc}(c), we truncate the horizontal axes of \figureref{fig:all_ilvs_gc} at \qty{1.72}{\angstrom} which we determine to be the 99-th displacement percentile of the data, thus excluding any points with larger displacements from our presentation due to the sparse data. Examining \figureref{fig:all_ilvs_gc}(a), we see that the distribution of $\hat{\bs\omega}_1^0$ has two peaks, one at a positive displacement and one at a negative displacement, as seen previously for the single run in \sectionref{sec:firstilv_gc}. As we move to the distribution of $\hat{\bs\omega}_{4}^0$ in panel (b) of \figureref{fig:all_ilvs_gc}, the peak at \qty{0.2}{\angstrom} seen in (a) appears to split into two peaks in (b) (which is similar to the progression seen in \figureref{fig:all_ilvs} for the AT case), while the negative displacement peak from (a) has almost vanished in (b). In \figureref{fig:all_ilvs_gc}(c) where the distribution of $\hat{\bs\omega}_{8}^0$ is plotted, we see that the two positive peaks from (b) persist but move further away from \qty{0.2}{\angstrom}, and the same can be said for panel (d) for $\hat{\bs\omega}_{15}^0$. While the small negative displacement peak persists in panel (e) for $\hat{\bs\omega}_{30}^0$, this peak has shifted slightly rightwards since the previous panels, while one of the positive displacement peaks has vanished entirely. Finally for $\hat{\bs\omega}_{40}^0$ in panel (f), it appears that the two remaining peaks from (e) have merged into a single peak around \qty{-0.08}{\angstrom} in (f). The main qualitative difference between the pure AT and GC cases seen in Figs.~\ref{fig:all_ilvs} and \ref{fig:all_ilvs_gc}, respectively, is that the ILV distributions in the GC case each have an additional peak at negative displacements which is not present in the AT case.

As in the AT case, we offer a heuristic explanation for most of these observations regarding the ILV distributions for GC. Recall from \figureref{fig:ile_gc} that the first ILE $\tilde{\chi}_1^0$ as a function of displacement $x$ for a single GC oscillator has a local maximum at \qty{0.2}{\angstrom} and increases without bound as $x\to-\infty$. Of course, only finite negative displacements are obtained in our simulations because of the finite energy used, and large negative displacements are rare due to the shape of the Morse potential, hence we see two peaks in \figureref{fig:all_ilvs_gc}(a), one at the smallest displacements available and one at $x_i=\qty{0.2}{\angstrom}$. However, once the first few ILVs have claimed all lattice sites with these displacement values, then the largest attainable $\tilde{\chi}_1^0$ occurs at \textit{three} displacement values.\footnote{This can be seen easily by drawing a horizontal line at, say, $\tilde{\chi}_1^0=16.5$ of \figureref{fig:ile_gc}(b), which clearly intersects the curve at three points.} Eventually, what remains for the higher-indexed ILVs are sites which individually have small $\tilde{\chi}_1^0$ values due to their displacements, and we see from \figureref{fig:ile_gc}(a) that $\tilde{\chi}_1^0$ is minimised at approximately \qty{-0.08}{\angstrom}. Qualitatively, we see this behaviour in the progression of \figureref{fig:all_ilvs_gc} through ILVs of increasing index.

\begin{figure}[htbp]
	\centering
	\includegraphics[width=\linewidth]{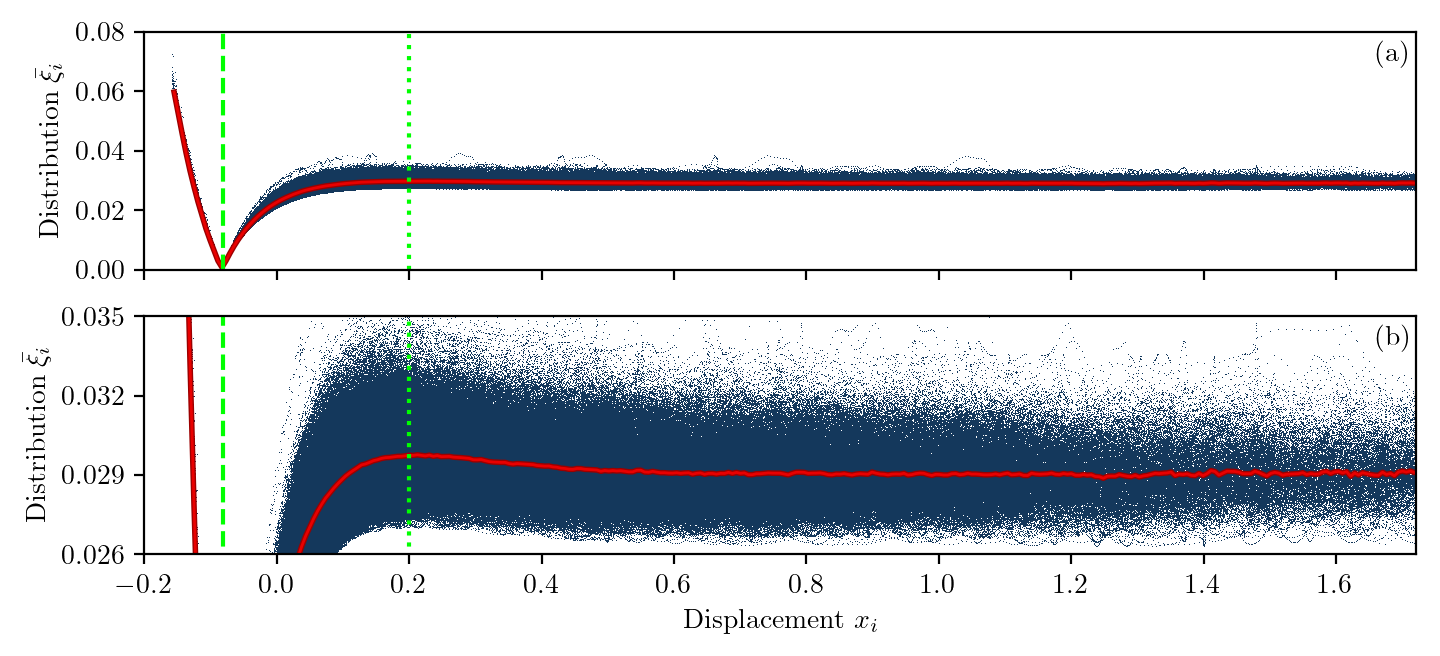}
	\caption{Similar to \figureref{fig:weight}, but using data from 10 simulations of a 40-site pure GC lattice over a \qty{50}{ps} interval. The green dashed line denotes $x_i=\qty{-0.08}{\angstrom}$ and the green dotted line denotes $x_i=\qty{0.2}{\angstrom}$.}
	\label{fig:weight_gc}
\end{figure}

Finally, we compute the weighted ILV distribution $\bar\xi_i$ \eqref{eq:weight}, $i=1,\dots,40$, using all 40 ILVs and ILEs over a \qty{50}{ps} interval for 10 different runs, and the relationship between $\bar\xi_i$ and displacement $x_i$ is presented in \figureref{fig:weight_gc}, truncating the displacement axis at the 99-th percentile as usual. From \figureref{fig:weight_gc}(a), we see that $\bar\xi_i$ is largest for the smallest attained displacements, but quickly decreases to a minimum at \qty{-0.08}{\angstrom} (indicated by the green dashed line) before increasing and saturating to an approximately constant value for displacements greater than $x_i\approx\qty{0.1}{\angstrom}$. However, we see from \figureref{fig:weight_gc}(b) that $\bar\xi_i$ is not in fact constant, but has a local maximum at \qty{0.2}{\angstrom} (indicated by the green dotted line). Using a completely analogous argument to the one in \sectionref{sec:character} when we were discussing the results of \figureref{fig:weight} for AT, the closeness of $\bar\xi_i$ values for medium-to-large displacements in \figureref{fig:weight_gc} is explained by the near-degeneracy of the first 15 ILEs. Qualitatively, \figureref{fig:weight_gc} is the same as its AT counterpart (\figureref{fig:weight}) except for the interval to the left of \qty{-0.08}{\angstrom} wherein $\bar\xi_i$ is decreasing with respect to displacement, which does not appear in the AT case since the small displacements required for this behaviour to be seen do not occur in our simulations. Therefore, $\bar\xi_i$ is maximised at the smallest attained displacements for GC, but it is worth remembering that these very small displacements are rare, as seen in the histogram of \figureref{fig:scatter_gc}(c) for a single run.

\begin{figure}[htbp]
	\centering
	\includegraphics[width=\linewidth]{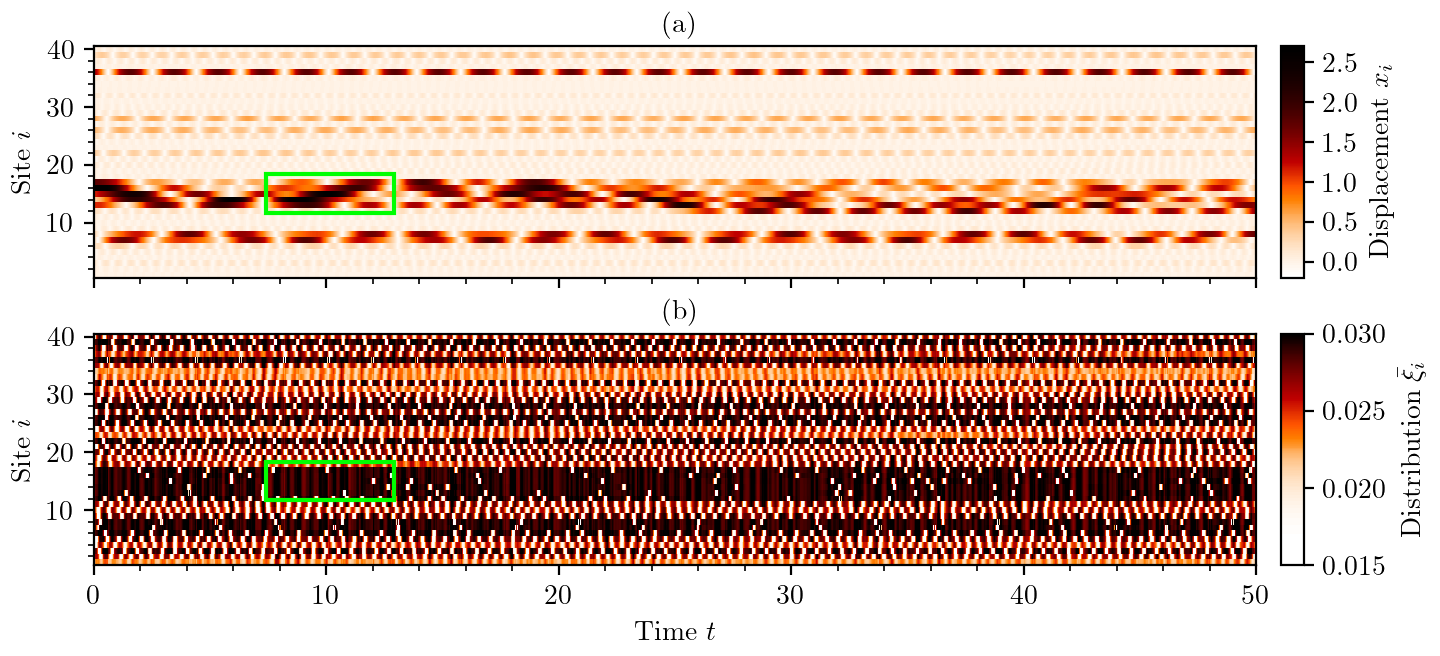}
	\caption{Similar to \figureref{fig:weighted_heat}, but for a 40-site pure GC lattice over a \qty{50}{ps} interval using the same initial condition as in \figureref{fig:heat_gc}, and the green rectangle highlights a different region in this case.}
	\label{fig:weighted_heat_gc}
\end{figure}

Taking the same approach used to generate \figureref{fig:weighted_heat}, we use the simulation from \figureref{fig:heat_gc} and present the spatiotemporal evolution of the displacement $x_i$ and the weighted ILV distribution $\bar{\xi_i}$ in \figureref{fig:weighted_heat_gc}(b). Comparing the location of bubbles in \figureref{fig:weighted_heat_gc}(a), such as the one inside the green rectangle, to the value of the weighted ILV distribution $\bar{\xi_i}$ in \figureref{fig:weighted_heat_gc}(b), it is clear that large values of $\bar{\xi_i}$ typically occur at sites with large displacements, including those inside bubbles. As in the case of \figureref{fig:weighted_heat} for AT, this result is the opposite of our expectation that bubbles linearise the potential of \eqref{eq:dnahamiltonian} and hence do not contribute to chaos, giving further evidence that more work is needed in order to properly interpret the weighted distribution $\bar{\xi}_i$ and its link to chaos.

\section{Summary}
In this chapter, we numerically studied the dynamical behaviour of homogeneous PBD lattices consisting of either pure AT or pure GC base pairs. In particular, we tried to understand the extent to which sites inside bubbles contribute to the chaoticity of the overall system. We began by looking at the distribution of the first CLV over the lattice, but no clear relationship between this quantity and the displacement of individual sites could be determined. As a new alternative approach, we introduced the notions of ILVs and ILEs in the context of Hamiltonian systems by defining them in a similar manner to the CLVs and LEs (see \sectionref{sec:ilv}), just over an infinitesimal time interval instead of an infinite one. Analogous to the LEs, ILEs are a spectrum of values which quantify the growth rates of perturbations in different directions, and ILVs are those directions. By measuring the spread of ILV distributions for chaotic orbits in 40-site pure AT/GC lattices, we determined that these distributions are spatially localised. Furthermore, we found that the distribution of the first ILV is clearly maximised at sites of particular displacements, which we explained from the shape of the on-site Morse potential. Due to the near degeneracy in the ILE spectra observed, we expanded our study to include ILVs other than just the first. We defined and computed an ILE-weighted average of all the ILV distributions, which we found to be almost constant for sites with medium-to-large displacements. Our interpretation of this result, that sites inside bubbles typically contribute strongly to the chaoticity of the system, was unexpected, and more work needs to be done in order to determine the validity of our interpretation and the link (if any exists) between these ILV distributions and chaos.

%% file: conclusion.tex
\chapter{Summary and conclusions}\label{ch:conclusion}
In this thesis, we have provided a holistic overview of Lyapunov exponents and vectors, discussing their theory, computation, and application to various autonomous Hamiltonian systems. With the introduction of several CLV algorithms in recent years (e.g.\ \cites{WolfeSamelson2007}{GinelliEtAl2007}{KuptsovParlitz2012}), there has been an increase in interest in the subject, which led to our investigation of the G\&C algorithm for computing CLVs along chaotic orbits, and our exploration of the ability of CLVs and related quantities to provide insight into the dynamics of nonlinear systems.

We began our study in \chapterref{ch:intro} with a survey of CLVs and their applications in the literature. We then proceeded to \chapterref{ch:theory} with a thorough overview of the numerous mathematical and numerical notions used in this thesis. The QR decomposition and SVD were discussed as these are fundamental to many of the algorithms we used. We defined a notion of distance $\Delta$ between linear subspaces, which we later used to quantify convergence rates. A brief introduction to continuous dynamical systems was then given before defining LEs and CLVs and discussing their properties in great detail. The numerical integration of autonomous Hamiltonian systems using symplectic integrators was also addressed. Finally, we explained the G\&C algorithm for computing CLVs, pseudocode for which was provided.

The dynamical behaviour of the H\'enon-Heiles system was investigated numerically in \chapterref{ch:hh}. We studied the convergence properties of the G\&C algorithm using an efficient technique of computing two sets of the relevant vectors/subspaces and measuring their mutual convergence via the distance $\Delta$. Our results were found to be in agreement with the convergence rates given in the literature related to the spectral gaps between LEs \cites{LegrasVautard1996}{GinelliEtAl2013}{Noethen2019}. Repeating this investigation of convergence for a system with three degrees of freedom, we provided substantial evidence of the validity of our approach to measuring exponential convergence rates, which we propose as an efficient method for determining the transient lengths needed when using the G\&C algorithm in practice.

Further investigating the H\'enon-Heiles system, we observed that the two CLVs in the centre subspace computed by the G\&C algorithm tend to converge, and we found that they converge by shearing along the direction of the flow. Due to this dynamics of the centre subspace, we observed that the this subspace cannot be computed very accurately by implementing the traditional form of the G\&C algorithm, since such an estimate becomes increasingly inaccurate over time. To remedy this issue, we introduced the centre correction technique. Applying the G\&C algorithm with the centre correction, we then used a sticky orbit to study the angles between the three splitting subspaces of the H\'enon-Heiles system and found that these subspaces become nearly tangent during regular regimes of motion, indicating that more frequent violations of hyperbolicity occur in such regimes.

We studied the dynamical behaviour of the PBD model of DNA in \chapterref{ch:dna}, which is a Hamiltonian lattice model with many degrees of freedom. After defining the spatial distribution of deviation vectors over the lattice, we used notions from circular statistics to quantify the spread of these distributions, which was needed because of the periodic lattices used in our simulations. Computing the first CLV, we found no clear relationship between its distribution at a site and the displacement of that site. However, after defining ILEs and ILVs as instantaneous analogues of LEs and CLVs, we observed a definite correlation between ILV distributions and displacements at individual sites, which we explained directly from the on-site Morse potential modelling the dynamics of individual base pairs. Finally, we introduced the weighted ILV distribution, defined as the ILE-weighted average of the ILV distributions, and found that sites inside bubbles typically have high weighted ILV distribution values. However, the correct interpretation of this weighted ILV distribution is still unclear to us.

A possibility for future work related to this thesis is the application of our analysis in \sectionref{sec:cent} of the dynamics of centre subspace in the H\'enon-Heiles system to other autonomous Hamiltonian systems, as this might provide insight into the tangent dynamics of these systems in general. Additionally, to further extend our work on DNA in \chapterref{ch:dna}, we propose using ILVs to probe heterogeneous DNA lattices, consisting of both AT and GC base pairs, particularly sequences of base pairs which appear in biology.

%% file: appendix.tex
\chapter*{Appendix}
\addcontentsline{toc}{chapter}{Appendix}
\markboth{Appendix}{Appendix}
\renewcommand{\thechapter}{A}
\section{Are LEs exponential growth rates?}\label{app:rate}
Recall from equation \eqref{eq:le_filtration} that if $\bs w(0)\in\Gamma_i^{\pm}\setminus\Gamma_{i\pm1}^{\pm}$ then
\begin{align}
	\lim_{t\to\pm\infty}\frac{1}{|t|}\ln\frac{\|\bs w(t)\|}{\|\bs w(0)\|}=\pm\lambda_i,\label{eq:simple}
\end{align}
where we have dropped the $\bs x$ dependence to simplify the notation. Focusing on the $t\to\infty$ case, we rearrange this equation to get
\begin{align}
	\lim_{t\to\infty}\frac{\ln\frac{\|\bs w(t)\|}{\|\bs w(0)\|}-\lambda_i t}{t}=0.\label{eq:dom}
\end{align}
The numerator in \eqref{eq:dom} is asymptotically dominated by $t$, so the numerator is said to be $o(t)$ (see e.g.\ \cite{BalcazarGabarro1989}), that is,
\begin{align}
	\ln\frac{\|\bs w(t)\|}{\|\bs w(0)\|}-\lambda_i t = o(t).
\end{align}
Exponentiating both sides of this equation and solving for $\|\bs w(t)\|$, we get
\begin{align}
	\|\bs w(t)\| = \|\bs w(0)\|\exp(\lambda_it + o(t)).\label{eq:obey}
\end{align}
Based on \eqref{eq:simple}, some authors (e.g.\ \cite{KuptsovParlitz2012}) refer to $\lambda_i$ as the exponential growth rate of $\bs w(t)$ and write $\|\bs w(t)\|\sim \exp(\lambda_i t)$, which means that $\|\bs w(t)\|$ and $\exp(\lambda_i t)$ are \textit{asymptotically equivalent}, i.e.
\begin{align}
	\lim_{t\to\infty}\frac{\|\bs w(t)\|}{\exp(\lambda_i t)}=1
\end{align}
(see e.g.\ \cite[p.~4]{Murray1984}). But this asymptotic equivalence is not strictly true given \eqref{eq:obey}. By way of counter-example, consider $\|\bs w(t)\|=\exp(\lambda_i t+\sqrt[3]{t})$ which satisfies \eqref{eq:simple} but is not asymptotically equivalent to $\exp(\lambda_i t)$ because
\begin{align}
	\lim_{t\to\infty}\frac{\|\bs w(t)\|}{\exp(\lambda_i t)}=\lim_{t\to\infty}\frac{\exp(\lambda_i t+\sqrt[3]{t})}{\exp(\lambda_i t)}=\lim_{t\to\infty}\exp(\sqrt[3]{t})=\infty.
\end{align}
Nevertheless, it is easily shown that for large enough $t$, $\|\bs w(t)\|$ is squeezed between two exponential functions with exponential growth rates $\lambda_i\pm\epsilon$ for arbitrarily small $\epsilon>0$. In particular, it is clear that for all $\epsilon>0$, a large enough $t$ will yield $-\epsilon t\leq \sqrt[3]{t}\leq\epsilon t$, from which it follows that
\begin{align}
	\exp((\lambda_i-\epsilon) t)\leq \exp(\lambda_i t+\sqrt[3]{t})\leq \exp((\lambda_i+\epsilon) t).
\end{align}
A similar result holds for the $t\to-\infty$ case. We conclude that any exponential function with a larger (smaller) growth rate than $\lambda_i$ will grow faster (slower) than $\|\bs w(t)\|$ asymptotically. Despite the lack of rigour in referring to $\lambda_i$ as an asymptotic exponential growth rate, the description can nonetheless be used meaningfully in practice with the understanding that the actual growth is bounded by two exponential functions with growth rates arbitrarily close to $\lambda_i$. We therefore use this exponential growth rate characterisation of LEs in the text.

\section{Matrix of the linear propagator restricted to the 2-D centre subspace}\label{app:basis}
Repeating \eqref{eq:bases} here, we claim that
\begin{align}
	\overline{M}(k\tau,(k+1)\tau)=(\hat{\bs c}_2'\,\ \hat{\bs c}_3^{\prime\perp})\transpose R((k+1)\tau)(\hat{\bs c}_2\,\ \hat{\bs c}_3^{\perp}),\label{eq:bases2}
\end{align}
where $\overline{M}(k\tau,(k+1)\tau)$ is the matrix of $\smash{\mathrm{d}_{\bs x(k\tau)}\Phi^{\tau}|_{\Omega}^{\Omega'}}$ \eqref{eq:cool} with respect to the bases $\overline{\mathcal B}$ and $\overline{\mathcal B}'$ \eqref{eq:particularbasis}, $R((k+1)\tau)$ is the matrix of $\mathrm{d}_{\bs x(k\tau)}\Phi^{\tau}$ with respect to the GSV bases, $\hat{\bs c}_2$ and $\smash{\hat{\bs c}_3^{\perp}}$ are computed orthonormal basis vectors of $\Omega$ (the 2-D centre subspace of $T_{\bs x(k\tau)}\mathcal{M}$) expressed in the GSV basis, and similarly for $\hat{\bs c}_2'$ and $\smash{\hat{\bs c}_3^{\prime\perp}}$ with respect to the subspace $\Omega'$ (the 2-D centre subspace of $T_{\bs x((k+1)\tau)}\mathcal{M}$). For a reminder of the notation used here, refer to the discussion around \figureref{fig:restrict}. Using the bracket notation \eqref{eq:bracketdefinition} to denote the matrix of a function, we can rewrite \eqref{eq:bases2} as follows with explicit reference to the underlying linear maps and bases:
\begin{align}
	\big[\mathrm{d}_{\bs x(k\tau)}\Phi^{\tau}|_{\Omega}^{\Omega'}\big]_{\overline{\mathcal B}}^{\overline{\mathcal B}'}={\overline{P}'}\transpose\big[\mathrm{d}_{\bs x(k\tau)}\Phi^{\tau}\big]_{\mathcal A}^{\mathcal A'}\overline{P},\label{eq:basesop}
\end{align}
where $\mathcal A$ and $\mathcal A'$ are the GSV bases, $\overline P=(\hat{\bs c}_2\,\ \hat{\bs c}_3^{\perp})$, and $\overline{P}'=(\hat{\bs c}_2'\,\ \hat{\bs c}_3^{\prime\perp})$. Recall from the discussion around the bases \eqref{eq:particularbasis} that the vectors $\hat{\bs c}_2,\hat{\bs c}_3^{\perp}$ are computed estimates of $\hat{\bs w}_{\mathrm{flow}},\hat{\bs w}_{\mathrm{flow}}^{\perp}$ in the GSV basis, and similarly for the vectors $\hat{\bs c}_2',\hat{\bs c}_3^{\prime\perp}$ and $\hat{\bs w}_{\mathrm{flow}}',\hat{\bs w}_{\mathrm{flow}}^{\prime\perp}$. Therefore, the columns of $\overline P$ and $\smash{\overline{P}'}$ are estimates for the orthonormal basis vectors constituting the bases $\overline{\mathcal B}$ and $\overline{\mathcal B}'$ represented in the GSV bases $\mathcal A$ and $\mathcal A'$, respectively. We will show that \eqref{eq:basesop} holds, which is equivalent to showing that \eqref{eq:bases2} holds. For the purpose of this exercise, we simplify the notation by letting $S=\smash{\mathrm{d}_{\bs x(k\tau)}\Phi^{\tau}}$ and $\overline{S}=\smash{\mathrm{d}_{\bs x(k\tau)}\Phi^{\tau}|_{\Omega}^{\Omega'}}$. Hence, we need to show that
\begin{align}
	\big[\raisebox{-1pt}{$\overline S$}\big]_{\overline{\mathcal B}}^{\overline{\mathcal B}'}={\overline{P}'}\transpose\big[S\big]_{\mathcal A}^{\mathcal A'}\overline{P}.
\end{align}

As a first step, we extend the orthonormal bases $\overline{\mathcal B}=(\beta_1,\beta_2)$ and $\overline{\mathcal B}'=(\beta_1',\beta_2')$ to the orthonormal bases $\mathcal B=(\beta_1,...,\beta_N)$ and $\mathcal B'=(\beta_1',...,\beta_N')$ of $T_{\bs x(0)}\mathcal {M}$ and $T_{\bs x(t)}\mathcal {M}$, respectively.\footnote{Recall from \eqref{eq:particularbasis} that we have chosen $\beta_1,\beta_2,\beta_1',\beta_2'$ to be $\hat{\bs w}_{\mathrm{flow}},\hat{\bs w}_{\mathrm{flow}}^{\perp},\hat{\bs w}_{\mathrm{flow}}',\hat{\bs w}_{\mathrm{flow}}^{\prime\perp}$, which in the GSV basis are approximated by $\hat{\bs c}_2,\hat{\bs c}_3^{\perp},\hat{\bs c}_2',\hat{\bs c}_3^{\prime\perp}$, respectively. We use the $\beta$ symbols here for convenience.} The matrix of the linear propagator $S$ with respect to bases $\mathcal B$ and $\mathcal B'$ has entries $S_{ij}$, defined by
\begin{align}
	S(\beta_j)=S_{1j}\beta_1'+\cdots+S_{Nj}\beta_N'.\label{eq:S}
\end{align}
for $j=1,2,\dots,N$ (see e.g.\ \cite[p.~70]{Axler2015}). Likewise, the matrix entries $\overline{S}_{ij}$ of the restricted linear propagator $\overline{S}$ are defined by
\begin{align}
	\begin{split}
		\overline{S}(\beta_1)&=\overline{S}_{11}\beta_1'+\overline{S}_{21}\beta_2',\\
		\overline{S}(\beta_2)&=\overline{S}_{12}\beta_1'+\overline{S}_{22}\beta_2'.\label{eq:S1}
	\end{split}
\end{align}
Now, recall from \eqref{eq:covariance} that $S(\Omega)=\Omega'$ due to the covariance of the centre subspace and that we chose our bases such that $\beta_1,\beta_2\in \Omega$ and $\beta_1',\beta_2'\in \Omega'$. Therefore, the $j=1$ and $j=2$ equations in \eqref{eq:S} reduce to
\begin{align}
	\begin{split}
		S(\beta_1)&=S_{11}\beta_1'+S_{21}\beta_2',\\
		S(\beta_2)&=S_{12}\beta_1'+S_{22}\beta_2'.\label{eq:S2}
	\end{split}
\end{align}
Since $\beta_1,\beta_2\in \Omega$, it follows from the definition \eqref{eq:cool} of the restriction of the linear propagator $S$ to $\Omega$ and $\Omega'$ that $\overline S(\beta_1)=S(\beta_1)$ and $\overline S(\beta_2)=S(\beta_2)$. We therefore combine equations \eqref{eq:S1} and \eqref{eq:S2} to get
\begin{align}
	\begin{split}
		0 &= (\overline{S}_{11}-S_{11})\beta_1' + (\overline{S}_{21}-S_{21})\beta_2',\\
		0 &= (\overline{S}_{12}-S_{12})\beta_1' + (\overline{S}_{22}-S_{22})\beta_2'.\label{eq:S3}
	\end{split}
\end{align}
Due to the linear independence of $\mathcal B'$, equations \eqref{eq:S3} imply $\overline S_{ij}=S_{ij}$ for all $i,j\in\{1,2\}$. Hence, $\smash{\big[\raisebox{-1pt}{$\overline S$}\big]_{\raisebox{-0.5pt}{$\scriptstyle\overline{\mathcal B}$}}^{\raisebox{-2pt}{$\scriptstyle\overline{\mathcal B}'$}}}$ is simply the $2\times2$ upper-left block of $\big[S\big]_{\raisebox{-0.5pt}{$\scriptstyle\mathcal B$}}^{\raisebox{-2pt}{$\scriptstyle\mathcal B'$}}$. We note that this result is briefly discussed in \cite[p.~200]{HoffmanKunze1971} in the special case where $S$ is a linear operator.

Given the matrix $\big[S\big]_{\raisebox{-0.5pt}{$\scriptstyle\mathcal A$}}^{\raisebox{-2pt}{$\scriptstyle\mathcal A'$}}$, we change bases to get
\begin{align}
	\big[S\big]_{\mathcal B}^{\mathcal B'}={P'}\transpose\big[S\big]_{\mathcal A}^{\mathcal A'}P,\label{eq:S4}
\end{align}
where $P$ ($P'$) is an orthogonal matrix whose columns are the ordered basis vectors in $\mathcal B$ ($\mathcal B'$) represented in the orthonormal basis $\mathcal A$ ($\mathcal A'$) \cite[p.~42]{Johnston2021}. To obtain the $2\times2$ upper-left block of $\big[S\big]_{\raisebox{-0.5pt}{$\scriptstyle\mathcal B$}}^{\raisebox{-2pt}{$\scriptstyle\mathcal B'$}}$, we need to remove the last $N-2$ rows and the last $N-2$ columns from the matrix, which is easily done by removing the last $N-2$ columns from both $P$ and ${P'}$. Therefore,
\begin{align}
	\big[\raisebox{-1pt}{$\overline S$}\big]_{\overline{\mathcal B}}^{\overline{\mathcal B}'}={\overline{P}'}\transpose\big[S\big]_{\mathcal A}^{\mathcal A'}\overline{P},
\end{align}
where $\overline{P}$ ($\overline{P}'$) is the matrix consisting of the first two columns of $P$ ($P'$).

\section{Convergence of vectors under a shear transformation}\label{app:conv}
Repeating the claimed asymptotic equivalence from \eqref{eq:showequiv} here,
\begin{align}
	\sqrt{2-\frac{2kt}{\sqrt{k^2t^2+1}}}\sim \frac{1}{kt},\label{eq:toshow}
\end{align}
we now show that this holds for positive $k$. To do so, we need to take the limit
\begin{align}
	L=\lim_{t\to\infty}\frac{\sqrt{2-\frac{2kt}{\sqrt{k^2t^2+1}}}}{\frac{1}{kt}}
\end{align}
and show that $L=1$ \cite[p.~4]{Murray1984}. We begin with a change of variables, $s=kt$, so
\begin{align}
	L=\lim_{s\to\infty}\frac{\sqrt{2-\frac{2s}{\sqrt{s^2+1}}}}{\frac{1}{s}}=\lim_{s\to\infty}\sqrt{2s^2-\frac{2s^3}{\sqrt{s^2+1}}}
	=\sqrt{\lim_{s\to\infty}\left(2s^2-\frac{2s^3}{\sqrt{s^2+1}}\right)},
\end{align}
where the last step follows from the continuity of the square root function. With some algebra, we get
\begin{align}
	L=\sqrt{\lim_{s\to\infty}\frac{2s^2\sqrt{s^2+1}-2s^3}{\sqrt{s^2+1}}}
	=\sqrt{\lim_{s\to\infty}\frac{2s^2\left(\sqrt{s^2+1}-s\right)\left(\sqrt{s^2+1}+s\right)}{\sqrt{s^2+1}\left(\sqrt{s^2+1}+s\right)}}.
\end{align}
Finally, we can simplify and evaluate the limit,
\begin{align}
	L=\sqrt{\lim_{s\to\infty}\frac{2s^2}{s^2+1+s\sqrt{s^2+1}}}
	=\sqrt{\lim_{s\to\infty}\frac{2}{1+\frac{1}{s^2}+\sqrt{1+\frac{1}{s^2}}}}.\label{eq:showfinal}
\end{align}
The $1/s^2$ terms in this final expression for $L$ all vanish as $s\to\infty$, hence $L=1$ and the asymptotic equivalence of \eqref{eq:toshow} holds.

\section{Symmetry of ILEs and ILV distributions}\label{app:appsymmetry}
Consider an autonomous Hamiltonian system of $n$ degrees of freedom with a Hamiltonian of the form
\begin{align}
	H(\bs q,\bs p)=T(\bs p)+V(\bs q),\label{eq:sepham}
\end{align}
where $\bs q$ and $\bs p$ contain the $n$ position and $n$ momentum coordinates, respectively. The Jacobian matrix for a general time-independent Hamiltonian function can be found, for example, in \cite[p.~70]{Skokos2010}, but for an autonomous Hamiltonian system with a Hamiltonian function of the form \eqref{eq:sepham}, the general form of the Jacobian matrix $J$ reduces to
\begin{align}
	J = \begin{pmatrix}
		0 & D^2 T(\bs p)\\
		-D^2 V(\bs q) & 0
	\end{pmatrix},
\end{align}
where $D^2T(\bs p)$ and $D^2V(\bs q)$ denote the Hessian matrices of $T(\bs p)$ and $V(\bs q)$, respectively, the entries of which are given by
\begin{align}
	D^2T(\bs p)_{ij}=\frac{\partial^2 T}{\partial p_i\partial p_j},\quad D^2V(\bs q)_{ij}=\frac{\partial^2 V}{\partial q_i\partial q_j},
\end{align}
evaluated at the relevant state $\bs x=\smash{\big(\begin{smallmatrix}\raisebox{0.8pt}{$\scriptstyle\bs q$}\\\raisebox{-0.8pt}{$\scriptstyle\bs p$}\end{smallmatrix}\big)}$. The symmetric part of the Jacobian is then
\begin{align}
	\frac{J\transpose + J}{2} = \begin{pmatrix}
		0 & Y\\
		Y & 0
	\end{pmatrix},\label{eq:symjac}
\end{align}
where $Y=(D^2 T(\bs p) - D^2 V(\bs q))/2$. Now, let $\lambda$ be an eigenvalue of the matrix \eqref{eq:symjac} and let $\smash{\big(\begin{smallmatrix}\raisebox{-1.5pt}{$\scriptstyle\bs v_q$}\\\raisebox{1.5pt}{$\scriptstyle\bs v_p$}\end{smallmatrix}\big)}$ be a corresponding eigenvector. It follows that
\begin{align}
	\lambda\begin{pmatrix}\bs v_q\\\bs v_p\end{pmatrix} = \begin{pmatrix}0&Y\\Y&0\end{pmatrix}\begin{pmatrix}\bs v_q\\\bs v_p\end{pmatrix} = \begin{pmatrix}Y\bs v_p\\Y\bs v_q\end{pmatrix},
\end{align}
so $\lambda\bs v_q = Y\bs v_p$ and $\lambda\bs v_p = Y\bs v_q$. Using this result,
\begin{align}
	\begin{pmatrix}0&Y\\Y&0\end{pmatrix}\begin{pmatrix}\bs v_p\\-\bs v_q\end{pmatrix} = \begin{pmatrix}-Y\bs v_q\\Y\bs v_p\end{pmatrix} = \begin{pmatrix}-\lambda \bs v_p\\\lambda\bs v_q\end{pmatrix} = -\lambda\begin{pmatrix}\bs v_p\\-\bs v_q\end{pmatrix},\label{eq:resultfrom}
\end{align}
from which we see that $-\lambda$ is also an eigenvalue of \eqref{eq:symjac} with a corresponding eigenvector of $\smash{\big(\begin{smallmatrix}\raisebox{-1.5pt}{$\scriptstyle\bs v_p$}\\\raisebox{1.5pt}{$\scriptstyle-\bs v_q$}\end{smallmatrix}\big)}$.

For autonomous Hamiltonian systems with a Hamiltonian function of the form \eqref{eq:sepham}, ILEs are the eigenvalues of matrix \eqref{eq:symjac}, as discussed in \sectionref{sec:ilv}. Therefore, the result from \eqref{eq:resultfrom} implies that the ILE spectrum has the symmetry
\begin{align}
	\chi_j^0=-\chi_{N-j+1}^0.
\end{align}
Furthermore, since the ILVs are the eigenvectors corresponding to these ILEs, it follows from the relationship between the components of $\hat{\bs\omega}_j^0$ and $\hat{\bs\omega}_{N-j+1}^0$ that their distributions $\xi_i$ defined in \eqref{eq:dvd} have the symmetry
\begin{align}
	\xi_i(\hat{\bs\omega}_j^0)=\xi_i(\hat{\bs\omega}_{N-j+1}^0).
\end{align}